RICE UNIVERSITY

# Constrained Counting and Sampling: Bridging the Gap between Theory and Practice

by

**Kuldeep Singh Meel**

A THESIS SUBMITTED
IN PARTIAL FULFILLMENT OF THE
REQUIREMENTS FOR THE DEGREE

**Doctor of Philosophy**

APPROVED, THESIS COMMITTEE:

Supratik Chakraborty
Bajaj Chair Professor in Department of Computer Science and Engineering at IIT Bombay

Swarat Chaudhuri
Associate Professor of Computer Science at Rice University

Leonardo Dueñas-Osorio
Associate Professor of Civil and Environmental Engineering at Rice University

Sanjit A. Seshia
Professor of Electrical Engineering and Computer Sciences at University of California, Berkeley

Moshe Y. Vardi
Karen Ostrum George Distinguished Service Professor in Computational Engineering at Rice University

Houston, Texas
August, 2017

To my parents: Ranjeet and Bhanwari

ABSTRACT

Constrained Counting and Sampling: Bridging the Gap between Theory and Practice

by

Kuldeep Singh Meel

Constrained counting and sampling are two fundamental problems in Computer Science with numerous applications, including network reliability, privacy, probabilistic reasoning, and constrained-random verification. In constrained counting, the task is to compute the total weight, subject to a given weighting function, of the set of solutions of the given constraints. In constrained sampling, the task is to sample randomly, subject to a given weighting function, from the set of solutions to a set of given constraints. Consequently, constrained counting and sampling have been subject to intense theoretical and empirical investigations over the years. Prior work, however, offered either heuristic techniques with poor guarantees of accuracy or approaches with proven guarantees but poor performance in practice.

In this thesis, we introduce a novel hashing-based algorithmic framework for constrained sampling and counting that combines the classical algorithmic technique of universal hashing with the dramatic progress made in combinatorial reasoning tools, in particular, SAT and SMT, over the past two decades. The resulting frameworks for counting ( ApproxMC2) and sampling (UniGen) can handle formulas with up to million variables representing a significant boost up from the prior state of the art tools' capability to handle few hundreds of variables. If the initial set of constraints is expressed as Disjunctive Normal Form (DNF), ApproxMC2 is the only known Fully Polynomial Randomized Approximation Scheme (FPRAS) that does not involve Monte Carlo steps. By exploiting the connection between definability of formulas and variance of the distribution of solutions in a cell defined by 3-universal hash functions, we introduced an algorithmic technique, MIS, that reduced the size of XOR constraints employed in the underlying universal hash functions by as much as two orders of magnitude.

We demonstrate the utility of the above techniques on real-world applications including probabilistic inference, design verification and estimating the reliability of critical infrastructure networks during natural disasters. The high parallelizability of our approach opens up new directions for development of artificial intelligence tools that can effectively leverage high-performance computing resources.

# List of Publications

This thesis is based on the following publications. Except for [2], the names of authors are sorted alphabetically, and the order does not reflect contribution.

1. **Counting-based Reliability Estimation for Power-Transmission Grids**
   Leonardo Duenas-Osorio, Kuldeep S. Meel, Roger Paredes, and Moshe Y. Vardi
   In Proceedings of AAAI Conference on Artificial Intelligence (AAAI) 2017

2. **Constrained Sampling and Counting: Universal Hashing Meets SAT Solving**
   Kuldeep S. Meel, Moshe Vardi, Supratik Chakraborty, Daniel J. Fremont, Sanjit A. Seshia, Dror Fried, Alexander Ivrii and Sharad Malik
   In Proceedings of AAAI-16 Workshop on Beyond NP (BNP) 2016

3. **Algorithmic Improvements in Approximate Counting for Probabilistic Inference: From Linear to Logarithmic SAT Calls**
   Supratik Chakraborty, Kuldeep S. Meel, and Moshe Y. Vardi
   In Proceedings of International Joint Conference on Artificial Intelligence (IJCAI) 2016

4. **On Computing Minimal Independent Support and its Applications to Sampling and Counting**
   Alexander Ivrii, Sharad Malik, Kuldeep S. Meel and Moshe Y. Vardi
   Constraints 21(1), 2016

5. **Approximate Probabilistic Inference via Word-Level Counting**
   Supratik Chakraborty, Kuldeep S. Meel, Rakesh Mistry and Moshe Y. Vardi
   In Proceedings of AAAI Conf. on Artificial Intelligence (AAAI) 2016

6. **On Computing Minimal Independent Support and its applications to sampling and counting**
   ***Best Student Paper Award***
   Alexander Ivrii, Sharad Malik, Kuldeep S. Meel and Moshe Y. Vardi
   In Proc of International Conference on Principles and Practice of Constraint Programming (CP) 2015

7. **From Weighted to Unweighted Model Counting**
   Supratik Chakraborty, Dror Fried, Kuldeep S. Meel, and Moshe Y. Vardi
   In Proceedings of International Joint Conference on Artificial Intelligence (IJ-CAI) 2015, pages 304-319

8. **On Parallel Scalable Uniform SAT Witness Generation**
   Supratik Chakraborty, Daniel J. Fremont, Kuldeep S. Meel, Sanjit A. Seshia, and Moshe Y. Vardi
   In Proceedings of International Conference on Tools and Algorithms for the Construction and Analysis of Systems (TACAS) 2015

9. **Distribution-Aware Sampling and Weighted Model Counting for SAT**
   Supratik Chakraborty, Daniel J. Fremont, Kuldeep S. Meel, Sanjit A. Seshia, and Moshe Y. Vardi
   In Proceedings of AAAI Conf. on Artificial Intelligence (AAAI) 2014, pages 1722-1730

10. **Balancing Scalability and Uniformity in SAT Witness Generator**
    Supratik Chakraborty, Kuldeep S. Meel, and Moshe Y. Vardi
    In Proceedings of Design Automation Conference (DAC) 2014, pages 60:1-60:6

11. **A Scalable Approximate Model Counter**

    Supratik Chakraborty, Kuldeep S. Meel, and Moshe Y. Vardi

    In Proc. of International Conf. on Principles and Practice of Constraint Programming (CP), 2013, pages 200-216

12. **A Scalable and Nearly Uniform Generator of SAT-Witnesses**

    Supratik Chakraborty, Kuldeep S. Meel, and Moshe Y. Vardi

    In Proc. of International Conf. on Computer-Aided Verification (CAV) 2013, pages 608-623

# Acknowledgements

The completion of Ph.D. signals the completion of the formal education. As I stand at the finish line, I allow myself to look back at the journey and appreciate how lucky I have been. To be reminded that my family had to borrow money even for footwear a year before I was born, I do not think my words can do justice to sacrifices my family has made over the past 22 years. While money was scarce during the early years, love was in abundance, and I was incredibly lucky to grow up in a big joint family. Growing up, I rallied the support of my grandparents (Rameshwar, Mohini, Ramchandra, and Banarasi) to rationalize my naughtiness. My uncles (Ashok, Manoj, and Ramshwaroop Meena) and my aunts (Rajni and Pratibha) have always treated me as their child. Despite being the eldest among siblings, I have found myself more at the receiving end of support and love from my siblings (Viru, Amit, Nirmal, Taramani, Santosh, Shruti, and Hardik). Despite his hectic schedule and in addition to fulfilling his societal responsibilities, Viru has filled in for me all these years.

Through sheer luck, I had excellent teachers who went far beyond their job descriptions and cared about my career path more than myself. In particular, had it not been for Madan Kajla, Mahal Singh, Nemichand Fageria, and Sudarshan Guha, I would have never been aware of the doors that led me here.

Not everyone who was at the start line is with me now, and I dearly miss my great-grandparents, grandfather Ramchandra Bhanwaria, teachers Mahal Singh and Nemichand Fageria, and uncle Sripal Jakhar every day.

I thank my advisers, Prof. Supratik Chakraborty and Prof. Moshe Vardi, for their constant guidance, encouragement, and freedom to pursue my ideas. I would not have pursued graduate school had I not met Moshe in Mumbai at FSTTCS 2011.

I hope I have learned a tiny fraction of all that Moshe has tried to teach me in the past four and half years. Ever since I began my undergraduate studies, Supratik has been an adviser in one capacity or the other. While I have had the opportunity to watch him from close quarters, I have not been able to decode how he was able to be available whenever I requested a meeting. He was patient enough to hear out my rumblings thoughts and highly non-rigorous sketches of algorithms, theorems, and proofs. His positive criticism was crucial in the design of the algorithmic frameworks of UniGen and ApproxMC2. But most importantly, I treasure friendship with Supratik and can never forget his help during times of personal crisis.

I am thankful to Prof. Swarat Chaudhuri for his mentorship during Ph.D. and for being just a phone call away during the job search. I am grateful to Prof. Sanjit Seshia and Prof. Leonardo Dueñas-Osorio for serving on thesis committee and their excellent suggestions.

I spent the whole final year of Ph.D. in faculty job search, and I would not have navigated this stressful period without the constant mentorship of Prof. Supratik Chakraborty and Prof. Krishna Palem. I am thankful to Prof. Devika Subramanian for several chats in her office throughout past four years. I am grateful to Prof. Adnan Darwiche, Prof. Sharad Malik and Prof. Bart Selman for their guidance and recommendation letters.

During Ph.D., I have been fortunate to collaborate with some of the amazing researchers around the world: (listed alphabetically) Suguman Bansal, Dr. Raj Barik, Prof. Swarat Chaudhuri, Prof. Sourav Chakraborty, Prof. Supratik Chakraborty, Dr. Tiago Cogumbreiro, Jeffrey Dudek, Prof. Leonardo Dueñas-Osorio, Dr. Alexander Ivrii, Daniel Fremont, Dr. Dror Fried, Dr. William Hung, Dr. Deepak Majeti, Prof. Sharad Malik, Dr. Dmitry Malioutov, Prof. John Mellor-Crummey, Rakesh Mistry, Karthik Murthy, Dr. Aditya Nori, Roger Parades, Sri Raj Paul, Dr. Sriram Rajamni, Prof. Vivek Sarkar, Prof. Sanjit Seshia, Dr. Mate Soos, Aditya Shrotri, and Prof. Moshe Vardi. This thesis would not have a detailed empirical evaluation without Mate Soos' unwavering dedication to CryptoMiniSAT. "*Mate, Thank You!*".

While I could never come to terms with lonely corridors in Duncan Hall, it was bearable thanks to Jianwen Li, Aditya Shrotri, Sumit Nain, Giuseppe Perelli, Sonali Dutta, Milind Chabbi, Deepak Majeti, Niketan Pansare, Sailesh Prabhu, Ryan Luna, Morteza Lahijanian, Afsane Rahbar, and Kevin Smith.

My social life in Houston revolved around Dror Fried, Mehul Chadha, Rakesh Malladi, Vaideesh Loganathan, Amit Bhatia, and Rho Zou. Dror and his wife, Sagit, along with their three cute kids have been generous to let me be part of their family in Houston. I will forever remember the email that Dror sent just before I traveled to Israel "Of course, you can call me 24/7." Besides being an amazing roommate during the eventful year of 2014-15, Rakesh was always there to hear out my work-related frustrations, discuss love relationships (although not always providing good advice), and weekend beers and dinners. Vaideesh has been my sole connection to undergrad in Houston and therefore, a partner in some of the most memorable trips. I am thankful to Amit for his counsel and introducing me to the world of cooking.

I owe special acknowledgments to my undergraduate friends who inspire me and on whom I rely on in the moments of self-doubt: Ashish Gupta, Rahul Rekhi, Neeraj Salhotra, Devendra Shelar, Ravi Bhoraskar, Prashant Sachdeva, Harshit Mittal, Shaunak Chhaparia, Smeet Bhatt, Hrishikesh Karmarkar, Sanket Patel, Shubham Agrawal, and Saurabh Agrawal.

Falling in love with Suguman has been one of the best things that have happened to me. From Luna Pizzeria to campus visits, it has been a crazy ride with a fair share of lows, high and perfect imperfections. I could not be more excited about the future.

Finally and most importantly, my parents, Bhanwari and Ranjeet, have been the source of my energy all these years. They faced years of financial hardship without negotiating a small cut in the educational expenses of my siblings and me. They have never pushed me, but they have always been right beside me to catch if I would stumble. Their love and support of my decisions have always been unconditional.
"Mummy and Papa: *We did it. This thesis is for you*"

Kuldeep Singh Meel
September 29, 2017

# Contents

# List of Algorithms

# List of Illustrations

# List of Tables

# Part I

# Prologue

# Chapter 1

# Introduction

The paradigmatic NP-complete problem of Boolean satisfiability (SAT) solving is a central problem in Computer Science [47]. While the mention of SAT can be traced to early 19th century, efforts to develop practically successful SAT solvers go back to 1950s. The past 20 years have witnessed a "SAT revolution" with the development of *conflict-driven clause-learning* (CDCL) solvers [22]. Such solvers combine a classical backtracking search with a rich set of effective heuristics. While 20 years ago SAT solvers were able to solve instances with at most a few hundred variables, modern SAT solvers solve instances with up to *millions* of variables in a reasonable time [121]. Motivated by "SAT revolution", this thesis seeks to develop algorithmic foundations for two widely useful extensions of SAT: constrained counting and sampling.

## 1.1 Constrained Counting

In constrained counting, the task is to compute the total weight, subject to a given weighting function, of the set of solutions of the given constraints. In the field of machine learning and artificial intelligence, the problem of constrained counting is popularly referred as the problem of discrete integration [71]. If the weight function assigns equal weight to every assignment, the problem is referred as unweighted counting. Also, if the underlying formulas are propositional formulas, then unweighted counting problem is known as #SAT [150].

The earliest investigations of constrained counting were primarily based on understanding the complexity of the problem. In his seminal paper, Valiant showed that #SAT is #P-complete, where #P is the set of counting problems associated with NP decision problems [150]. Theoretical investigations of #P have led to the discovery of deep connections in complexity theory, and there is strong evidence for its hardness [9, 148]. In particular, Toda showed that every problem in the polynomial hierarchy could be solved by just one call to a #P oracle; more formally, $PH \subseteq P^{\#\mathsf{P}}$ [148].

The earliest practical approaches to constrained counting focused on algorithmic procedures motivated by Davis-Putnam-Logemann-Loveland (DPLL) algorithm [55, 54]. These approaches, e.g. CDP [24], incrementally counted the number of solutions by introducing appropriate multiplication factors for each partial solution found, eventually covering the entire solution space. Subsequent counters such as Relsat [14], Cachet [138], and sharpSAT [147] improved upon this by using several optimizations such as component caching, clause learning, and the like. Techniques based on Binary Decision Diagrams (BDD) and their variants [125, 52], have also been used to compute exact counts. Although exact counters have been successfully used in small- to medium-sized problems, scaling to the larger problem instances have posed significant challenges. Consequently, a large class of practical applications has remained beyond the reach of exact counters [123].

Owing to the hardness of constrained counting, efforts have focused on studying the complexity of approximate variants of counting. In a breakthrough, Stockmeyer provided a randomized approximation scheme for counting that makes polynomially many invocations of NP oracle. The procedure, however, is computationally prohibitive in practice and no practical tools exist based on Stockmeyer's proposed algorithmic framework. The large majority of approximate counters used in practice are *bounding counters*, which provide lower or upper bounds but do not offer guarantees on the tightness of these bounds. Examples include SampleCount [84], BPCount [110], MBound and Hybrid-MBound [87], and MiniCount [110]. Another category of counters is called *guarantee-less counters* such as ApproxCount [156], SearchTreeSampler [73], SE [136], and SampleSearch [82]. These counters are based on a large plethora of sampling techniques ranging from rejection sampling, Gibbs sampling, MCMC-based sampling techniques to variational techniques. While these counters may be efficient, they provide no guarantees and the computed estimates may differ from the exact counts by several orders of magnitude [85].

The problem of constrained counting has numerous applications in disciplines ranging from machine learning and privacy to biology and physics. In particular, the algorithmic framework developed in this thesis has been applied to problems arising from three application domains:

**Probabilistic Inference** Probabilistic inference is key to reason about uncertain and large data sets arising from medical diagnostics, weather modeling, computer vision and the like. The problem of probabilistic inference requires us to determine the probability of an event of interest given observed evidence. This problem has been the subject of intense investigations by both theoreticians and practitioners over the last few decades. A promising approach that has emerged over the last few years is to reduce probabilistic inference to constrained counting queries on a finite domain knowledge base. In this thesis, we demonstrate the effectiveness of constrained counting techniques to answer probabilistic inference queries.

**Network Reliability** Modern society is increasingly reliant on the availability of critical facilities and utility services, such as power, telecommunications, water, gas, and transportation among others [146]. One of the key challenging problems is network reliability, wherein the input to the problem consists of a network, represented as a graph, arising out of the distribution of water, power, transportation routes and the like. The network reliability problem seeks to measure the likelihood of two points of interest being reachable under conditions such as natural disasters. In this thesis, we demonstrate the effectiveness of the approach of reducing network reliability queries to constrained counting.

**Quantified Information Flow** Quantitative information flow (QIF) computation [45] is a powerful quantitative technique to detect information leakage directly at the code level. A specific fragment of the program (e.g., a function, or the whole program) is modeled as an information-theoretic channel from its input to its output. To compute the maximum amount of information that can leak from the program fragment of interest, a *constrained counter* is used to determine the number of distinct outputs of the fragment (e.g., the return values of the function, or the outputs of the program). Consequently, the techniques developed in this provide a scalable and accurate approach to detecting information leakage.

## 1.2 Constrained Sampling

In constrained sampling, the task is to sample randomly, subject to a given weight function, from the set of solutions of input constraints. If the weight function assigns equal weight to assignments, then the problem is referred to as uniform sampling.

Early theoretical investigations of constrained sampling led to the design of ran-domized polynomial time schemes given access to exact constrained counters. Owing to the hardness of exact counters, approximate variants of sampling were studied. Of particular interest was to understand the complexity of the problem of sampling solutions *almost-uniformly*; a relaxed notion of the problem of uniform sampling. Jerrum, Valiant, and Vazirani showed that for all *self-reducible* problems, generating solutions *almost uniformly* is inter-reducible with approximate counting; hence, they have similar complexity [100]. In a breakthrough, Bellare, Goldreich, and Petrank [15] later showed that in fact, a NP-oracle suffices for generating solutions of NP problems *exactly uniformly* in randomized polynomial time. Unfortunately, these deep theoretical results have not been successfully reduced to practice. Our experience in implementing these techniques indicates that they do not scale in practice even to the small problem instances involving few tens of variables [123].

Industrial approaches to constrained sampling [129] either rely on Binary Decision Diagram (BDD)-based techniques [161], which scale rather poorly, or use heuristics that offer no guarantee of performance or uniformity when applied to large problem instances [109]. In prior academic works [74, 107, 88, 155], the focus is on heuristic techniques including Markov Chain Monte Carlo (MCMC) methods and techniques based on the random seeding of combinatorial solvers. These methods scale to large problem instances, but either offer very weak or no guarantees on the uniformity of sampling, or require the user to provide hard-to-estimate problem-specific parameters that crucially affect the performance and uniformity of sampling [70, 71, 81, 109].

The problem of constrained sampling has numerous applications in disciplines ranging from machine learning and verification to program synthesis. In particular, the algorithmic framework developed in this thesis has been applied to problems arising from three application domains:

**Functional Verification** Functional verification constitutes one of the most challenging and time-consuming steps in the design of modern digital systems. The state of simulation technology today is mature enough to allow simulation of large designs within a reasonable time using modest computational resources. The verification engineer declaratively specifies a set of constraints on the values of circuit inputs. A constraint solver is then used to generate random values for the circuit inputs satisfying the constraints. Since the distribution of errors in the design's behavior space is not known *a priori*, every solution to the set of constraints is as likely to discover a bug as any other solution. It is therefore important to sample the space of all solutions uniformly or almost-uniformly at random.

**Pattern Sampling** Given the deluge of data, providing concise representations, also known as patterns, of the underlying dataset has been the holy grail in the field of data mining. Often, finding a single concise representation for a real-world data is not possible, and as a result, researchers focus on finding a set of patterns, often known as pattern mining. Recently, pattern sampling has emerged as a promising alternative that supports a broad class of quality measures and constraints while providing strong guarantees regarding sampling accuracy [64]. The core of state of the art pattern sampling techniques rely on constraint sampler, and tools such as FLEXICS employ the constrained sampling techniques proposed in this thesis [64].

**Program Synthesis** The problem of program synthesis is to synthesize programs from the specification, which finds applications in many disciplines ranging from computer-aided programming, aiding students in introductory program-ming courses to industrial tools such as FlashFill [67, 91, 140]. The runtime of state of the art synthesis techniques crucially depends on the quality of solutions returned by the underlying constraint solver; sometimes, leading to about two orders of magnitude runtime variation depending on the quality of underlying solver. The sampling techniques developed in this thesis has resulted in significant improvement in the efficiency of the state of the art synthesis tools [67].

## 1.3 Contributions

Despite intense theoretical and empirical investigations over the years, the prior work for counting and sampling offered either techniques with poor guarantees of accuracy or approaches with proven guarantees but poor performance in practice. The contribution of this thesis is a novel hashing-based framework that combines the classical algorithmic technique of universal hashing with the dramatic progress made in Boolean reasoning over the past two decades. The proposed framework has yielded significant progress in bridging the gap between theory and practice for constrained counting and sampling. In particular, this thesis contributes the following key results:

**Constrained Counting** We present hashing-based scalable approximate unweighted counter, ApproxMC2, in which the number of oracle invocations grows logarithmically in the number of variables and provides rigorous $(\varepsilon, \delta)$ guarantees, i.e., the estimates computed by ApproxMC2 are within $(1 + \varepsilon)$ multiplicative factor of the true count with confidence at least $1 - \delta$, where both $\varepsilon$ and $\delta$ are supplied by the user. If the initial set of constraints is expressed as Disjunctive Normal Form (DNF), ApproxMC2 is a Fully Polynomial Randomized Approximation Scheme (FPRAS) – the only known FPRAS scheme for DNF formulas, which does not involve Monte Carlo steps.

We extend the hashing-based paradigm to handle weighted distributions and define a novel parameter, *tilt*, to capture the hardness of weighted counting. We extend the hashing-based approach to bit-vector formulas and present the first word-level approximate unweighted counter, SMTApproxMC. We apply our hashing-based framework to construct a scalable reliability estimation framework, RelNet, which, unlike the previous state of the art techniques, can scale to real-world networks arising from cities across U.S.

**Constrained Sampling** We present the first scalable almost-uniform sampler, UniGen, which requires only one call to an approximate counter vis-a-vis linear calls in prior work. UniGen is highly parallelizable and achieves near-linear speedup in practice. We then present WeightGen, an adaptation of UniGen, to handle the problem of weighted sampling.

**Efficient Hash Functions** The performance of hashing-based techniques for constrained counting and sampling is primarily affected by the runtime of combinatorial solvers for the queries based on constraints from hash functions. We describe a new construction of universal hash functions based on the Independent support of the formulas and present the first algorithmic procedure and corresponding tool, MIS, to determine minimal independent support. The hash functions constructed using MIS achieves up to two orders of magnitude runtime improvement in our counting and sampling techniques.

Furthermore, the algorithmic frameworks for counting and sampling have been implemented as open source tools, which can handle formulas with up to 1 million variables, representing a significant boost up from the prior state of the art tools' ability to handle few tens of variables.

## 1.4 Tools

The following open source tools have been developed as part of this thesis:

**Constrained Counting**

ApproxMC2 https://bitbucket.org/kuldeepmeel/approxmc

SMTApproxMC https://bitbucket.org/kuldeepmeel/smtapproxmc

WeightMC https://bitbucket.org/kuldeepmeel/weightmc

WeightCount https://bitbucket.org/kuldeepmeel/weightcount

**Constrained Sampling**

UniGen https://bitbucket.org/kuldeepmeel/unigen

WeightGen https://bitbucket.org/kuldeepmeel/weightgen

**Efficient Hash Functions**

MIS https://bitbucket.org/kuldeepmeel/mis

## 1.5 Outline

This thesis is divided into four parts. The next Chapter, i.e., Chapter 2, introduces notations and definitions and should be treated as an index for standard concepts. Chapter 3 discusses several applications of constrained counting and sampling in detail.

We then move to Part II where we discuss the hashing-based framework for constrained counting. The first chapter of this part, i.e. Chapter 4, focuses on un-weighted constrained counting, i.e., UMC and presents the core algorithmic framework, ApproxMC2, which is primarily based on results in [40]. In Chapter 5, we discuss techniques to handle weighted distributions. We then discuss in Chapter 6 how hashing-based paradigm introduced in Chapter 4 can be extended to handle bit-vector formulas. Most of the results in this chapter appear in the paper [36]. The final chapter of this part, Chapter 7, presents a case study where we apply techniques introduced in this part to compute reliability estimates of the power grids arising from several cities in the USA. A preliminary version of this chapter appeared in [62].

We then move to Part III where we discuss the hashing-based framework for constrained sampling. The first chapter of this part, i.e., Chapter 8, focuses on the uniform generation and introduces the core hashing-based algorithmic framework, UniGen, which is primarily based on results in [39]. Chapter 9 then discusses how UniGen can trade off independence for performance gains and demonstrate high parallelizability of our framework. This chapter is based on results reported in [35]. Chapter 10 discusses an adaptation of UniGen, called WeightGen, to handle general weight functions. A preliminary version of this chapter appeared in [34].

And finally, we have the closing act of this thesis: Part IV. The algorithmic techniques described in Part II and III crucially relies on the hash functions. In Chapter 11, we describe an efficient construction of hash functions by exploiting the connection between defianibility of formulas and the distribution of solutions. Most of the results in this chapter appear in the paper [97]. A preliminary version of this paper appeared in 21st International Conference on Principles and Practice of Constraint Programming (CP-2015) and was awarded Best Student Paper. Chapter 12 summarizes the thesis and presents an assorted list of possible directions for future work.

# Chapter 2

# Background

## 2.1 Standard Probability Results

We state some standard probability results that are used throughout this work. Standard textbooks [153, 68] can be consulted for detailed information.

### r-wise Independence

A Set $V$ of random variables is said to exhibit $r-$wise independence *iff* for every subset of $V$ size $r$ or less, the joint probability distribution function of the subset is equal to product of individual marginal distributions.

We write $\mathsf{Pr}\left[Z : \mathcal{P}\right]$ to denote the probability of outcome $Z$ when sampling from a probability space $\mathcal{P}$. For brevity, we omit $\mathcal{P}$ when it is clear from the context. The expected value of $Z$ is denoted $\mathsf{E}\left[Z\right]$ and its variance is denoted by $\mathsf{V}\left[X\right]$ or $\sigma^2\left[Z\right]$. We now state three basic inequalities that are repeatedly used in this thesis.

### Markov Inequality

Let $Z$ be a nonnegative random variables and let $a > 0$, then

$$\mathsf{Pr}[Z > a] \leq \frac{\mathsf{E}[Z]}{a} \tag{2.1}$$

**Chebyshev Inequality**

Let $Z$ be a nonnegative random variable and let $\beta > 0$, then

$$\mathsf{Pr}[|Z - \mathsf{E}[X]| \geq \beta\sigma^2[Z]] \leq \frac{1}{\beta^2} \tag{2.2}$$

**Paley-Zygmund Inequality**

Let $Z$ be a random variable with finite variance, and for $0 \leq \beta \leq 1$, then

$$\mathsf{Pr}(Z \geq \beta\mathsf{E}[Z]) \geq (1-\beta)^2 \frac{\mathsf{E}[Z]^2}{\mathsf{E}[Z^2]} \tag{2.3}$$

## 2.2 Boolean Formulas

Let $F$ be a Boolean formula and let $X$ be the set of variables appearing in $F$, also referred to as the support of $F$. For a variable $x \in X$, we denote the assignment of $x$ to *true* by $x^1$ (also referred to as positive literal) and the assignment of $x$ to *false* by $x^0$ (also referred to as negative literal). We say that a formula $F$ is in Conjunctive Normal Form (CNF) if $F$ is expressed as:

$$F = C_1 \wedge C_2 \wedge \cdots C_m \tag{2.4}$$

where each $C_i$ is expressed as disjunct of literals, i.e, $C_i = (l_i^1 \vee l_i^2 \cdots)$. Similarly, a formula $F$ is in Disjunctive Normal Form (DNF) if $F$ is expressed as:

$$F = D_1 \vee D_2 \vee \cdots D_m \tag{2.5}$$

where each $D_i$ is expressed as conjunct of literals, i.e. $D_i = (l_i^1 \wedge l_i^2 \cdots)$.

A *satisfying assignment* or a *witness* of $F$ is an assignment of variables in $X$ that makes $F$ evaluate to *true*. We denote the set of all witnesses of $F$ by $R_F$. If $\sigma$ is an assignment of variables in $X$ and $x \in X$, we use $\sigma(x)$ to denote the value assigned to $x$ in $\sigma$.

For a formula $F$ over $X$ variables, we say that $G$ is a $\Sigma_1^1$ formula if $G$ is expressed as:

$$G := \exists S F(X) \tag{2.6}$$

where $S \subseteq X$. Note that there is a many to one mapping between satisfying assignments of $F$ and $G$. In particular, for every satisfying assignment $\sigma$ of $F$, the projection of $\sigma$ onto $X - S$ is a satisfying assignment of $G$. Consequently, we interchangeably represent a $\Sigma_1^1$ formula $G$ as a tuple $(F, S)$ where $S$ is referred to as the sampling set. We denote the set of all witnesses of $F$ by $R_F$ and the projection of $R_F$ onto $S$ by $R_{F\downarrow S}$. For $G := \exists S F(X)$, we have $R_G = R_{F\downarrow S}$. If $S = X$, then $R_G = R_F$.

## 2.3 Independent Support

For a given Boolean formula, Independent support is a subset of variables whose values uniquely determine the values of the remaining variables in any satisfying assignment to the formula. Formally, let $\mathcal{I} \subseteq X$ be a subset of the support such that if two satisfying assignments $\sigma_1$ and $\sigma_2$ agree on $\mathcal{I}$, then $\sigma_1 = \sigma_2$. In other words, in every satisfying assignment, the truth values of variables in $\mathcal{I}$ uniquely determine the truth value of every variable in $X \setminus \mathcal{I}$. The set $\mathcal{I}$ is called an *independent support* of $F$, and $\mathcal{D} = X \setminus \mathcal{I}$ is referred to as *dependent support*. There may be more than one independent support: $(a \vee \neg b) \wedge (\neg a \vee b)$ has three, namely $\{a\}$, $\{b\}$ and $\{a, b\}$. Clearly, if $\mathcal{I}$ is an independent support of $F$, so is every superset of $\mathcal{I}$. Note that there is a one-to-one correspondence between $R_F$ and $R_{F\downarrow I}$. The following lemma formalizes the above discussion.

**Lemma 1.** *Let $F(X)$ be a Boolean function with support $X$, and let $\mathcal{I}$ be an independent support of $F$. Then there exist Boolean functions $g_0, g_1, \ldots g_{n-k}$, each with support $S$ such that*

$$F(X) \leftrightarrow \left( g_0(\mathcal{I}) \wedge \bigwedge_{j=1}^{n-k} (x_{k+j} \leftrightarrow g_j(\mathcal{I})) \right)$$

*Proof.* Since $S$ is an independent support of $F$, we have $D = X \setminus S$ is a dependent support of $F$. From the definition of a dependent support, there exist Boolean functions $g_1, \ldots g_k$, each with support $S$, such that $F(\vec{X}) \rightarrow \bigwedge_{j=1}^{n-k}(x_{k+j} \leftrightarrow g_j(\mathcal{I}))$.

Let $g_0(\mathcal{I})$ be the characteristic function of the projection of $R_F$ on $S$. More formally, $g_0(\mathcal{I}) \equiv \bigvee_{(x_{k+1},\ldots x_n)\in\{0,1\}^{n-k}} F(\vec{X})$. It follows that $F(\vec{X}) \rightarrow g_0(\mathcal{I})$. Combining this with the result from the previous paragraph, we get the implication $F(\vec{X}) \rightarrow \left(g_0(\mathcal{I}) \wedge \bigwedge_{j=1}^{n-k}(x_{k+j} \leftrightarrow g_j(\mathcal{I}))\right)$

From the definition of $g_0(\mathcal{I})$ given above, we have $g_0(\mathcal{I}) \rightarrow F(\mathcal{I}, x_{k+1}, \ldots x_n)$, for some values of $x_{k+1}, \ldots x_n$. However, we also know that $F(\vec{X}) \rightarrow \bigwedge_{j=1}^{n-k}(x_{k+j} \leftrightarrow g_j(\mathcal{I}))$. It follows that $\left(g(\mathcal{I}) \wedge \bigwedge_{j=1}^{n-k}(x_{k+j} \leftrightarrow g_j(\mathcal{I}))\right) \rightarrow F(\vec{X})$. □

## 2.4 Group-oriented Unsatisfiable Subformulas and Subsets

In the problem of *group-oriented minimization* of unsatisfiable subsets [116, 127], we are given an unsatisfiable formula $\Psi$ of the form $\Psi = H_1 \wedge \cdots \wedge H_m \wedge \Omega$, and the task is to find a subset $\{H_{i_1}, \ldots, H_{i_k}\}$ of $\{H_1, \ldots, H_m\}$ so that $H_{i_1} \wedge \cdots \wedge H_{i_k} \wedge \Omega$ remains unsatisfiable. The subformulas $H_1, \ldots, H_m$ are called *groups* (or *high-level constraints*) and $\Omega$ is called the *remainder*. The remainder plays a special role – it consists of *non-interesting* constraints that do not need to be minimized and are always part of the formula.

If $H_{i_1} \wedge \cdots \wedge H_{i_k} \wedge \Omega$ is unsatisfiable, we say that $\{H_{i_1}, \ldots, H_{i_k}\}$ is a (group-oriented) unsatisfiable subset, or equivalently that $H_{i_1} \wedge \cdots \wedge H_{i_k} \wedge \Omega$ is an unsatisfiable subformula of $\Psi$. In addition, when $\{H_{i_1}, \ldots, H_{i_k}\}$ is *minimal* (removal of any $H_{i_j}$ renders the formula satisfiable), we say that $\{H_{i_1}, \ldots, H_{i_k}\}$ is a *group-oriented minimal unsatisfiable subset* (GMUS), or equivalently that $H_{i_1} \wedge \cdots \wedge H_{i_k} \wedge \Omega$ is a minimal unsatisfiable subformula of $\Psi$. (If $\{H_{i_1}, \ldots, H_{i_k}\}$ is of minimum size, that is there is no smaller unsatisfiable subset, then we call it a *minimum* unsatisfiable subset (SGMUS).)

## 2.5 Bit-Vector Formulas

A *word* (or *bit-vector*) is an array of bits. The size of the array is called the *width* of the word. We consider here *fixed-width* words, whose width is a constant. It is easy to see that a word of width $k$ can be used to represent elements of a set of size $2^k$. The first-order theory of fixed-width words has been extensively studied (see [111, 30] for an overview). The vocabulary of this theory includes interpreted predicates and functions, whose semantics are defined over words interpreted as signed integers, unsigned integers, or vectors of propositional constants (depending on the function or predicate). When a word of width $k$ is treated as a vector, we assume that the component bits are indexed from 0 through $k-1$, where index 0 corresponds to the rightmost bit.

## 2.6 Weight Function

Given a weight function $W : \{0,1\}^n \mapsto [0,1]$, we use $W(\sigma)$ to denote the weight of an assignment $\sigma$. To avoid notational clutter, we overload $W(\cdot)$ to denote the weight of an assignment or formula, depending on the context. Given a set $Y$ of assignments, we use $W(Y)$ to denote $\sum_{\sigma \in Y} W(\sigma)$. Given a formula $F$ and sampling set $S$, we $W(F \downarrow S)$ to denote $\sum_{\sigma \in R_{F\downarrow S}} W(\sigma)$. If the sampling set $S$ is an Independent support, we use $W(F)$ to avoid notational clutter. For example, the formula $F = (x_1 \leftrightarrow \neg x_2)$ has two satisfying assignments: $\sigma_1 = (x_1 : \mathit{true}, x_2 : \mathit{false})$, and $\sigma_2 = (x_1 : \mathit{false}, x_2 : \mathit{true})$. Thus, we have $W(\sigma_1) = W(x_1^1) \cdot W(x_2^0)$ and $W(\sigma_2) = W(x_1^0) \cdot W(x_2^1)$. The weight of $F$, or $W(F)$, is then $W(\sigma_1) + W(\sigma_2)$.

There are several representations of weight function over assignments. Of particular interest to us is literal-weight representations, in which weights are assigned to literals, and the weight of an assignment is the product of weights of its literals. We are yet again overloading the $W(\cdot)$ to represent weight of literal as well as we will see, the context matters. For a variable $x$ of $F$ and a weight function $W(\cdot)$, we use $W(x^1)$ and $W(x^0)$ to denote the weights of the positive and negative literals, respectively. Adopting terminology used in [139], we assume that every variable $x$ either has an *indifferent* weight, i.e. $W(x^0) = W(x^1) = 1$, or a *normal* weight, i.e. $W(x^0) = 1 - W(x^1)$, where $0 \leq W(x^1) \leq 1$. Note that having a variable with a normal weight of $W(x^1) = 1$ (resp. $W(x^1) = 0$) makes the variable $x$ redundant in computation of the weight of $F$, since in this case only assignments $\sigma$ with $\sigma(x) = \mathit{true}$ (resp. $\sigma(x) = \mathit{false}$) can contribute to the weight of $F$. Thus, we can assign *true* (resp. *false*) to $x$ without changing the overall weight of models of the formula. This suggests that we can further assume $0 < W(x^1) < 1$ for every variable $x$ with a normal weight.

For every variable $x_i$ with normal weight, we assume that $W(x_i^1)$, which is a positive fraction, is specified in binary using $m_i$ bits. Without loss of generality, the least significant bit in the binary representation of $W(x_i^1)$ is always taken to be 1. Thus, the rational decimal representation of $W(x_i^1)$ is $k_i/2^{m_i}$, where $k_i$ is an odd integer in $\{1, \ldots 2^{m_i} - 1\}$. It follows that $W(x_i^0)$ is $(2^{m_i} - k_i)/2^{m_i}$. Let $N_F$ denote the set of indices of variables in $X$ that have normal weights, and let $\widehat{m} = \Sigma_{i \in N_F} m_i$. Let $C_F = \prod_{i \in N_F} 2^{-m_i} = 2^{-\widehat{m}}$. Note that $W(\sigma)/C_F$ is a natural number for every assignment $\sigma$; hence $W(F)/C_F$ is a natural number as well.

## 2.7 Universal Hash Functions

The concept of universal hash functions is central to this thesis. For positive integers $n$, $m$, and $r$, we write $H(n,m,r)$ to denote a family of $r$-universal hash functions mapping $\{0,1\}^n$ to $\{0,1\}^m$. We use $h \xleftarrow{R} H(n,m,r)$ to denote the probability space obtained by choosing a hash function $h$ uniformly at random from $H(n,m,r)$. The property of $r$-universality guarantees that:

$$\forall \alpha_1, \alpha_2, \cdots \alpha_r \text{ and distinct } y_1, \ldots, y_r \in \{0,1\}^n$$

$$\mathsf{Pr}[h(y_i) = \alpha_i] = \frac{1}{2^m} \tag{2.7}$$

$$\mathsf{Pr}\left[h(y_1) = h(y_2) = \cdots = h(y_r) : h \xleftarrow{R} H(n,m,r)\right] \leq \left(\frac{1}{2^m}\right)^r. \tag{2.8}$$

We use a particular class of such hash functions, denoted by $H_{xor}(n,m)$, which is defined as follows. Let $h(y)[i]$ denote the $i^{th}$ component of the vector $h(y)$. This family of hash functions is then defined as $\{h \mid h(y)[i] = a_{i,0} \oplus (\bigoplus_{k=1}^{n} a_{i,k} \cdot y[k]), a_{i,k} \in \{0,1\}, 1 \leq i \leq m,\ 0 \leq k \leq n\}$, where $\oplus$ denotes the XOR operation. By choosing values of $a_{i,k}$ randomly and independently, we can effectively choose a random hash function from $H_{xor}(n,m)$. It was shown in [88] that this family is 3-universal.

In several of the algorithms presented in this thesis, we randomly choose one function $h$ from $H_{xor}(n,n-1)$, and one vector $\alpha$ from $\{0,1\}^{n-1}$. Thereafter, we use

"prefix-slices" of $h$ and $\alpha$ to obtain $h_m$ and $\alpha_m$ for all other values of $m$. Formally, for every $m \in \{1, \ldots |S|-1\}$, the $m^{th}$ prefix-slice of $h$, denoted $h^{(m)}$, is a map from $\{0,1\}^{|S|}$ to $\{0,1\}^m$, such that $h^{(m)}(y)[i] = h(y)[i]$, for all $y \in \{0,1\}^{|S|}$ and for all $i \in \{1, \ldots m\}$. Similarly, the $m^{th}$ prefix-slice of $\alpha$, denoted $\alpha^{(m)}$, is an element of $\{0,1\}^m$ such that $\alpha^{(m)}[i] = \alpha[i]$ for all $i \in \{1, \ldots m\}$. The randomness in the choices of $h$ and $\alpha$ induces randomness in the choices of $h_m$ and $\alpha_m$. However, the $(h_m, \alpha_m)$ pairs chosen for different values of $m$ are no longer independent. Specifically, $h_j(y)[i] = h_k(y)[i]$ and $\alpha_j[i] = \alpha_k[i]$ for $1 \le j < k < |S|$ and for all $i \in \{1, \ldots j\}$. This lack of independence is a fundamental departure from previous design of hashing-based algorithms and crucial for the theoretical as well as practical performance of our algorithms.

## 2.8 Constrained Counting and Sampling

Given a formula $F$, sampling set $S$, and a weight function $W(\cdot)$, the *constrained counting problem*, also referred to as weighted model counting (denoted as WMC), is to determine $W(F \downarrow S)$. If the weight function assigns equal weight to all the assignments, then the problem is called unweighted counting, also referred to as unweighted model counting (denoted as UMC).

An *approximate counter* is a probabilistic algorithm $\mathsf{ApproxCount}(\cdot,\cdot,\cdot,\cdot,\cdot)$ that, given a formula $F$, sampling set $S$, weight function $W(\cdot)$, tolerance $\varepsilon > 0$, and confidence parameter $\delta \in (0,1]$, guarantees that

$$\mathsf{Pr}\Big[W(F \downarrow S)/(1+\varepsilon) \le \mathsf{ApproxCount}(F, S, W(\cdot), \varepsilon, \delta) \le (1+\varepsilon)W(F \downarrow S)\Big] \ge 1 - \delta. \tag{2.9}$$

The *constrained-sampling problem* is to sample a witness $y$ randomly from $R_{F\downarrow S}$ with probability proportional to its weight, i.e. $W(y)$. Formally, a constrained sam-pler, $\mathcal{G}(\cdot,\cdot,\cdot)$ ensures:

$$\mathsf{Pr}\left[\mathcal{G}^{u}(F,S,W(\cdot))=y\right]\propto W(y) \tag{2.10}$$

An *almost-sampler* $\mathcal{G}^{au}(\cdot,\cdot,\cdot,\cdot)$ guarantees that for every $y\in R_F$, we have:

$$\frac{1}{(1+\varepsilon)W(F\downarrow S)}\leq \mathsf{Pr}\left[\mathcal{G}^{au}(F,S,W(\cdot)\,,\varepsilon)=y\right]\leq \frac{1+\varepsilon}{W(F\downarrow S)} \tag{2.11}$$

, where $\varepsilon>0$ is a specified *tolerance*. Probabilistic generators are allowed to occasionally "fail" in the sense that no solution may be returned even if $R_F$ is non-empty. The failure probability for such generators must be bounded by a constant strictly less than 1.

# Chapter 3

# Applications of Counting and Sampling

In this Chapter, we discuss several applications of constrained counting and sampling. The techniques developed in this thesis have been applied to the application domains discussed in this Chapter.

## 3.1 Probabilistic Inference

Probabilistic inference is key to reason about uncertain and large data sets arising from diverse applications including medical diagnostics, weather modeling, computer vision and the like [10, 60, 138, 159]. In the domain of probabilistic reasoning, we typically have a probabilistic model capturing dependencies between variables in a system, and evidence described as a valuation of a subset of variables. The problem of probabilistic inference requires us to determine the probability of an event of interest, (valuations of a subset of variables) given observed evidence. (valuations of some other subset of variables). , i.e., valuation to the variables of interest. This problem has been the subject of intense investigations by both theoreticians and practitioners over the last few decades.

Not surprisingly, probabilistic inference in its exact form is intractable due to the curse of dimensionality, and it has been shown to be #P-complete for variables with finite domains [135]. As a result, researchers have investigated approximate techniques to solve real-world instances of this problem. Of these, the most popular ones are those based on Monte Carlo Markov Chain (MCMC) methods and variational approximations. While these techniques scale to large problem instances, they fail to provide rigorous approximation guarantees in practice [72]. Interval propagation and techniques based on the random seeding of combinatorial reasoning tools have also been used by researchers to tame the problem. Unfortunately, these approaches also suffer from the same drawback – the formal guarantees provided are either very weak or non-existent [109].

A promising alternative approach that has emerged over the years is to reduce the probabilistic inference problem to discrete integration or constrained counting [135, 42] on a finite domain knowledge base. For example, given a Bayesian approach, the constrained counting approach encodes the Bayesian network into a knowledge base in conjunctive normal form and maps the weights to literals of CNF formula to elements in conditional probability tables. For a detailed discussion of various reductions of probabilistic inference to constrained counting, we refer the reader to [42].

## 3.2 Network Reliability

Modern society is increasingly reliant on the availability of critical facilities and utility services, such as power, telecommunications, water, gas, and transportation among others [146]. To ensure adequate service, it is imperative to quantify system reliability, or the probability of the system to remain functional, as well as system resilience, or the ability of the system to quickly return to normalcy when failure is unavoidable [29]. While resilience assessment requires human decision-making principles, it also heavily depends on intrinsic system reliability. Hence, the recent focus on community resilience and sustainability has spurred significant activity in engineering reliability [165].

One of the key challenging problems in the area of engineering reliability is network reliability, wherein the input to the problem consists of a network, represented as a graph, arising out of the distribution of water, power, transportation routes and the like. The problem of the network reliability seeks to measure the likelihood of two points of interest being reachable under conditions such as natural disasters. Early theoretical investigations showed that the problem of network reliability is #P complete [150]. Although graph contraction strategies combined with DNF counting provide a Fully Polynomial Randomized Approximation Scheme (FPRAS) with error guarantees [103], implementation on practical systems does not scale well due to the requirement of a large number of Monte Carlo steps. Consequently, recent investigations have focused on advancing algorithmic strategies that build upon advanced Monte Carlo simulation [166] and analytical approaches [117, 63]. Furthermore, inventive sampling methods, such as line sampling and variance reduction schemes [76], along with graphical models, especially Bayesian networks, provide versatile strategies to quantify the reliability of complex engineered systems and their dynamics [21].

Despite significant progress, most techniques remain computationally expensive. As an alternative, when invoking approximations, most methods are unable to guarantee the quality of the reliability estimation a priori, barring small instances where exact methods do not time out. Therefore, the design of techniques that offer strong theoretical guarantees on the quality of estimates and can scale to large real-world instances remains an unattained goal across multiple disciplines.

In this thesis, we demonstrate the effectiveness of the approach of reducing network reliability queries to constrained counting. For a detailed discussion of our approach, we refer the reader to Chapter 7.

## 3.3 Quantified Information Flow

The remarkable progress in artificial intelligence has led to data being the key component for the new economy. This has led to unprecedented increase in the storage of personal data, which is often accessed by software to provide a variety of services to the end user ranging from the recommendation for consumable products to personalized mortgage interest rates. Finding vulnerabilities in the programs that access data is fundamental for guaranteeing user security and data confidentiality. Due to the increasing complexity of software systems, automated techniques have to be deployed to assist architects and engineers in verifying the quality of their code. Among these, *quantitative* techniques have been shown to effectively detect complex vulnerabilities.

*Quantitative information flow* (QIF) computation [45] is a powerful quantitative technique to detect information leakage directly at the code level. QIF leverages information theory to measure the flow of information between different functions of the program. An unexpectedly large flow of information may characterize a potential leakage of information. In practice, this technique relies on the following: the maximum amount of information that can leak from a function (known as *channel capacity*) is the logarithm of the number of distinct outputs that the function can produce [49].

Recently, QIF computation based on program analysis and constrained counting has effectively analyzed codebases of tens of thousands of lines of C code [149]. This technique proceeds as follows. A specific fragment of the program (e.g., a function, or the whole program) is modeled as an information-theoretic channel from its input to its output. Program analysis techniques such as symbolic execution or model checking are used to explore the possible executions of the fragment. Program analysis produces a set of constraints that characterize these executions. Afterwards, a *con-* *strained counter* is used to determine the number of distinct outputs of the fragment (e.g., the return values of the function, or the outputs of the program). Finally, the base-2 logarithm of the number of possible outputs gives us the channel capacity in bits, which corresponds to the maximum amount of information that can flow through the channel modeling the fragment.

## 3.4 Functional Verification

Functional verification constitutes one of the most challenging and time-consuming steps in the design of modern digital systems. The primary objective of functional verification is to expose design bugs early in the design cycle. Among various techniques available for this purpose, those based on simulation overwhelmingly dominate industrial practice. In a typical simulation-based functional verification exercise, a gate-level or RTL model of the circuit is simulated for a large number of cycles with specific input patterns. The values at observable outputs, as computed by the simulator, are then compared against their expected values, and any discrepancy is flagged as a manifestation of a bug. The state of simulation technology today is mature enough to allow simulation of large designs within reasonable time using modest computational resources. Generating input patterns that exercise diverse corners of the design's behavior space, however, remains a challenging problem [20].

In recent years, constrained-random simulation (also called constrained-random verification, or CRV) [129] has emerged as a practical approach to address the problem of simulating designs with "random enough" input patterns. In CRV, the verification engineer declaratively specifies a set of constraints on the values of circuit inputs. Typically, these constraints are obtained from usage requirements, environmental constraints, constraints on operating conditions and the like. A constraint solver is then used to generate random values for the circuit inputs satisfying the constraints. Since the distribution of errors in the design's behavior space is not known *a priori*, every solution to the set of constraints is as likely to discover a bug like any other solution. It is therefore important to sample the space of all solutions uniformly or almost-uniformly (defined formally below) at random. Unfortunately, guaranteeing uniformity poses significant technical challenges when scaling to large problem sizes. This has repeatedly been noted in the literature (see, for example, [58, 133, 108]) and also confirmed by industry practitioners*.

Given the important of constrained-random simulation in the hardware design process, our benchmark suite for constrained sampling has a significant fraction of benchmarks arising from constrained-random simulation domain (See experimental evaluation in Chapters 8 and 9).

## 3.5 Pattern Sampling

Given the deluge of data, providing concise representations, also known as patterns, of the underlying dataset has been the holy grail in the field of data mining. Often, finding a single concise representation for a real-world data is not possible, and as a result, researchers focus on finding a set of patterns, often known as pattern mining.

The earliest approaches to pattern mining focused on the enumeration of all the patterns, but this hindered scalability of such approaches due to a large number of patterns. As a result, the alternate approaches such as *Condensed representations, top-k mining, pattern set mining* have been proposed [27, 33, 164]. However, these approaches either result in too few patterns or too similar patterns. Recently, Dzyuba,

*Private communication: R. Kurshan

van Leeuwen, and De Raedt proposed FLEXICS, the first flexible pattern sampler that supports a broad class of quality measures and constraints while providing strong guarantees regarding sampling accuracy [64]. The core of FLEXICS relies on constraint sampler, and it employs the constrained sampling techniques proposed in this thesis. Empirically, FLEXICS is shown to be highly accurate and efficient, thus leading to being the state of the art tool for pattern-based data exploration. For a detailed discussion, we refer the reader to [64].

## 3.6 Program Synthesis

The problem of program synthesis is to synthesize programs from specification, which finds applications in many disciplines ranging from computer-aided programming, aiding students in introductory programming courses to industrial tools such as FleshFill [67, 91, 140]. Recent breakthrough successes in program synthesis owes to the paradigm of counter example guided inductive synthesis (CEGIS). The CEGIS paradigm involves (i) representing space of programs using set of constraints, (ii) modeling the learned concepts as additional constraints, and (iii) employing a constraint sampler to generate the learned program [142]. The runtime of CEGIS-based techniques crucially depends on the quality of solutions returned by the underlying constraint solver; sometimes, leading to about two orders of magnitude runtime variation depending on the quality of underlying solver. Therefore, constraint sampling is a key step in CEGIS based techniques and the sampling techniques developed in this thesis has resulted into significant improvement in the efficiency of CEGIS-based tools [67].

# Part II

# Constrained Counting

This part will focus on constrained counting. The first chapter of this part, i.e. Chapter 4, will focus on unweighted constrained counting, i.e., UMC and then in. Chapter 5, we discuss techniques to handle weighted distributions. We then discuss in Chapter 6 how hashing-based paradigm introduced in Chapter 4 can be extended to handle bit-vector formulas. The final chapter of this part, Chapter 7, presents a case study where we apply techniques introduced in this part to compute reliability estimates of the power grids arising from several cities in USA.

# Chapter 4

# Hashing-based Scalable Unweighted Counter

Complexity theoretic studies of unweighted counting, also referred to as UMC, were initiated by Valiant, who showed that the problem is #P-complete [150]. The earliest approaches to UMC were based on DPLL-style SAT solvers and computed exact counts. These approaches, e.g. CDP [24], incrementally counted the number of solutions by introducing appropriate multiplication factors for each partial solution found, eventually covering the entire solution space. Subsequent counters such as Relsat [14], Cachet [138], and sharpSAT [147] improved upon this by using several optimizations such as component caching, clause learning, and the like. Techniques based on Binary Decision Diagrams (BDDs) and their variants [125], or d-DNNF formulas [52], have also been used to compute exact counts.

Although exact counters have been successfully used in small- to medium-sized problems, scaling to the larger problem instances have posed significant challenges. Consequently, a large class of practical applications has remained beyond the reach of exact counters [123]. The study of approximate model counting has therefore been an important topic of research for several decades. Approximate counting was shown to lie in the third level of the polynomial hierarchy in [144]. Stockmeyer's approximate counter crucially relies on Sipser's technique for estimating the size of a set using universal hash functions [141], and requires access to a $\Sigma_2^p$ oracle. For DNF formulas, Karp, Luby and Madras gave a fully polynomial randomized approximation scheme for counting models [104]. One can build on [144] and design a hashing-based probably approximately correct counting algorithm that makes polynomially many calls to an NP oracle [83]. Unfortunately, this does not lend itself to a scalable implementation because every invocation of the NP oracle (a SAT solver in practice) must reason about a formula with significantly large, viz. $\mathcal{O}(n/\varepsilon)$, support.

To overcome the scalability challenge, relaxed variants of counting have been pursued. Among them of note are *bounding counters*, which provide lower or upper bounds but do not offer guarantees on the tightness of these bounds. Examples include SampleCount [84], BPCount [110], MBound and Hybrid-MBound [87], and MiniCount [110]. Another category of counters is called *guarantee-less counters* such as ApproxCount [156], SearchTreeSampler [73], SE [136], and SampleSearch [82].. While these counters may be efficient, they provide no guarantees and the computed estimates may differ from the exact counts by several orders of magnitude [85].

In [38], a new hashing-based strongly probably approximately correct counting algorithm, called ApproxMC, was shown to scale to formulas with hundreds of thousands of variables, while providing rigorous PAC-style $(\varepsilon, \delta)$ guarantees. The core idea of ApproxMC is to use 2-universal hash functions to randomly partition the solution space of the original formula into "small" enough cells. The sizes of sufficiently many randomly chosen cells are then determined using calls to a specialized SAT solver (CryptoMiniSAT [143]), and a scaled median of these sizes is used to estimate the desired model count. Finding the right parameters for the hash functions is crucial to the success of this technique. ApproxMC uses a linear search for this purpose, where each search step invokes the specialized SAT solver, viz. CryptoMiniSAT, $\mathcal{O}(1/\varepsilon^2)$ times. Overall, ApproxMC makes a total of $\mathcal{O}(\frac{n \log(1/\delta)}{\varepsilon^2})$ calls to CryptoMiniSAT. Significantly, and unlike the algorithm in [83], each call of CryptoMiniSAT reasons about a formula with only $n$ variables.

The works of [70, 34, 35, 17] have subsequently extended the ApproxMC approach to finite domain discrete integration. The work of [17] had experimental evaluation on an implementation inconsistent with the algorithms presented in the work and the revised evaluation did not support earlier claims. Furthermore, approaches based on ApproxMC form the core of various sampling algorithms proposed recently [69, 39, 34, 35]. Therefore, any improvement in the core algorithmic structure of ApproxMC can potentially benefit several other algorithms. in sampling and discrete integration. Recently, Zhu and Ermon [163] proposed an approximate algorithm, named RP-InfAlg, for approximate probabilistic inference. This algorithm does not provide $(\varepsilon, \delta)$ approximation guarantees, and requires the use of hard-to-estimate parameters. The computational effort required in identifying the right values of the parameters is not addressed in their work. Furthermore, their experiments were done with setting length of xor-constraints to 1, 2 and 4 for which the proofs are known not to hold [162].

Prior work on improving the scalability of hashing-based approximate counting algorithms has largely focused on improving the efficiency of 2-universal linear (xor-based) hash functions. It is well-known that long xor-based constraints make SAT solving significantly hard in practice [86]. Researchers have therefore investigated theoretical and practical aspects of using short xors [86, 39, 72, 162].

Recently, Ermon et al. [72] and Zhao et al. [162] attempted to show how short xor constraints (even logarithmic in the number of variables) can be used for approximate counting with certain theoretical guarantees. The resulting algorithms, however, do not provide PAC-style $(\varepsilon, \delta)$ guarantees. Furthermore, their proposed technique to obtain lower and upper bounds does not have upper bound on failure probability. In addition, the experimental results were performed with an implementation inconsistent with the algorithm for which theoretical analysis was performed in the work.

Given the promise of hashing-based counting techniques in bridging the gap between scalability and providing rigorous guarantees for constrained counting, there have been several recent efforts to design efficient universal hash functions [97, 36]. While these efforts certainly help push the scalability frontier of hashing-based techniques for probabilistic inference, the structure of the underlying algorithms has so far escaped critical examination. For example, all recent approaches to probabilistic inference via hashing-based counting use a linear search to identify the right values of parameters for the hash functions. As a result, the number of calls to the NP oracle (SAT solver in practice) increases linearly in the number of variables, $n$, in the input constraint. Since SAT solver calls are by far the computationally most expensive steps in these algorithms [124], this motivates us to ask: *Can we design a hashing-based approximate counting algorithm that requires sub-linear (in $n$) calls to the* SAT *solver, while providing strong theoretical guarantees?*

In this chapter, we provide a positive answer to the above question. In particular:

1. We present a new hashing-based approximate counting algorithm, called ApproxMC2, for CNF formulas, that reduces the number of SAT solver calls *from linear in $n$ to logarithmic in $n$* while still providing rigorous $(\varepsilon, \delta)$ guarantees.

2. Furthermore, for DNF formulas, ApproxMC2 gives a fully polynomial randomized approximation scheme (FPRAS), which differs fundamentally from the only known FPRAS for DNF formulas [104].

3. Extensive experiments demonstrate that ApproxMC2 outperforms the prior state of the art tool, ApproxMC, by 1-2 orders of magnitude in running time, when using the same family of hash functions.

We also discuss how the framework and analysis of ApproxMC2 can be lifted to

**Algorithm 1** ApproxMC2$(F, S, \varepsilon, \delta)$

thresh $\leftarrow 1 + 9.84\left(1 + \frac{\varepsilon}{1+\varepsilon}\right)\left(1 + \frac{1}{\varepsilon}\right)^2$;
$Y \leftarrow$ BoundedSAT$(F,$ thresh$, S)$;
**if** ($|Y| <$ thresh) **then return** $|Y|$;
$t \leftarrow \lceil 17 \log_2(3/\delta) \rceil$;
nCells $\leftarrow 2$; $C \leftarrow$ emptyList; iter $\leftarrow 0$;
**repeat**
    iter $\leftarrow$ iter $+ 1$;
    (nCells, nSols) $\leftarrow$ ApproxMC2Core$(F, S,$ thresh, nCells$)$;
    **if** (nCells $\neq \perp$) **then** AddToList$(C,$ nSols $\times$ nCells$)$;
**until** (iter $< t$);
finalEstimate $\leftarrow$ FindMedian$(C)$;
**return** finalEstimate

other hashing-based probabilistic inference algorithms [34]. Significantly, the algorithmic improvements of ApproxMC2 are orthogonal to recent advances in the design of hash functions [97], permitting the possibility of combining ApproxMC2-style algorithms with efficient hash functions to boost the performance of hashing-based probabilistic inference even further.

The remainder of the chapter is organized as follows. In Section 4.1, we present ApproxMC2 and its analysis. We discuss our experimental methodology and present experimental results in Section 4.2. Finally, we conclude in Section 4.3.

## 4.1 The Algorithm

We now present ApproxMC2: a hashing-based approximate counting algorithm. Algorithm 1 shows the pseudocode for ApproxMC2. It takes as inputs a formula $F$, a sampling set $S$, a tolerance $\varepsilon$ ($> 0$), and a confidence $1 - \delta \in (0, 1]$. It returns an estimate of $|R_{F\downarrow S}|$ within tolerance $\varepsilon$, with confidence at least $1 - \delta$. Note that although ApproxMC2 draws on several ideas from ApproxMC, the original algorithm in [38] computed an estimate of $|R_F|$ (and not of $|R_{F\downarrow S}|$). Nevertheless, the idea of using sampling sets, as described in [39], can be trivially extended to ApproxMC. Therefore, whenever we refer to ApproxMC in this chapter, we mean the algorithm in [38] extended in the above manner.

There are several high-level similarities between ApproxMC2 and ApproxMC. Both algorithms start by checking if $|R_{F\downarrow S}|$ is smaller than a suitable threshold (called pivot in ApproxMC and thresh in ApproxMC2). This check is done using subroutine BoundedSAT, that takes as inputs a formula $F$, a threshold thresh, and a sampling set $S$, and returns a subset $Y$ of $R_{F\downarrow S}$, such that $|Y| = \min(\mathsf{thresh}, |R_{F\downarrow S}|)$. The thresholds used in invocations of BoundedSAT lie in $O(1/\varepsilon^2)$ in both ApproxMC and ApproxMC2, although the exact values used are different. If $|Y|$ is found to be less than thresh, both algorithms return $|Y|$ for the size of $|R_{F\downarrow S}|$. Otherwise, a core subroutine, called ApproxMCCore in ApproxMC and ApproxMC2Core in ApproxMC2, is invoked. This subroutine tries to randomly partition $R_{F\downarrow S}$ into "small" cells using hash functions from $H_{xor}(|S|, m)$, for suitable values of $m$. There is a small probability that this subroutine fails and returns $(\bot, \bot)$. Otherwise, it returns the number of cells, nCells, into which $R_{F\downarrow S}$ is partitioned, and the count of solutions, nSols, in a randomly chosen small cell. The value of $|R_{F\downarrow S}|$ is then estimated as $\mathsf{nCells} \times \mathsf{nSols}$. In order to achieve the desired confidence of $(1 - \delta)$, both ApproxMC2 and ApproxMC invoke

their core subroutine repeatedly, collecting the resulting estimates in a list $C$. The number of such invocations lies in $O(\log(1/\delta))$ in both cases. Finally, both algorithms compute the median of the estimates in $C$ to obtain the desired estimate of $|R_{F\downarrow S}|$.

Despite these high-level similarities, there are key differences in the ways ApproxMC and ApproxMC2 work. These differences stem from: (i) the use of dependent hash functions when searching for the "right" way of partitioning $R_{F\downarrow S}$ within an invocation of ApproxMC2Core, and (ii) the lack of independence between successive invocations of ApproxMC2Core. We discuss these differences in detail below.

Subroutine ApproxMC2Core lies at the heart of ApproxMC2. Functionally, ApproxMC2Core serves the same purpose as ApproxMCCore; however, it works differently. To understand this difference, we briefly review the working of ApproxMCCore. Given a formula $F$ and a sampling set $S$, ApproxMCCore finds a triple $(m, h_m, \alpha_m)$, where $m$ is an integer in $\{1, \ldots |S| - 1\}$, $h_m$ is a hash function chosen randomly from $H_{xor}(|S|, m)$, and $\alpha_m$ is a vector chosen randomly from $\{0,1\}^m$, such that $|R_{\langle F, h_m, \alpha_m \rangle \downarrow S}| < \mathsf{thresh}$ and $|R_{\langle F, h_{m-1}, \alpha_{m-1} \rangle \downarrow S}| \geq \mathsf{thresh}$. In order to find such a triple, ApproxMCCore uses a linear search: it starts from $m = 1$, chooses $h_m$ and $\alpha_m$ randomly and independently from $H_{xor}(|S|, m)$ and $\{0,1\}^m$ respectively, and checks if $|R_{\langle F, h_m, \alpha_m \rangle \downarrow S}| \geq \mathsf{thresh}$. If so, the partitioning is considered too coarse, $h_m$ and $\alpha_m$ are discarded, and the process repeated with the next value of $m$; otherwise, the search stops. Let $m^*, h_{m^*}$ and $\alpha_{m^*}$ denote the values of $m$, $h_m$ and $\alpha_m$, respectively, when the search stops. Then ApproxMCCore returns $|R_{\langle F, h_{m^*}, \alpha_{m^*} \rangle \downarrow S}| \times 2^{m^*}$ as the estimate of $|R_{F\downarrow S}|$. If the search fails to find $m$, $h_m$ and $\alpha_m$ with the desired properties, we say that ApproxMCCore fails.

Every iteration of the linear search above invokes BoundedSAT once to check if $|R_{\langle F, h_m, \alpha_m \rangle \downarrow S}| \geq \mathsf{thresh}$. A straightforward implementation of BoundedSAT makes up

to thresh calls to a SAT solver to answer this question. Therefore, an invocation of ApproxMCCore makes $\mathcal{O}(\mathsf{thresh}.|S|)$ SAT solver calls. A key contribution of this chapter is a new approach for choosing hash functions that allows ApproxMC2Core to make at most $\mathcal{O}(\mathsf{thresh}.\log_2|S|)$ calls to a SAT solver. Significantly, the sizes of formulas fed to the solver remain the same as those used in ApproxMCCore; hence, the reduction in number of calls comes without adding complexity to the individual calls.

A salient feature of ApproxMCCore is that it randomly and independently chooses $(h_m, \alpha_m)$ pairs for different values of $m$, as it searches for the right partitioning of $R_{F\downarrow S}$. In contrast, in ApproxMC2Core, we randomly choose one function $h$ from $H_{xor}(|S|, |S|-1)$, and one vector $\alpha$ from $\{0,1\}^{|S|-1}$. Thereafter, we use "prefix-slices" of $h$ and $\alpha$ to obtain $h_m$ and $\alpha_m$ for all other values of $m$. Formally, for every $m \in \{1, \ldots |S|-1\}$, the $m^{th}$ prefix-slice of $h$, denoted $h^{(m)}$, is a map from $\{0,1\}^{|S|}$ to $\{0,1\}^m$, such that $h^{(m)}(y)[i] = h(y)[i]$, for all $y \in \{0,1\}^{|S|}$ and for all $i \in \{1, \ldots m\}$. Similarly, the $m^{th}$ prefix-slice of $\alpha$, denoted $\alpha^{(m)}$, is an element of $\{0,1\}^m$ such that $\alpha^{(m)}[i] = \alpha[i]$ for all $i \in \{1, \ldots m\}$. Once $h$ and $\alpha$ are chosen randomly, ApproxMC2Core uses $h^{(m)}$ and $\alpha^{(m)}$ as choices of $h_m$ and $\alpha_m$, respectively. The randomness in the choices of $h$ and $\alpha$ induces randomness in the choices of $h_m$ and $\alpha_m$. However, the $(h_m, \alpha_m)$ pairs chosen for different values of $m$ are no longer independent. Specifically, $h_j(y)[i] = h_k(y)[i]$ and $\alpha_j[i] = \alpha_k[i]$ for $1 \leq j < k < |S|$ and for all $i \in \{1, \ldots j\}$. This lack of independence is a fundamental departure from ApproxMCCore.

Algorithm 2 shows the pseudo-code for ApproxMC2Core. After choosing $h$ and $\alpha$ randomly, ApproxMC2Core checks if $|R_{\langle F,h,\alpha\rangle\downarrow S}| < \mathsf{thresh}$. If not, ApproxMC2Core fails and returns $(\bot, \bot)$. Otherwise, it invokes sub-routine LogSATSearch to find

**Algorithm 2** $\mathsf{ApproxMC2Core}(F, S, \mathsf{thresh}, \mathsf{prevNCells})$

1: Choose $h$ at random from $H_{xor}(|S|, |S|-1)$;
2: Choose $\alpha$ at random from $\{0,1\}^{|S|-1}$;
3: $Y \leftarrow \mathsf{BoundedSAT}(F \wedge h(S) = \alpha, \mathsf{thresh}, S)$;
4: **if** $(|Y| \geq \mathsf{thresh})$ **then return** $(\bot, \bot)$;
5: $\mathsf{mPrev} \leftarrow \log_2 \mathsf{prevNCells}$;
6: $m \leftarrow \mathsf{LogSATSearch}(F, S, h, \alpha, \mathsf{thresh}, \mathsf{mPrev})$;
7: $\mathsf{nSols} \leftarrow |\mathsf{BoundedSAT}(F \wedge h^{(m)}(S) = \alpha^{(m)}, \mathsf{thresh}, S)|$;
8: **return** $(2^m, \mathsf{nSols})$;

a value of $m$ (and hence, of $h^{(m)}$ and $\alpha^{(m)}$) such that $|R_{\langle F,h^{(m)},\alpha^{(m)}\rangle\downarrow S}| < \mathsf{thresh}$ and $|R_{\langle F,h^{(m-1)},\alpha^{(m-1)}\rangle\downarrow S}| \geq \mathsf{thresh}$. This ensures that nSols computed in line 7 is $|R_{\langle F,h^{(m)},\alpha^{(m)}\rangle\downarrow S}|$. Finally, ApproxMC2Core returns $(2^m, \mathsf{nSols})$, where $2^m$ gives the number of cells into which $R_{F\downarrow S}$ is partitioned by $h^{(m)}$.

An easy consequence of the definition of prefix-slices is that for all $m \in \{1, \ldots |S|-1\}$, we have $R_{\langle F,h^{(m)},\alpha^{(m)}\rangle\downarrow S} \subseteq R_{\langle F,h^{(m-1)},\alpha^{(m-1)}\rangle\downarrow S}$. This linear ordering is exploited by sub-routine LogSATSearch (see Algorithm 3), which uses a galloping search to zoom down to the right value of $m$, $h^{(m)}$ and $\alpha^{(m)}$. LogSATSearch uses an array, BigCell, to remember values of $m$ for which the cell $\alpha^{(m)}$ obtained after partitioning $R_{F\downarrow S}$ with $h^{(m)}$ is large, i.e. $|R_{\langle F,h^{(m)},\alpha^{(m)}\rangle\downarrow S}| \geq \mathsf{thresh}$. As boundary conditions, we set $\mathsf{BigCell}[0]$ to 1 and $\mathsf{BigCell}[|S|-1]$ to 0. These are justified because (i) if $R_{F\downarrow S}$ is partitioned into $2^0$ (i.e. 1) cell, line 3 of Algorithm 1 ensures that the size of the cell (i.e. $|R_{F\downarrow S}|$) is at least thresh, and (ii) line 4 of Algorithm 2 ensures that $|R_{\langle F,h^{|S|-1},\alpha^{|S|-1}\rangle\downarrow S}| < \mathsf{thresh}$. For every other $i$, $\mathsf{BigCell}[i]$ is initialized to $\bot$ (unknown value). Subsequently, we set $\mathsf{BigCell}[i]$ to 1 (0) whenever we find that $|R_{\langle F,h^{(i)},\alpha^{(i)}\rangle\downarrow S}|$ is at least as large as (smaller than) thresh.

**Algorithm 3** LogSATSearch$(F, S, h, \alpha, \text{thresh}, \text{mPrev})$

loIndex $\leftarrow 0$; hiIndex $\leftarrow |S| - 1$; $m \leftarrow$ mPrev;
BigCell$[0] \leftarrow 1$; BigCell$[|S| - 1] \leftarrow 0$;
BigCell$[i] \leftarrow \perp$ for all $i$ other than 0 and $|S| - 1$;
**while** true **do**
    $Y \leftarrow$ BoundedSAT$(F \wedge (h^{(m)}(S) = \alpha^{(m)}), \text{thresh}, S)$;
    **if** $(|Y| \geq \text{thresh})$ **then**
        **if** (BigCell$[m + 1] = 0$) **then return** $m + 1$;
        BigCell$[i] \leftarrow 1$ for all $i \in \{1, \ldots m\}$;
        loIndex $\leftarrow m$;
        **if** $(|m - \text{mPrev}| < 3)$ **then** $m \leftarrow m + 1$;
        **else if** $(2.m < |S|)$ **then** $m \leftarrow 2.m$;
        **else** $m \leftarrow (\text{hiIndex} + m)/2$;
    **else**
        **if** (BigCell$[m - 1] = 1$) **then return** $m$;
        BigCell$[i] \leftarrow 0$ for all $i \in \{m, \ldots |S|\}$;
        hiIndex $\leftarrow m$;
        **if** $(|m - \text{mPrev}| < 3)$ **then** $m \leftarrow m - 1$;
        **else** $m \leftarrow (m + \text{loIndex})/2$;

In the context of probabilistic hashing-based counting algorithms like ApproxMC, it has been observed [123] that the "right" values of $m$, $h_m$ and $\alpha_m$ for partitioning $R_{F\downarrow S}$ are often such that $m$ is closer to 0 than to $|S|$. In addition, repeated invocations of a hashing-based probabilistic counting algorithm with the same input formula $F$ often terminate with similar values of $m$. To optimize LogSATSearch using these observations, we provide mPrev, the value of $m$ found in the last invocation of ApproxMC2Core, as an input to LogSATSearch. This is then used in LogSATSearch to linearly search a small neighborhood of mPrev, viz. when $|m - \mathsf{mPrev}| < 3$, before embarking on a galloping search. Specifically, if LogSATSearch finds that $|R_{\langle F,h^{(m)},\alpha^{(m)}\rangle\downarrow S}| \geq \mathsf{thresh}$ after the linear search, it keeps doubling the value of $m$ until either $|R_{\langle F,h^{(m)},\alpha^{(m)}\rangle\downarrow S}|$ becomes less than thresh, or $m$ overshoots $|S|$. Subsequently, binary search is done by iteratively bisecting the interval between loIndex and hiIndex. This ensures that the search requires $\mathcal{O}(\log_2 m^*)$ calls (instead of $\mathcal{O}(\log_2 |S|)$ calls) to BoundedSAT, where $m^*$ (usually $\ll |S|$) is the value of $m$ when the search stops. Note also that a galloping search inspects much smaller values of $m$ compared to a naive binary search, if $m^* \ll |S|$. Therefore, the formulas fed to the SAT solver have fewer xor clauses (or number of components of $h^{(m)}$) conjoined with $F$ than if a naive binary search was used. This plays an important role in improving the performance of ApproxMC2.

In order to provide the right value of mPrev to LogSATSearch, ApproxMC2 passes the value of nCells returned by one invocation of ApproxMC2Core to the next invocation (line 8 of Algorithm 1), and ApproxMC2Core passes on the relevant information to LogSATSearch (lines 5–6 of Algorithm 2). Thus, successive invocations of ApproxMC2Core in ApproxMC2 are *no longer independent* of each other. Note that the independence of randomly chosen $(h_m, \alpha_m)$ pairs for different values of $m$, and the independence of successive invocations of ApproxMCCore, are features of ApproxMC that are exploited in its analysis [38]. Since these independence no longer hold in ApproxMC2, we must analyze ApproxMC2 afresh.

### 4.1.1 Analysis

**Lemma 2.** *For* $1 \leq i < |S|$, *let* $\mu_i = R_{F\downarrow S}/2^i$. *For every* $\beta > 0$ *and* $0 < \varepsilon < 1$, *we have the following:*

1. $\mathsf{Pr}\left[|R_{\langle F,h^{(i)},\alpha^{(i)}\rangle\downarrow S}| - \mu_i| \geq \frac{\varepsilon}{1+\varepsilon}\mu_i\right] \leq \frac{(1+\varepsilon)^2}{\varepsilon^2\mu_i}$

2. $\mathsf{Pr}\left[|R_{\langle F,h^{(i)},\alpha^{(i)}\rangle\downarrow S}| \leq \beta\mu_i\right] \leq \frac{1}{1+(1-\beta)^2\mu_i}$

*Proof.* For every $y \in \{0,1\}^{|S|}$ and for every $\alpha \in \{0,1\}^i$, define an indicator variable $\gamma_{y,\alpha,i}$ which is 1 iff $h^{(i)}(y) = \alpha$. Let $\Gamma_{\alpha,i} = \sum_{y \in R_{F\downarrow S}} (\gamma_{y,\alpha,i})$, $\mu_{\alpha,i} = \mathsf{E}[\Gamma_{\alpha,i}]$ and $\sigma^2_{\alpha,i} = \mathsf{V}[\Gamma_{\alpha,i}]$. Clearly, $\Gamma_{\alpha,i} = |R_{\langle F,h^{(i)},\alpha\rangle\downarrow S}|$ and $\mu_{\alpha,i} = 2^{-i}|R_{F\downarrow S}|$. Note that $\mu_{\alpha,i}$ is independent of $\alpha$ and equals $\mu_i$, as defined in the statement of the Lemma. From the pairwise independence of $h^{(i)}(y)$ (which, effectively, is a randomly chosen function from $H_{xor}(|S|, i)$), we also have $\sigma^2_{\alpha,i} \leq \mu_{\alpha,i} = \mu_i$. Statements 1 and 2 of the lemma then follow from Chebhyshev inequality and Paley-Zygmund inequality, respectively. □

Let $B$ denote the event that ApproxMC2Core either returns $(\perp, \perp)$ or returns a pair $(2^m, \mathsf{nSols})$ such that $2^m \times \mathsf{nSols}$ does not lie in the interval $\left[\frac{|R_{F\downarrow S}|}{1+\varepsilon}, |R_{F\downarrow S}|(1+\varepsilon)|\right]$. We wish to bound $\mathsf{Pr}[B]$ from above. Towards this end, let $T_i$ denote the event $\left(|R_{\langle F,h^{(i)},\alpha^{(i)}\rangle\downarrow S}| < \mathsf{thresh}\right)$, and let $L_i$ and $U_i$ denote the events $\left(|R_{\langle F,h^{(i)},\alpha^{(i)}\rangle\downarrow S}| < \frac{|R_{F\downarrow S}|}{(1+\varepsilon)2^i}\right)$ and $\left(|R_{\langle F,h^{(i)},\alpha^{(i)}\rangle\downarrow S}| > \frac{|R_{F\downarrow S}|}{2^i}(1+\frac{\varepsilon}{1+\varepsilon})\right)$, respectively. Furthermore, let $m^*$ denote the integer $\lfloor \log_2 |R_{F\downarrow S}| - \log_2\left(4.92\left(1+\frac{1}{\varepsilon}\right)^2\right)\rfloor$.

**Lemma 3.** *The following bounds hold:*

1. $\mathsf{Pr}[T_{m^*-3}] \leq \frac{1}{62.5}$

2. $\mathsf{Pr}[L_{m^*-2}] \leq \frac{1}{20.68}$

3. $\mathsf{Pr}[L_{m^*-1}] \leq \frac{1}{10.84}$

4. $\mathsf{Pr}[L_{m^*} \cup U_{m^*}] \leq \frac{1}{4.92}$

The proofs follow from the definitions of $m^*$, thresh, $\mu_i$, and from applications of Lemma 2 with appropriate values of $\beta$.

**Lemma 4.** $\mathsf{Pr}\,[B] \leq 0.36$

*Proof.* For any event $E$, let $\overline{E}$ denote its complement. For notational convenience, we use $T_0$ and $U_{|S|}$ to denote the empty (or impossible) event, and $T_{|S|}$ and $L_{|S|}$ to denote the universal (or certain) event. It then follows from the definition of $B$ that $\mathsf{Pr}\,[B] \leq \mathsf{Pr}\left[\bigcup_{i\in\{1,\ldots|S|\}}\left(\overline{T_{i-1}} \cap T_i \cap (L_i \cup U_i)\right)\right]$.

We now wish to simplify the upper bound of $\mathsf{Pr}\,[B]$ obtained above. In order to do this, we use three observations, labeled O1, O2 and O3 below, which follow from the definitions of $m^*$, thresh and $\mu_i$, and from the linear ordering of $R_{\langle F,h^{(m)},\alpha^{(m)}\rangle\downarrow S}$.

O1: $\forall i \leq m^* - 3$, $T_i \cap (L_i \cup U_i) = T_i$ and $T_i \subseteq T_{m^*-3}$,

O2: $\mathsf{Pr}[\bigcup_{i\in\{m^*,\ldots|S|\}} \overline{T_{i-1}} \cap T_i \cap (L_i \cup U_i)] \quad \leq \quad \mathsf{Pr}[\overline{T_{m^*-1}} \cap (L_{m^*} \cup U_{m^*})] \leq \mathsf{Pr}[L_{m^*} \cup U_{m^*}]$,

O3: For $i \in \{m^* - 2, m^* - 1\}$, since $\mathsf{thresh} \leq \mu_i(1 + \frac{\varepsilon}{1+\varepsilon})$, we have $T_i \cap U_i = \emptyset$.

Using O1, O2 and O3, we get $\mathsf{Pr}[B] \leq \mathsf{Pr}[T_{m^*-3}] + \mathsf{Pr}[L_{m^*-2}] + \mathsf{Pr}[L_{m^*-1}] + \mathsf{Pr}[L_{m^*} \cup U_{m^*}]$. Using the bounds from Lemma 3, we finally obtain $\mathsf{Pr}\,[B] \leq 0.36$. □

Note that Lemma 4 holds regardless of the order in which the search in LogSATSearch proceeds. Our main theorem now follows from Lemma 4 and from the count $t$ of invocations of ApproxMC2Core in ApproxMC2 (see lines 4-10 of Algorithm 1).

**Theorem 5.** *Suppose* ApproxMC2$(F, S, \varepsilon, \delta)$ *returns* $c$ *after making* $k$ *calls to a* SAT *solver. Then* $\Pr[|R_{F\downarrow S}|/(1+\varepsilon) \leq c \leq (1+\varepsilon)|R_{F\downarrow S}|] \geq 1-\delta$, *and* $k \in \mathcal{O}(\frac{\log(|S|)\log(1/\delta)}{\varepsilon^2})$.

Note that the number of SAT solver calls in ApproxMC [38] lies in $\mathcal{O}(\frac{|S|\log(1/\delta)}{\varepsilon^2})$, which is exponentially worse than the number of calls in ApproxMC2, for the same $\varepsilon$ and $\delta$. Furthermore, if the formula $F$ fed as input to ApproxMC2 is in DNF, the subroutine BoundedSAT can be implemented in PTIME, since satisfiability checking of DNF + XOR is in PTIME. This gives us the following result.

**Theorem 6.** ApproxMC2 *is a fully polynomial randomized approximation scheme (FPRAS) for* #DNF.

Note that this is fundamentally different from FPRAS for $\#DNF$ described in earlier work, viz. [104].

### 4.1.2 Generalizing beyond ApproxMC

So far, we have shown how ApproxMC2 significantly reduces the number of SAT solver calls vis-a-vis ApproxMC, without sacrificing theoretical guarantees, by relaxing independence requirements. Since ApproxMC serves as a paradigmatic representative of several hashing-based counting and probabilistic inference algorithms, the key ideas of ApproxMC2 can be used to improve these other algorithms too.

PAWS [69] is a hashing-based sampling algorithm for high dimensional probability spaces. Similar to ApproxMC, the key idea of PAWS is to find the "right" number and set of constraints that divides the solution space into appropriately sized cells. To do this, PAWS iteratively adds independently chosen constraints, using a linear search. An analysis of the algorithm in [69] shows that this requires $\mathcal{O}(n \log n)$ calls to an NP oracle, where $n$ denotes the size of the support of the input constraint. Our approach based on dependent constraints can be used in PAWS to search out-of-order, and reduce the number of NP oracle calls from $\mathcal{O}(n \log n)$ to $\mathcal{O}(\log n)$, while retaining the same theoretical guarantees.

## 4.2 Evaluation

To evaluate the runtime performance and quality of approximations computed by ApproxMC2, we implemented a prototype in C++ and conducted experiments on a wide variety of publicly available benchmarks. Specifically, we sought answers to the following questions:

1. How does runtime performance and number of SAT invocations of ApproxMC2 compare with that of ApproxMC ?

2. How far are the counts computed by ApproxMC2 from the exact counts?

Our benchmark suite consisted of problems arising from probabilistic inference in grid networks, synthetic grid-structured random interaction Ising models, plan recognition, DQMR networks, bit-blasted versions of SMTLIB benchmarks, ISCAS89 combinational circuits, and program synthesis examples.

We used a high-performance cluster to conduct experiments in parallel. Each node of the cluster had a 12-core 2.83 GHz Intel Xeon processor, with 4GB of main memory, and each experiment was run on a single core. For all our experiments, we used $\varepsilon = 0.8$ and $\delta = 0.2$, unless stated otherwise. To further optimize the running time, we used improved estimates of the iteration count $t$ required in ApproxMC2 by following an analysis similar to that in [37].

### 4.2.1 Performance comparison

Table 4.1 presents the performance of ApproxMC2 vis-a-vis ApproxMC over a subset of our benchmarks*. Column 1 of this table gives the benchmark name, while columns 2

*The complete table is available in Appendix as Table A1

| Benchmark | Vars | Clauses | ApproxMC2 Time | ApproxMC Time | ApproxMC2 SATCalls | ApproxMC SATCalls |
|---|---|---|---|---|---|---|
| tutorial3 | 486193 | 2598178 | 12373.99 | – | 1744 | – |
| case204 | 214 | 580 | 166.2 | – | 1808 | – |
| case205 | 214 | 580 | 300.11 | – | 1793 | – |
| case133 | 211 | 615 | 18502.44 | – | 2043 | – |
| s953a_15_7 | 602 | 1657 | 161.41 | – | 1648 | – |
| llreverse | 63797 | 257657 | 1938.1 | 4482.94 | 1219 | 2801 |
| lltraversal | 39912 | 167842 | 151.33 | 450.57 | 1516 | 4258 |
| karatsuba | 19594 | 82417 | 23553.73 | 28817.79 | 1378 | 13360 |
| enqueueSeqSK | 16466 | 58515 | 192.96 | 2036.09 | 2207 | 23321 |
| progsyn_20 | 15475 | 60994 | 1778.45 | 20557.24 | 2308 | 34815 |
| progsyn_77 | 14535 | 27573 | 88.36 | 1529.34 | 2054 | 24764 |
| sort | 12125 | 49611 | 209.0 | 3610.4 | 1605 | 27731 |
| LoginService2 | 11511 | 41411 | 26.04 | 110.77 | 1533 | 10653 |
| progsyn_17 | 10090 | 27056 | 100.76 | 4874.39 | 1810 | 28407 |
| progsyn_29 | 8866 | 31557 | 87.78 | 3569.25 | 1712 | 28630 |
| LoginService | 8200 | 26689 | 21.77 | 101.15 | 1498 | 12520 |
| doublyLinkedList | 6890 | 26918 | 17.05 | 75.45 | 1615 | 10647 |

Table 4.1 : Performance comparison of ApproxMC2 vis-a-vis ApproxMC. The runtime is reported in seconds and "–" in a column reports timeout after 8 hours.

and 3 list the number of variables and clauses, respectively. Columns 4 and 5 list the runtime (in seconds) of ApproxMC2 and ApproxMC respectively, while columns 6 and 7 list the number of SAT invocations for ApproxMC2 and ApproxMC respectively. We use "–" to denote timeout after 8 hours. Table 4.1 clearly demonstrates that ApproxMC2 outperforms ApproxMC by 1-2 orders of magnitude. Furthermore, ApproxMC2 is able to compute counts for benchmarks that are beyond the scope of ApproxMC. The runtime improvement of ApproxMC2 can be largely attributed to the reduced (by almost an order of magnitude) number of SAT solver calls vis-a-vis ApproxMC.

There are some large benchmarks in our suite for which both ApproxMC and ApproxMC2 timed out; hence, we did not include these in Table 4.1. Importantly, for a significant number of our experiments, whenever ApproxMC or ApproxMC2 timed out, it was because the algorithm could execute *some, but not all* required iterations of ApproxMCCore or ApproxMC2Core, respectively, within the specified time limit. In all such cases, we obtain a model count within the specified tolerance, but with reduced confidence. This suggests that it is possible to extend ApproxMC2 to obtain an anytime algorithm.

### 4.2.2 Approximation Quality

To measure the quality of approximation, we compared the approximate counts returned by ApproxMC2 with the counts computed by an exact model counter, viz. sharpSAT [147]. Figure 4.1 shows the model counts computed by ApproxMC2, and the bounds obtained by scaling the exact counts with the tolerance factor ($\varepsilon = 0.8$) for a small subset of benchmarks. Since sharpSAT can not handle $\Sigma_1^1$ formulas, we ensured that sampling set $S$ for these subset of benchmarks is an independent support. The $y$-axis represents model counts on log-scale while the $x$-axis represents benchmarks ordered in ascending order of model counts. We observe that for *all* the benchmarks, ApproxMC2 computed counts within the tolerance. Furthermore, for each instance, the observed tolerance ($\varepsilon_{obs}$) was calculated as $\max(\frac{\text{AprxCount}}{|R_{F\downarrow S}|} - 1, 1 - \frac{|R_{F\downarrow S}|}{\text{AprxCount}})$, where AprxCount is the estimate computed by ApproxMC2. We observe that the geometric mean of $\varepsilon_{obs}$ across all benchmarks is 0.021 – far better than the theoretical guarantee of 0.8. In comparison, the geometric mean of the observed tolerance obtained from ApproxMC running on the same set of benchmarks is 0.036.

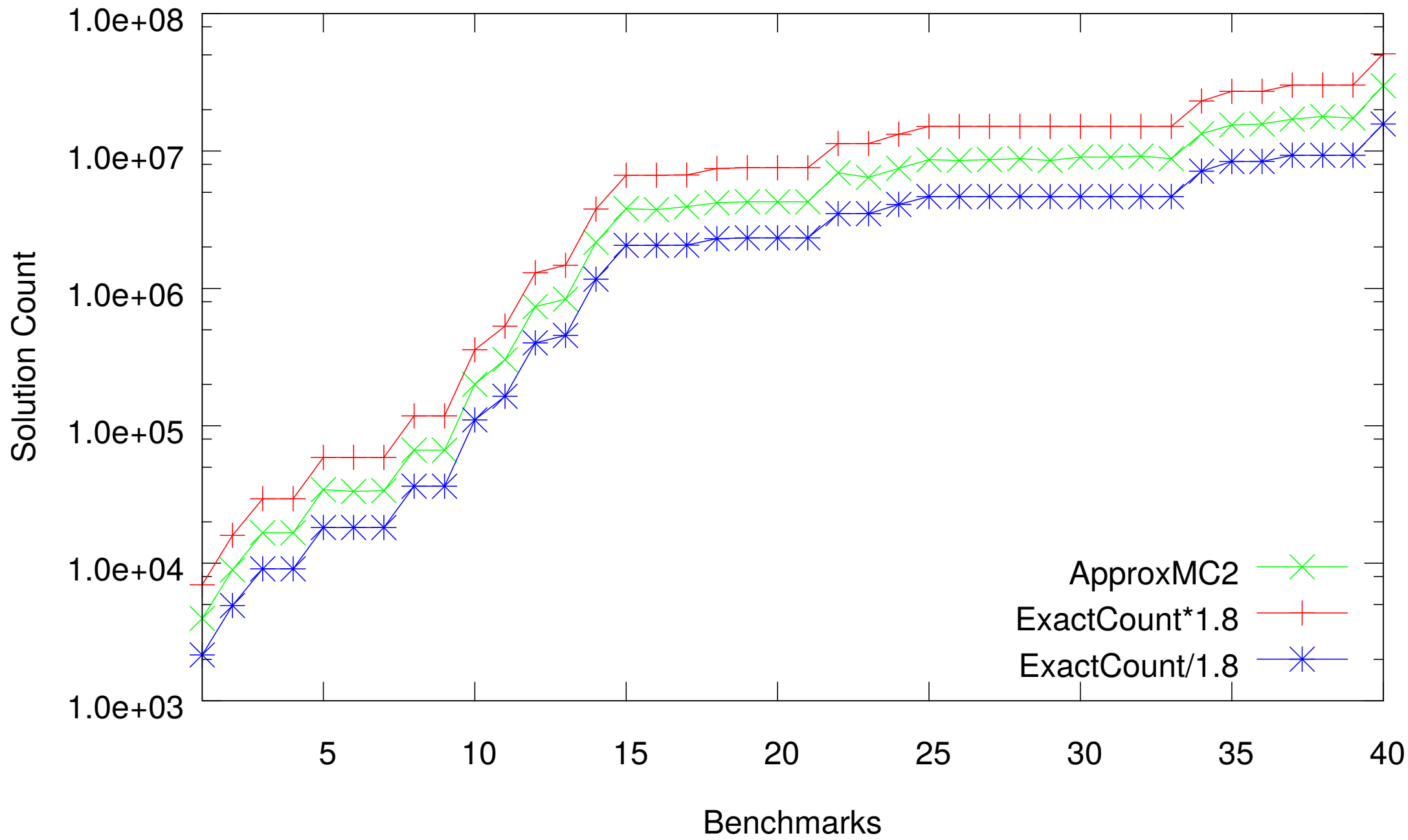

Figure 4.1 : Quality of counts computed by ApproxMC2

**Discussion on Quality**

While the observation that estimates computed by ApproxMC2 are better in practice than theoretical guarantees is certainly more relieving than the alternate scenario. These results, however, do not imply that to obtain results within a tolerance of 0.04, one could supply $\varepsilon$ of 0.8 to ApproxMC2. In fact, if one desired tolerance of 0.04 instead of 0.8, there is no free lunch: there is an associated cost in terms of increase in the number of SAT calls. Note that the number of invocations to SAT oracle is $\mathcal{O}(\frac{1}{\varepsilon^2})$. We discuss possible future directions of research related to dependence of ApproxMC-esque techniques on $\varepsilon$ in Chapter 12.

## 4.3 Chapter Summary

In this chapter, we presented a new approach to hashing-based counting, which allows out-of-order-search with dependent hash functions, dramatically reducing the number of SAT solver calls from linear to logarithmic in the size of the support of interest. This is achieved while retaining strong theoretical guarantees and without increasing the complexity of each SAT solver call. Extensive experiments demonstrate the practical benefits of our approach vis-a-vis state-of-the art techniques. Combining our approach with more efficient hash functions promises to push the scalability horizon of approximate counting further.

# Chapter 5

# Handling Weighted Distributions for Counting

In the previous chapter, we discussed hashing-based paradigm for unweighted variant of constrained counting (UMC), i.e. the weight function assigns weight of 1 to every assignment. However, many applications of constrained counting, including probabilistic inference and network reliability, arising from real world are naturally expressed as weighted counting problem (WMC).

Theoretical investigations into WMC were led by Roth who asserted that probabilistic inference is #P-complete, which by a known connection with WMC [48, 42], implies that WMC is also #P-complete for both CNF and DNF formulas [135]. On the practical side, the earliest efforts at WMC such as CDP [24] were inspired from DPLL-style SAT solvers and consisted of incrementally counting the number of solutions after a partial solution was found. Subsequently, heuristics such as component caching, clause learning, no-good learning and the like improved upon the initial approach and ensuing counters such as Relsat [14], Cachet [138], and sharpSAT [147] have been shown to scale to larger formulas. These approaches were later *manually* adapted for WMC in Cachet [139].Again, alternative approaches based on BDDs and their variants [118, 159] have also been proposed for UMC. Similar to SAT-based approaches, the resulting solvers have also been *manually* adapted for WMC, resulting in solvers like SDD [3].

In this chapter, we discuss two complementary approaches to handle WMC. In the first half of this chapter, we discuss how hashing-based techniques introduced in Chapter 4 for UMC can be lifted to handle WMC. Prior hashing-based approaches to WMC employed computationally expensive MPE oracle. In contrast, we only employ SAT oracle. In this half of the chapter, we do not make any assumption on the weight function. In the second half of this chapter, we discuss a complementary approach wherein we propose an efficient reduction of WMC to UMC if the weight function is expressed using *literal-weighted* representation.

## 5.1 Lifting Hashing-based Techniques to Weighted Counting

Let $W(\cdot)$ be a function that takes as input an assignment $\sigma$ and yields a real number $W(\sigma) \in (0, 1]$ called the *weight* of $\sigma$. Given a set $Y$ of assignments, we use $W(Y)$ to denote $\Sigma_{\sigma \in Y} W(\sigma)$. In this section, we make no assumptions about the nature of the weight function, treating it as a black-box function. Three important quantities derived from the weight function are $w_{max} = \max_{\sigma \in R_F} W(\sigma)$, $w_{min} = \min_{\sigma \in R_F} W(\sigma)$, and the *tilt* $\rho = w_{max}/w_{min}$. Our hashing-based algorithm requires an upper bound on the tilt, denoted $r$, which is provided by the user. As tight a bound as possible is desirable to maximize the efficiency of the algorithms. the tilt concerns weights of only satisfying assignments, our assumption about it being bounded by a small number is reasonable in several practical situations. For example, when solving probabilistic inference with evidence by reduction to weighted model counting [42], every satisfying assignment of the CNF formula corresponds to an assignment of values to variables in the underlying probabilistic graphical model that is consistent with the evidence. Furthermore, the weight of a satisfying assignment is the joint probability of the corresponding assignment of variables in the probabilistic graphical model. A large tilt would therefore mean existence of two assignments that are consistent with the evidence, but one of which is overwhelmingly more likely than the other.

In several real-world problems (see, e.g. Sec 8.3 of [59]), this is considered unlikely given that numerical conditional probability values are often obtained from human experts providing qualitative and rough quantitative data. The algorithms presented in this section require an upper bound for $\rho$ as the input. It is worth noting that although better estimation of upper bounds improve the performance, the algorithms are sound with respect to any upper bound estimate. While an algorithm solution to estimation of upper bound for $\rho$ is beyond the scope of this work, such an estimate can be easily obtained from the designers of probabilistic models. It is easy for designers to estimate upper bound for $\rho$ than accurate estimation of $\mathrm{w}_{\max}$ as the former does not require precise knowledge of probabilities of all the models.

**Algorithm 4** $\mathsf{WeightMC}(F, S, \varepsilon, \delta, r)$

1: counter $\leftarrow 0$; $C \leftarrow$ $\mathsf{emptyList}$; $\mathrm{w}_{\max} \leftarrow 1$;
2: pivot $\leftarrow 2 \times \lceil e^{3/2} \left(1 + \frac{1}{\varepsilon}\right)^2 \rceil$;
3: $t \leftarrow \lceil 35 \log_2(3/\delta) \rceil$;
4: **repeat**
5: $\quad (c, \mathrm{w}_{\max}) \leftarrow \mathsf{WeightMCCore}(F, S, \mathrm{pivot}, r, \mathrm{w}_{\max})$;
6: $\quad$ counter $\leftarrow$ counter $+ 1$;
7: $\quad$ **if** $c \neq \bot$ **then**
8: $\quad\quad \mathsf{AddToList}(C, c \cdot \mathrm{w}_{\max})$;
9: **until** counter $< t$
10: finalCount $\leftarrow \mathsf{FindMedian}(C)$;
11: **return** finalCount;

Our weighted counting algorithm, called WeightMC, is best viewed as an adaptation of the ApproxMC algorithm. The key idea in ApproxMC is to partition the set

**Algorithm 5** WeightMCCore$(F, S, \text{pivot}, r, \text{w}_{\max})$

$(Y, \text{w}_{\max}) \leftarrow$ BoundedWeightSAT$(F, \text{pivot}, r, \text{w}_{\max}, S)$;
**if** $W(Y)/\text{w}_{\max} \leq \text{pivot}$ **then**
  **return** $W(Y)$;
**else**
  $i \leftarrow 0$;
  **repeat**
    $i \leftarrow i + 1$;
    Choose $h$ at random from $H_{xor}(|S|, i, 3)$;
    Choose $\alpha$ at random from $\{0,1\}^i$;
    $(Y, \text{w}_{\max}) \leftarrow$ BoundedWeightSAT$(F \wedge (h(x_1, \ldots x_{|S|}) = \alpha), \text{pivot}, \rho, \text{w}_{\max}, S)$;
  **until** $(0 < W(Y)/\text{w}_{\max} \leq \text{pivot})$ or $i = n$
  **if** $W(Y)/\text{w}_{\max} > \text{pivot}$ **or** $W(Y) = 0$ **then return** $(\perp, \text{w}_{\max})$;
  **elsereturn** $(\frac{W(Y) \cdot 2^{i-1}}{\text{w}_{\max}}, \text{w}_{\max})$;

of satisfying assignments into "cells" containing roughly equal numbers of satisfying assignments, by employing 2-universal hash functions. For weighted counting, the primary modification that needs to be done to ApproxMC is that instead of requiring "cells" to have roughly equal numbers of satisfying assignments, we now require them to have roughly equal weights of satisfying assignments. To ensure that all weights lie in $[0, 1]$, we scale weights by a factor of $\frac{1}{\text{w}_{\max}}$. Unlike earlier works [70, 71], however, we do not require a MPE-oracle to get $\text{w}_{\max}$; instead we estimate $\text{w}_{\max}$ online without incurring any additional performance cost.

WeightMC assumes access to a subroutine called BoundedWeightSAT that takes a

**Algorithm 6** BoundedWeightSAT($F$, pivot, $r$, $\mathrm{w_{max}}$, $S$)

1: $\mathrm{w_{min}} \leftarrow \mathrm{w_{max}}/r; \mathrm{w_{total}} \leftarrow 0; Y = \{\};$
2: **repeat**
3: $\quad y \leftarrow$ SolveSAT($F$);
4: $\quad$ **if** y == UNSAT **then**
5: $\quad\quad$ break;
6: $\quad Y = Y \cup y;$
7: $\quad F =$ AddBlockClause($F, y|_S$);
8: $\quad \mathrm{w_{total}} \leftarrow \mathrm{w_{total}} + W(y);$
9: $\quad \mathrm{w_{min}} \leftarrow min(\mathrm{w_{min}}, W(y));$
10: **until** $\mathrm{w_{total}}/(\mathrm{w_{min}} \cdot r) > pivot;$
11: **return** $(Y, \mathrm{w_{min}} \cdot r);$

CNF formula $F$, a "pivot", an upper bound $r$ of the tilt and an upper bound $w_{max}$ of the maximum weight of an assignment in $R_{F\downarrow S}$. It returns a set of satisfying assignments of $F$ such that the total weight of the returned assignments scaled by $1/w_{max}$ exceeds pivot. It also updates the minimum weight of a satisfying assignment seen so far and returns the same. BoundedWeightSAT accesses a subroutine AddBlockClause that takes as inputs a formula $F$ and a projected assignment $\sigma|_S$, computes a blocking clause for $\sigma|_S$, and returns the formula $F'$ obtained by conjoining $F$ with the blocking clause thus obtained. Finally, the algorithms assume access to an NP-oracle, which in particular can decide SAT. Both algorithms also accept as input a positive real-valued parameter $r$ which is an upper bound on $\rho$.

### 5.1.1 WeightMC Algorithm

The pseudo-code for WeightMC is shown in Algorithm 4. The algorithm takes a CNF formula $F$, sampling set $S$, tolerance $\varepsilon \in (0, 1)$, confidence parameter $\delta \in (0, 1)$, and tilt upper bound $r$, and returns an approximate weighted model count. WeightMC invokes an auxiliary procedure WeightMCCore that computes an approximate weighted model count by randomly partitioning the space of satisfying assignments using hash functions from the family $H_{xor}(|S|, m, 3)$.. WeightMC first computes two parameters: pivot, which quantifies the size of "small" cell and $t$ which determines the number of invocation of WeightMC. The particular choice of expressions to compute these parameters is motivated by technical reasons. After invoking WeightMCCore sufficiently many times, WeightMC returns the median of the non-$\perp$ counts returned by WeightMCCore.

The pseudo-code for subroutine WeightMCCore is presented in 5. WeightMCCore takes in a CNF formula $F$, sampling set $S$, parameter to quantify size of "small" cell pivot, tilt upper bound $r$ and current estimate of upper bound on $\mathrm{w}_{\max}$ and returns an approximate weighted model count and revised estimate of upper bound on $\mathrm{w}_{\max}$. WeightMCCore first handles the easy case of total weighted count of $F$ being less than pivot in lines 1–3. Otherwise, in every iteration of the loop 6–12, WeightMCCore randomly partitions the solution space of $F$ using $H_{xor}(|S|, i, 3)$ until a randomly chosen cell is "small" i.e. the total weighted count of the "cell" is less than pivot. We also refine the estimate for $\mathrm{w}_{\max}$ in every iteration of the loop 6–12 using the minimum weight of solutions seen so far (computed in calls to BoundedWeightSAT) and the *tilt*. In the event a chosen cell is "small", ApproxMCthe weighted count of "cell" is multiplied by total number of cells to obtain the estimated total weighted count. The estimated total weighted count along with refined estimate of $\mathrm{w}_{\max}$ is returned in line 13.

### Implementation Details

In our implementations of WeightMC, BoundedWeightSAT is implemented using CryptoMiniSAT [1], a SAT solver that handles xor clauses efficiently. CryptoMiniSAT uses *blocking clauses* to prevent already generated witnesses from being generated again. Since we are only interested in determining —$R_{F\downarrow S}$—, blocking clauses can be restricted to only variables in the set $S$. We implemented this optimization in CryptoMiniSAT, leading to significant improvements in performance. We used "random_device" implemented in C++11 as source of pseudo-random numbers to make random choices in WeightMC.

### 5.1.2 Analysis of WeightMC

In this section we denote the quantity $\log_2 \mathcal{W}(R_F) - \log_2 pivot + 1$ by $m$. For simplicity of exposition, we assume henceforth that $m$ is an integer. A more careful analysis removes this restriction with only a constant factor scaling of the probabilities.

**Lemma 7.** *Let algorithm* WeightMCCore*, when invoked from* WeightMC*, return* $c$ *with* $i$ *being the final value of the loop counter in* WeightMCCore*. Let* $p_i$ *be short hand for 1-* $\mathsf{Pr}\left[(1+\varepsilon)^{-1} \cdot W(F \downarrow S) \leq c \ \leq (1+\varepsilon) \cdot W(F \downarrow S)\right]$*. Then* $p_i \leq \frac{e^{-3/2}}{2^{m-i}}$

*Proof.* Referring to the pseudocode of WeightMCCore, the lemma is trivially satisfied if $W(F \downarrow S) \leq pivot$. Therefore, the only non-trivial case to consider is when $W(F \downarrow S) > pivot$ and WeightMCCore returns from line 13. In this case, the count returned is $2^i \cdot \mathcal{W}(R_{F,h,\alpha})$, where $\alpha, i$ and $h$ denote (with abuse of notation) the values of the corresponding variables and hash functions in the final iteration of the repeat-until loop in lines 6–11 of the pseudocode. From the pseudocode of WeightMCCore, we know that $pivot = \lceil e^{3/2}(1+1/\varepsilon)^2 \rceil$. The lemma is now proved by showing that for every $i$ in $\{0, \ldots m\}$, $h \in H(n,i,3)$, and $\alpha \in \{0,1\}^i$, we have $\mathsf{Pr}\left[(1+\varepsilon)^{-1} \cdot W(F\downarrow S) \le 2^i \mathcal{W}\left(R_{F,h,\alpha}\right) \le (1+\varepsilon) \cdot W(F \downarrow S)\right] \ge 1 - \frac{e^{-3/2}}{2^{m-i}}$.

For every $y \in \{0,1\}^n$ and $\alpha \in \{0,1\}^i$, define an indicator variable $\gamma_{y,\alpha}$ as follows: $\gamma_{y,\alpha} = \mathcal{W}(y)$ if $h(y) = \alpha$, and $\gamma_{y,\alpha} = 0$ otherwise. Let us fix $\alpha$ and $y$ and choose $h$ uniformly at random from $H(n,i,3)$. The random choice of $h$ induces a probability distribution on $\gamma_{y,\alpha}$ such that $\mathsf{Pr}\left[\gamma_{y,\alpha} = \mathcal{W}(y)\right] = \mathsf{Pr}\left[h(y) = \alpha\right] = 2^{-i}$, and $\mathsf{E}\left[\gamma_{y,\alpha}\right] = \mathcal{W}(y)\,\mathsf{Pr}\left[\gamma_{y,\alpha} = \mathcal{W}(y)\right] = 2^{-i}\mathcal{W}(y)$. In addition, the 3-wise independence of hash functions chosen from $H(n,i,3)$ implies that for every distinct $y_a, y_b, y_c \in R_F$, the random variables $\gamma_{y_a,\alpha}$, $\gamma_{y_b,\alpha}$ and $\gamma_{y_c,\alpha}$ are 3-wise independent.

Let $\Gamma_\alpha = \sum_{y \in R_F} \gamma_{y,\alpha}$ and $\mu_\alpha = \mathsf{E}\left[\Gamma_\alpha\right]$. Clearly, $\Gamma_\alpha = \mathcal{W}\left(R_{F,h,\alpha}\right)$ and $\mu_\alpha = \sum_{y\in R_F} \mathsf{E}\left[\gamma_{y,\alpha}\right] = 2^{-i} W(F\downarrow S)$. Therefore, using Chebyshev's Inequality, we have $\mathsf{Pr}\left[W(F\downarrow S)\left(1-\frac{\varepsilon}{1+\varepsilon}\right) \le 2^i \mathcal{W}\left(R_{F,h,\alpha}\right) \le (1+\frac{\varepsilon}{1+\varepsilon}) W(F\downarrow S)\right] \ge 1 - \frac{e^{-3/2}}{2^{m-i}}$. Simplifying and noting that $\frac{\varepsilon}{1+\varepsilon} < \varepsilon$ for all $\varepsilon > 0$, we obtain $\mathsf{Pr}\left[(1+\varepsilon)^{-1} \cdot W(F\downarrow S) \le 2^i \mathcal{W}\left(R_{F,h,\alpha}\right) \le (1+\varepsilon) \cdot W(F\downarrow S)\right] \ge 1 - \frac{e^{-3/2}}{2^{m-i}}$. □

**Lemma 8.** *Let an invocation of* WeightMCCore *from* WeightMC *return $c$. Then* $\mathsf{Pr}\left[c \ne \bot \wedge (1+\varepsilon)^{-1} \cdot W(F\downarrow S) \le c \cdot \mathrm{w}_{\max} \le (1+\varepsilon) \cdot W(F\downarrow S)\right] \ge 0.6$.

*Proof.* It is easy to see that the required probability is at least as large as $\mathsf{Pr}\left[c \ne \bot \wedge i \le m \wedge (1+\varepsilon)^{-1} W(F\downarrow S) \le c \cdot \mathrm{w}_{\max} \le (1+\varepsilon)\cdot W(F\downarrow S)\right]$. Dividing by $\mathrm{w}_{\max}$ and applying Lemma 7 this probability is $\ge 1 - p_m - p_{m-1} - p_{m-2} \ge 1 - e^{-3/2} - \frac{e^{-3/2}}{2} - \frac{e^{-3/2}}{4} \ge 0.6$. □

We now turn to proving that the confidence can be raised to at least $1-\delta$ for $\delta \in (0,1]$ by invoking WeightMCCore $\mathcal{O}(\log_2(1/\delta))$ times, and by using the median of the non-$\bot$ counts thus returned. For convenience of exposition, we use $\eta(t,m,p)$ in the following discussion to denote the probability of at least $m$ heads in $t$ independent tosses of a biased coin with $\mathsf{Pr}\left[heads\right] = p$. Clearly, $\eta(t,m,p) = \sum_{k=m}^{t} \binom{t}{k} p^k (1-p)^{t-k}$.

**Theorem 9.** *Given a propositional formula $F$ and parameters $\varepsilon$ $(0 < \varepsilon \leq 1)$ and $\delta$ $(0 < \delta \leq 1)$, suppose* WeightMC$(F, \varepsilon, \delta, X, r)$ *returns $c$. Then* $\mathsf{Pr}\left[(1+\varepsilon)^{-1} W(F \downarrow S)) \leq c \leq (1+\varepsilon) \cdot W(F \downarrow S))\right] \geq 1-\delta$.

*Proof.* Throughout this proof, we assume that WeightMCCore is invoked $t$ times from WeightMC, where $t = \lceil 35 \log_2(3/\delta) \rceil$ (see pseudocode for ComputeIterCount in Section 6.3). Referring to the pseudocode of WeightMC, the final count returned is the median of the non-$\perp$ counts obtained from the $t$ invocations of WeightMCCore. Let $Err$ denote the event that the median is not in $[(1+\varepsilon)^{-1} \cdot W(F \downarrow S), (1+\varepsilon) \cdot W(F \downarrow S)]$. Let "$\#non\perp = q$" denote the event that $q$ (out of $t$) values returned by WeightMCCore are non-$\perp$. Then, $\mathsf{Pr}\left[Err\right] = \sum_{q=0}^{t} \mathsf{Pr}\left[Err \mid \#non\perp = q\right] \cdot \mathsf{Pr}\left[\#non\perp = q\right]$.

In order to obtain $\mathsf{Pr}\left[Err \mid \#non\perp = q\right]$, we define a 0-1 random variable $Z_i$, for $1 \leq i \leq t$, as follows. If the $i^{th}$ invocation of WeightMCCore returns $c$, and if $c$ is either $\perp$ or a non-$\perp$ value that does not lie in the interval $[(1+\varepsilon)^{-1} \cdot W(F \downarrow S), (1+\varepsilon) \cdot W(F \downarrow S)]$, we set $Z_i$ to 1; otherwise, we set it to 0. From Lemma 8, $\mathsf{Pr}\left[Z_i = 1\right] = p < 0.4$. If $Z$ denotes $\sum_{i=1}^{t} Z_i$, a necessary (but not sufficient) condition for event $Err$ to occur, given that $q$ non-$\perp$s were returned by WeightMCCore, is $Z \geq (t - q + \lceil q/2 \rceil)$. To see why this is so, note that $t-q$ invocations of WeightMCCore must return $\perp$. In addition, at least $\lceil q/2 \rceil$ of the remaining $q$ invocations must return values outside the desired interval. To simplify the exposition, let $q$ be an even integer. A more careful analysis removes this restriction and results in an additional constant scaling factor for $\mathsf{Pr}\left[Err\right]$. With our simplifying assumption, $\mathsf{Pr}\left[Err \mid \#non\perp = q\right] \leq \mathsf{Pr}[Z \geq (t - q + q/2)] = \eta(t, t-q/2, p)$. Since $\eta(t,m,p)$ is a decreasing function of $m$ and since $q/2 \leq t - q/2 \leq t$, we have $\mathsf{Pr}\left[Err \mid \#non\perp = q\right] \leq \eta(t, t/2, p)$. If $p < 1/2$, it is easy to verify that $\eta(t, t/2, p)$ is an increasing function of $p$. In our case, $p < 0.4$; hence, $\Pr[Err \mid \#non\bot = q] \leq \eta(t, t/2, 0.4)$.

It follows from the above that $\Pr[Err] = \sum_{q=0}^{t} \Pr[Err \mid \#non\bot = q] \cdot \Pr[\#non\bot = q] \leq \eta(t, t/2, 0.4) \cdot \sum_{q=0}^{t} \Pr[\#non\bot = q] = \eta(t, t/2, 0.4)$. Since $\binom{t}{t/2} \geq \binom{t}{k}$ for all $t/2 \leq k \leq t$, and since $\binom{t}{t/2} \leq 2^t$, we have $\eta(t, t/2, 0.4) = \sum_{k=t/2}^{t} \binom{t}{k} (0.4)^k (0.6)^{t-k} \leq \binom{t}{t/2} \sum_{k=t/2}^{t} (0.4)^k (0.6)^{t-k} \leq 2^t \sum_{k=t/2}^{t} (0.6)^t (0.4/0.6)^k \leq 2^t \cdot 3 \cdot (0.6 \times 0.4)^{t/2} \leq 3 \cdot (0.98)^t$. Since $t = \lceil 35 \log_2(3/\delta) \rceil$, it follows that $\Pr[Err] \leq \delta$. □

**Theorem 10.** *Given an oracle for* SAT*,* WeightMC$(F, \varepsilon, \delta, S, r)$ *runs in time polynomial in* $\log_2(1/\delta), r, |F|$ *and* $1/\varepsilon$ *relative to the oracle.*

*Proof.* Referring to the pseudocode for WeightMC, lines 1–3 take $\mathcal{O}(1)$ time. The repeat-until loop in lines 4–9 is repeated $t = \lceil 35 \log_2(3/\delta) \rceil$ times. The time taken for each iteration is dominated by the time taken by WeightMCCore. Finally, computing the median in line 10 takes time linear in $t$. The proof is therefore completed by showing that WeightMCCore takes time polynomial in $|F|, r$ and $1/\varepsilon$ relative to the SAT oracle.

Referring to the pseudocode for WeightMCCore, we find that BoundedWeightSAT is called $\mathcal{O}(|F|)$ times. Observe that when the loop in BoundedWeightSAT terminates, $\mathrm{w}_{\min}$ is such that each $y \in R_F$ whose weight was added to $\mathrm{w}_{\mathrm{total}}$ has weight at least $\mathrm{w}_{\min}$. Thus since the loop terminates when $\mathrm{w}_{\mathrm{total}}/\mathrm{w}_{\min} > r \cdot pivot$, it can have iterated at most $(r \cdot pivot) + 1$ times. Therefore each call to BoundedWeightSAT makes at most $(r \cdot pivot) + 1$ calls to the SAT oracle, and takes time polynomial in $|F|$, $r$, and $pivot$ relative to the oracle. Since $pivot$ is in $\mathcal{O}(1/\varepsilon^2)$, the number of calls to the SAT oracle, and the total time taken by all calls to BoundedWeightSAT in each invocation of WeightMCCore is polynomial in $|F|, r$ and $1/\varepsilon$ relative to the oracle. The random choices in lines 8 and 9 of WeightMCCore can be implemented in time polynomial in $n$ (hence, in $|F|$) if we have access to a source of random bits. Constructing $F \wedge h(z_1, \ldots z_n) = \alpha$ in line 10 can also be done in time polynomial in $|F|$. □

### 5.1.3 Experimental Results

To evaluate the performance of WeightMC, we built prototype implementations and conducted an extensive set of experiments. The suite of benchmarks was made up of problems arising from various practical domains as well as problems of theoretical interest. Specifically, we used bit-level unweighted versions of constraints arising from grid networks, plan recognition, DQMR networks, bounded model checking of circuits, bit-blasted versions of SMT-LIB [2] benchmarks, and ISCAS89 [26] circuits with parity conditions on randomly chosen subsets of outputs and next-state variables [139, 102]. While our algorithms are agnostic to the weight oracle, other tools that we used for comparison require the weight of an assignment to be the product of the weights of its literals. Consequently, to create weighted problems with tilt at most some bound $r$, we randomly selected $m = \max(15, n/100)$ of the variables and assigned them the weight $w$ such that $(w/(1-w))^m = r$, their negations the weight $1 - w$, and all other literals the weight 1. To illustrate agnostic nature of our algorithms w.r.t. to weight oracle, we also evaluated WeightMC with non-factored representation of the weights. In our implementation of weight oracle without factored representation, we first randomly chose a range of minimum ($w_{min}$) and maximum ($w_{max}$) possible weights and then randomly selected 20 variables of the input formula. We now compute weight of an assignment as $w_{min}+(w_{max} - w_{min} * \frac{x}{2^20})$, where $x$ is the integer value of binary representation of assignment to our randomly selected 20 variables. Unless mentioned otherwise, our experiments for WeightMC used $r = 5$, $\epsilon = 0.8$, and $\delta = 0.2$.

To facilitate performing multiple experiments in parallel, we used a high performance cluster, each experiment running on its own core. Each node of the cluster had two quad-core Intel Xeon processors with 4GB of main memory. We used 2500 seconds as the timeout of each invocation of BoundedWeightSAT and 20 hours as the overall timeout for WeightMC. If an invocation of BoundedWeightSAT timed out in line 10 (WeightMC), we repeated the execution of the corresponding loops without incrementing the variable $i$ (in both algorithms). With this setup, WeightMC was able to successfully return weighted counts and generate weighted random instances for formulas with close to 64,000 variables.

We compared the performance of WeightMC with the SDD Package [3], a state-of-the-art tool which can perform exact weighted model counting by compiling CNF formulae into Sentential Decision Diagrams [44]. We also tried to compare our tools against Cachet, WISH and PAWS but the current versions of the tools made available to us were broken and we are yet, at the time of submission, to receive working tools. If we get access to working tools in future, we will update our full version with the corresponding comparisons. Our results are shown in Table 5.1, where column 1 lists the benchmarks and columns 2 and 3 give the number of variables and clauses for each benchmark. Column 4 lists the time taken by WeightMC, while column 5 lists the time taken by SDD. "T" and "mem" indicate that an experiment exceeded our imposed 20-hour and 4GB-memory limits, respectively. While SDD was generally superior for small problems, WeightMC was significantly faster for all benchmarks with more than 1,000 variables.

To evaluate the quality of the approximate counts returned by WeightMC, we computed exact weighted model counts using the SDD tool for our benchmarks. Since SDD could not compute counts for all the benchmark, we list results for the subset of benchmarks for which SDD returned an answer. Figure 5.1 shows the counts returned by WeightMC, and the exact counts from SDD scaled up and down by $(1+\varepsilon)$. The weighted model counts are represented on the y-axis, while the x-axis represents benchmarks arranged in increasing order of counts. We observe, for all our experiments, that the weighted counts returned by WeightMC lie within the tolerance of the exact counts. Over all of the benchmarks, the $L_1$ norm of the relative error was 0.036, demonstrating that in practice WeightMC is substantially more accurate than the theoretical guarantees provided by Theorem 54.

Table 5.1 : WeightMC and SDD runtimes in seconds.

| **Benchmark** | **vars** | **#clas** | **Weight-MC** | **SDD** |
|---|---:|---:|---:|---:|
| or-50 | 100 | 266 | 15 | 0.38 |
| or-70 | 140 | 374 | 771 | 0.83 |
| s526_3_2 | 365 | 943 | 62 | 29.54 |
| s526a_3_2 | 366 | 944 | 81 | 12.16 |
| s953a_3_2 | 515 | 1297 | 11978 | 355.7 |
| s1238a_7_4 | 704 | 1926 | 3519 | mem |
| s1196a_15_7 | 777 | 2165 | 3087 | 2275 |
| Squaring9 | 1434 | 5028 | 34942 | mem |
| Squaring7 | 1628 | 5837 | 39367 | mem |
| ProcessBean | 4768 | 14458 | 53746 | mem |
| LoginService2 | 11511 | 41411 | 322 | mem |
| Sort | 12125 | 49611 | 19303 | T |
| EnqueueSeq | 16466 | 58515 | 8620 | mem |
| Karatsuba | 19594 | 82417 | 4962 | mem |
| TreeMax | 24859 | 103762 | 34 | T |
| LLReverse | 63797 | 257657 | 1496 | mem |

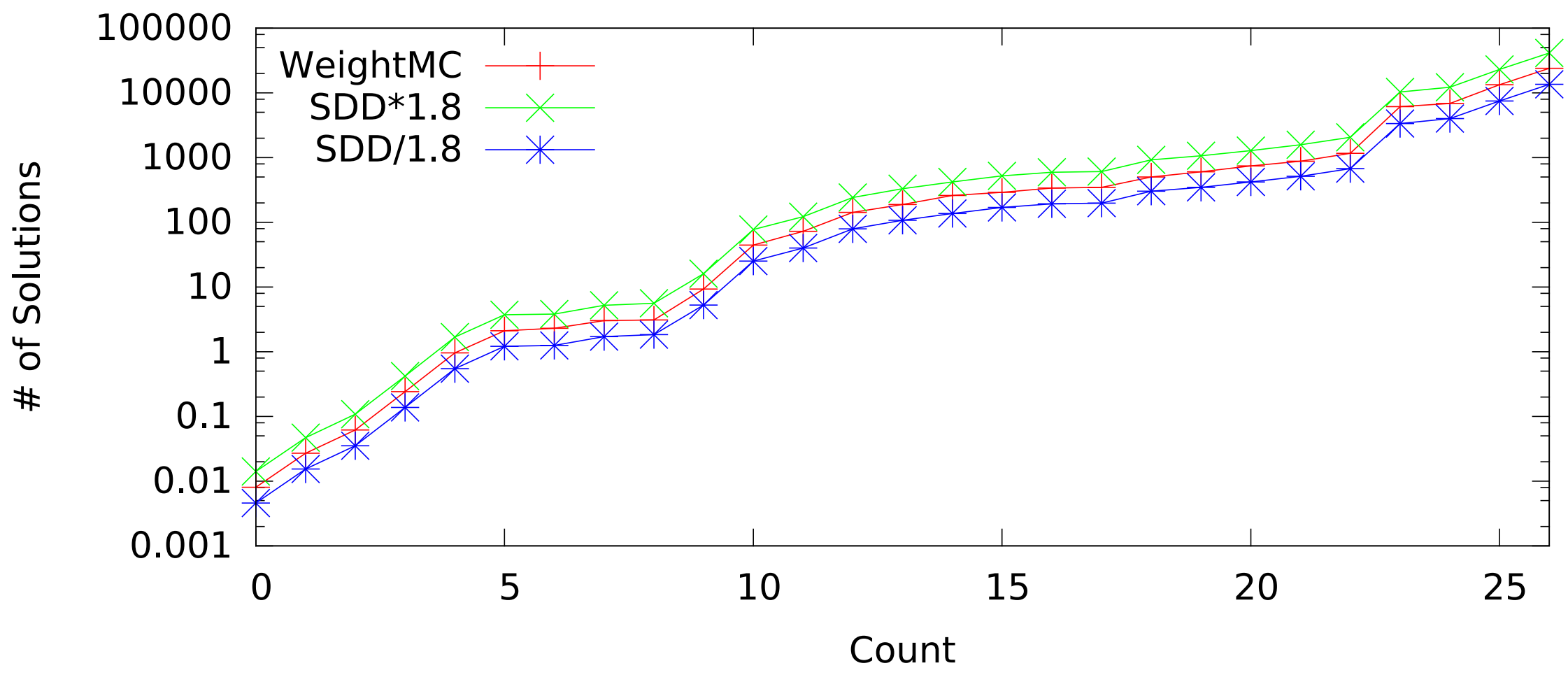

Figure 5.1 : Quality of counts computed by WeightMC. The benchmarks are arranged in increasing order of weighted model counts.

In another experiment, we studied the effect of different values of the tilt bound $r$ on the runtime of WeightMC. Runtime as a function $r$ is shown for several benchmarks in Figure 5.2, where times have been normalized so that at the lowest tilt $(r = 1)$ each benchmark took one time unit. Each runtime is an average over five runs on the same benchmark. The theoretical linear dependence on the tilt shown in Theorem 10 can be seen to roughly occur in practice.

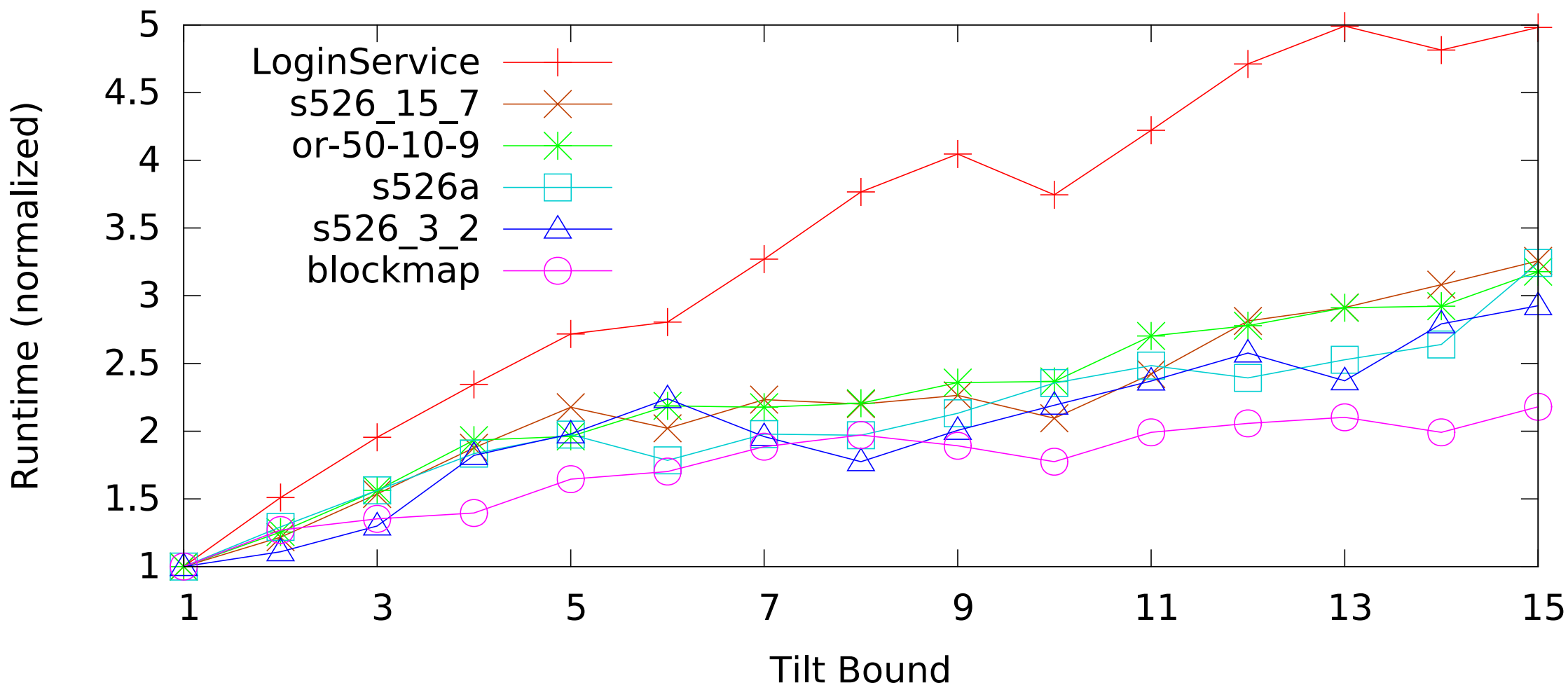

Figure 5.2 : Runtime of WeightMC as a function of *tilt bound*.

## 5.2 Handling Literal-Weighted Representation

Many applications, including probabilistic inference, of WMC arising from real-world can be expressed by a *literal-weighted* representation, in which the weight of an assignment is the product of weights of its literals [42]. We use this representation throughout this chapter, and use *literal-weighted* WMC to denote the corresponding WMC problem. Note that literal-weighted WMC problems for both CNF and DNF formulas arise in real-life applications; e.g., DNF formulas are used in problems arising from probabilistic databases [51], while CNF is the de-facto form of representation for probabilistic-inference problems [42].

Recent approaches to WMC have focused on adapting UMC techniques to work in the weighted setting [139, 44, 34]. Such adaption requires intimate understanding of the implementation details of the UMC techniques, and on-going maintenance, since some of these techniques evolve over time. In this chapter, we flip this approach and present an efficient reduction of literal-weighted WMC to UMC. The reduction preserves the normal form of the input formula, i.e. it provides the UMC formula in the same normal form as the input WMC formula. Therefore, an important contribution of our reduction is to provide a WMC-to-UMC module that allows any UMC solver, viewed as a *black box*, to be converted to a WMC solver. This enables the automatic leveraging of progress in UMC solving to progress in WMC solving.

We have implemented our WMC-to-UMC module on top of state-of-the-art exact unweighted model counters to obtain exact weighted model counters for CNF formulas with literal-weighted representation. Experiments on a suite of benchmarks indicate that the resulting counters scale to significantly larger problem instances than what can be handled by a state-of-the-art exact weighted model counter [44]. Our results suggest that we can leverage powerful techniques developed for SAT and related domains in recent years to handle probabilistic inference queries for graphical models encoded as WMC instances. Furthermore, we demonstrate that our techniques can be extended to more general representations where weights are associated with constraints instead of individual literals.

In this section, we adopt a different approach and propose to solve WMC by reducing it to UMC. Our key contribution lies in showing that this reduction is efficient and effective, thereby making it possible to solve weighted model counting problems using *any* unweighted model counter as a black-box. Our reduction makes use of chain formulas to encode each weighted variable. Interestingly, these formulas can be viewed as adaptations of switching circuits proposed by [157] in the context of stochastic switching networks. Chain formulas are also reminiscent of the log-encoding approach of encoding variables with bounded domains in the CSP literature [66, 77, 79, 94, 154]. Indeed, a chain formula encoding a variable with weight $k/2^m$ is logically equivalent to the constraint $(X \geq 2^m - k)$, where $X$ is an unsigned integer represented using $m$ boolean variables, as described in [79, 154]. The use of log-encoding for exact counting weighted models of Boolean formulas is novel, to the best of our knowledge.

The remainder of the section is organized as follows. We present our the polynomial-time reduction from WMC to UMC in Section 5.2.1. Using our reduction, we have implemented a literal-weighted exact model counter module called sharpWeightSAT. In Section 5.2.6, we present results of our experiments using sharpWeightSAT on top of state-of-the-art UMC solvers, and compare them with SDD – a state-of-the-art exact weighted model counter. We then demonstrate, in Section 5.3, that our reduction can be extended to more general representation of associating weights with constraints.

### 5.2.1 From Literal-weighted WMC to UMC

In this section, we first show how chain formulas can be used to represent normal weights of variables. We then present two polynomial-time reductions from WMC to UMC using chain formulas. These reductions, though related, are motivated by the need to preserve different normal forms (CNF and DNF) of the input formula. Finally, we discuss the optimality of our reductions with respect to number of variables in the unweighted formula.

### 5.2.2 Representing Weights using Chain Formulas

The central idea of our reduction is the use of chain formulas to represent weights. Let $m > 0$ be a natural number, and $k < 2^m$ be a positive odd number. Let $c_1 c_2 \cdots c_m$ be the $m$-bit binary representation of $k$, where $c_m$ is the least significant bit. We then construct a chain formula $\varphi_{k,m}(\cdot)$ on $m$ variables $a_1, \ldots a_m$ as follows. For every $j$ in $\{1, \ldots m-1\}$, let $C_j$ be the connector "$\vee$" if $c_j = 1$, and the connector "$\wedge$" if $c_j = 0$. Define

$$\varphi_{k,m}(a_1, \cdots a_m) = a_1\, C_1\, (a_2\, C_2(\cdots(a_{m-1}\, C_{m-1}\, a_m)\cdots))$$

For example, consider $k = 5$ and $m = 4$. The binary representation of 5 using 4 bits is 0101. Therefore, $\varphi_{5,4}(a_1, a_2, a_3, a_4) = a_1 \wedge (a_2 \vee (a_3 \wedge a_4))$. We first show in Lemma 11 that $\varphi_{k,m}(\cdot)$ has exactly $k$ satisfying assignments. Next, as a simple application of the distributive laws of Boolean algebra, Lemma 12 shows that every chain formula can be efficiently represented in both CNF and DNF.

**Lemma 11.** *Let $m > 0$ be a natural number, $k < 2^m$ , and $\varphi_{k,m}$ as defined above. Then $|\varphi_{k,m}|$ is linear in $m$ and $\varphi_{k,m}$ has exactly $k$ satisfying assignments.*

*Proof.* By construction, $\varphi_{k,m}(a_1, \cdots a_m)$ is of size linear in $m$. To prove that $\varphi_{k,m}(a_1, \cdots a_m)$ has exactly $k$ satisfying assignments, we use induction on $m$. The base case $(m = 1)$ is trivial. For $m \geq 1$, let $c_2 \cdots c_m$ represent the number $k'$ in binary, and assume that $a_2C_2(\cdots(a_{m-1}C_{m-1}a_m)\cdots)$ has exactly $k'$ satisfying assignments. If $c_1$ is 0, then on one hand $k = k'$, and on the other hand $C_1$ is the connector "$\wedge$". Therefore, $\varphi_{k,m}(a_1, \cdots a_m)$ is $a_1 \wedge (a_2C_2(\cdots(a_{m-1}C_{m-1}a_m)\cdots))$, which has $k' = k$ satisfying assignments. Otherwise, if $c_1$ is 1, then on one hand $k = 2^{m-1} + k'$, and on the other hand $C_1$ is the connector "$\vee$". Therefore, $\varphi_{k,m}(a_1, \cdots a_m)$ is $a_1 \vee (a_2C_2(\cdots(a_{m-1}C_{m-1}a_m)\cdots))$, which has $2^{m-1} + k' = k$ satisfying assignments. This completes the induction. □

Recall from Section 2.6 that $N_F$ denotes the set of indices of normal-weighted variables in $F$. For $i$ in $N_F$, let $W(x_i^1) = k_i/2^{m_i}$, where $k_i$ is a positive odd number less than $2^{m_i}$. Additionally, let $\{x_{i,1}, \ldots x_{i,m_i}\}$ be a set of $m_i$ "fresh" variables (i.e. variables that were not used before) for each $i$ in $N_F$. We call the chain formula $\varphi_{k_i,m_i}(x_{i,1} \cdots x_{i,m_i})$, the *representative formula of* $x_i$. For notational clarity, we simply write $\varphi_{k_i,m_i}$ when the arguments of the representative formula are clear from the context.

**Lemma 12.** *Every chain formula $\psi$ on $n$ variables is equivalent to a CNF (resp., $\mathsf{DNF}$) formula $\psi^{\mathsf{CNF}}$ (resp., $\psi^{\mathsf{DNF}}$) having at most $n$ clauses. In addition, $|\psi^{\mathsf{CNF}}|$ (resp., $|\psi^{\mathsf{DNF}}|$) is in $O(n^2)$.*

*Proof.* We first prove the CNF case and then obtain a similar proof for $\mathsf{DNF}$. The proof is by induction on $n$. The base case ($n = 1$) is trivial. To prove the induction step, we consider two cases. First, assume that $\psi$ is $l_i \vee \phi$, where $\phi$ is a chain formula on $n-1$ variables. By the induction hypothesis, $\phi$ is equivalent to a CNF formula $\phi^{\mathsf{CNF}}$. Let $\phi^{\mathsf{CNF}}$ be given by $(\phi_1 \wedge \cdots \phi_{n-1})$, where each $\phi_j$ is a disjunction of literals. Then, $\psi$ is equivalent to $l_i \vee (\phi_1 \wedge \cdots \phi_{n-1})$. Distributing "$\vee$" over "$\wedge$", we get the equivalent formula $(l_i \vee \phi_1) \wedge \cdots (l_i \vee \phi_{n-1})$. Since each $\phi_j$ is a disjunction of literals, so is $(l_i \vee \phi_j)$. Therefore, $(l_i \vee \phi_1) \wedge \cdots (l_i \vee \phi_{n-1})$ is the desired CNF formula $\psi^{\mathsf{CNF}}$. Next, assume that $\psi$ is $l_i \wedge \phi$, where $\phi$ is a chain formula on $n-1$ variables. By the induction hypothesis, $\phi$ is equivalent to a CNF formula $\phi^{\mathsf{CNF}}$. It follows immediately that $l_i \wedge \phi^{\mathsf{CNF}}$ is the desired CNF formula $\psi^{\mathsf{CNF}}$. To see why $|\psi^{\mathsf{CNF}}|$ is in $O(n^2)$, recall that a variable can appear only once (in negated or un-negated form) in a chain formula. Therefore, $\psi^{\mathsf{CNF}}$ as constructed above has at most $n$ clauses, each with at most $n$ literals.

The proof for $\mathsf{DNF}$ is very similar to the above one. Again, the proof is by induction on $n$. The base case ($n = 1$) is trivial. To prove the induction step, we consider two cases. First, assume that $\psi$ is $l_i \wedge \phi$, where $\phi$ is a chain formula on $n-1$ variables. By the induction hypothesis, $\phi$ is equivalent to a DNF formula $\phi^{\mathsf{DNF}}$. Let $\phi^{\mathsf{DNF}}$ be given by $(\phi_1 \vee \cdots \phi_{n-1})$, where each $\phi_j$ is a conjunction of literals. Then, $\psi$ is equivalent to $l_i \wedge (\phi_1 \vee \cdots \phi_{n-1})$. Distributing "$\wedge$" over "$\vee$", we get the equivalent formula $(l_i \wedge \phi_1) \vee \cdots (l_i \wedge \phi_{n-1})$. Since each $\phi_j$ is a conjunction of literals, so is $(l_i \wedge \phi_j)$. Therefore, $(l_i \wedge \phi_1) \vee \cdots (l_i \wedge \phi_{n-1})$ is the desired DNF formula $\psi^{\mathsf{DNF}}$. Next, assume that $\psi$ is $l_i \vee \phi$, where $\phi$ is a chain formula on $n-1$ variables. By the induction hypothesis, $\phi$ is equivalent to a DNF formula $\phi^{\mathsf{DNF}}$. It follows immediately that $l_i \wedge \phi^{\mathsf{DNF}}$ is the desired DNF formula $\psi^{\mathsf{DNF}}$. Again, as every variable can appear only once (in negated or un-negated form) in a chain formula, $\psi^{\mathsf{DNF}}$ as constructed above has at most $n$ clauses, each with at most $n$ literals. Therefore $|\psi^{\mathsf{DNF}}|$ is in $O(n^2)$. □

### 5.2.3 Polynomial-time Reductions

We now present two reductions from literal-weighted WMC to UMC. Since weighted model count is a real number in general, while unweighted model count is a natural number, any reduction from WMC to UMC must use a normalization constant. Given that all literal weights are of the form $k_i/2^{m_i}$, a natural choice for the normalization constant is $C_F = \prod_{i \in N_F} 2^{-m_i}$. Theorem 13a gives a transformation of an instance $(F, W(\cdot))$ of literal-weighted WMC to an unweighted Boolean formula $\widehat{F}$ such that $W(F) = C_F \cdot |R_{\widehat{F}}|$. This reduction is motivated by the need to preserve CNF form of the input formula. We may also allow an additive correction term when doing the reduction. Theorem 13b provides a transformation of $(F, W(\cdot))$ to an unweighted Boolean formula $\check{F}$ such that $W(F) = C_F \cdot |R_{\check{F}}| - 2^n + 2^{n-|N_F|}$. The motivation for this reduction comes from the need to preserve DNF form of the input formula. Both reductions take time linear in the size of $F$ and in the number of bits required to represent the weights of normal-weighted variables in $F$.

Note that since $C_F = 2^{-\widehat{m}}$, computing $C_F \cdot |R_{\widehat{F}}|$ (respectively, $C_F \cdot |R_{\check{F}}|$) amounts to computing $|R_{\widehat{F}}|$ (respectively, $|R_{\widehat{F}}|$), which is an instance of UMC, and shifting the radix point in the binary representation of the result to the left by $\widehat{m}$ positions.

**Theorem 13.** *Let $(F, W(\cdot))$ be an instance of literal-weighted* WMC*, where $F$ has $n$ variables. Then, we can construct in linear time the following unweighted Boolean formulas, each of which has $n + \widehat{m}$ variables and is of size linear in $|F| + \widehat{m}$.*

*(a) $\widehat{F}$ such that $W(F) = C_F \cdot |R_{\widehat{F}}|$.*

*(b) $\check{F}$ such that $W(F) = C_F \cdot |R_{\check{F}}| - 2^n \cdot (1 - 2^{-|N_F|})$*

*Proof.* Let $X = \{x_1, \cdots, x_n\}$ be the set of variables of $F$. Without loss of generality, let $N_F = \{1, \cdots r\}$ be the set of indices of the normal-weighted variables of $F$. For each normal-weighted variable $x_i$, let $\varphi_{k_i,m_i}(x_{i,1}, \cdots x_{i,m_i})$ be the representative formula, as defined above. Let $\Omega = (x_1 \leftrightarrow \varphi_{k_1,m_1}) \wedge \cdots \wedge (x_r \leftrightarrow \varphi_{k_r,m_r})$.

*Proof of part (a):* We define the formula $\widehat{F}$ as follows.

$$\widehat{F} = F \wedge \Omega$$

Recalling $\widehat{m} = \sum_{i \in N_F} m_i$, it is easy to see that $\widehat{F}$ has $n + \widehat{m}$, variables. From Lemma 11, we know that $\varphi_{k_i,m_i}(x_{i,1}, \cdots x_{i,m_i})$ is of size linear in $m_i$, for every $i$ in $N_F$. Therefore, the size of $\Omega$ is linear in $\widehat{m}$, and the size of $\widehat{F}$ is linear in $|F| + \widehat{m}$.

We now show that $W(F) = C_F \cdot |R_{\widehat{F}}|$. Let $W'(\cdot)$ be a new weight function, defined over the literals of $X$ as follows. If $x_i$ has indifferent weight, then $W'(x_i^0) = W'(x_i^1) = 1$. If $x_i$ has normal weight with $W(x_i^1) = k_i/2^{m_i}$, then $W'(x_i^1) = k_i$ and $W'(x_i^0) = 2^{m_i} - k_i$. By extending the definition of $W'(\cdot)$ in a natural way (as was done for $W(\cdot)$) to assignments, sets of assignments and formulas, it is easy to see that $W(F) = W'(F) \cdot \prod_{i \in N_F} 2^{-m_i} = W'(F) \cdot C_F$.

Next, for every assignment $\sigma$ of variables in $X$, let $\sigma^1 = \{i \in N_F \mid \sigma(x_i) = \mathit{true}\}$ and $\sigma^0 = \{i \in N_F \mid \sigma(x_i) = \mathit{false}\}$. Then, we have $W'(\sigma) = \prod_{i \in \sigma^1} k_i \prod_{i \in \sigma^0} (2^{m_i} - k_i)$. Let $\widehat{\sigma}$ be an assignment of variables appearing in $\widehat{F}$. We say that $\widehat{\sigma}$ is *compatible* with $\sigma$ if for all variables $x_i$ in $X$, we have $\widehat{\sigma}(x_i) = \sigma(x_i)$. Observe that $\widehat{\sigma}$ is compatible with exactly one assignment, viz. $\sigma$, of variables in $X$. Let $S_\sigma$ denote the set of all satisfying assignments of $\widehat{F}$ that are compatible with $\sigma$. Then $\{S_\sigma | \sigma \in R_F\}$ is a partition of $R_{\widehat{F}}$. From Lemma 11, we know that there are $k_i$ witnesses of $\varphi_{k_i,m_i}$ and $2^{m_i} - k_i$ witnesses of $\neg\varphi_{k_i,m_i}$. Since the representative formula of every normal-weighted variable uses a fresh set of variables, we have from the structure of $\widehat{F}$ that if $\sigma$ is a witness of $F$, then $|S_\sigma| = \prod_{i \in \sigma^1} k_i \prod_{i \in \sigma^0} (2^{m_i} - k_i)$. Therefore $|S_\sigma| = W'(\sigma)$. Note that if $\sigma$ is not a witness of $F$, then there are no compatible satisfying assignments of $\widehat{F}$; hence $S_\sigma = \emptyset$ in this case. Overall, this gives

$$|R_{\widehat{F}}| = \sum_{\sigma \in R_F} |S_\sigma| + \sum_{\sigma \notin R_F} |S_\sigma| = \sum_{\sigma \in R_F} |S_\sigma| + 0 = W'(F)\,.$$

It follows that $W(F) = C_F \cdot W'(F) = C_F \cdot |R_{\widehat{F}}|$. This completes the proof of part (a).

*Proof of part (b):* We define the formula $\breve{F}$ as follows.

$$\breve{F} = \Omega \rightarrow F$$

Clearly, $\breve{F}$ has $n + \widehat{m}$ variables. Since the size of $\Omega$ is linear in $\widehat{m}$, the size of $\breve{F}$ is linear in $|F| + \widehat{m}$.

We now show that $W(F) = C_F \cdot |R_{\breve{F}}| - 2^n \cdot (1 - 2^{-|N_F|})$. First, note that $\breve{F}$ is logically equivalent to $\neg\Omega \vee (F \wedge \Omega) = \neg\Omega \vee \widehat{F}$, where $\widehat{F}$ is as defined in part (a) above. Since $\widehat{F}$ and $\neg\Omega$ are mutually inconsistent, it follows that $|R_{\breve{F}}|$ is the sum of $|R_{\widehat{F}}|$ and the number of satisfying assignments (over all variables in $\breve{F}$) of $\neg\Omega$.

By definition, $\Omega$ does not contain any variable in $X \setminus N_F$. Hence, the number of satisfying assignments (over all variables in $\breve{F}$) of $\neg\Omega$ is $2^{n-|N_F|} \cdot |R_{\neg\Omega}|$. To calculate $|R_{\neg\Omega}|$, observe that $|R_{(x_i \leftrightarrow \varphi_{k_i,m_i})}| = 2^{m_i}$, and the sub-formulas $(x_i \leftrightarrow \varphi_{k_i,m_i})$ and $(x_j \leftrightarrow \varphi_{k_j,m_j})$ have disjoint variables for $i \neq j$. Therefore, $|R_\Omega| = \prod_{i \in N_F} 2^{m_i} = 2^{\widehat{m}}$, and $|R_{\neg\Omega}| = 2^{\widehat{m}+|N_F|} - 2^{\widehat{m}} = 2^{\widehat{m}} \cdot (2^{|N_F|} - 1)$. From part (a) above, we also know that $|R_{\widehat{F}}| = W(F)\,/C_F$. Hence, $|R_{\breve{F}}| = |R_{\widehat{F}}| + 2^{n-|N_F|} \cdot |R_{\neg\Omega}| = W(F)\,/C_F + 2^{n+\widehat{m}} \cdot (1 - 2^{-|N_F|})$. Rearraging terms, we get $W(F) = C_F \cdot \left(|R_{\breve{F}}| - 2^{n+\widehat{m}} \cdot (1 - 2^{-|N_F|})\right)$. Since $C_F = 2^{-\widehat{m}}$, we obtain $W(F) = C_F \cdot |R_{\breve{F}}| - 2^n \cdot (1 - 2^{-|N_F|})$. This completes the proof of part (b). □

### 5.2.4 Preservation of Normal Forms

The representative formula of a normal-weighted variable is a chain formula, which is generally neither in CNF nor in DNF. Therefore, even if the input formula $F$ is in a normal form (CNF/DNF), the formulas $\widehat{F}$ and $\breve{F}$ in Theorem 13 may be neither in CNF nor in DNF. We ask if our reductions can be adapted to preserve the normal form (CNF/DNF) of $F$. Theorem 14 answers this question affirmatively.

**Theorem 14.** *Let $(F, W(\cdot))$ be an instance of literal-weighted* WMC*, where $F$ is in* CNF *(resp.,* DNF*) and has $n$ variables. We can construct in polynomial time a* CNF *(resp.,* DNF*) formula $F^\star$ such that $W(F) = C_F \cdot |R_{F^\star}|$ (resp., $C_F \cdot |R_{F^\star}| - 2^n \cdot (1 - 2^{-|N_F|})$). Moreover, $F^\star$ has $n + \widehat{m}$ variables and its size is linear in $(|F| + \sum_{i \in N_F} m_i^2)$.*

*Proof.* We first prove the case of $F$ in CNF. To this end, we first show that $\Omega$ obtained in the proof of Theorem 13 can be transformed to a CNF formula $\Omega^{\mathsf{CNF}}$. Transform $\Omega$ by replacing every sub-formula $(x_i \leftrightarrow \varphi_{k_i,m_i})$ in $\widehat{F}$ with the equivalent sub-formula $(\neg x_i \vee \varphi^{\mathsf{CNF}}_{k_i,m_i}) \wedge (x_i \vee (\neg\varphi_{k_i,m_i})^{\mathsf{CNF}})$. Note that since $\varphi_{k_i,m_i}$ is a chain formula, so is $\neg\varphi_{k_i,m_i}$. Hence, by Lemma 12, $\neg\varphi_{k_i,m_i}$ can be transformed into an equivalent CNF formula $(\neg\varphi_{k_i,m_i})^{\mathsf{CNF}}$. We can obtain $\Omega^{\mathsf{CNF}}$ (in CNF) by distributing $\vee$ over $\wedge$ in each of $(\neg x_i \vee \varphi_{k_i,m_i}^{\mathsf{CNF}})$ and $(x_i \vee (\neg\varphi_{k_i,m_i})^{\mathsf{CNF}})$. Finally $F^\star$ is simply $F \wedge \Omega^{\mathsf{CNF}}$. Since $F^\star$ is semantically equivalent to $\widehat{F}$, we have $|R_{F^\star}| = |R_{\widehat{F}}|$. From Theorem 13, we also know that $W(F) = C_F \cdot |R_{\widehat{F}}|$. Therefore, $W(F) = C_F \cdot |R_{F^\star}|$. From the above construction, and from Lemma 12 and Theorem 13, it is also easy to see that $|F^\star|$ is linear in $(|F| + \sum_{i \in N_F} m_i^2)$. Moreover, $F^\star$ has exactly the same variables as $\widehat{F}$. Hence, $F^\star$ has $n + \hat{m}$ variables.

Next, we show how to construct in polynomial time a $\mathsf{DNF}$ formula if $F$ is in $\mathsf{DNF}$. We first observe that $\breve{F}$ obtained in Theorem 13 can be rewritten as $\neg\Omega \vee F$. Since $\Omega$ can be transformed to $\Omega^{\mathsf{CNF}}$, we have $\neg\Omega^{\mathsf{CNF}}$ in DNF. Therefore $F^\star$ is simply $(\neg\Omega^{\mathsf{CNF}}) \vee F$. Since $F^\star$ is semantically equivalent to $\breve{F}$, we have $|R_{F^\star}| = |R_{\breve{F}}|$. From Theorem 13, we know that $W(F) = C_F(|R_{\breve{F}}| - 2^{\hat{m}+n} + 2^{\hat{m}})$. Therefore, $W(F) = C_F(|R_{F^\star}| - 2^{\hat{m}+n} + 2^{\hat{m}})$. Again, from the above construction, and from Lemma 12 and Theorem 13, it is also easy to see that $|F^\star|$ is linear in $(|F| + \sum_{i \in N_F} m_i^2)$. Moreover, $F^\star$ has exactly the same variables as $\breve{F}$. Hence, $F^\star$ has $n + \hat{m}$ variables. □

### 5.2.5 Optimality of Reductions

We now ask if there exists an algorithm that reduces literal-weighted $\mathsf{WMC}$ to $\mathsf{UMC}$ and gives unweighted Boolean formulas that have significantly fewer variables than $\widehat{F}$ or $\breve{F}$. We restrict our discussion to reductions that use $C_F$ as a normalization constant, and perhaps use an additive correction term $D(n, W(\cdot))$ that is agnostic to $F$, and depends only on the number of variables in $F$ and on the weight function. An example of such a term is $-2^n \cdot (1 - 2^{-|N_F|})$, used in Theorem 13b, where $N_F$ can be determined from $W(\cdot)$ by querying the weights of individual literals.

**Theorem 15.** *Let* $\mathsf{Reduce}(\cdot)$ *be an algorithm that takes as input an instance* $(F, W(\cdot))$

*of literal-weighted* WMC*, and returns an unweighted Boolean formula* $\widetilde{F}$ *such that* $W(F) = C_F \cdot |R_{\widetilde{F}}| + D(n, W(\cdot))$*, where* $D(\cdot,\cdot)$ *is a real-valued function and* $n$ *is the number of variables in* $F$*. Then* $\widetilde{F}$ *has at least* $n - 1 + \widehat{m} - 2|N_F|$ *variables.*

Observe that the number of variables in $\widehat{F}$ and $\breve{F}$ in Theorem 13 differ from the lower bound given by Theorem 15 by $2|N_F| + 1$, which is independent of $n$ as well as the number of bits ($\widehat{m}$) used to represent weights.

*Proof.* We first show that $D(\cdot,\cdot)$ must always be non-positive. Otherwise, suppose $D(n, W(\cdot)) > 0$. Consider the instance $(G, W(\cdot))$ of literal-weighted UMC, where $G = G_1 \wedge G_2$, where $G_1 = x_1 \wedge \neg x_1$ and $G_2 = x_2 \wedge x_3 \wedge \ldots \wedge x_n$. Since $G$ is unsatisfiable, $W(G) = 0$. However, $G_F \cdot |R_{\widetilde{G}}| + D(n, W(\cdot))$ is positive for every $\widetilde{G}$ that Reduce$(G, W(\cdot))$ may generate. This gives a contradiction; hence $D(\cdot,\cdot)$ must be non-positive.

Now, let $F$ be the formula $(x_1 \wedge x_2 \cdots \wedge x_r) \wedge \neg(x_{r+1} \wedge \cdots x_n)$, where $N_F = \{1, 2, \ldots r\}$. Furthermore, let $W(x_i^1) = \frac{2^{m_i}-1}{2^{m_i}}$, for every $i$ in $N_F$. Clearly, $W(F) = (2^{n-r} - 1) \cdot \prod_{i \in N_F} \frac{2^{m_i}-1}{2^{m_i}}$. Factoring out $C_F$, i.e. $\prod_{i \in N_F} 2^{-m_i}$, we get $W(F) = C_F \cdot (2^{n-r} - 1) \cdot \prod_{i \in N_F}(2^{m_i} - 1)$. In order to have $W(F) = C_F \cdot |R_{\widetilde{F}}| + D(n, W(\cdot))$, we must have $(2^{n-r} - 1) \cdot \prod_{i \in N_F}(2^{m_i} - 1) - \frac{D(n,W(\cdot))}{C_F}$ witnesses of $\widetilde{F}$. Since $D(\cdot,\cdot) \leq 0$, we need at least $(2^{n-r} - 1) \cdot \prod_{i \in N_F}(2^{m_i} - 1)$ witnesses of $\widetilde{F}$. In other words, $\widetilde{F}$ must have at least $\lceil \log_2((2^{n-r} - 1) \prod_{i \in N_F}(2^{m_i} - 1)) \rceil$ variables. Noting that $r = |N_F|$ and $2^m - 1 \geq 2^{m-1}$ for all $m \geq 1$, we conclude that $\widetilde{F}$ must have at least $n - |N_F| - 1 + \sum_{i \in N_F}(m_i - 1)$ variables. Rearranging terms, we get the desired lower bound on the number of variables.

Every Boolean formula in $\sum_{i \in N_F}(m_i - 1)$ variables is trivially of size $\Omega(\sum_{i \in N_F} m_i)$. Now, assume that the algorithm Reduce$(F, W(\cdot))$ uses the input formula $F$ as a black box. Then to ensure that $W(F) = C_F \cdot |R_{\widetilde{F}}| + D(n, W(\cdot))$, the formula $\widetilde{F}$ generated by $\mathsf{Reduce}(F, W(\cdot))$ must have $F$ as a sub-formula. Otherwise, $C_F \cdot |R_{\widetilde{F}}| + D(n, W(\cdot))$ will be independent of $R_F$. However, this cannot happen since $W(F) = C_F \cdot |R_{\widetilde{F}}| + D(n, |N_F|)$. Hence $\widetilde{F}$ must have $F$ as a sub-formula, and the size of $\widetilde{F}$ is at least as large as that of $F$. Putting the above arguments together, the size of $\widetilde{F}$ is in $\Omega(|F| + \sum_{i \in N_F} m_i)$. □

### 5.2.6 Experimental Analysis

The construction outlined in the proof of Theorem 13 naturally suggests an algorithm for solving WMC using a UMC solver as a black-box. This is particularly important in the context of weighted model counting, since state-of-the-art unweighted model counters (viz. sharpSAT [147]) scale to much larger problem sizes than existing state-of-the-art weighted model counters (viz. SDD [53]). To investigate the practical advantages of using the reduction based approach, we developed a literal-weighted model counter module called sharpWeightSAT, that takes as input an instance $(F, W(\cdot))$ of literal-weighted WMC and reduces it to an instance $F^\star$ of UMC, as outlined in Theorem 13 and 14. The sharpWeightSAT module then invokes an underlying state-of-the-art exact UMC solver, to count the witnesses of $F^\star$. Finally, sharpWeightSAT computes $C_F \cdot |R^\star_F|$ as the weighted model count of $(F, W(\cdot))$. In our experiments we employed both sharpSAT and DSharp as the underlying exact UMC solver.

We conducted experiments on a suite of diverse CNF benchmarks to compare the performance of sharpWeightSAT with that of SDD. We also tried to compare our tool with the weighted variant of Cachet [139], but despite extensive efforts, we have not been able to run this tool on our system. We focused on CNF formulas because of the availability of CNF model counters and the lack of DNF model counters in the public-domain. The suite of benchmarks used in our experiments consisted of problems arising from probablistic inference in grid networks, synthetic grid-structured random interaction Ising models, plan recognition, DQMR networks, bit-blasted versions of SMTLIB benchmarks, ISCAS89 combinational circuits with weighted inputs, and program synthesis examples. Note that normal weights of variables in our benchmarks typically correspond to (conditional) probabilities of events in the original problem from which the benchmark is derived. To allow specification of probabilities with a precision of up to to two decimal places, we rounded off the weights such that all weights were of the form $k/2^i$ $(1 \leq i \leq 7)$. A uniform timeout of 5 hours was used for all tools in our experiments.

| | | | sharpWeightSAT | | | | | SDD |
|---|---|---|---|---|---|---|---|---|
| **Benchmark** | **Orig #vars** | **Orig #clas** | **Final #vars** | **Final #claus** | **Transform time (s)** | **sharpSAT counting time (s)** | **DSharp counting time (s)** | **Overall time (s)** |
| case_1_b11_1 | 340 | 1026 | 550 | 1266 | 0.03 | 92.16 | 1059.82 | 64.3 |
| s1196a_15_7 | 777 | 2165 | 867 | 2285 | 0.06 | 0.54 | 8.88 | – |
| case_2_b12_2 | 827 | 2725 | 917 | 2845 | 0.06 | 34.11 | 714.37 | 735.68 |
| squaring1 | 891 | 2839 | 981 | 2959 | 0.04 | 10.02 | 97.86 | – |
| cliquen30 | 930 | 1800 | 2517 | 3821 | 0.11 | 300.86 | – | – |
| BN_63 | 1112 | 2661 | 1272 | 2853 | 0.04 | 0.68 | 8.68 | – |
| BN_55 | 1154 | 2692 | 1314 | 2884 | 0.1 | 1.11 | – | – |
| BN_47 | 1336 | 3376 | 1406 | 3460 | 0.11 | 0.11 | 1.49 | 170.92 |
| BN_61 | 1348 | 3388 | 1418 | 3472 | 0.05 | 0.2 | 1.77 | 157.88 |
| squaring9 | 1434 | 5028 | 1524 | 5148 | 0.07 | 32.68 | 721.14 | – |
| squaring16 | 1627 | 5835 | 1723 | 5963 | 0.07 | – | 2623.12 | – |
| BN_43 | 1820 | 3806 | 2240 | 4286 | 0.34 | 8393.12 | – | – |
| BN_108 | 2289 | 8218 | 11028 | 19105 | 0.27 | 2.14 | 8.66 | 270.31 |
| smokers_20 | 2580 | 3740 | 6840 | 8860 | 0.33 | 224.25 | – | – |
| treemax | 24859 | 103762 | 26353 | 105754 | 1.5 | 3.93 | 338.16 | – |
| BN_26 | 50470 | 93870 | 276675 | 352390 | 244.29 | 68.99 | 259.42 | 693.09 |

Table 5.2 : Performance comparison of sharpWeightSAT vis-a-vis SDD

Table 5.2 presents the results of comparing the performances of sharpWeightSAT and SDD on a subset of our benchmarks *. In this table, the benchmarks are listed in Column 1. Columns 2 and 3 list the number of variables and clauses, respectively, for each benchmark. Columns 4 through 8 present our experimental observations on running sharpWeightSAT via either sharpSAT or DSharp as UMC solvers. Specifically, columns 4 and 5 give the total number of variables and clauses of the unweighted formula obtained after applying our reduction. Note that these numbers are larger than the corresponding numbers in the original problem, since all normal-weighted variables in the original problem have been replaced by their respective representative formulas. The run-time of sharpWeightSAT via sharpSAT is the sum of the transform time taken to reduce a WMC instance to an instance of UMC, as presented in Column 6, and the counting time taken by sharpSAT to solve an instance of UMC, as presented in Column 7. The run-time of sharpWeightSAT via DSharp is the sum of the transform time as presented in Column 6, and and the counting time taken by DSharp to solve an instance of UMC, as presented in Column 8. Finally, run-time for SDD to solve the same instance of WMC is presented in column 9. A "-" in a column indicates that the corresponding experiment either did not complete within 5 hours or ran out of memory.

Overall, out of 79 benchmarks for which the weighted model count could be computed by either SDD or sharpWeightSAT, SDD timed/spaced out on 30 benchmarks, sharpWeightSAT via sharpSAT timed out on 2 benchmarks, and sharpWeightSAT via DSharp timed out on 11 benchmarks . Table 5.2 clearly shows that on most benchmarks sharpWeightSAT via either sharpSAT or DSharp outperformed SDD in terms of running time by 1 to 3 orders of magnitude. Moreover, sharpWeightSAT could gener-

*The full version of Table 5.2 is available in Appendix as Table A2

ate weighted counts for a large class of benchmarks for which SDD timed out. Thus, our reduction helps in solving instances of literal-weighted WMC that are otherwise beyond the reach of a state-of-the-art weighted model counter. Significantly, column 6 of Table 5.2 demonstrates that the overhead for reducing a WMC problem to a UMC instance is very small. The comparison between sharpWeightSAT via sharpSAT and sharpWeightSAT via DSharp is interesting but beyond the scope of this work.

To empirically study the effect of using our reduction in the context of approximate WMC, we also augmented ApproxMC2 – a state-of-the-art approximate unweighted model counter – to obtain an approximate weighted model counter called ApproxWeightMC. It is important to note that ApproxMC2 uses random hash functions under-the-hood, as do several recent approximate model counters [72] that provide strong guarantees. Surprisingly, our experiments showed that ApproxWeightMC faced serious performance bottlenecks when run on the benchmarks in Table 5.2. A crucial step in ApproxMC2 is the use of random xor clauses over the set of *independent* variables (see Chapter 4 for details) to constrain the space of witnesses of the original problem before sampling a witness from the constrained space. The count of independent variables crucially affects the expected size of a random xor clause, which in turn has a significant bearing on the performance of SAT solvers used to generate a witness of the constrained problem. Therefore, a likely explanation for the inability of ApproxWeightMC to scale as well as sharpWeightSAT is the explosion in the size of random xor clauses over the independent variables when we conjoin representative formulas for all normal-weighted variables. To see why this explosion happens, recall from Section 5.2.1 that the representative formula of a variable with weight $k/2^m$ introduces $m$ fresh variables; indeed, these are independent variables in the formula obtained using our reduction.

Overall, our experiments demonstrate that state-of-the-art UMC solvers can be augmented with an implementation of our reduction to obtain literal-weighted model counts on formulas with tens of thousands of variables – problems that are clearly beyond the reach of existing weighted model counters. Significantly, our approach requires no modification of the implementation of the UMC solver, which can therefore be treated as a black-box. Approximate literal-weighted WMC, however, does not seem to benefit from our reduction due to the significant increase in the size of random xor clauses. This underlines the need for further research on understanding the intricacies of the reduction of literal-weighted WMC to UMC.

## 5.3 Beyond Literal Weights

While literal-weighted representation is typically employed in applications of WMC, richer forms of representations of weights are increasingly used in a wide variety of applications. Of these, associating weights to constraints instead of literals has been widely employed in probabilistic programming, verification, and the like [4, 131]. For example, Figaro, a popular probabilistic programming framework, contains a construct called *setConstraint* that associates weights with constraints. We now demonstrate that our techniques can be generalized to handle such representations as well.

Define ConstraintWMC to be a variant of WMC, wherein the weight of an assignment is specified using a set of constraints. Specifically, given a formula $F$, a set $\boldsymbol{G} = (G_1, \cdots G_r)$ of Boolean constraints, and a weight function $W(\cdot)$ over $\boldsymbol{G}$, the weight of an assignment $\sigma$ is defined as the product of the weights of constraints in $\boldsymbol{G}$ that are satisfied by $\sigma$. The weight of every constraint $G_i$ is assumed to be of the form $k_i/2^{m_i}$, where $k_i$ is an odd integer between 1 and $2^{m_i} - 1$. In case $\sigma$ satisfies none of the constraints in $\boldsymbol{G}$, the weight of $\sigma$ is defined to be 1. The ConstraintWMC problem is to compute the sum of the weights of all witnesses of $F$.

By an extension of the reasoning used in the proof of Theorem 13a, we can obtain an efficient reduction from ConstraintWMC to UMC. We do not yet know how to preserve the normal form of the input formula.

**Theorem 16.** *Let $(F, \boldsymbol{G}, W(\cdot))$ be an instance of* ConstraintWMC*, where $|\boldsymbol{G}| = r$ and $\varphi_{k_i,m_i}(x_{i,1}, \cdots x_{i,m_i})$ is the chain formula that describes $W(G_i)$. Then by defining $\widehat{F} = F \wedge (G_1 \rightarrow \varphi_{k_1,m_1}) \wedge \cdots \wedge (G_r \rightarrow \varphi_{k_r,m_r})$, we get a linear-time reduction from* ConstraintWMC *to* UMC*, such that $W(F) = C_{\boldsymbol{G}} \cdot |R_{\widehat{F}}|$, where $C_{\boldsymbol{G}} = \prod_{i=1}^{r} 2^{-m_i}$.*

*Proof.* The proof is almost identical to the proof of Theorem 13. Clearly, $\widehat{F}$ has $n + \sum_{i \leq r}(n_i + m_i)$, variables. From Lemma 11, we know that $\varphi_{k_i,m_i}(x_{i,1}, \cdots x_{i,m_i})$ is of size linear in $m_i$, for every $i$ in $N_F$. Therefore, the size of $\widehat{F}$ is linear in $(|F| + \sum_{i \leq r}(|G_i| + m_i)$.

We now show that $W(F) = C_G \cdot |R_{\widehat{F}}|$. Let $W'(\cdot)$ be a new weight function, defined over the constraints $G_i$ of $\boldsymbol{G}$ as follows. If $W(G_i) = k_i/2^{m_i}$, then $W'(G_i) = k_i$, and $W'(G_i) = 2^{m_i}$. We extend the definition of $W'(\cdot)$ in a natural way (as was done for $W(\cdot)$) to assignments, sets of assignments and formulas. Note that for every assignment $\sigma$ for the variables in $X$ we have $W(\sigma) = W'(\sigma) \cdot \prod_{i \leq r} 2^{-m_i} = W'(\sigma) \cdot C_G$, and therefore we have $W(F) = W'(F) \cdot C_G$. In addition, for every assignment $\sigma$ of variables in $X$, Denote by $G(\sigma)$ the set of indices of the constraints in $\boldsymbol{G}$ that satisfy $\sigma$. Then we also have $W'(\sigma) = \prod_{i \in G(\sigma)} k_i \prod_{i \notin G(\sigma)} 2^{m_i}$. Let $\widehat{\sigma}$ be an assignment of variables appearing in $\widehat{F}$. We say that $\widehat{\sigma}$ is *compatible* with $\sigma$ if for all variables $x_i$ in $X$, we have $\widehat{\sigma}(x_i) = \sigma(x_i)$. Observe that $\widehat{\sigma}$ is compatible with exactly one assignment, viz. $\sigma$, of variables in $X$. Let $S_\sigma$ denote the set of all satisfying assignments of $\widehat{F}$ that are compatible with $\sigma$. Then $\{S_\sigma | \sigma \in R_F\}$ is a partition of $R_{\widehat{F}}$. From Lemma 11, we know that there are $k_i$ witnesses of $\varphi_{k_i,m_i}$. Since the representative formula of every weighted constraint uses a fresh set of variables, we have from the structure of $\widehat{F}$ that if $\sigma$ is a witness to $F$ then $|S_\sigma| = \prod_{i \in G(\sigma)} k_i \prod_{i \notin G(\sigma)} 2^{m_i}$. Therefore $|S_\sigma| = W'(\sigma)$. Note that if $\sigma$ is not a witness of $F$, then there are no compatible satisfying assignments of $\widehat{F}$; hence $S_\sigma = \emptyset$ in this case. Overall, this gives

$$|R_{\widehat{F}}| = \sum_{\sigma \in R_F} |S_\sigma| + \sum_{\sigma \notin R_F} |S_\sigma| = \sum_{\sigma \in R_F} |S_\sigma| + 0 = W'(F)\,.$$

It follows that $W(F) = C_G \cdot W'(F) = C_G \cdot |R_{\widehat{F}}|$. Finally, note that $C_G = \prod_{i=1}^{r} 2^{-m_i} = 2^{-\sum_{1 \le i \le r} m_i}$. Therefore, computing $C_G \cdot |R_{\widehat{F}}|$ amounts to computing $|R_{\widehat{F}}|$ (an instance of UMC) and shifting the radix point in the binary representation of $|R_{\widehat{F}}|$ left by $\left(\sum_{1 \le i \le r} m_i\right)$ positions. □

## 5.4 Chapter Summary

In this chapter, we discussed two complementary approaches to handle WMC. In the first half of this chapter, we discussed how hashing-based techniques introduced in Chapter 4 for UMC can be lifted to handle WMC. Prior hashing-based approaches to WMC employed computationally expensive MPE oracle. In contrast, we only employ SAT oracle. In this half of the chapter, we do not make any assumption on the weight function. We introduced a novel parameter, *t*ilt, to capture the hardness of benchmarks with respect to hashing-based approach. In the second half of this chapter, we discussed a complementary approach wherein we propose an efficient reduction of WMC to UMC if the weight function is expressed using *literal-weighted* representation.

# Chapter 6

# Handling Bit-Vector Formulas

In a large class of probabilistic inference problems, an important case being lifted inference on first order representations [106], the values of variables come from finite but large (exponential in the size of the representation) domains. Data values coming from such domains are naturally encoded as fixed-width words, where the width is logarithmic in the size of the domain. Conditions on observed values are, in turn, encoded as word-level constraints, and the corresponding model-counting problem asks one to count the number of solutions of a word-level constraint. It is therefore natural to ask if the success of approximate propositional model counters can be replicated at the word-level.

The balance between efficiency and strong guarantees of hashing-based algorithms for constrained counting for Boolean formulas crucially depends on two factors: (i) use of XOR-based 2-universal bit-level hash functions, and (ii) use of state-of-the-art propositional satisfiability solvers, viz. CryptoMiniSAT [143], that can efficiently reason about formulas that combine disjunctive clauses with XOR clauses.

In recent years, the performance of SMT (Satisfiability Modulo Theories) solvers has witnessed spectacular improvements [12]. Indeed, several highly optimized SMTsolvers for fixed-width words are now available in the public domain [28, 101, 92, 56]. Nevertheless, 2-universal hash functions for fixed-width words that are also amenable to efficient reasoning by SMT solvers have hitherto not been studied. The reasoning power of SMTsolvers for fixed-width words has therefore remained untapped for word-level model counting. Thus, it is not surprising that all existing work on probabilistic inference using model counting (viz. [43, 16, 71]) effectively reduce the problem to propositional model counting. Such approaches are similar to "bit blasting" in SMT solvers [111].

The primary contribution of this chapter is an efficient word-level approximate model counting algorithm SMTApproxMC that can be employed to answer inference queries over high-dimensional discrete domains. Our algorithm uses a new class of word-level hash functions that are 2-universal and can be solved by word-level SMTsolvers capable of reasoning about linear equalities on words. Therefore, unlike previous works, SMTApproxMC is able to leverage the power of sophisticated SMT solvers.

To illustrate the practical utility of SMTApproxMC, we implemented a prototype and evaluated it on a suite of benchmarks. Our experiments demonstrate that SMTApproxMC can significantly outperform the prevalent approach of bit-blasting a word-level constraint and using an approximate propositional model counter that employs XOR-based hash functions. Our proposed word-level hash functions embed the domain of all variables in a large enough finite domain. Thus, one would not expect our approach to work well for constraints that exhibit a hugely heterogeneous mix of word widths, or for problems that are difficult for word-level SMT solvers. Indeed, our experiments suggest that the use of word-level hash functions provides significant benefits when the original word-level constraint is such that (i) the words appearing in it have long and similar widths, and (ii) the SMTsolver can reason about the constraint at the word-level, without extensive bit-blasting.

## 6.1 Related Work

Over the last two decades, there has been tremendous progress in the development of decision procedures, called Satisfiability Modulo Theories (or SMT) solvers, for combinations of first-order theories, including the theory of fixed-width words [13, 11]. An SMT solver uses a core propositional reasoning engine and decision procedures for individual theories, to determine the satisfiability of a formula in the combination of theories. It is now folklore that a well-engineered word-level SMT solver can significantly outperform the naive approach of *blasting* words into component bits and then using a propositional satisfiability solver [56, 101, 31]. The power of word-level SMT solvers stems from their ability to reason about words directly (e.g. $a+(b-c) = (a-c)+b$ for every word $a$, $b$, $c$), instead of *blasting* words into component bits and using propositional reasoning.

The work of [43] tried to extend ApproxMC [38] to non-propositional domains. A crucial step in their approach is to propositionalize the solution space (e.g. bounded integers are equated to tuples of propositions) and then use XOR-based bit-level hash functions. Unfortunately, such propositionalization can significantly reduce the effectiveness of theory-specific reasoning in an SMT solver. The work of [16] used bit-level hash functions with the propositional abstraction of an SMT formula to solve the problem of *weighted model integration*. This approach also fails to harness the power of theory-specific reasoning in SMT solvers.

Recently, [25] proposed SGDPLL($T$), an algorithm that generalizes SMT solving to do lifted inferencing and model counting (among other things) modulo background theories (denoted $T$). A fixed-width word model counter, like the one proposed in this Chapter, can serve as a theory-specific solver in the SGDPLL($T$) framework. In addition, it can also serve as an alernative to SGDPLL($T$) when the overall problem is simply to count models in the theory $T$ of fixed-width words, There have also been other attempts to exploit the power of SMT solvers in machine learning. For example, [145] used optimizing SMT solvers for structured relational learning using Support Vector Machines. This is unrelated to our approach of harnessing the power of SMT solvers for probabilistic inference via model counting.

## 6.2 Word-level Hash Function

The performance of hashing-based techniques for approximate model counting depends crucially on the underlying family of hash functions used to partition the solution space. A popular family of hash functions used in propositional model counting is $\mathcal{H}_{xor}$, defined as the family of functions obtained by XOR-ing a random subset of propositional variables, and equating the result to either 0 or 1, chosen randomly. The family $\mathcal{H}_{xor}$ enjoys important properties like 2-independence and easy implementability, which make it ideal for use in practical model counters for propositional formulas [71, 38]. Unfortunately, word-level universal hash families that are 2-independent, easily implementable and amenable to word-level reasoning by SMT solvers, have not been studied thus far. In this section, we present $\mathcal{H}_{SMT}$, a family of word-level hash functions that fills this gap.

As discussed earlier, let $\mathsf{sup}(F) = \{x_0, \ldots x_{n-1}\}$, where each $x_i$ is a word of width $k$. We use $\mathbf{X}$ to denote the $n$-dimensional vector $(x_0, \ldots x_{n-1})$. The space of all assignments to words in $\mathbf{X}$ is $\{0,1\}^{n.k}$. Let $p$ be a prime number such that $2^k \leq p < 2^{n.k}$. Consider a family $\mathcal{H}$ of hash functions mapping $\{0,1\}^{n.k}$ to $\mathbb{Z}_p$, where each hash function is of the form $h(\mathbf{X}) = (\sum_{j=0}^{n-1} a_j * x_j + b) \mod p$, and the $a_j$'s and $b$ are elements of $\mathbb{Z}_p$, represented as words of width $\lceil \log_2 p \rceil$. Observe that every $h \in \mathcal{H}$ partitions $\{0,1\}^{n.k}$ into $p$ bins (or cells). Moreover, for every $\xi \in \{0,1\}^{n.k}$ and $\alpha \in \mathbb{Z}_p$, $\mathsf{Pr}\left[h(\xi) = \alpha : h \xleftarrow{R} \mathcal{H}\right] = p^{-1}$. For a hash function chosen uniformly at random from $\mathcal{H}$, the expected number of elements per cell is $2^{n.k}/p$. Since $p < 2^{n.k}$, every cell has at least 1 element in expectation. Since $2^k \leq p$, for every word $x_i$ of width $k$, we also have $x_i \mod p = x_i$. Thus, distinct words are not aliased (or made to behave similarly) because of modular arithmetic in the hash function.

Suppose now we wish to partition $\{0.1\}^{n.k}$ into $p^c$ cells, where $c > 1$ and $p^c < 2^{n.k}$. To achieve this, we need to define hash functions that map elements in $\{0,1\}^{n.k}$ to a tuple in $(\mathbb{Z}_p)^c$. A simple way to achieve this is to take a $c$-tuple of hash functions, each of which maps $\{0,1\}^{n.k}$ to $\mathbb{Z}_p$. Therefore, the desired family of hash functions is simply the iterated Cartesian product $\mathcal{H} \times \cdots \times \mathcal{H}$, where the product is taken $c$ times. Note that every hash function in this family is a $c$-tuple of hash functions. For a hash function chosen uniformly at random from this family, the expected number of elements per cell is $2^{n.k}/p^c$.

An important consideration in hashing-based techniques for approximate model counting is the choice of a hash function that yields cells that are neither too large nor too small in their expected sizes. Since increasing $c$ by 1 reduces the expected size of each cell by a factor of $p$, it may be difficult to satisfy the above requirement if the value of $p$ is large. At the same time, it is desirable to have $p > 2^k$ to prevent aliasing of two distinct words of width $k$. This motivates us to consider more general classes of word-level hash functions, in which each word $x_i$ can be split into thinner slices, effectively reducing the width $k$ of words, and allowing us to use smaller values of $p$. We describe this in more detail below.

Assume for the sake of simplicity that $k$ is a power of 2, and let $q$ be $\log_2 k$. For every $j \in \{0, \ldots q-1\}$ and for every $x_i \in \mathbf{X}$, define $\mathbf{x_i}^{(j)}$ to be the $2^j$-dimensional vector of slices of the word $x_i$, where each slice is of width $k/2^j$. For example, the two slices in $\mathbf{x_1}^{(1)}$ are $\mathsf{extract}(x_1, 0, k/2-1)$ and $\mathsf{extract}(x_1, k/2, k-1)$. Let $\mathbf{X}^{(j)}$ denote the $n.2^j$-dimensional vector $(\mathbf{x_0}^{(j)}, \mathbf{x_1}^{(j)}, \ldots \mathbf{x_{n-1}}^{(j)})$. It is easy to see that the $m^{th}$ component of $\mathbf{X}^{(j)}$, denoted $\mathbf{X}_m^{(j)}$, is $\mathsf{extract}(x_i, s, t)$, where $i = \lfloor m/2^j \rfloor$, $s = (m \mod 2^j) \cdot (k/2^j)$ and $t = s + (k/2^j) - 1$. Let $p_j$ denote the smallest prime larger than or equal to $2^{(k/2^j)}$. Note that this implies $p_{j+1} \leq p_j$ for all $j \geq 0$. In order to obtain a family of hash functions that maps $\{0,1\}^{n.k}$ to $\mathbb{Z}_{p_j}$, we split each word $x_i$ into slices of width $k/2^j$, treat these slices as words of reduced width, and use a technique similar to the one used above to map $\{0,1\}^{n.k}$ to $\mathbb{Z}_p$. Specifically, the family $\mathcal{H}^{(j)} = \left\{h^{(j)} : h^{(j)}(\mathbf{X}) = \left(\sum_{m=0}^{n.2^j-1} a_m^{(j)} * \mathbf{X}_m^{(j)} + b^{(j)}\right) \mod p_j\right\}$ maps $\{0,1\}^{n.k}$ to $\mathbb{Z}_{p_j}$, where the values of $a_m^{(j)}$ and $b^{(j)}$ are chosen from $\mathbb{Z}_{p_j}$, and represented as $\lceil \log_2 p_j \rceil$-bit words.

In general, we may wish to define a family of hash functions that maps $\{0,1\}^{n.k}$ to $\mathcal{D}$, where $\mathcal{D}$ is given by $(\mathbb{Z}_{p_0})^{c_0} \times (\mathbb{Z}_{p_1})^{c_1} \times \cdots \left(\mathbb{Z}_{p_{q-1}}\right)^{c_{q-1}}$ and $\prod_{j=0}^{q-1} p_j^{c_j} < 2^{n.k}$. To achieve this, we first consider the iterated Cartesian product of $\mathcal{H}^{(j)}$ with itself $c_j$ times, and denote it by $\left(\mathcal{H}^{(j)}\right)^{c_j}$, for every $j \in \{0, \ldots q-1\}$. Finally, the desired family of hash functions is obtained as $\prod_{j=0}^{q-1} \left(\mathcal{H}^{(j)}\right)^{c_j}$. Observe that every hash function $h$ in this family is a $\left(\sum_{l=0}^{q-1} c_l\right)$-tuple of hash functions. Specifically, the $r^{th}$ component of $h$, for $r \leq \left(\sum_{l=0}^{q-1} c_l\right)$, is given by $\left(\sum_{m=0}^{n.2^j-1} a_m^{(j)} * \mathbf{X}_m^{(j)} + b^{(j)}\right) \mod p_j$, where $\left(\sum_{i=0}^{j-1} c_i\right) < r \leq \left(\sum_{i=0}^{j} c_i\right)$, and the $a_m^{(j)}$s and $b^{(j)}$ are elements of $\mathbb{Z}_{p_j}$.

The case when $k$ is not a power of 2 is handled by splitting the words $x_i$ into slices of size $\lceil k/2 \rceil$, $\lceil k/2^2 \rceil$ and so on. Note that the family of hash functions defined above depends only on $n$, $k$ and the vector $C = (c_0, c_1, \ldots c_{q-1})$, where $q = \lceil \log_2 k \rceil$. Hence, we call this family $\mathcal{H}_{SMT}(n, k, C)$. Note also that by setting $c_i$ to 0 for all $i \neq \lfloor \log_2(k/2) \rfloor$, and $c_i$ to $r$ for $i = \lfloor \log_2(k/2) \rfloor$ reduces $\mathcal{H}_{SMT}$ to the family $\mathcal{H}_{xor}$ of XOR-based bit-wise hash functions mapping $\{0,1\}^{n.k}$ to $\{0,1\}^r$. Therefore, $H_{SMT}$ strictly generalizes $\mathcal{H}_{xor}$.

We summarize below important properties of the $\mathcal{H}_{SMT}|S|, k, C)$ class. Let $\mathcal{D}$ denote $(\mathbb{Z}_{p_0})^{c_0} \times (\mathbb{Z}_{p_1})^{c_1} \times \cdots (\mathbb{Z}_{p_{q-1}})^{c_{q-1}}$, where $\prod_{j=0}^{q-1} p_j^{c_j} < 2^{n.k}$. Let $C$ denote the vector $(c_0, c_1, \ldots c_{q-1})$.

**Lemma 17.** *For every* $\mathbf{X} \in \{0,1\}^{n.k}$ *and every* $\alpha \in \mathcal{D}$, $\mathsf{Pr}[h(\mathbf{X}) = \alpha \mid h \xleftarrow{R} \mathcal{H}_{SMT}|S|, k, C)] = \prod_{j=0}^{|C|-1} {p_j}^{-c_j}$

*Proof.* Let $h_r$, the $r^{th}$ component of $h$, for $r \leq \left(\sum_{j=0}^{|C|-1} c_j\right)$, be given by $\left(\sum_{m=0}^{n.2^j-1} a_m^{(j)} * \mathbf{X}_m^{(j)} + b^{(j)}\right)$ mod $p_j$, where $\left(\sum_{i=0}^{j-1} c_i\right) < r \leq \left(\sum_{i=0}^{j} c_i\right)$, and the $a_m^{(j)}$s and $b^{(j)}$ are randomly and independently chosen elements of $\mathbb{Z}_{p_j}$, represented as words of width $\lceil \log_2 p_j \rceil$. Let $\mathcal{H}^{(j)}$ denote the family of hash functions of the form $\left(\sum_{m=0}^{n.2^j-1} u_m^{(j)} * \mathbf{X}_m^{(j)} + v^{(j)}\right)$ mod $p_j$, where $u_m^{(j)}$ and $v^{(j)}$ are elements of $\mathbb{Z}_{p_j}$. We use $\alpha_r$ to denote the $r$th component of $\alpha$. For every choice of $\mathbf{X}$, $a_m^{(j)}$s and $\alpha_r$, there is exactly one $b^{(j)}$ such that $h_r(\mathbf{X}) = \alpha_r$. Therefore, $\mathsf{Pr}[h_r(\mathbf{X}) = \alpha_r | h_r \xleftarrow{R} \mathcal{H}^{(j)}] = p_i^{-1}$.

Recall that every hash function $h$ in $\mathcal{H}_{SMT}(n, k, C)$ is a $\left(\sum_{j=0}^{q-1} c_j\right)$-tuple of hash functions. Since $h$ is chosen uniformly at random from $\mathcal{H}_{SMT}(n, k, C)$, the $\left(\sum_{j=0}^{q-1} c_j\right)$ components of $h$ are effectively chosen randomly and independently of each other. Therefore, $\mathsf{Pr}[h(\mathbf{X}) = \alpha \mid h \xleftarrow{R} \mathcal{H}_{SMT}(n, k, C)] = \prod_{i=0}^{|C|-1} {p_i}^{-c_i}$ □

**Theorem 18.** *For every* $\alpha_1, \alpha_2 \in \mathcal{D}$ *and every distinct* $\mathbf{X}_1, \mathbf{X}_2 \in \{0,1\}^{n.k}$, $\mathsf{Pr}[(h(\mathbf{X}_1) = \alpha_1 \wedge h(\mathbf{X}_2) = \alpha_2) \mid h \xleftarrow{R} \mathcal{H}_{SMT}|S|, k, C)] = \prod_{j=0}^{|C|-1} (p_j)^{-2.c_j}$. *Therefore,* $\mathcal{H}_{SMT}|S|, k, C)$ *is pairwise independent.*

*Proof.* We know that $\mathsf{Pr}[(h(\mathbf{X}_1) = \alpha_1 \wedge h(\mathbf{X}_2) = \alpha_2)] = \mathsf{Pr}[h(\mathbf{X}_2) = \alpha_2 \mid h(\mathbf{X}_1) = \alpha_1] \times \mathsf{Pr}[h(\mathbf{X}_1) = \alpha_1]$. Theorem 18 implies that in order to prove pairwise independence of $\mathcal{H}_{SMT}(n, k, C)$, it is sufficient to show that $\mathsf{Pr}[h(\mathbf{X}_2) = \alpha_2 \mid h(\mathbf{X}_1) = \alpha_1] = \mathsf{Pr}[h(\mathbf{X}_2) = \alpha_2]$.

Since $h(\mathbf{X}) = \alpha$ can be viewed as conjunction of $\left(\sum_{j=0}^{q-1} c_j\right)$ ordered and *independent* constraints, it is sufficient to prove 2-wise independence for every ordered constraint. We now prove 2-wise independence for one of the ordered constraints below. Since the proof for the other ordered constraints can be obtained in exactly the same way, we omit their proofs.

We formulate a new hash function based on the first constraint as $g(\mathbf{X}) = ((\sum_{m=0}^{n.2^j-1} a_m^{(0)} * \mathbf{X}_m^{(0)} +$ mod $p_0$, where the $a_m^{(0)}$'s and $b^{(0)}$ are randomly and independently chosen elements of $\mathbb{Z}_{p_0}$, represented as words of width $\lceil \log_2 p_0 \rceil$. It is sufficient to show that $g(\mathbf{X})$ is 2-universal. This can be formally stated as $\mathsf{Pr}[g(\mathbf{X}_2) = \alpha_{2,0} \mid g(\mathbf{X}_1) = \alpha_{1,0}] = \mathsf{Pr}[g(\mathbf{X}_2) = \alpha_{2,0}]$, where $\alpha_{2,0}, \alpha_{1,0}$ are the $0^{th}$ components of $\alpha_2$ and $\alpha_1$ respectively. We consider two cases based on linear independence of $\mathbf{X}_1$ and $\mathbf{X}_2$.

- **Case 1:** $\mathbf{X}_1$ and $\mathbf{X}_2$ are linearly dependent. Without loss of generality, let $\mathbf{X}_1 = (0, 0, 0, \ldots 0)$ and $\mathbf{X}_2 = (r_1, 0, 0, \ldots 0)$ for some $r_1 \in \mathbb{Z}_{p_0}$, represented as a word. From $g(\mathbf{X}_1)$ we can deduce $b^{(0)}$. However for $g(\mathbf{X}_2) = \alpha_{2,0}$ we require $a_1^{(0)} * r_1 + b^{(0)} = \alpha_{2,0} \mod p_0$. Using Fermat's Little Theorem, we know that there exists a unique $a_1^{(0)}$ for every $r_1$ that satisfies the above equation. Therefore, therefore $\mathsf{Pr}[g(\mathbf{X}_2) = \alpha_{2,0} | g(\mathbf{X}_1) = \alpha_{1,0}] = \mathsf{Pr}[g(\mathbf{X}_2) = \alpha_{2,0}] = \frac{1}{p_0}$.

- **Case 2:** $\mathbf{X}_1$ and $\mathbf{X}_2$ are linearly independent. Since $2^k < p_0$, every component of $\mathbf{X}_1$ and $\mathbf{X}_2$ (i.e. an element of $\{0,1\}^k$) can be treated as an element of $\mathbb{Z}_{p_0}$. The space $\{0,1\}^{n.k}$ can therefore be thought of as lying within the vector space $(\mathbb{Z}_{p_0})^n$, and any $\mathbf{X} \in \{0,1\}^{n.k}$ can be written as a linear combination of the set of basis vectors over $(\mathbb{Z}_{p_0})^n$. It is therefore sufficient to prove pairwise independence when $\mathbf{X}_1$ and $\mathbf{X}_2$ are basis vectors. Without loss of generality, let $\mathbf{X}_1 = (r_1, 0, 0, \dots 0)$ and $\mathbf{X}_2 = (0, r_2, 0, 0, \dots 0)$ for some $r_1, r_2 \in \mathbb{Z}_{p_0}$. From $g(\mathbf{X}_1)$, we can deduce $\left(a_1^{(0)} * r_1 + b^{(0)} = \alpha_{1,0}\right) \mod p_0$. But since $a_1^{(0)}$ is randomly chosen, therefore $\mathsf{Pr}[g(\mathbf{X}_2) = \alpha_{2,0} \mid g(\mathbf{X}_1) = \alpha_{1,0}] = \mathsf{Pr}[(a_2^{(0)} * r_2 + \alpha_{1,0} - a_1^{(0)} * r_1 = \alpha_{2,0}) \mod p_0] = \mathsf{Pr}[(a_2^{(0)} * r_2 - a_1^{(0)} * r_1 = \alpha_{2,0} - \alpha_{1,0}) \mod p_0]$, where $-a$ refers to the additive inverse of $a$ in the field $\mathbb{Z}_{p_0}$. Using Fermat's Little Theorem, we know that for every choice $a_1^{(0)}$ there exists a unique $a_2^{(0)}$ that satisfies the above requirement, given $\alpha_{1,0}$, $\alpha_{2,0}$, $r_1$ and $r_2$. Therefore $\mathsf{Pr}[g(\mathbf{X}_2) = \alpha_{2,0} \mid g(\mathbf{X}_1) = \alpha_{1,0}] = \frac{1}{p_0} = \mathsf{Pr}[g(\mathbf{X}_2) = \alpha_{2,0}]$.

□

### 6.2.1 Gaussian Elimination

The practical success of XOR-based bit-level hashing techniques for propositional model counting owes a lot to solvers like CryptoMiniSAT [143] that use Gaussian Elimination to efficiently reason about XOR constraints. It is significant that the constraints arising from $\mathcal{H}_{SMT}$ are linear modular equalities that also lend themselves to efficient Gaussian Elimination. We believe that integration of Gaussian Elimination engines in SMT solvers will significantly improve the performance of hashing-based word-level model counters.

## 6.3 Algorithm

We now present SMTApproxMC, a word-level hashing-based approximate model counting algorithm. SMTApproxMC takes as inputs a formula $F$ in the theory of fixed-width words, sampling set $S$, a tolerance $\varepsilon$ ($> 0$), and a confidence $1 - \delta \in (0, 1]$. It returns an estimate of $|R_{F\downarrow S}|$ within the tolerance $\varepsilon$, with confidence $1 - \delta$. The formula $F$ is assumed to have $n$ variables, each of width $k$, in its support. The central idea of SMTApproxMC is to randomly partition the solution space of $F$ into "small" cells of roughly the same size, using word-level hash functions from $\mathcal{H}_{SMT}(|S|, k, C)$, where $C$ is incrementally computed. The check for "small"-ness of cells is done using a word-level SMT solver. The use of word-level hash functions and a word-level SMT solver allows us to directly harness the power of SMT solving in model counting.

The pseudocode for SMTApproxMC is presented in Algorithm 7. Lines 1– 3 initialize the different parameters. Specifically, pivot determines the maximum size of a "small" cell as a function of $\varepsilon$, and $t$ determines the number of times SMTApproxMCCore must be invoked, as a function of $\delta$. The value of $t$ is determined by technical arguments in the proofs of our theoretical guarantees, and is not based on experimental observations Algorithm SMTApproxMCCore lies at the heart of SMTApproxMC. Each invocation of SMTApproxMCCore either returns an approximate model count of $F$, or $\perp$ (indicating a failure). In the former case, we collect the returned value, $m$, in a list $M$ in line 8. Finally, we compute the median of the approximate counts in $M$, and return this as FinalCount.

The pseudocode for SMTApproxMCCore is shown in Algorithm 8. This algorithm takes as inputs a word-level SMT formula $F$, a threshold pivot, and the width $k$ of words in $\mathsf{sup}(F)$. We assume access to a subroutine BoundedSMT that accepts a word-level SMT formula $\varphi$, sampling set $S$, and a threshold pivot as inputs, and returns pivot+1 solutions of $\varphi$ if $|R_{\varphi\downarrow S}| >$ pivot; otherwise it returns $R_\varphi$. In lines 1– 2 of Algorithm 8, we return the exact count if $|R_F| \leq$ pivot. Otherwise, we initialize $C$ by setting $C[0]$ to 0 and $C[1]$ to 1, where $C[i]$ in the pseudocode refers to $c_i$ in the previous section's discussion. This choice of initialization is motivated by our experimental observations. We also count the number of cells generated by an arbitrary

**Algorithm 7** SMTApproxMC$(F, \varepsilon, \delta, k)$

1: counter $\leftarrow 0$; $M \leftarrow$ emptyList;

2: pivot $\leftarrow 2 \times \lceil e^{-3/2} \left(1 + \frac{1}{\varepsilon}\right)^2 \rceil$;

3: $t \leftarrow \lceil 35 \log_2(3/\delta) \rceil$;

4: **repeat**

5: $\quad m \leftarrow$ SMTApproxMCCore$(F, \text{pivot}, k)$;

6: $\quad$ counter $\leftarrow$ counter $+ 1$;

7: $\quad$ **if** $m \neq \perp$ **then**

8: $\quad\quad$ AddToList$(M, m)$;

9: **until** (counter $< t$)

10: FinalCount $\leftarrow$ FindMedian$(M)$;

11: **return** FinalCount;

hash function from $\mathcal{H}_{SMT}(|S|, k, C)$ in numCells. The loop in lines 6–20 iteratively partitions $R_F$ into cells using randomly chosen hash functions from $\mathcal{H}_{SMT}(|S|, k, C)$. The value of $i$ in each iteration indicates the extent to which words in the support of $F$ are sliced when defining hash functions in $\mathcal{H}_{SMT}(|S|, k, C)$ – specifically, slices that are $\lceil k/2^i \rceil$-bits or more wide are used. The iterative partitioning of $R_F$ continues until a randomly chosen cell is found to be "small" (i.e. has $\geq 1$ and $\leq$ pivot solutions), or the number of cells exceeds $2^{n.k}$, rendering further partitioning meaningless. The random choice of $h$ and $\alpha$ in lines 7 and 8 ensures that we pick a random cell. It is important to note that while the choice of $h$ and $\alpha$ is random but not independent for different iterations of the loop. In particular, if $(h^1, \alpha^1)$, and $(h^2, \alpha^2)$ are choice of $h$ and $\alpha$ in two different iterations, then either $h^1$ is a prefix of $h^2$ and $\alpha^1$ is a prefix of $\alpha^2$ or $h^2$ is a prefix of $h^1$ and $\alpha^2$ is a prefix of $\alpha^1$. The call to BoundedSMT returns at most pivot + 1 solutions of $F$ within the chosen cell in the set $Y$. If $|Y| >$ pivot, the cell is deemed to be large, and the algorithm partitions each cell further into $p_i$ parts. This is done by incrementing $C[i]$ in line 11, so that the hash function chosen from $\mathcal{H}_{SMT}(|S|, k, C)$ in the next iteration of the loop generates $p_i$ times more cells than in the current iteration. On the other hand, if $Y$ is empty and $p_i > 2$, the cells are too small (and too many), and the algorithm reduces the number of cells by a factor of $p_{i+1}/p_i$ (recall $p_{i+1} \leq p_i$) by setting the values of $C[i]$ and $C[i+1]$ accordingly (see lines15 –17). If $Y$ is non-empty and has no more than pivot solutions, the cells are of the right size, and we return the estimate $|Y| \times$ numCells. In all other cases, SMTApproxMCCore fails and returns $\perp$.

**Algorithm 8** SMTApproxMCCore$(F, \text{pivot}, k)$

$Y \leftarrow \mathsf{BoundedSMT}(F, \text{pivot})$;
**if** $|Y| \leq \text{pivot})$ **then return** $|Y|$;
**else**
    $C \leftarrow \text{emptyVector}$; $C[0] \leftarrow 0$; $C[1] \leftarrow 1$;
    $i \leftarrow 1$; numCells $\leftarrow p_1$;
    **repeat**
        Choose $h$ at random from $\mathcal{H}_{SMT}(|S|, k, C)$;
        Choose $\alpha$ at random from $\prod_{j=0}^{i} \left(\mathbb{Z}_{p_j}\right)^{C[j]}$;
        $Y \leftarrow \mathsf{BoundedSMT}(F \wedge (h(\mathbf{X}) = \alpha), \text{pivot})$;
        **if** $(|Y| > \text{pivot})$ **then**
            $C[i] \leftarrow C[i] + 1$;
            numCells $\leftarrow$ numCells $\times\, p_i$;
        **if** $(|Y| = 0)$ **then**
            **if** $p_i > 2$ **then**
                $C[i] \leftarrow C[i] - 1$;
                $i \leftarrow i + 1$; $C[i] \leftarrow 1$;
                numCells $\leftarrow$ numCells $\times\, (p_{i+1}/p_i)$;
            **else**
                break;
    **until** $((0 < |Y| \leq \text{pivot})$ or $(\text{numCells} > 2^{n.k}))$
    **if** $((|Y| > \text{pivot})$ **or** $(|Y| = 0))$ **then return** $\perp$;
    **else return** $|Y| \times$ numCells;

## 6.4 Analysis of SMTApproxMC

Similar to the analysis of ApproxMC [38], the current theoretical analysis of SMTApproxMC assumes that for some $C$ during the execution of SMTApproxMCCore, $\log|R_F| - \log(\text{numCells}) + 1 = \log(\text{pivot})$. We leave analysis of SMTApproxMC without above assumption to future work.

For a given $h$ and $\alpha$, we use $R_{F,h,\alpha}$ to denote the set $R_F \cap h^{-1}(\alpha)$, i.e. the set of solutions of $F$ that map to $\alpha$ under $h$. Let $\mathsf{E}[Y]$ and $\mathsf{V}[Y]$ represent expectation and variance of a random variable $Y$ respectively. The analysis below focuses on the random variable $|R_{F,h,\alpha}|$ defined for a chosen $\alpha$. We use $\mu$ to denote the expected value of the random variable $|R_{F,h,\alpha}|$ whenever $h$ and $\alpha$ are clear from the context. The following lemma based on pairwise independence of $\mathcal{H}_{SMT}(|S|, k, C)$ is key to our analysis.

**Lemma 19.** *The random choice of $h$ and $\alpha$ in* SMTApproxMCCore *ensures that for each $\varepsilon > 0$, we have* $\mathsf{Pr}\left[(1 - \frac{\varepsilon}{1+\varepsilon})\mu \leq |R_{F,h,\alpha}| \leq (1 + \frac{\varepsilon}{1+\varepsilon})\mu\right] \geq 1 - \frac{(1+\varepsilon)^2}{\varepsilon^2 \quad \mu}$*, where* $\mu = \mathsf{E}[|R_{F,h,\alpha}|]$

*Proof.* For every $y \in \{0,1\}^{n.k}$ and for every $\alpha \in \prod_{i=0}^{|C|-1}(\mathbb{Z}_{p_i})^{C[i]}$, define an indicator variable $\gamma_{y,\alpha}$ as follows: $\gamma_{y,\alpha} = 1$ if $h(y) = \alpha$, and $\gamma_{y,\alpha} = 0$ otherwise. Let us fix $\alpha$ and $y$ and choose $h$ uniformly at random from $\mathcal{H}_{SMT}(|S|, k, C)$. The 2-wise independence $\mathcal{H}_{SMT}(|S|, k, C)$ implies that for every distinct $y_1, y_2 \in R_F$, the random variables $\gamma_{y_1}, \gamma_{y_2}$ are 2-wise independent. Let $|R_{F,h,\alpha}| = \sum_{y \in R_F} \gamma_{y,\alpha}$, $\mu = \mathsf{E}\left[|R_{F,h,\alpha}|\right]$ and $\mathsf{V}[|R_{F,h,\alpha}|] = \mathsf{V}[\sum_{y \in R_F} \gamma_{y,\alpha}]$. The pairwise independence of $\gamma_{y,\alpha}$ ensures that $\mathsf{V}[|R_{F,h,\alpha}|] = \sum_{y \in R_F} \mathsf{V}[\gamma_{y,\alpha}] \leq \mu$. The result then follows from Chebyshev's inequality. □

Let $Y$ be the set returned by $\mathsf{BoundedSMT}(F \wedge (h(\mathbf{X}) = \alpha), \text{pivot})$ where pivot is as calculated in Algorithm 7.

**Lemma 20.** *Let an invocation of* SMTApproxMCCore *from* SMTApproxMC *return* $m$. *Then* $\mathsf{Pr}\left[(1+\varepsilon)^{-1}|R_F| \leq m \leq (1+\varepsilon)|R_F|\right] \geq 0.6$

*Proof.* For notational convenience, we use $(\text{numCells}_l)$ to denote the value of numCells when $i = l$ in the loop in SMTApproxMCCore. As noted earlier, we assume, for some $i = \ell^*$, $\log|R_F| - \log(\text{numCells}_{\ell^*}) + 1 = \log(\text{pivot})$. Furthermore, note that for all $i \neq j$ and $\text{numCells}_i > \text{numCells}_j$, $\text{numCells}_i/\text{numCells}_j \geq 2$. Let $F_l$ denote the event that $|Y| < \text{pivot}$ and $(|Y| > (1+\varepsilon)|R_F| \vee |Y| < \frac{|R_F|}{(1+\varepsilon)})$ for $i = l$. Let $\ell_1$ be the value of $i$ such that $\text{numCells}_{\ell_1} < \text{numCells}_{\ell^*}/2 \wedge \forall j, \text{numCells}_j < \text{numCells}_{\ell^*}/2 \implies \text{numCells}_{\ell_1} \geq \text{numCells}_j$. Similarly, let $\ell_2$ be the value of $i$ such that $\text{numCells}_{\ell_2} < \text{numCells}_{\ell^*}/4 \wedge \forall j, \text{numCells}_j < \text{numCells}_{\ell^*}/4 \implies \text{numCells}_{\ell_2} \geq \text{numCells}_j$

Then, $\forall_{i|\text{numCells}_i < \text{numCells}_{\ell^*}/4}$, $F_i \subseteq F_{\ell_2}$. Therefore, the probability of $\mathsf{Pr}\left[(1+\varepsilon)^{-1}|R_F| \leq m \leq (1+\varepsilon)|R_F|\right]$ is at least $1 - \mathsf{Pr}[F_{\ell_2}] - \mathsf{Pr}[F_{\ell_1}] - \mathsf{Pr}[F_{\ell^*}] = 1 - \frac{e^{-3/2}}{4} - \frac{e^{-3/2}}{2} - e^{-3/2} \geq 0.6$ .

□

Now, we apply standard combinatorial analysis on repetition of probabilistic events and prove that SMTApproxMC is $(\varepsilon, \delta)$ model counter.

**Theorem 21.** *Suppose an invocation of* $\mathsf{SMTApproxMC}(F, \varepsilon, \delta, k)$ *returns* FinalCount. *Then* $\mathsf{Pr}\left[(1+\varepsilon)^{-1}|R_F| \leq \text{FinalCount} \leq (1+\varepsilon)|R_F|\right] \geq 1 - \delta$

*Proof.* Throughout this proof, we assume that SMTApproxMCCore is invoked $t$ times from SMTApproxMC, where $t = \lceil 35\log_2(3/\delta)\rceil$ in Section 6.3). Referring to the pseudocode of SMTApproxMC, the final count returned by SMTApproxMC is the median of non-$\bot$ counts obtained from the $t$ invocations of SMTApproxMCCore. Let $Err$ denote the event that the median is not in $[(1+\varepsilon)^{-1} \cdot |R_F|, (1+\varepsilon) \cdot |R_F|]$. Let "$\#non\bot = q$" denote the event that $q$ (out of $t$) values returned by SMTApproxMCCore are non-$\bot$. Then, $\mathsf{Pr}\left[Err\right] = \sum_{q=0}^{t} \mathsf{Pr}\left[Err \mid \#non\bot = q\right] \cdot \mathsf{Pr}\left[\#non\bot = q\right]$.

In order to obtain $\mathsf{Pr}\left[Err \mid \#non\bot = q\right]$, we define a 0-1 random variable $Z_i$, for $1 \leq i \leq t$, as follows. If the $i^{th}$ invocation of SMTApproxMCCore returns $c$, and if $c$ is either $\bot$ or a non-$\bot$ value that does not lie in the interval $[(1+\varepsilon)^{-1} \cdot |R_F|, (1+\varepsilon) \cdot |R_F|]$, we set $Z_i$ to 1; otherwise, we set it to 0. From Lemma 20, $\mathsf{Pr}\left[Z_i = 1\right] = p < 0.4$. If $Z$ denotes $\sum_{i=1}^{t} Z_i$, a necessary (but not sufficient) condition for event $Err$ to occur, given that $q$ non-$\bot$s were returned by SMTApproxMCCore, is $Z \geq (t - q + \lceil q/2 \rceil)$. To see why this is so, note that $t-q$ invocations of SMTApproxMCCore must return $\bot$. In addition, at least $\lceil q/2 \rceil$ of the remaining $q$ invocations must return values outside the desired interval. To simplify the exposition, let $q$ be an even integer. A more careful analysis removes this restriction and results in an additional constant scaling factor for $\mathsf{Pr}\left[Err\right]$. With our simplifying assumption, $\mathsf{Pr}\left[Err \mid \#non\bot = q\right] \leq \mathsf{Pr}[Z \geq (t - q + q/2)] = \eta(t, t - q/2, p)$. Since $\eta(t, m, p)$ is a decreasing function of $m$ and since $q/2 \leq t - q/2 \leq t$, we have $\mathsf{Pr}\left[Err \mid \#non\bot = q\right] \leq \eta(t, t/2, p)$. If $p < 1/2$, it is easy to verify that $\eta(t, t/2, p)$ is an increasing function of $p$. In our case, $p < 0.4$; hence, $\mathsf{Pr}\left[Err \mid \#non\bot = q\right] \leq \eta(t, t/2, 0.4)$.

It follows from above that $\mathsf{Pr}\left[Err\right] = \sum_{q=0}^{t} \mathsf{Pr}\left[Err \mid \#non\bot = q\right] \cdot \mathsf{Pr}\left[\#non\bot = q\right] \leq \eta(t, t/2, 0.4) \cdot \sum_{q=0}^{t} \mathsf{Pr}\left[\#non\bot = q\right] = \eta(t, t/2, 0.4)$. Since $\binom{t}{t/2} \geq \binom{t}{k}$ for all $t/2 \leq k \leq t$, and since $\binom{t}{t/2} \leq 2^t$, we have $\eta(t, t/2, 0.4) = \sum_{k=t/2}^{t} \binom{t}{k} (0.4)^k (0.6)^{t-k} \leq \binom{t}{t/2} \sum_{k=t/2}^{t} (0.4)^k (0.6)^{t-k} \leq 2^t \sum_{k=t/2}^{t} (0.6)^t (0.4/0.6)^k \leq 2^t \cdot 3 \cdot (0.6 \times 0.4)^{t/2} \leq 3 \cdot (0.98)^t$. Since $t = \lceil 35 \log_2(3/\delta) \rceil$, it follows that $\mathsf{Pr}\left[Err\right] \leq \delta$. □

**Theorem 22.** SMTApproxMC$(F, \varepsilon, \delta, k)$ *runs in time polynomial in* $|F|$, $1/\varepsilon$ *and* $\log_2(1/\delta)$ *relative to an* NP-*oracle.*

*Proof.* Referring to the pseudocode for SMTApproxMC, lines 1– 3 take time no more than a polynomial in $\log_2(1/\delta)$ and $1/\varepsilon$. The repeat-until loop in lines 4– 9 is repeated $t = \lceil 35 \log_2(3/\delta) \rceil$ times. The time taken for each iteration is dominated by the time taken by SMTApproxMCCore. Finally, computing the median in line 10 takes time linear in $t$. The proof is therefore completed by showing that SMTApproxMCCore takes time polynomial in $|F|$ and $1/\varepsilon$ relative to the SAT oracle.

Referring to the pseudocode for SMTApproxMCCore, we find that BoundedSMT is called $\mathcal{O}(|F|)$ times. Each such call can be implemented by at most $pivot + 1$ calls to a NP oracle (SMT solver in case), and takes time polynomial in $|F|$ and $pivot + 1$ relative to the oracle. Since $pivot + 1$ is in $\mathcal{O}(1/\varepsilon^2)$, the number of calls to the NP oracle, and the total time taken by all calls to BoundedSMT in each invocation of SMTApproxMCCore is a polynomial in $|F|$ and $1/\varepsilon$ relative to the oracle. The random choices in lines 7 and 8 of SMTApproxMCCore can be implemented in time polynomial in $|S|.k$ (hence, in $|F|$) if we have access to a source of random bits. Constructing $F \wedge (h(\mathbf{X}) = \alpha)$ in line 9 can also be done in time polynomial in $|F|$. □

## 6.5 Experimental Methodology and Results

To evaluate the performance and effectiveness of SMTApproxMC, we built a prototype implementation and conducted extensive experiments. Our suite of benchmarks consisted of more than 150 problems arising from diverse domains such as reasoning about circuits, planning, program synthesis and the like.

For purposes of comparison, we also implemented a state-of-the-art bit-level hashing-based approximate model counting algorithm for bounded integers, proposed by [43]. Henceforth, we refer to this algorithm as CDM, after the authors' initials. Both model counters used an overall timeout of 12 hours per benchmark, and a BoundedSMT timeout of 2400 seconds per call. Both used Boolector, a state-of-the-art SMT solver for fixed-width words [28]. Note that Boolector (and other popular SMT solvers for fixed-width words) does not yet implement Gaussian elimination for linear modular equalities; hence our experiments did not enjoy the benefits of Gaussian elimination. We employed the Mersenne Twister to generate pseudo-random numbers, and each thread was seeded independently using the Python `random` library. All experiments used $\varepsilon = 0.8$ and $\delta = 0.2$. Similar to ApproxMC, we determined value of $t$ based on tighter analysis offered by proofs. For detailed discussion, we refer the reader to Section 6 in [38]. Every experiment was conducted on a single core of high-performance computer cluster, where each node had a 20-core, 2.20 GHz Intel Xeon processor, with 3.2GB of main memory per core.

| Benchmark | Total Bits | Variable Types | # of Operations | SMTApproxMC time(s) | CDM time(s) |
|---|---|---|---|---|---|
| squaring27 | 59 | {1: 11, 16: 3} | 10 | – | 2998.97 |
| squaring51 | 40 | {1: 32, 4: 2} | 7 | 3285.52 | 607.22 |
| 1160877 | 32 | {8: 2, 16: 1} | 8 | 2.57 | 44.01 |
| 1160530 | 32 | {8: 2, 16: 1} | 12 | 2.01 | 43.28 |
| 1159005 | 64 | {8: 4, 32: 1} | 213 | 28.88 | 105.6 |
| 1160300 | 64 | {8: 4, 32: 1} | 1183 | 44.02 | 71.16 |
| 1159391 | 64 | {8: 4, 32: 1} | 681 | 57.03 | 91.62 |
| 1159520 | 64 | {8: 4, 32: 1} | 1388 | 114.53 | 155.09 |
| 1159708 | 64 | {8: 4, 32: 1} | 12 | 14793.93 | – |
| 1159472 | 64 | {8: 4, 32: 1} | 8 | 16308.82 | – |
| 1159115 | 64 | {8: 4, 32: 1} | 12 | 23984.55 | – |
| 1159431 | 64 | {8: 4, 32: 1} | 12 | 36406.4 | – |
| 1160191 | 64 | {8: 4, 32: 1} | 12 | 40166.1 | – |

Table 6.1 : Runtime performance of SMTApproxMC vis-a-vis CDM

We sought answers to the following questions from our experimental evaluation:

1. How does the performance of SMTApproxMC compare with that of a bit-level

hashing-based counter like CDM?

2. How do the approximate counts returned by SMTApproxMC compare with exact counts?

Our experiments show that SMTApproxMC significantly outperforms CDM for a large class of benchmarks. Furthermore, the counts returned by SMTApproxMC are highly accurate and the observed geometric tolerance($\varepsilon_{obs}$) = 0.04.

### 6.5.1 Performance Comparison

Table 6.1 presents the result of comparing the performance of SMTApproxMC vis-a-vis CDM on a subset of our benchmarks*. In Table 6.1, column 1 gives the benchmark identifier, column 2 gives the sum of widths of all variables, column 3 lists the number of variables (numVars) for each corresponding width (w) in the format {w : numVars}. To indicate the complexity of the input formula, we present the number of operations in the original SMT formula in column 4. The runtimes for SMTApproxMC and CDM are presented in columns 5 and column 6 respectively. We use "–" to denote timeout after 12 hours. Table 6.1 clearly shows that SMTApproxMC significantly outperforms CDM (often by 2-10 times) for a large class of benchmarks. In particular, we observe that SMTApproxMC is able to compute counts for several cases where CDM times out.

Benchmarks in our suite exhibit significant heterogeneity in the widths of words, and also in the kinds of word-level operations used. Propositionalizing all word-level variables eagerly, as is done in CDM, prevents the SMT solver from making full use of word-level reasoning. In contrast, our approach allows the power of word-level reasoning to be harnessed if the original formula $F$ and the hash functions are such

*An extended version of Table 6.1 is available in Appendix as Table A3

that the SMT solver can reason about them without bit-blasting. This can lead to significant performance improvements, as seen in Table 6.1. Some benchmarks, however, have heterogenous bit-widths and heavy usage of operators like $\mathsf{extract}(x, n_1, n_2)$ and/or word-level multiplication. It is known that word-level reasoning in modern SMT solvers is not very effective for such cases, and the solver has to resort to bit-blasting. Therefore, using word-level hash functions does not help in such cases. We believe this contributes to the degraded performance of SMTApproxMC vis-a-vis CDM in a subset of our benchmarks. This also points to an interesting direction of future research: to find the right hash function for a benchmark by utilizing SMT solver's architecture.

### 6.5.2 Quality of Approximation

To measure the quality of the counts returned by SMTApproxMC, we selected a subset of benchmarks that were small enough to be bit-blasted. We set the sampling set was set to all the variables in the formula and fed to sharpSAT [147] – a state-of-the-art exact model counter. Figure 6.1 compares the model counts computed by SMTApproxMC with the bounds obtained by scaling the exact counts (from sharpSAT) with the tolerance factor ($\varepsilon = 0.8$). The y-axis represents model counts on log-scale while the x-axis presents benchmarks ordered in ascending order of model counts. We observe that for *all* the benchmarks, SMTApproxMC computes counts within the tolerance. Furthermore, for each instance, we computed observed tolerance ( $\varepsilon_{obs}$) as $\frac{\text{count}}{|R_F|} - 1$, if count $\geq |R_F|$, and $\frac{|R_F|}{\text{count}} - 1$ otherwise, where $|R_F|$ is computed by sharpSAT and count is computed by SMTApproxMC. We observe that the geometric mean of $\varepsilon_{obs}$ across all the benchmarks is only 0.04 – far less (i.e. closer to the exact count) than the theoretical guarantee of 0.8.

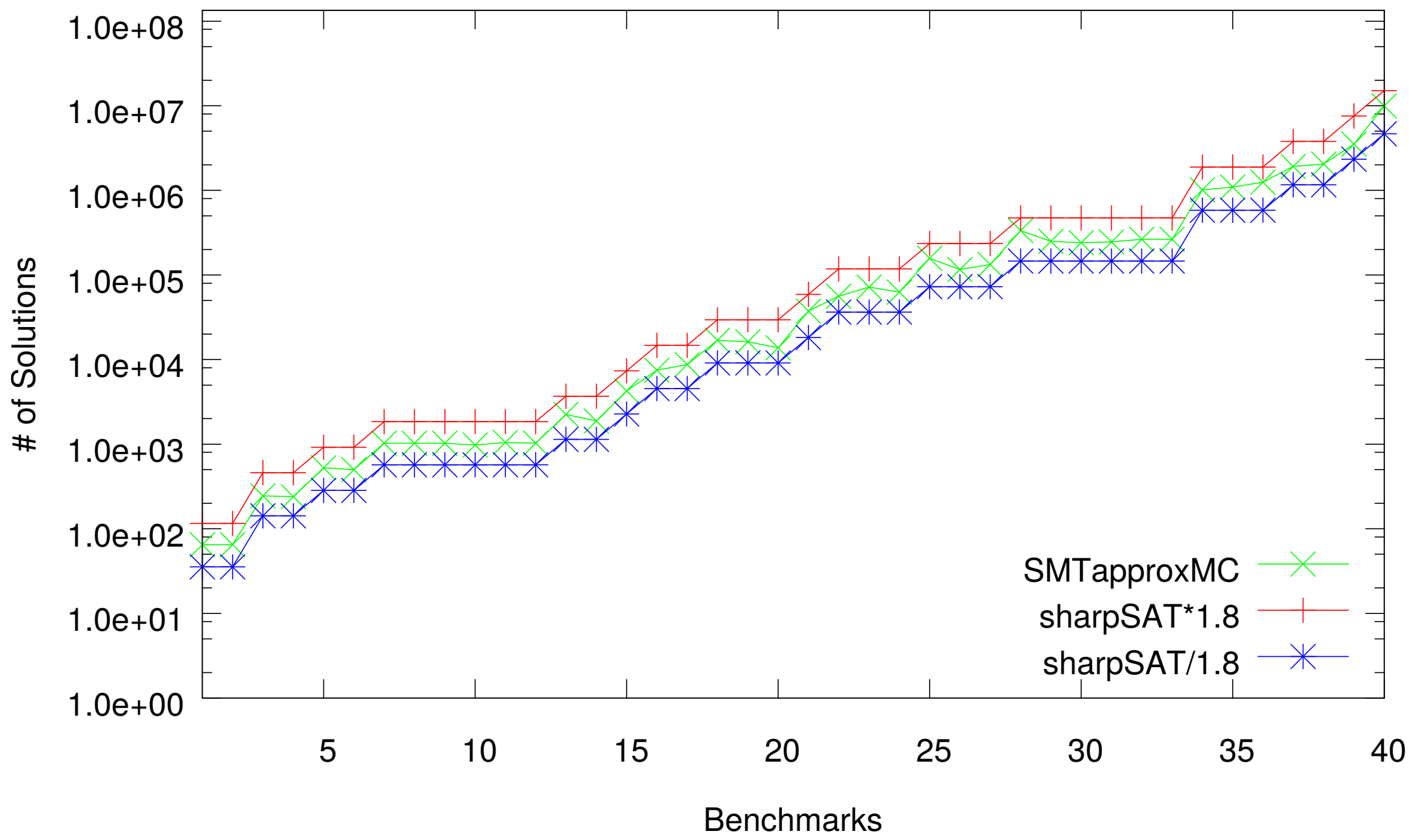

Figure 6.1 : Quality of counts computed by SMTApproxMC vis-a-vis exact counts

## 6.6 Chapter Summary

Hashing-based constrained counting has emerged as a promising approach for probabilistic inference on graphical models. While real-world examples naturally have word-level constraints, state-of-the-art approximate model counters effectively reduce the problem to propositional model counting due to lack of non-bit-level hash functions. In this work, we presented, $\mathcal{H}_{SMT}$, a word-level hash function and used it to build SMTApproxMC, an approximate word-level model counter. Our experiments show that SMTApproxMC can significantly outperform techniques based on bit-level hashing.

Our study also presents interesting directions for future work. For example, the performance of SMTApproxMC seems to be closely related to how the SMT solver handles the original constraint and the hashing constraints. adapting SMTApproxMC to be aware of SMT solving strategies, and augmenting SMT solving strategies to efficiently reason about hash functions used in counting, are exciting directions of future work. Second, the performance of SMTApproxMC is expected to improve significantly with the integration of Gaussian Elimination in SMT solvers.

Our work goes beyond serving as a replacement for other approximate counting techniques. SMTApproxMC can also be viewed as an efficient building block for more sophisticated inference algorithms [25]. The development of SMT solvers has so far been primarily driven by the verification and static analysis communities. Our work hints that probabilistic inference could well be another driver for SMT solver technology development.

# Chapter 7

# Case Study: Reliability for Power-Transmission Networks

Modern society is increasingly reliant on the availability of critical facilities and utility services, such as power, telecommunications, water, gas, and transportation among others [146]. To ensure adequate service, it is imperative to quantify system reliability, or the probability of the system to remain functional, as well as system resilience, or the ability of the system to quickly return to normalcy when failure is unavoidable [29]. While resilience assessment requires human decision making principles, it also heavily depends on intrinsic system reliability. Hence, the recent focus on community resilience and sustainability has spurred significant activity in engineering reliability [165].

One of the key challenging problems in the area of engineering reliability is network reliability, wherein the input to the problem consists of a network, represented as a graph, arising out of distribution of water, power, transportation routes and the like. The problem of the network reliability seeks to measure the likelihood of two points of interest being reachable under conditions such as natural disasters. Early theoretical investigations showed that the problem of network reliability is #P complete [150]. Although graph contraction strategies combined with DNF counting provide a Fully Polynomial Randomized Approximation Scheme (FPRAS) with error guarantees [103], implementation on practical systems does not scale well due to the requirement of a large number of Monte Carlo steps. Consequently, recent investi-gations have focused on advancing algorithmic strategies that build upon advanced Monte Carlo simulation [166] and analytical approaches [117, 63]. In addition, inventive sampling methods, such as line sampling and variance reduction schemes [76], along with graphical models, especially Bayesian networks, provide versatile strategies to quantify the reliability of complex engineered systems and their dynamics [21].

Despite significant progress, most techniques remain computationally expensive. As an alternative, when invoking approximations, most methods are unable to guarantee the quality of the reliability estimation a priori, barring small instances where exact methods do not time out. Therefore, design of techniques that offer strong theoretical guarantees on the quality of estimates and can scale to large real world instances remains an unattained goal across multiple disciplines.

A promising alternative approach to answer #P queries is to reduce a #P problem to a #SAT problem, where #SAT denotes the problem of computing the number of solutions for a given SAT formula. This motivates us to ask: *Can we design a counting-based framework that can take advantage of progress in hashing-based techniques in this thesis to provide theoretically sound estimates for the network reliability problem?*

In this Chapter, we provide a positive answer to the above question. We present a counting-based framework, called RelNet, that reduces the problem of computing reliability for a given network to counting the number of satisfying assignments of a $\Sigma_1^1$ formula, which is amenable to recent hashing-based techniques developed for counting satisfying assignments of SAT formula. RelNet significantly outperforms state of the art techniques and in particular, allowed us to obtain the first theoretically sound estimates of reliability for ten networks representing different cities in the U.S.

## 7.1 Preliminaries

For a set $A$, $\bar{A}$ denotes the complement of the set $A$.Let $G = (V, E)$ be a graph, where $V$ is set of the vertices, also referred as nodes, and $E$ is set of edges. For every edge $e \in E$ from $u$ to $v$, we define $\mathsf{start}(e) = u$ and $\mathsf{end}(e) = v$. Note that we allow multiple edges between pairs of nodes.

We say that $\pi = (u, w_1, \cdots w_{k-1}, v)$ is a path of length $k$ that connects $u$ and $v$ if $\forall i < k-1, w_i \in V$ and $\exists e\ (u = \mathsf{start}(e) \wedge w_1 = \mathsf{end}(e)) \wedge$ $\exists e\ (w_{k-1} = \mathsf{start}(e) \wedge v = \mathsf{end}(e)) \wedge$ $\forall i < k-2, \exists e\ (w_i = \mathsf{start}(e) \wedge w_{i+1} = \mathsf{end}(e))$. We use $T_\pi$ to denote set of all edges in $\pi$. For every subset $\sigma \subseteq E$, we say $u$ and $v$ are connected under $\sigma$, denoted by $(u, v) \models \sigma$, if $\exists \pi, k$ such that $\pi$ is a path of length $k$ that connects $u$ and $v$ and $T_\pi \subseteq \sigma$. For a given graph $G$, we use $\Gamma_{G,u,v}$ to denote the set of all subsets $\sigma$ of $E$ that make $u$ and $v$ connected, i.e $\Gamma_{G,u,v} = \{\sigma \subseteq E | (u, v) \models \sigma\}$.

For a given graph $G = (V, E)$ and nodes $u$ and $v$, we use $e(u, v) \cup G$ to denote the augmented graph $G'$ obtained by putting an edge $e$ such that $u = \mathsf{start}(e)$ and $v = \mathsf{end}(e)$. Note that if $G$ has $i$ edges from $u$ to $v$, then $G'$ has $i + 1$ edges from $u$ to $v$. In this Chapter, we focus on probabilistic variant of graphs, where probability function is associated to edges in $E$. For every edge $e \in E$, we use $e^1$ to denote the event that edge $e$ does not fail and $e^0$ to denote the the event that edge $e$ fails. We have $\mathsf{Pr}[e^0] + \mathsf{Pr}[e^1] = 1$. As discussed in Section 7.3, the failure of edge corresponds to event in real life when an existing edge is broken due to events such as natural disasters. We assume all $e_i^1$ to be independent. Without loss of generality, the least significant bit in the representation of $\mathsf{Pr}[e_i^1]$ is always taken to be 1. We call a graph as unweighted if for all edges $e \in E$, we have $\mathsf{Pr}[e^0] = 1/2$, otherwise the graph is called weighted. Therefore for $\sigma \subseteq E$, $\mathsf{Pr}[\sigma] = \prod_{e_i \in \sigma} \mathsf{Pr}(e_i^1) \times \prod_{e_j \notin \sigma} \mathsf{Pr}(e_j^0)$. Furthermore, we have $\mathsf{Pr}[\Gamma_{G,u,v}] = \sum_{\sigma \in \Gamma_{G,u,v}} \mathsf{Pr}[\sigma]$. For a given graph $G$, source node $u$ and terminal node $v$, the reliability of $u \to v$ is defined as $\mathsf{Pr}[\Gamma_{G,u,v}]$. In this Chapter, we consider the problem of estimating $r(u, v) = 1 - \mathsf{Pr}[\Gamma_{G,u,v}]$

In this chapter, we employ a special class of graphs, called chain graphs, which are inspired from the work on chain formulas discussed in Chapter 5. Similar to every edge, every chain graph has start and end node defined as follows. Every edge $e$ is a chain graph, say $G$, such that $u = \mathsf{start}(G)$ if $u = \mathsf{start}(e)$ and $v = \mathsf{end}(G)$ if $v = \mathsf{end}(e)$, and we represented $G$ as $G := (u \vee v)$. In addition, if $G = (V, E)$ is a chain graph and $e$ is an edge such that (i) $u = \mathsf{start}(e) = \mathsf{start}(G) \in V$ and $v = \mathsf{end}(e) = \mathsf{end}(G) \in V$, we say that $e \cup G$ is a chain formula, represented by $(u \vee G)$ or (ii) $u = \mathsf{start}(e) \notin V$ and $v = \mathsf{end}(e) = \mathsf{start}(G) \in V$, then $e \cup G$ is a chain formula, represented by $(u \wedge G)$. Every chain graph $G$ over nodes $a_1, a_2, ...a_m$ and $n$ edges can be represented as $(b_1 C_1 (b_2 C_2 (\cdots (b_n C_n b_{n+1}) \cdots)))$, where $C_i = \vee$ or $\wedge$ and performing a many to one mapping from $\{b_1, \cdots b_{n+1}\}$ to $\{a_1, a_2, \cdots a_m\}$ such that (i) $b_1 \mapsto a_1 \wedge b_{n+1} \mapsto a_m$, and (ii) $\forall i < m-1, b_i \mapsto a_j \wedge b_{i+1} \mapsto a_l \to j < l$ if $C_i = \wedge$ and $j = l$, otherwise.

## 7.2 Prior Work

The problem of computing $r(u, v)$ for a given graph $G$ was shown to be #P-complete by Valiant [150]. Consequently, there has been focus on development of approximate techniques for $r(u, v)$. In his seminal Chapter, Karger [103] provided the first Fully Polynomial Randomized Approximation Scheme (FPRAS) such that returned estimate satisfies $(\varepsilon, \delta)$ guarantees while the runtime of algorithm (referred as Karger's algorithm in rest of the Chapter) is polynomial in the $|G|$, $\log(1/\delta)$, $1/\varepsilon$. Our experiments demonstrate that the high requirement of Monte Carlo samples in the above algorithm is a major bottleneck and for our benchmarks, Karger's algorithm times out.

The recent investigations into network reliability have focused on advancing algorithmic strategies that build upon advanced Monte Carlo simulation [166] and analytical approaches [117, 63]. In particular, statistical learning techniques when combined with numerical simulation afford the reliability assessment of complex engineered systems, while unraveling component importance and sensitivities [95]. Also, successful strategies in data science, such as hierarchical clustering, provide novel tools for reliability and risk assessment [160, 89]. Also, state space partition strategies and optimization allow for analytical modeling of system reliability, which also offers, as a by-product, insights on the geometry of the failure space [8, 61]. Classical universal generating functions but combined with optimization also offer fresh alternatives to quantify system reliability approximately [41]. Besides, inventive sampling methods, such as line sampling and variance reduction schemes [76], along with graphical models, especially Bayesian networks, provide versatile strategies to quantify the reliability of complex engineered systems and their dynamics [21].

With the advent of resilience engineering, analytical methods are highly regarded in engineering reliability as they provide accurate estimates or, in more challenging instances, they yield lower and upper bounded estimates with 100% confidence. Furthermore, we can classify analytical network reliability methods in two groups based on their algorithmic approach. The first uses prior enumeration of cut sets (or path sets) or boolean algebra to account for non-disjoint events [7, 5], whereas the latter uses recursive or iterative decompositions of disjoint events [61, 134, 130]. The latter group has proven more practical due to its online decomposition capabilities while not relying on the prior cut (or path) set enumeration and applications of the inclusion-exclusion principle, both NP-hard problems. In particular, research that builds upon the work by Dotson et al. has found wide technical application for medium-size networks [113, 117] and in this Chapter we use the Selective path based Recursive Decomposition Algorithm (S-RDA) as a representative approach of state-of-the-art analytical reliability methods for civil infrastructure systems. Herein, we refer to the gap between upper and lower bound estimates of reliability as the gap error. S-RDA aims at shrinking the gap error as much as possible by finding disjoint path sets that contain the shortest path of maximum likelihood at every decomposition step while prioritizing partitioning subsets of larger likelihood as well allowing it to provide anytime approximation guarantee.

## 7.3 Datasets

In this Chapter we use as benchmark 10 power-transmission networks powering small to medium size cities in the states of Texas (TX), Florida (FL), California (CA), Tennessee (TN), Georgia (GA), and South Carolina(SC). Such states are susceptible to extreme natural disasters such as flooding, hurricanes, or earthquakes. These cities have populations in the order of tens to hundreds of thousands and the grids connect generators and substations with 110-765 kV transmission-level power lines. Also, as shown in Table 7.3, networks' size go from 47 to 112 nodes and the number of edges are of the same order. The raw network data was obtained in GIS format from the "Platts" repository for maps and geospatial data *.

Transmission-line outages due to random failures are not uncommon in power transmission systems during regular operation. The annualized probability of such failures depends on technical characteristics such as length of lines, supply/demand, temperature, etc. Typical values for ten-hour line outages, based on their annual occurrence rate, range from 60% to 98% for lines of length 50 and 200 kilometers respectively [23]. Although these values may appear high, such contingencies can be managed relatively easily. In contrast, extensive and complete damage due to natural disasters have smaller occurrence probabilities but are much more difficult to manage due to increased time of repairs. Even though the likelihood of such extreme natural events is small, conditioned on their occurrence, the probability of failures with significant damage for power transmission lines and facilities can be much larger as is typically depicted in fragility curves that encode probabilities of failure conditioned on some hazard intensity level (Fig. 7.1, source: [93]). For our experiments, we consider failure probability of 0.125 – a value that is attainable in practice by wide range of extreme natural events.

*http://www.platts.com/products/gis-data.

| Index | City Name | $\|V\|$ | $\|E\|$ |
|---|---|---|---|
| G1 | Amarillo, TX | 47 | 62 |
| G2 | Lakeland, FL | 50 | 69 |
| G3 | El Paso, TX | 52 | 65 |
| G4 | San Luis Obispo, CA | 57 | 69 |
| G5 | Eureka, CA | 61 | 70 |
| G6 | Bulls Gap, TN | 62 | 91 |
| G8 | Memphis, TN | 66 | 83 |
| G12 | Lubbock, TX | 85 | 106 |
| G22 | Athens, GA | 103 | 116 |
| G27 | Sumter, SC | 112 | 139 |

Table 7.1 : Test power networks.

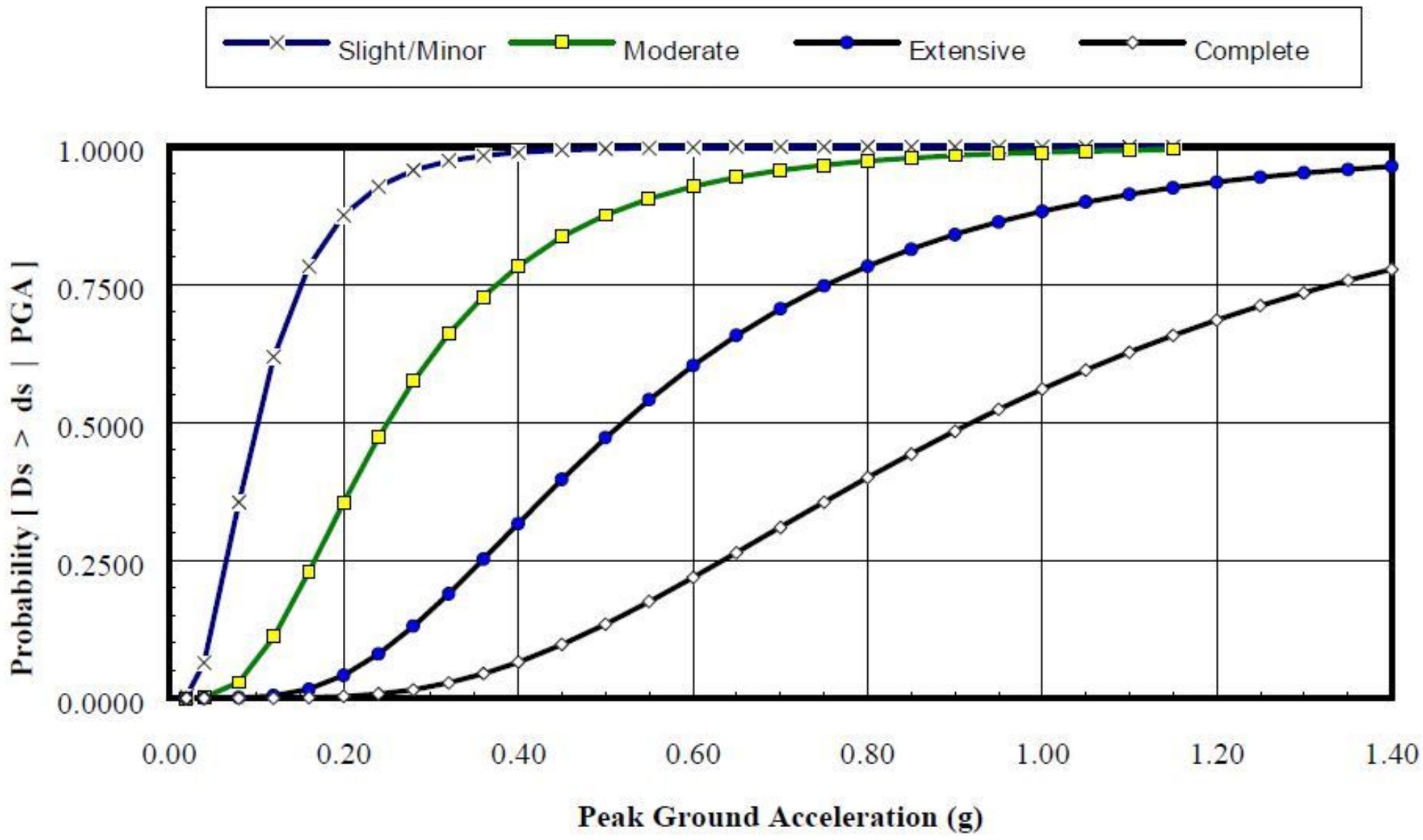

Figure 7.1 : Probability of exceeding a given damage state (DS) for Medium/Large Generation Facilities with Anchored Components as a function of the peak ground acceleration intensity after an earthquake.

## 7.4 From Network Reliability to Constrained Counting

In this section, we first discuss how weighted graphs can be reduced to unweighted graphs. We then discuss how the problem of computing reliability for an unweighted graph can be reduced to constrained unweighted counting. We then discuss our proposed framework, RelNet, that combines the two reductions and employs hashing-based techniques to compute reliability for arbitrary graphs.

### 7.4.1 From Weighted to Unweighted Graph

The central idea of our reduction is usage of chain graphs to represent weights, which is closely related to usage of chain formulas for weighted counting(c.f., Chapter 5). Let $m > 0$ be a natural number, and $k < 2^m$ be a positive odd number. Let $c_1 c_2 \cdots c_m$ be the $m$-bit binary representation of $k$, where $c_m$ is the least significant bit. Let $z$ be the number of zeros in the representation of $k$. Define

$$\psi_{k,m}(b_1, \cdots b_{m+1}) = (b_1 C_1 (b_2 C_2 \cdots (b_m C_m b_{m+1}) \cdots))$$

where $C_i = \vee$ if $c_i = 1$ and $\wedge$ otherwise. We now construct chain graph $\phi_{k,m}(a_1, \cdots a_{z+2})$ by performing a many to one mapping between $\{b_1, \cdots b_{m+1}\}$ and $\{a_1, a_2, \cdots a_{z+2}\}$ such that (i) $b_1 \mapsto a_1 \wedge b_{m+1} \mapsto a_{z+2}$, and (ii) $\forall i < m-1, (b_i \mapsto a_j \wedge b_{i+1} \mapsto a_l) \rightarrow j < l$ if $C_i = \wedge$ and $j = l$, otherwise. Note that there is one to one correspondence between $\psi_{k,m}(b_1, \cdots b_{m+1})$ and $\phi_{k,m}(a_1, \cdots a_{z+2})$ For example, consider $k = 3$ and $m = 3$. The binary representation of 3 using 3 bits is 011 and $z = 1$. Therefore, we have $\psi_{3,3}(b_1, b_2, b_3, b_4) = ( b_1 \ \wedge \ (b_2 \vee \ (b_3 \vee \ b_4))))$, which gives us $\varphi_{3,3}(a_1, a_2, a_3) = (a_1 \ \wedge \ (a_2 \vee (a_2 \vee a_3))))$. We now first show that $|\varphi_{k,m}|$ is of linear size and then discuss the relationship between $k$, $m$ and $\Gamma_{G,a_1,a_{z+2}}$.

**Lemma 23.** *Let $m > 0$ be a natural number, $k < 2^m$ , $z$ and $\varphi_{k,m}$ as defined above. Then $|\varphi_{k,m}|$ is linear in $m$. Furthermore $|\Gamma_{\varphi_{k,m},a_1,a_{z+2}}| = k$*

*Proof.* By construction, $\varphi_{k,m}(a_1, \cdots a_{z+2})$ is of size linear in $m$. To prove that $|\Gamma_{\varphi_{k,m},a_1,a_{z+2}}|$ is of exactly size $k$, we use induction on $m$. We apply induction on $\psi_{k,m}$ since $\psi_{k,m}$ and $\varphi_{k,m}$ have 1-1 correspondence. The base case ($m = 1$) is trivial. For $m \geq 1$, let $c_2 \cdots c_m$ represent the number $k'$ in binary, and assume that $\psi_{k',m-1} = (b_2 \cdots C_m b_{m+1}) \cdots)$ has corresponding chain graph $\varphi_{k',m-1}$ such that $|\Gamma_{\varphi_{k',m-1},u,v}| = k'$, where $u = \mathsf{start}(\varphi_{k',m-1})$ and $v = \mathsf{end}(\varphi_{k',m-1})$. If $c_1$ is 0, then on one hand $k = k'$, and on the other hand we have, $\varphi_{k,m} \equiv e \cup \varphi_{k',m-1}$, where $a_1 = \mathsf{start}(e), \mathsf{end}(e) = \mathsf{start}(\varphi_{k'm-1})$ which has $|\Gamma_{\varphi_{k,m},a_1,a_{z+2}}| = k$ . Otherwise, if $c_1$ is 1, then on one hand $k = 2^{m-1} + k'$, and on the other hand $C_1$ is the connector "$\vee$". Therefore, $\varphi_{k,m} \equiv e \cup \varphi_{k',m-1}$ where $a_1 = \mathsf{start}(e)$, $\mathsf{end}(e) = \mathsf{end}(\varphi_{k'm-1})$, which has

$|\Gamma_{\varphi_{k,m},a_1,a_{z+2}}| = 2^{m-1} + k' = k$. This completes the induction.

□

### 7.4.2 From Graphs to $\Sigma_1^1$ Formulas

In this section, we discuss how for a given graph $G = (V, E)$ and nodes $u$ and $v$, and associated probability function such that $\mathsf{Pr}[e^1|e \in E] = 1/2$, we can reduce the problem of computing $r(u, v)$ to the problem of computing $|R_F|$ wherein $F$ is a $\Sigma_1^1$ formula.

The central idea of our reduction is based on usage of transitive closure for connectivity. Our reduction has close connection to previously proposed formulations for s-t connectivity (See [46] for related survey). Let $R(u, v)$ denote the event that $\exists$ path $\pi$ such that $\pi$ connects $u$ and $v$. If $R(u, v)$ occurs and there exists an edge $e \in E$, such that $v = \mathsf{start}(e) \wedge w = \mathsf{end}(e)$, then $R(u, w)$ must occur. For a given graph $G = (V, E)$ and pair of nodes $u$ and $v$, the goal is to create a $\Sigma_1^1$ formula $F$ such that every satisfying assignment to $F$ has one to one correspondence with $\sigma \subseteq E$ such that $u$ and $v$ are <u>not</u> connected under $\sigma$. To this end, we define a propositional variables $p_u$ and $q_e$ for every node $u \in V$ and every edge $e \in E$ respectively. Define,

$$C_e = (p_u \wedge q_e \rightarrow p_v)$$

$$S = \{p_u | u \in V\}$$

$$F_{u,v} = \exists S(p_u \wedge \neg p_v \wedge \bigwedge_{e \in E} C_e)$$

**Lemma 24.** *For a given graph $G = (V, E)$ and nodes $u$ and $v$, let $F_{u,v}$ be as defined above. Then, $|R_{F_{u,v}}| = |\overline{\Gamma_{G,u,v}}|$. Furthermore if $\forall e \in E$, we have $\mathsf{Pr}[e^1] = \frac{1}{2}$, then $r(u, v) = \frac{|R_{F_{u,v}}|}{2^{|E|}}$*

*Proof.* We prove $|R_{F_{u,v}}| = |\overline{\Gamma_{G,u,v}}|$ by constructing a bijective mapping from $\Gamma_{G,u,v}$ to $\overline{R_{F_{u,v}}}$. Note that for $A, B$ such that $|A| + |\overline{A}| = |B| + |\overline{B}|$, we have $|A| = |\overline{B}|$ iff there is a bijective mapping from $\overline{A}$ to $B$. (Notice the change in complement signs). For every $\sigma \subseteq E$, if $(u, v) \models \sigma$, let $\pi = uw_1w_2 \cdots w_{k-1}v$ be path of length $k$ between $u$ and $v$ under $\sigma$. Define truth assignment $\tau_\sigma : q_e|e \in E \rightarrow \{0, 1\}$ as follows: $\tau_\sigma(q_e) = 1$ if $e \in \sigma$ and 0 otherwise. Note that $p_u = 1$ and we have $\tau_\sigma(e_1) = 1$, where $e_1$ is edge between $u$ and $w_1$. Furthermore, constraint $C_{e_1}$ forces $p_{w_1} = 1$. By inductively applying this implication for every node appearing in the graph, we observe that $p(v)$ is forced to be 1, which is a contradiction. Note that definition of $F_{u,v}$ has unit clause $(\neg p_v)$. Therefore, $\tau_\sigma$ is not a satisfying assignment of $F_{u,v}$. Similarly, if $\tau$ is not a satisfying assignment, then we have $\sigma_\tau = \{e|\tau(q_e) = 1\}$. Following similar arguments as above, we can show that if $\tau$ is not a satisfying assignment, $\sigma_\tau \notin \Gamma_{G,u,v}$.

□

### 7.4.3 RelNet

We now describe how the above reductions can be employed to design a counting-based framework, called RelNet, for the problem of network reliability. For a given graph $G = (V, E)$, source node $u$ and sink node $v$ and a probability space $\Omega$ over the edges, RelNet consists of the following three steps:

Step 1: We obtain a transformed graph $G'$ by replacing every $e_i \in E$ with $\phi_{k,m}$ if $\mathsf{Pr}[e_i^1] = \frac{k_i}{2^{m_i}}$. Let $M = \sum_{e_i \in E} m_i$ where $\mathsf{Pr}[e_i^1] = \frac{k_i}{2^{m_i}}$.

Step 2: Construct $F_{u,v}$ as described above for the transformed graph $G'$, source node $u$ and sink node $v$

Step 3: Invoke ApproxMC2 estimate $|R_{F_{u,v}}|$

The following theorem proves the correctness of our framework

**Theorem 25.** *For a given Graph $G$, source node $u$ and sink node $v$, and probability space $\Omega$ over the edges,* $r(u, v) = \frac{|R_{F_{u,v}}|}{2^M}$

*Proof.* The proof follows directly from Lemmas 23 and 24. □

## 7.5 Evaluation

Since the primary objective of this project was to compute connectivity reliability of power transmission grid networks across different cities in U.S., we compared the effectiveness of RelNet vis-a-vis state of the art techniques. Specifically, we sought to answer the following questions:

1. How does the runtime performance of RelNet compare to that of the state-of-the art techniques on real world power transmission networks?

2. How do estimates computed by RelNet compare to the exact estimates of reliability for networks that could be handled by exact techniques?

### 7.5.1 Experimental Methodology

We sought to compute reliability between every pair of nodes for all the ten cities discussed in Section 7.3. We implemented a Python prototype of RelNet, which invokes ApproxMC2 to perform counting over $\Sigma_1^1$ formulas as required by Step 3 of the RelNet. For all our experiments, we used $\varepsilon = 0.8$ and $\delta = 0.2$ as parameters for ApproxMC2, which is consistent with previously reported studies of using hashing-based counting techniques.

For comparison purposes, we considered: (i) Karger's FPRAS algorithm [103], (ii) a recently proposed MCMC-based technique [166] and (iii) selective path based RDA (S-RDA), one of the current state of the art techniques employed by the reliability engineering community. For all our benchmarks, S-RDA outperformed Karger's FPRAS algorithm and the above stated MCMC technique, therefore we omit further discussion of these two techniques in the rest of the section.

Each experiment consisted of running a given tool on a given graph for a pair of nodes termed as source and sink. The timeout for each experiment was set to 1,000 seconds.

### 7.5.2 Results

The analysis of runtime performance of S-RDA and RelNet shows that RelNet dramatically outperforms S-RDA. First of all, RelNet can compute $r(u, v)$ for each pair of source ($u$) and terminal ($v$) for all the ten cities while S-RDA could handle only G5 and G27 and timed out for almost every pair for rest of the cities. It is worth reiterating before RelNet, no theoretically sound estimates were, to the best of our knowledge, a priori available for rest of the eight cities. Figure 7.2 presents heat-maps for both S-RDA and RelNet for cities G1, G2, and G3. For every city Gi, the corresponding heatmap is labeled by either Gi (S-RDA) if it presents runtime results for S-RDA or Gi (RelNet), otherwise. For every heat-map, the $y$-axis represent source node while the $x$-axis represents terminal node. For every pair of source and terminal, the runtime for the corresponding tool is represented by the color as specified by the scale next to each heat-map. Overall, the closer the color of the point is to blue, better the method is.

The heat-maps clearly show that while RelNet can compute estimates within few tens of seconds for each pair, S-RDA fails for almost every pair. In this context, it is worth mentioning that runtime of RelNet is very consistent across different pairs of

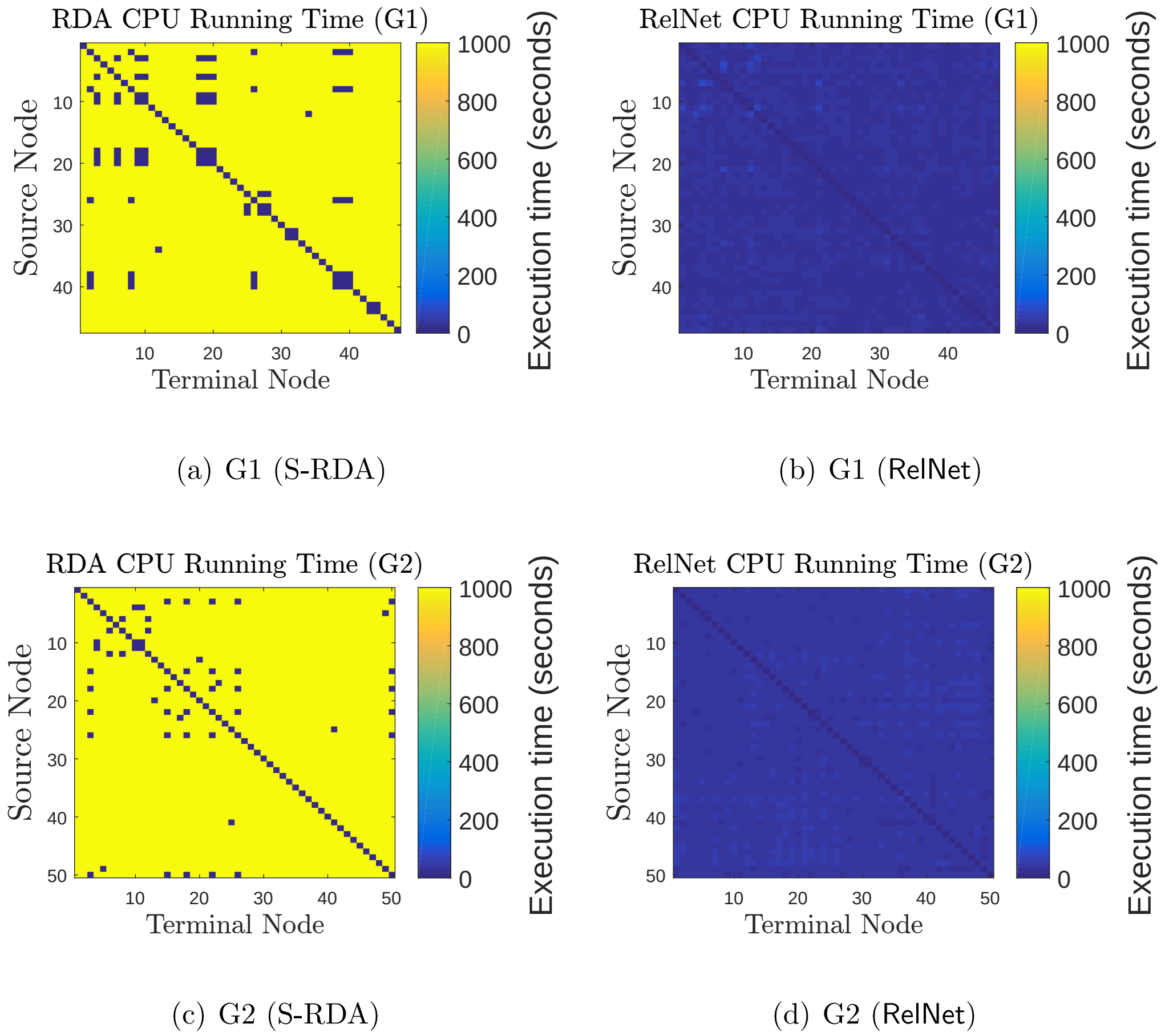

(a) G1 (S-RDA)

(b) G1 (RelNet)

(c) G2 (S-RDA)

(d) G2 (RelNet)

Figure 7.2 : CPU time in seconds using RDA and RelNet for every source and terminal pair

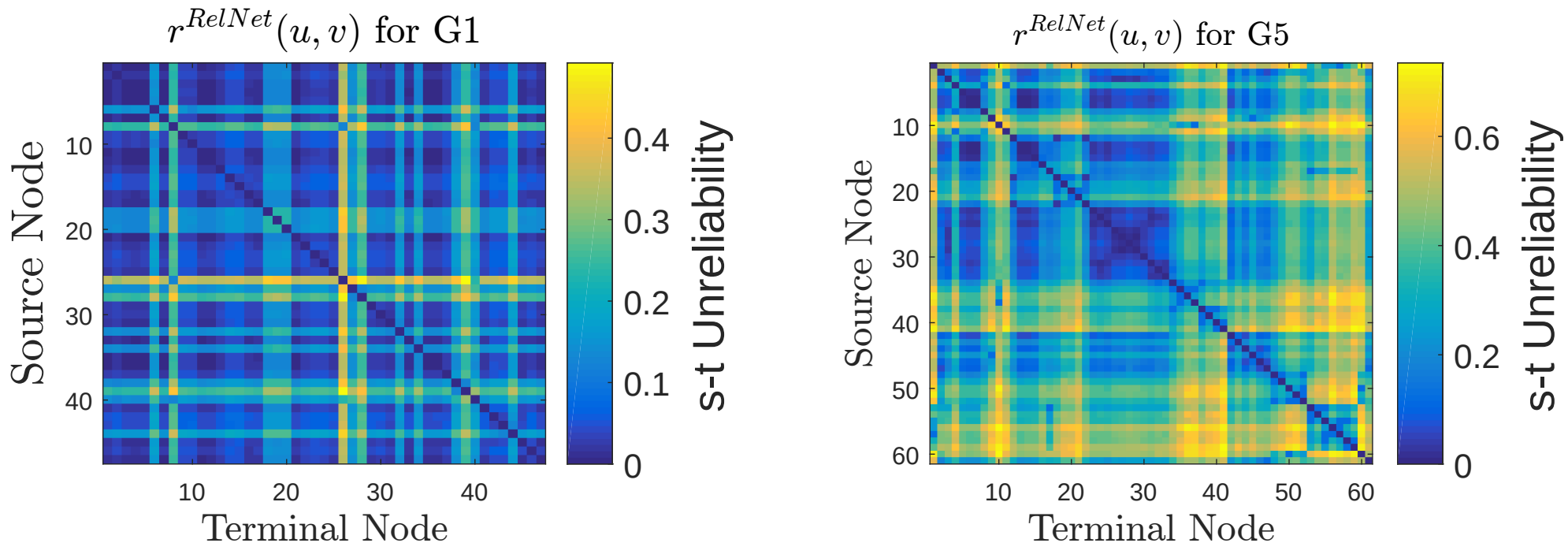

(a) G1 (b) G5

Figure 7.3 : s-t reliability estimates for G1 and G5 using RelNet for every pair.

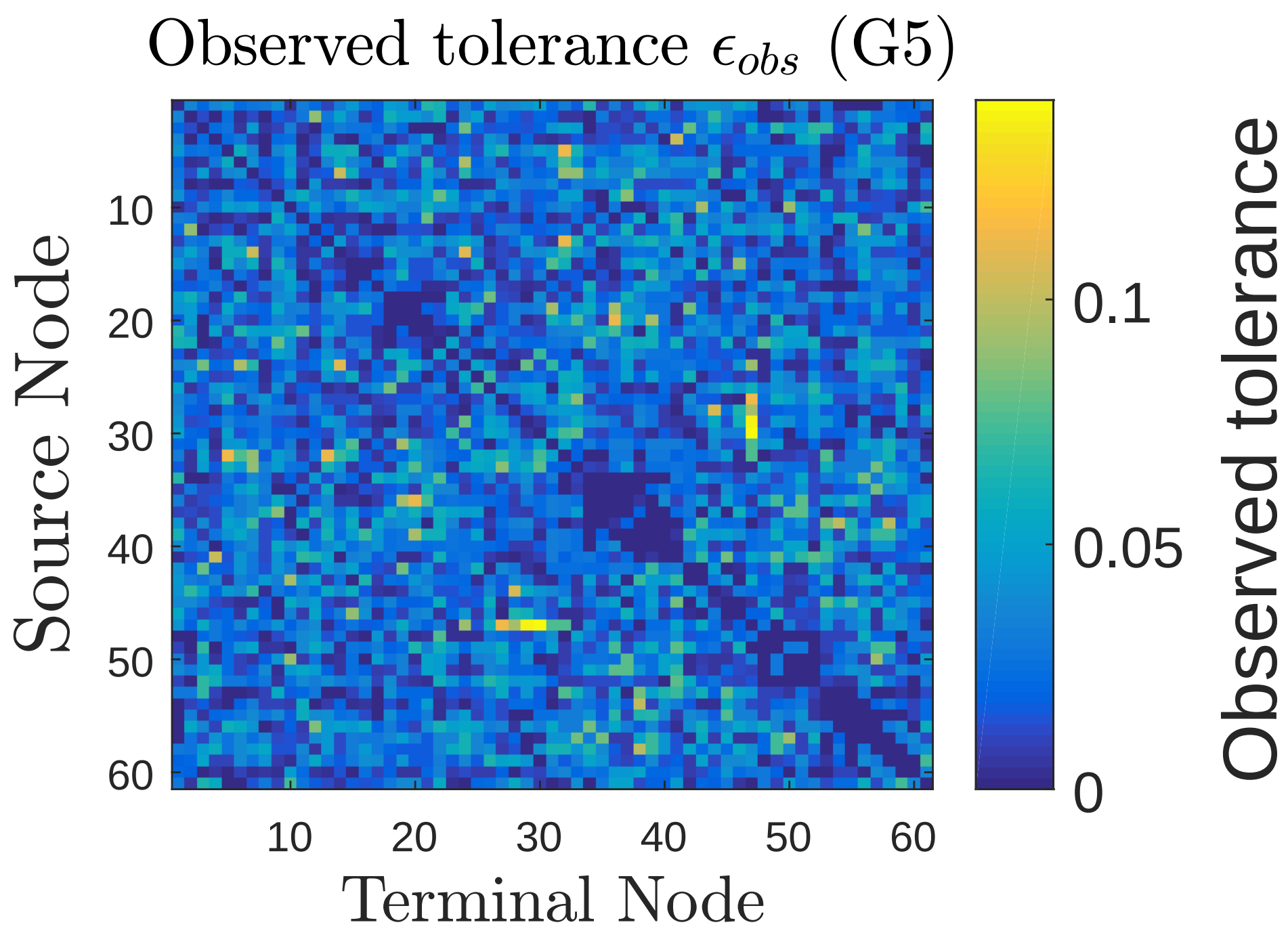

Figure 7.4 : Observed tolerance ($\varepsilon_{obs}$) for all pairs of city G5

source and sink nodes.

As an illustration, Figure 7.3 shows heat-maps of reliability estimates between all pairs of nodes for cities G1 and G5 as computed by RelNet. Similar to performance comparison heatmaps, the y-axis of every plot refers to source node while the x-axis refers to sink node. The reliability for $(u, v)$ is represented by the color as per the mapping presented on the right. Looking at these plots, one might wonder about the accuracy of reported results. While RelNet provides theoretical guarantees of accuracy, we sought to measure the quality of our estimates in practice. Given that S-RDA is an exact technique, we use the estimates from S-RDA on G5 to measure the quality of estimates of RelNet. For each pair, the observed tolerance $\varepsilon_{obs}$ was calculated as $\max(\frac{C}{f} - 1, \frac{f}{C} - 1)$ where $C$ is the estimate from RelNet and $f$ is the exact estimate computed by S-RDA. Figure 7.4 shows the heat-map of observed tolerance $\varepsilon_{obs}$ for each pair of G5. First of all, for every pair the observed tolerance is less than 0.14 – far better than the theoretical guarantee of 0.8. Furthermore, the geometric mean of observed tolerance is just 0.01951; almost an order of magnitude better than the theoretical guarantee. This highlights conservative nature of theoretical guarantees and the need to strengthen the analysis as part of future work. As this work is part of larger project, where estimates of reliability are required to support decision making for community resilience, the above observations are quite significant as they show how emerging computational algorithms could support analysis and management of infrastructure under uncertainty.

## 7.6 Chapter Summary

Estimation of network reliability is crucial for decision making to ensure availability and resilience of critical facilities. Despite significant interest and long history of prior work, the current state of the art techniques fail to either provide sound theoretical estimates or scale to large networks. In this chapter, we discussed how progress in the development of hashing-based techniques described in this Part can be utilized to construct a scalable reliability estimation framework, RelNet. Furthermore, unlike the current state of the art techniques, RelNet can scale to real world networks arising from cities across U.S., especially when exact reliability computations are not affordable.

# Part III

# Constrained Sampling

The first chapter of this part, i.e. Chapter 8, focuses on uniform generation and introduced the core hashing-based algorithmic framework, UniGen. Chapter 9 then discusses how UniGen can trade off independence for performance gains and demonstrate high parallelizability of our framework. Chapter 10 discusses adaptation of UniGen, called WeightGen, to handle general weight functions.

# Chapter 8

# Hashing-based Almost Uniform Generator

Marrying scalability with strong guarantees of uniformity has been the holy grail of algorithms that sample from solutions of constraint systems. The literature bears testimony to the significant tension between these objectives when designing random generators of SAT witnesses. Earlier work in this area either provide strong theoretical guarantees at the cost of scalability, or remedy the scalability problem at the cost of guarantees of uniformity. More recently, however, there have been efforts to bridge these two extremes.

Bellare, Goldreich and Petrank [15] showed that a provably uniform generator of SAT witnesses can be designed in theory to run in probabilistic polynomial time relative to an NP oracle. Unfortunately, it was shown in [37] that this algorithm does not scale beyond formulae with few tens of variables in practice. Weighted binary decision diagrams (BDD) have been used in [161] to sample uniformly from SAT witnesses. However, BDD-based techniques are known to suffer from scalability problems [108]. Adapted BDD-based techniques with improved performance were proposed in [112]; however, the scalability was achieved at the cost of guarantees of uniformity. Random seeding of DPLL SAT solvers [126] has been shown to offer performance, although the generated distributions of witnesses can be highly skewed [108].

Markov Chain Monte Carlo methods (also called MCMC methods) [108, 156] are widely considered to be a practical way to sample from a distribution of solutions. Several MCMC algorithms, such as those based on simulated annealing, Metropolis-Hastings algorithm and the like, have been studied extensively in the literature [119]. While MCMC methods guarantee eventual convergence to a target distribution under mild requirements, convergence is often impractically slow in practice. The work of [156, 108] proposed several such adaptations for MCMC-based sampling in the context of constrained-random verification. Unfortunately, most of these adaptations are heuristic in nature, and do not preserve theoretical guarantees of uniformity. constraints, thereby increasing constraint-solving time. Sampling techniques based on interval-propagation and belief networks have been proposed in [57, 80, 98]. The simplicity of these approaches lend scalability to the techniques, but the generated distributions can deviate significantly from the uniform distribution, as shown in [109].

Sampling techniques based on hashing were originally pioneered by Sipser [141], and have been used subsequently by several researchers [15, 88, 37]. The core idea in hashing-based sampling is to use $r$-wise independent hash functions (for a suitable value of $r$) to randomly partition the space of witnesses into "small cells" of roughly equal size, and then randomly pick a solution from a randomly chosen cell. The algorithm of Bellare et al. referred to above uses this idea with $n$-wise independent algebraic hash functions (where $n$ denotes the size of the support of $F$). As noted above, their algorithm scales very poorly in practice. Gomes, Sabharwal and Selman used 3-wise independent linear hash functions in [88] to design $\mathsf{XORSample}'$, a near-uniform generator of SAT witnesses. Nevertheless, to realize the guarantee of near-uniformity, their algorithm requires the user to provide difficult-to-estimate input parameters. Although $\mathsf{XORSample}'$ has been shown to scale to constraints involving a few thousand variables, Gomes et al. acknowledge the difficulty of scaling their algorithm to much larger problem sizes without sacrificing theoretical guarantees [88].

Recently, Chakraborty, Meel and Vardi [37] proposed a new hashing-based SAT

witness generator, called UniWit, that represents a small but significant step towards marrying the conflicting goals of scalability and guarantees of uniformity. Like XORSample′, the UniWit algorithm uses 3-wise independent linear hashing functions. Unlike XORSample′, however, the guarantee of near-uniformity of witnesses generated by UniWit does not depend on difficult-to-estimate input parameters. In [37], UniWit has been shown to scale to formulas with several thousand variables. In addition, Chakraborty et al proposed a heuristic called "leap-frogging" that allows UniWit to scale even further – to tens of thousands of variables [37]. Unfortunately, the guarantees of near-uniformity can no longer be established for UniWit with "leap-frogging". More recently, Ermon et al. [69] proposed a hashing-based algorithm called PAWS for sampling from a distribution defined over a discrete set using a graphical model but fails to provide guarantees of almost-uniformity. PAWS faces the same scalability hurdles as UniWit, and does not scale beyond a few thousand variables without heuristic adapatations that compromise its guarantees.

In this Chapter, we propose an algorithm called UniGen, which is the first algorithm to provide strong guarantees of almost-uniformity, while scaling to problems involving hundreds of thousands of variables. We also improve upon the success probability of the earlier algorithms significantly, both in theory and as evidenced by our experiments.

## 8.1 The UniGen Algorithm

The new algorithm, called UniGen, shares some features with earlier hashing-based algorithms such as XORSample′ [88], UniWit [37] and PAWS [69], but there are key differences that allow UniGen to significantly outperform these earlier algorithms, both in terms of theoretical guarantees and measured performance.

The effectiveness of a hashing-based probabilistic generator depends on its ability to quickly partition the set $R_F$ into "small" and "roughly equal" sized random cells. This, in turn, depends on the parameter $m$ used in the choice of the hash function family $H(n, m, r)$. A high value of $m$ leads to skewed distributions of sizes of cells, while a low value of $m$ leads to cells that are not small enough. The best choice of $m$ depends on $|R_F|$, which is not known *a priori*. Different algorithms therefore use different techniques to estimate a value of $m$. In XORSample′, this is achieved by requiring the user to provide some difficult-to-estimate input parameters. In UniWit, the algorithm sequentially iterates over values of $m$ until a good enough value is found. The approach of PAWS comes closest to our, although there are crucial differences. In both PAWS and UniGen, an approximate model counter is first used to estimate $|R_F|$ within a specified tolerance and with a specified confidence. This estimate, along with a user-provided parameter, is then used to determine a *unique* value of $m$ in PAWS. Unfortunately, this does not facilitate proving that PAWS is an almost-uniform generator. Instead, Ermon, et al. show that PAWS behaves like an almost-uniform generator with probability greater than $1-\delta$, for a suitable $\delta$ that depends on difficult-to-estimate input parameters. In contrast, we use the estimate of $|R_F|$ to determine a *small range* of candidate values of $m$. This allows us to prove that UniGen is almost-uniform generator with confidence 1.

The pseudocode for UniGen is shown in Algorithm 9. UniGen takes as inputs a Boolean CNF formula $F$, a tolerance $\varepsilon$ ($> 1.71$, for teachnical reasons explained in the Appendix) and a set $S$ of sampling variables. It either returns a random witness of $F$ or $\bot$ (indicating failure). The algorithm assumes access to a source of random binary numbers, and to two subroutines: (i) BoundedSAT$(F, N)$, which, for every $N > 0$, returns $\min(|R_F|, N)$ distinct witnesses of $F$, and (ii) an approximate model counter ApproxModelCounter$(F, \varepsilon', 1-\delta')$.

UniGen first computes two quantities, "pivot" and $\kappa$, that represent the expected size of a "small" cell and the tolerance of this size, respectively. The specific choices of expressions used to compute $\kappa$ and "pivot" in ComputeKappaPivot are motivated by technical reasons explained in the Appendix. The values of $\kappa$ and "pivot" are used to determine high and low thresholds (denoted "hiThresh" and "loThresh" respectively) for the size of each cell. Lines 5– 7 handle the easy case when $F$ has no more than "hiThresh" witnesses. Otherwise, UniGen invokes ApproxModelCounter to obtain an estimate, $C$, of $|R_F|$ to within a tolerance of 0.8 and with a confidence of 0.8. Once again, the specific choices of the tolerance and confidence parameters used in computing $C$ are motivated by technical reasons explained in the Appendix. The estimate $C$ is then used to determine a range of candidate values for $m$. Specifically, this range is $\{q-4, \ldots q\}$, where $q$ is determined in line 10 of the pseudocode. The loop in lines 14– 17 checks whether some value in this range is good enough for $m$, i.e., whether the number of witnesses in a cell chosen randomly after partitioning $R_F$ using $H_{xor}(|S|, m, 3)$, lies within "hiThresh" and "loThresh". If so, lines 21– 22 return a random witness from the chosen cell. Otherwise, the algorithm reports a failure in line 19.

An probabilistic generator is likely to be invoked multiple times with the same input constraint in constrained-random verification. Towards this end, note than lines 1–11 of the pseudocode need to executed only once for every formula $F$. Generating a new random witness requires executing afresh only lines 12–22. While this optimization appears similar to "leapfrogging" [37, 38], it is fundamentally different since it does not sacrifice any theoretical guarantees, unlike "leapfrogging".

**Algorithm 9** UniGen($F, \varepsilon, S$)

/*Assume $S = \{x_1, \ldots x_{|S|}\}$ is an independent support of $F$, and $\varepsilon > 1.71$ */

1: $(\kappa, \text{pivot}) \leftarrow \mathsf{ComputeKappaPivot}(\varepsilon)$;
2: $\text{hiThresh} \leftarrow 1 + (1+\kappa)\text{pivot}$;
3: $\text{loThresh} \leftarrow \frac{1}{1+\kappa}\text{pivot}$;
4: $Y \leftarrow \mathsf{BoundedSAT}(F, \text{hiThresh})$;
5: **if** $(|Y| \leq \text{hiThresh})$ **then**
6:     Let $y_1, \ldots y_{|Y|}$ be the elements of $Y$;
7:     Choose $j$ at random from $\{1, \ldots |Y|\}$; **return** $y_j$;
8: **else**
9:     $C \leftarrow \mathsf{ApproxModelCounter}(F, 0.8, 0.8)$;
10:     $q \leftarrow \lceil \log C + \log 1.8 - \log \text{pivot} \rceil$;
11:     $i \leftarrow q - 4$;
12:     Choose $h$ at random from $H_{xor}(|S|, n, 3)$;
13:     Choose $\alpha$ at random from $\{0,1\}^n$;
14:     **repeat**
15:         $i \leftarrow i + 1$;
16:         $Y \leftarrow \mathsf{BoundedSAT}(F \wedge (h_i(x_1, \ldots x_{|S|}) = \alpha_i), \text{hiThresh})$;
17:     **until** $(\text{loThresh} \leq |Y| \leq \text{hiThresh})$ or $(i = q)$
18:     **if** $(|Y| > \text{hiThresh})$ or $(|Y| < \text{loThresh})$ **then**
19:         **return** $\perp$
20:     **else**
21:         Let $y_1, \ldots y_{|Y|}$ be the elements of $Y$;
22:         Choose $j$ at random from $[|Y|]$ and **return** $y_j$;

**Algorithm 10** ComputeKappaPivot($t\varepsilon$)

1: Find $\kappa \in [0, 1)$ such that $\varepsilon = (1 + \kappa)(2.23 + \frac{0.48}{(1-\kappa)^2}) - 1$ ;

2: pivot $\leftarrow \lceil 3e^{1/2}(1 + \frac{1}{\kappa})^2 \rceil$;

3: **return** ($\kappa$, pivot)

## 8.2 Implementation Issues

In our implementation of UniGen, BoundedSAT is implemented using CryptoMiniSAT [1] – a SAT solver that handles xor clauses efficiently. CryptoMiniSAT uses *blocking clauses* to prevent already generated witnesses from being generated again. Since the independent support of $F$ determines every satisfying assignment of $F$, blocking clauses can be restricted to only variables in the set $S$. We implemented this optimization in CryptoMiniSAT, leading to significant improvements in performance. ApproxModelCounter is implemented using ApproxMC [38]. We disable "leapfrogging" optimization since it nullifies the theoretical guarantees of ApproxMC [38]. We use "random_device" implemented in C++ as the source of pseudo-random numbers in lines 7, 14, 15 and 22 of the pseudocode, and also as the source of random numbers in ApproxMC.

## 8.3 Analysis

Following notations introduced in Chapter 2, let $R_{F\downarrow S}$ denote set of witnesses of the Boolean formula $F$ projected on the sampling set $S$. For convenience of analysis, we assume that $\log(|R_{F\downarrow S}| - 1) - \log \mathit{pivot}$ is an integer, where *pivot* is the quantity computed by algorithm ComputeKappaPivot (see Section 8.1). A more careful analysis removes this assumption by scaling the probabilities by constant factors. Let us denote $\log(|R_F|-1) - \log pivot$ by $m$. The expression used for computing $pivot$ in algorithm ComputeKappaPivot ensures that pivot $\geq 17$. Therefore, if an invocation of UniGen does not return from line 7 of the pseudocode, then $|R_F| \geq 18$. Note also that the expression for computing $\kappa$ in algorithm ComputeKappaPivot requires $\varepsilon \geq 1.71$ in order to ensure that $\kappa \in [0,1)$ can always be found.

The following lemma shows that $q$, computed in line 10 of the pseudocode, is a good estimator of $m$.

**Lemma 26.** $\mathsf{Pr}[q-3 \leq m \leq q] \geq 0.8$

*Proof.* Recall that in line 9 of the pseudocode, an approximate model counter is invoked to obtain an estimate, $C$, of $|R_{F\downarrow S}|$ with tolerance 0.8 and confidence 0.8. By the definition of approximate model counting, we have $\mathsf{Pr}[\frac{C}{1.8} \leq |R_{F\downarrow S}| \leq (1.8)C] \geq 0.8$. Thus, $\mathsf{Pr}[\log C - \log(1.8) \leq \log|R_{F\downarrow S}| \leq \log C + \log(1.8)] \geq 0.8$. It follows that $\mathsf{Pr}[\log C - \log(1.8) - \log pivot - \log(\frac{1}{1-1/|R_{F\downarrow S}|}) \leq \log(|R_{F\downarrow S}|-1) - \log pivot \leq \log C - \log pivot + \log(1.8) - \log(\frac{1}{1-1/|R_{F\downarrow S}|})] \geq 0.8$. Substituting $q = \lceil \log C + \log 1.8 - \log pivot \rceil$, $m = \log(|R_{F\downarrow S}|-1) - \log pivot$, $log(1.8) = 0.85$ and $\log(\frac{1}{1-1/|R_{F\downarrow S}|}) \leq 0.12$ (since $|R_{F\downarrow S}| \geq 18$ on reaching line 10 of the pseudocode), we get $\mathsf{Pr}[q-3 \leq m \leq q] \geq 0.8$. □

The next lemma provides a lower bound on the probability of generation of a witness. Let $w_{i,y,\alpha}$ denote the probability $\mathsf{Pr}\left[\frac{\text{pivot}}{1+\kappa} \leq |R_{F,h,\alpha}| \leq 1 + (1+\kappa)\text{pivot} \text{ and } h(y) = \alpha : h \xleftarrow{R} H_{xor}(n,i,3)\right]$. The proof of the lemma also provides a lower bound on $w_{m,y,\alpha}$.

**Lemma 27.** *For every witness $y$ of $F$,* $\mathsf{Pr}[y \textit{ is output}] \geq \frac{0.8(1-e^{-1})}{(1.06+\kappa)(|R_{F\downarrow S}|-1)}$

*Proof.* If $|R_{F\downarrow S}| \leq 1 + (1+\kappa)$pivot, the lemma holds trivially (see lines 5–7 of the pseudocode). Suppose $|R_{F\downarrow S}| \geq 1 + (1+\kappa)$pivot and let $U$ denote the event that witness $y \in R_{F\downarrow S}$ is output by UniGen on inputs $F$, $\varepsilon$ and $X$. Let $p_{i,y}$ denote the probability that we return from line 17 for a particular value of $i$ with $y$ in $R_{F,h,\alpha}$, where $\alpha \in \{0,1\}^i$ is the value chosen in line 15. Then, $\mathsf{Pr}[U] = \cup_{i=q-3}^{q} \frac{1}{|Y|} p_{i,y}$, where $Y$ is the set of witnesses returned by BoundedSAT in line 16 of the pseudocode. Let $f_m = \mathsf{Pr}[q-3 \leq m \leq q]$. From Lemma 48, we know that $f_m \geq 0.8$. From the design of the algorithm, we also know that $\frac{1}{1+\kappa}\text{pivot} \leq |Y| \leq 1 + (1+\kappa)\text{pivot}$. Therefore, $\mathsf{Pr}[U] \geq \frac{1}{1+(1+\kappa)\text{pivot}} \cdot p_{m,y} \cdot f_m$. The proof is now completed by showing $p_{m,y} \geq \frac{1}{2^m}(1-e^{-1})$. This gives $\mathsf{Pr}[U] \geq \frac{0.8(1-e^{-1})}{(1+(1+\kappa)pivot)2^m} \geq \frac{0.8(1-e^{-1})}{(1.06+\kappa)(|R_{F\downarrow S}|-1)}$. The last inequality uses the observation that $1/\text{pivot} \leq 0.06$.

To calculate $p_{m,y}$, we first note that since $y \in R_{F\downarrow S}$, the requirement "$y \in R_{F,h,\alpha}$" reduces to "$y \in h^{-1}(\alpha)$". For $\alpha \in \{0,1\}^n$, we define $w_{m,y,\alpha}$ as $\mathsf{Pr}\left[\frac{\text{pivot}}{1+\kappa} \leq |R_{F,h,\alpha}| \leq 1 + (1+\kappa) \ \ pivot \text{ and } h(y) = \alpha : h \xleftarrow{R} H_{xor}(n,m,3)\right]$. Therefore, $p_{m,y} = \Sigma_{\alpha \in \{0,1\}^m} (w_{m,y,\alpha}.2^{-m})$. The proof is now completed by showing that $w_{m,y,\alpha} \geq (1-e^{-1})/2^m$ for every $\alpha \in \{0,1\}^m$ and $y \in \{0,1\}^n$.

Towards this end, let us first fix a random $y$. Now we define an indicator variable $\gamma_{z,\alpha}$ for every $z \in R_{F\downarrow S} \setminus \{y\}$ such that $\gamma_{z,\alpha} = 1$ if $h(z) = \alpha$, and $\gamma_{z,\alpha} = 0$ otherwise. Let us fix $\alpha$ and choose $h$ uniformly at random from $H_{xor}(n,m,3)$. The random choice of h induces a probability distribution on $\gamma_{z,\alpha}$ such that $E[\gamma_{z,\alpha}] = \mathsf{Pr}[\gamma_{z,\alpha} = 1] = 2^{-m}$. Since we have fixed $y$, and since hash functions chosen from $H_{xor}(n,m,3)$ are 3-wise independent, it follows that for every distinct $z_a, z_b \in R_{F\downarrow S} \setminus \{y\}$, the random variables $\gamma_{z_a,\alpha}, \gamma_{z_b,\alpha}$ are 2-wise independent. Let $\Gamma_\alpha = \sum_{z \in R_{F\downarrow S} \setminus \{y\}} \gamma_{z,\alpha}$ and $\mu_\alpha = E[\Gamma_\alpha]$. Clearly, $\Gamma_\alpha = |R_{F,h,\alpha}| - 1$ and $\mu_\alpha = \sum_{z \in R_{F\downarrow S} \setminus \{y\}} E[\gamma_{z,\alpha}] = \frac{|R_{F\downarrow S}|-1}{2^m}$. Also, $\mathsf{Pr}[\frac{\text{pivot}}{1+\kappa} \leq |R_{F,h,\alpha}| \leq 1 + (1+\kappa)\text{pivot}] = \mathsf{Pr}[\frac{\text{pivot}}{1+\kappa} - 1 \leq |R_{F,h,\alpha}| - 1 \leq (1+\kappa)\text{pivot}] \geq \mathsf{Pr}[\frac{\text{pivot}}{1+\kappa} \leq |R_{F,h,\alpha}| - 1 \leq (1+\kappa)\text{pivot}]$. Using the expression for pivot, we get $2 \leq \lfloor e^{-1/2}(1+1/\epsilon)^2 \cdot \frac{|R_{F\downarrow S}|-1}{2^m} \rfloor$. Therefore using Chebyshev's Inequality and substituting pivot $= (|R_{F\downarrow S}| - 1)/2^m$, we get $\mathsf{Pr}[\frac{\text{pivot}}{1+\kappa} \leq |R_{F,h,\alpha}| - 1 \leq (1+\kappa)\text{pivot}] \geq 1 - e^{-1}$. Therefore, $\mathsf{Pr}[\frac{\text{pivot}}{1+\kappa} \leq |R_{F,h,\alpha}| \leq 1 + (1+\kappa)\text{pivot}] \geq 1 - e^{-1}$ Since $h$ is chosen at random from $H_{xor}(n, m, 3)$, we also have $\mathsf{Pr}[h(y) = \alpha] = 1/2^m$. It follows that $w_{m,y,\alpha} \geq (1 - e^{-1})/2^m$. □

The next lemma provides an upper bound of $w_{i,y,\alpha}$ and $p_{i,y}$.

**Lemma 28.** *For $i < m$, both $w_{i,y,\alpha}$ and $p_{i,y}$ are bounded above by* $\frac{1}{|R_{F\downarrow S}|-1}\frac{1}{\left(1-\frac{1+\kappa}{2^{m-i}}\right)^2}$.

*Proof.* We will use the terminology introduced in the proof of Lemma 27. Clearly, $\mu_\alpha = \frac{|R_{F\downarrow S}|-1}{2^i}$. Since each $\gamma_{z,\alpha}$ is a 0-1 variable, $\mathsf{V}[\gamma_{z,\alpha}] \leq \mathsf{E}[\gamma_{z,\alpha}]$. Therefore, $\sigma^2_{z,\alpha} \leq \sum_{z \neq y, z \in R_{F\downarrow S}} \mathsf{E}[\gamma_{z,\alpha}] \leq \sum_{z \in R_{F\downarrow S}} \mathsf{E}[\gamma_{z,\alpha}] = \mathsf{E}[\Gamma_\alpha] = 2^{-m}(|R_{F\downarrow S}| - 1)$. So $\mathsf{Pr}[\frac{pivot}{1+\kappa} \leq |R_{F,h,\alpha}| \leq 1 + (1+\kappa)\text{pivot}] \leq \mathsf{Pr}[|R_{F,h,\alpha}| - 1 \leq (1+\kappa)\text{pivot}]$. From Chebyshev's inequality, we know that $\mathsf{Pr}[|\Gamma_\alpha - \mu_{z,\alpha}| \geq \kappa\sigma_{z,\alpha}] \leq 1/\kappa^2$ for every $\kappa > 0$. By choosing $\kappa = (1 - \frac{1+\kappa}{2^{m-i}})\frac{\mu_{z,\alpha}}{\sigma_{z,\alpha}}$, we have $\mathsf{Pr}[|R_{F,h,\alpha}| - 1 \leq (1+\kappa)\text{pivot}] \leq \mathsf{Pr}\left[|(|R_{F,h,\alpha}| - 1) - \frac{|R_{F\downarrow S}|-1}{2^i}| \geq (1 - \frac{1+\kappa}{2^{m-i}})\frac{|R_{F\downarrow S}|-1}{2^i}\right] \leq \frac{1}{\left(1-\frac{(1+\kappa)}{2^{m-i}}\right)^2} \cdot \frac{2^i}{|R_{F\downarrow S}|-1}$. Since $h$ is chosen at random from $H_{xor}(n, m, 3)$, we also have $\mathsf{Pr}[h(y) = \alpha] = 1/2^i$. It follows that $w_{i,y,\alpha} \leq \frac{1}{|R_{F\downarrow S}|-1}\frac{1}{\left(1-\frac{1+\kappa}{2^{m-i}}\right)^2}$. The bound for $p_{i,y}$ is easily obtained by noting that $p_{i,y} = \Sigma_{\alpha \in \{0,1\}^i}(w_{i,y,\alpha}.2^{-i})$. □

**Lemma 29.** *For every witness $y$ of $F$,* $\mathsf{Pr}[$*y is output*$] \leq \frac{1+\kappa}{|R_{F\downarrow S}|-1}(2.23 + \frac{0.48}{(1-\kappa)^2})$

*Proof.* We will use the terminology introduced in the proof of Lemma 27. $\mathsf{Pr}[U] = \cup_{i=q-3}^{q} \frac{1}{|Y|} p_{i,y} \leq \frac{1+\kappa}{\text{pivot}} \sum_{i=q-3}^{q} p_{i,y}$. We can sub-divide the calculation of $\mathsf{Pr}[U]$ into three cases based on the range of the values m can take.

**Case 1 :** $q - 3 \leq m \leq q$.

Now there are four values that $m$ can take.

1. $m = q-3$. We know that $p_{i,y} \leq \mathsf{Pr}[h(y) = \alpha] = \frac{1}{2^i}$. $\mathsf{Pr}[U|m = q-3] \leq \frac{1+\kappa}{\text{pivot}} \cdot \frac{1}{2^{q-3}}\frac{15}{8}$. Substituting the value of pivot and $m$, we get $\mathsf{Pr}[U|m = q-3] \leq \frac{15(1+\kappa)}{8(|R_{F\downarrow S}|-1)}$.

2. $m = q-2$. For $i \in [q-2, q]$ $p_{i,y} \leq \mathsf{Pr}[h(y) = \alpha] = \frac{1}{2^i}$ Using Lemma 39, we get $p_{q-3,y} \leq \frac{1}{|R_{F\downarrow S}|-1}\frac{1}{(1-\frac{1+\kappa}{2})^2}$. Therefore, $\mathsf{Pr}[U|m = q-2] \leq \frac{1+\kappa}{\text{pivot}}\frac{1}{|R_{F\downarrow S}|-1}(\frac{1}{1-\frac{1+\kappa}{2}}) + \frac{1+\kappa}{\text{pivot}}\frac{1}{2^{q-2}}\frac{7}{4}$. Noting that pivot $= \frac{|R_{F\downarrow S}|-1}{2^m} > 10$, $\mathsf{Pr}[U|m = q-2] \leq \frac{1+\kappa}{|R_{F\downarrow S}|-1}(\frac{7}{4} + \frac{0.4}{(1-\kappa)^2})$

3. $m = q-1$. For $i \in [q-1, q]$, $p_{i,y} \leq \mathsf{Pr}[h(y) = \alpha] = \frac{1}{2^i}$. Using Lemma 39, we get $p_{q-3,y} + p_{q-2,y} \leq \frac{1}{|R_{F\downarrow S}|-1}\left(\frac{1}{(1-\frac{1+\kappa}{2^2})} + \frac{1}{(1-\frac{1+\kappa}{2})^2}\right)$. Therefore, $\mathsf{Pr}[U|m = q-1] \leq \frac{1+\kappa}{\text{pivot}}\left(\frac{1}{|R_{F\downarrow S}|-1}\left(\frac{1}{(1-\frac{1+\kappa}{2^2})^2} + \frac{1}{(1-\frac{1+\kappa}{2})^2}\right) + \frac{1}{2^{q-1}}\frac{3}{2}\right)$. Noting that pivot $= \frac{|R_{F\downarrow S}|-1}{2^m} > 10$ and $\kappa \leq 1$, $\mathsf{Pr}[U|m = q-1] \leq \frac{1+\kappa}{|R_{F\downarrow S}|-1}(1.9 + \frac{0.4}{(1-\kappa)^2})$.

4. $m = q$, $p_{q,y} \leq \mathsf{Pr}[h(y) = \alpha] = \frac{1}{2^q}$. Using Lemma 39, we get $p_{q-3,y} + p_{q-2,y} + p_{q-1,y} \leq \frac{1}{|R_{F\downarrow S}|-1}\left(\frac{1}{(1-\frac{1+\kappa}{2^3})^2}\frac{1}{(1-\frac{1+\kappa}{2^2})^2} + \frac{1}{(1-\frac{1+\kappa}{2})^2}\right)$. Therefore, $\mathsf{Pr}[U|m = q] \leq \frac{1+\kappa}{\text{pivot}}\left(\frac{1}{|R_{F\downarrow S}|-1}\left(\frac{1}{(1-\frac{1+\kappa}{2^3})^2} + \frac{1}{(1-\frac{1+\kappa}{2^2})^2} + \frac{1}{(1-\frac{1+\kappa}{2})^2}\right) + 1\right)$. Noting that pivot $= \frac{|R_{F\downarrow S}|-1}{2^m} > 10$, $\mathsf{Pr}[U|m = q] \leq \frac{1+\kappa}{|R_{F\downarrow S}|-1}(1.58 + \frac{0.4}{(1-\kappa)^2})$.

$\mathsf{Pr}[U|q-3 \leq m \leq q] \leq \max_i(\mathsf{Pr}[U|m = i])$. Therefore, $\mathsf{Pr}[U|q-3 \leq m \leq q] \leq \mathsf{Pr}[U|m = q-1] \leq \frac{1+\kappa}{|R_{F\downarrow S}|-1}(1.9 + \frac{0.4}{(1-\kappa)^2})$.

**Case 2 :** $m < q-3$. $\mathsf{Pr}[U|m < q-3] \leq \frac{1+\kappa}{\text{pivot}} \cdot \frac{1}{2^{q-3}}\frac{15}{8}$. Substituting the value of pivot and maximizing $m - q + 3$, we get $\mathsf{Pr}[U|m < q-3] \leq \frac{15(1+\kappa)}{16(|R_{F\downarrow S}|-1)}$.

**Case 3 :** $m > q$. Using Lemma 39, we know that $\mathsf{Pr}[U|m > q] \leq \frac{1+\kappa}{|R_{F\downarrow S}|-1}\frac{2^m}{|R_{F\downarrow S}|-1}$ $\sum_{i=q-3}^{q}\frac{1}{1-\frac{1+\kappa}{2^{m-i}}}$. The R.H.S. is maximized when $m = q+1$. Hence $\mathsf{Pr}[U|m > q] \leq \frac{1+\kappa}{|R_{F\downarrow S}|-1}\frac{2^m}{|R_{F\downarrow S}|-1}\sum_{i=q-3}^{q}\frac{1}{1-\frac{1+\kappa}{2^{q+1-i}}}$. Noting that pivot $= \frac{|R_{F\downarrow S}|-1}{2^m} > 10$ and expanding the above summation $\mathsf{Pr}[U|m > q] \leq \frac{1+\kappa}{|R_{F\downarrow S}|-1}\frac{1}{10}\left(\frac{1}{(1-\frac{1+\kappa}{2^4})^2} + \frac{1}{(1-\frac{1+\kappa}{2^3})^2} + \right.$

$\frac{1}{(1-\frac{1+\kappa}{2^2})^2} + \frac{1}{(1-\frac{1+\kappa}{2^1})^2}\Bigg)$. Using $\kappa \leq 1$ for the first two summation terms, $\Pr[U|m > q] \leq \frac{1+\kappa}{|R_{F\downarrow S}|-1} \cdot \frac{1}{10} \cdot (7.1 + \frac{4}{(1-\kappa)^2})$

Summing up all the above cases, $\Pr[U] = \Pr[U|m < q-3] \times \Pr[m < q-3] + \Pr[U|q-3 \leq m \leq q] \times \Pr[q-3 \leq m \leq q] + \Pr[U|m > q] \times \Pr[m > q]$. Using $\Pr[m < q-1] \leq 0.2$, $\Pr[m > q] \leq 0.2$ and $\Pr[q-3 \leq m \leq q] \leq 1$. Therefore, $\Pr[U] \leq \frac{1+\kappa}{|R_{F\downarrow S}|-1}(2.23 + \frac{0.48}{(1-\kappa)^2})$

□

Combining Lemma 27 and 29, the following theorem is obtained.

**Theorem 30.** *For every witness $y$ of $F$, if $\varepsilon > 1.71$,*

$$\frac{1}{(1+\varepsilon)(|R_{F\downarrow S}|-1)} \leq \Pr\left[\mathsf{UniGen}(F,\varepsilon,X) = y\right] \leq (1+\varepsilon)\frac{1}{|R_{F\downarrow S}|-1}.$$

*Proof.* The proof is completed by using Lemmas 27 and 29 and substituting $(1+\varepsilon) = (1+\kappa)(2.23 + \frac{0.48}{(1-\kappa)^2})$. To arrive at the results, we use the inequality $\frac{1.06+\kappa}{0.8(1-e^{-1})} \leq (1+\kappa)(2.23 + \frac{0.48}{(1-\kappa)^2})$. □

**Theorem 31.** *Algorithm* UniGen *succeeds (i.e. does not return $\perp$) with probability at least* 0.62.

*Proof.* If $|R_{F\downarrow S}| \leq 1 + (1+\kappa)\mathsf{pivot}$, the theorem holds trivially. Suppose $|R_{F\downarrow S}| > 1 + (1+\kappa)\mathsf{pivot}$ and let $P_{\text{succ}}$ denote the probability that a run of the algorithm UniGen succeeds. Let $p_i$, such that $(q-3 \leq i \leq q)$ denote the conditional probability that UniGen $(F,\ \varepsilon,\ X)$ terminates in iteration $i$ of the repeat-until loop (line 11-16) with $\frac{\mathsf{pivot}}{1+\kappa} \leq |R_{F,h,\alpha}| \leq 1 + (1+\kappa)\mathsf{pivot}$, given $|R_{F\downarrow S}| > 1 + (1+\kappa)\mathsf{pivot}$. Therefore, $P_{\text{succ}} = \sum_{i=q-3}^{q} p_i \prod_{j=q-3}^{i}(1-p_j)$. Let $f_m = \Pr[q-3 \leq m \leq q]$. Therefore, $P_{\text{succ}} \geq p_m f_m \geq 0.8 p_m$. The theorem is now proved by using Chebyshev's Inequality to show

that $p_m \geq 1 - e^{-3/2} \geq 0.77$.

For every $y \in \{0,1\}^n$ and for every $\alpha \in \{0,1\}^m$, define an indicator variable $\nu_{y,\alpha}$ as follows: $\nu_{y,\alpha} = 1$ if $h(y) = \alpha$, and $\nu_{y,\alpha} = 0$ otherwise. Let us fix $\alpha$ and $y$ and choose $h$ uniformly at random from $H_{xor}(n,m,3)$. The random choice of $h$ induces a probability distribution on $\nu_{y,\alpha}$, such that $\mathsf{Pr}[\nu_{y,\alpha} = 1] = \mathsf{Pr}[h(y) = \alpha] = 2^{-m}$ and $\mathsf{E}[\nu_{y,\alpha}] = \mathsf{Pr}[\nu_{y,\alpha} = 1] = 2^{-m}$. In addition 3-wise independence of hash functions chosen from $H_{xor}(n,m,3)$ implies that for every distinct $y_a, y_b, y_c \in R_{F\downarrow S}$, the random variables $\nu_{y_a,\alpha}, \nu_{y_b,\alpha}$ and $\nu_{y_c,\alpha}$ are 3-wise independent.

Let $\Gamma_\alpha = \sum_{y \in R_{F\downarrow S}} \nu_{y,\alpha}$ and $\mu_\alpha = \mathsf{E}[\Gamma_\alpha]$. Clearly, $\Gamma_\alpha = |R_{F,h,\alpha}|$ and $\mu_\alpha = \sum_{y \in R_{F\downarrow S}} \mathsf{E}[\nu_{y,\alpha}] = 2^{-m}|R_{F\downarrow S}|$. Since $|R_{F\downarrow S}| > \mathit{pivot}$ and $i - l > 0$, using the expression for *pivot*, we get $3 \leq \left\lfloor e^{-1/2}(1+\frac{1}{\varepsilon})^{-2} \cdot \frac{|R_{F\downarrow S}|}{2^m} \right\rfloor$. Therefore, using Chebyshev's Inequality, $\mathsf{Pr}\left[\frac{|R_{F\downarrow S}|}{2^m}.(1-\frac{\kappa}{1+\kappa}) \leq |R_{F,h,\alpha}| \leq (1+\kappa)\frac{|R_{F\downarrow S}|}{2^m}\right] > 1 - e^{-3/2}$. Simplifying and noting that $\frac{\kappa}{1+\kappa} < \kappa$ for all $\kappa > 0$, we obtain $\mathsf{Pr}\left[(1+\kappa)^{-1} \cdot \frac{|R_{F\downarrow S}|}{2^m} \leq |R_{F,h,\alpha}| \leq (1+\kappa) \cdot \frac{|R_{F\downarrow S}|}{2^m}\right] > 1 - e^{-3/2}$. Also, $\frac{\text{pivot}}{1+\kappa} = \frac{1}{1+\kappa}\frac{|R_{F\downarrow S}|-1}{2^m} \leq \frac{|R_{F\downarrow S}|}{(1+\kappa)2^m}$ and $1 + (1+\kappa)\text{pivot} = 1 + \frac{(1+\kappa)(|R_{F\downarrow S}|-1)}{2^m} \geq \frac{(1+\kappa)|R_{F\downarrow S}|}{2^m}$. Therefore, $p_m = \mathsf{Pr}[\frac{\text{pivot}}{1+\kappa} \leq |R_{F,h,\alpha}| \leq 1+(1+\kappa)\text{pivot}] \geq \mathsf{Pr}\left[(1+\kappa)^{-1} \cdot \frac{|R_{F\downarrow S}|}{2^m} \leq |R_{F,h,\alpha}| \leq (1+\kappa) \cdot \frac{|R_{F\downarrow S}|}{2^m}\right] \geq 1 - e^{-3/2}$. □

The guarantees provided by Theorem 30 are significantly stronger than those provided by earlier generators that scale to large problem instances. Specifically, neither XORSample′ [88] nor UniWit [37] provide strong upper bounds for the probability of generation of a witness. PAWS [69] offers a *probabilistic* guarantee that the probability of generation of a witness lies within a tolerance factor of the uniform probability, while the guarantee of Theorem 30 is not prbabilistic. The success probability of PAWS, like that of XORSample′, is bounded below by an expression that depends on difficult-to-estimate input parameters. Interestingly, the same parameters also directly affect the tolerance of distribution of the generated witnesses. The success probability of UniWit is bounded below by 0.125, which is significantly smaller than the lower bound of 0.62 guaranteed by Theorem 30.

## 8.4 Trading scalability with uniformity

The tolerance parameter $\varepsilon$ provides a knob to balance scalability and uniformity in UniGen. Smaller values of $\varepsilon$ lead to stronger guarantees of uniformity (by Theorem 30). Note, however, that the value of "hiThresh" increases with decreasing values of $\varepsilon$, requiring BoundedSAT to find more witnesses. Thus, each invocation of BoundedSAT is likely to take longer as $\varepsilon$ is reduced.

## 8.5 Experimental Results

To evaluate the performance of UniGen, we built a prototype implementation and conducted an extensive set of experiments. Industrial constrained-random verification problem instances are typically proprietary and unavailable for published research. Therefore, we conducted experiments on CNF SAT constraints arising from several problems available in the public-domain. These included bit-blasted versions of constraints arising in bounded model checking of circuits and used in [37], bit-blasted versions of SMTLib benchmarks, constraints arising from automated program synthesis, and constraints arising from ISCAS89 circuits with parity conditions on randomly chosen subsets of outputs and next-state variables.

To facilitate running multiple experiments in parallel, we used a high-performance cluster and ran each experiment on a node of the cluster. Each node had two quad-core Intel Xeon processors with 4 GB of main memory. Recalling the terminology used in the pseudocode of UniGen (see Section 8.1), we set the tolerance $\varepsilon$ to 6, and the sampling set $S$ to an independent support of $F$ in all our experiments. Independent supports (not necessarily minimal ones) for all benchmarks were easily obtained from the providers of the benchmarks on request. We used $2,500$ seconds as the timeout for each invocation of BoundedSAT and 20 hours as the overall timeout for UniGen, for each problem instance. If an invocation of BoundedSAT timed out in line 16 of the pseudocode of UniGen, we repeated the execution of lines 14–16 without incrementing $i$. With this set-up, UniGen was able to successfully generate random witnesses for formulas having up to $486,193$ variables.

For performance comparisons, we also implemented and conducted experiments with UniWit – a state-of-art near-uniform generator [37]. Our choice of UniWit as a reference for comparison is motivated by several factors. First, UniGen and UniWit share some commonalities, and UniGen can be viewed as an improvement of UniWit. Second, XORSample′ is known to perform poorly vis-a-vis UniWit [37]; hence, comparing with XORSample′ is not meaningful. Third, the implementation of PAWS made available by the authors of [69] currently does not accept CNF formulae as inputs. It accepts only a graphical model of a discrete distribution as input, making a direct comparison with UniGen difficult. Since PAWS and UniWit share the same scalability problem related to large random xor-clauses, we chose to focus only on UniWit. Since the "leapfrogging" heuristic used in [37] nullifies the guarantees of UniWit, we disabled this optimization. For fairness of comparison, we used the same timeouts in UniWit as used in UniGen, i.e. $2,500$ seconds for every invocation of BoundedSAT, and 20 hours overall for every invocation of UniWit.

Table 8.1 presents the results of our performance-comparison experiments for a subset of benchmarks*. Column 1 lists the CNF benchmark, and columns 2 and 3 give the count of variables and size of independent support used, respectively. The

*The full version of Table 8.1 is available in Appedix as Table A4

results of experiments with UniGen are presented in the next 3 columns. Column 4 gives the observed probability of success of UniGen when generating $1,000$ random witnesses. Column 5 gives the average time taken by UniGen to generate one witness (averaged over a large number of runs), while column 6 gives the average number of variables per xor-clause used for randomly partitioning $R_F$. The next two columns give results of our experiments with UniWit. Column 7 lists the average time taken by UniWit to generate a random witness, and column 8 gives the average number of variables per xor-clause used to partition $R_F$. A "–" in any column means that the corresponding experiment failed to generate any witness in 20 hours.

It is clear from Table 8.1 that the average run-time for generating a random witness by UniWit can be two to three orders of magnitude larger than the corresponding run-time for UniGen. This is attributable to two reasons. The first stems from fewer variables in xor-clauses and blocking clauses when small independent supports are used. Benchmark "tutorial3" exemplifies this case. Here, UniWit failed to generate any witness because all calls to BoundedSAT in UniWit, with xor-clauses and blocking clauses containing numbers of variables, timed out. In contrast, the calls to BoundedSAT in UniGen took much less time, due to short xor-clauses and blocking clauses using only variables from the independent support. The other reason for UniGen's improved efficiency is that the computationally expensive step of identifying a a good range of values for $m$ (see Section 8.1 for details) needs to be executed only once per benchmark. Subsequently, whenever a random witness is needed, UniGen simply iterates over this narrow range of $m$. In contrast, generating every witness in UniWit (without leapfrogging) requires sequentially searching over all values afresh to find a good choice for $m$. Referring to Table 8.1, UniWit requires more than $20,000$ seconds on average to find a good value for $m$ and generate a random witness for

Table 8.1 : Runtime performance comparison of UniGen and UniWit

| | | | UniGen | | | UniWit | | |
|---|---|---|---|---|---|---|---|---|
| Benchmark | $\|X\|$ | $\|S\|$ | Succ Prob | Avg Run Time (s) | Avg XOR leng | Avg Run Time (s) | Avg XOR len | Succ Prob |
| Squaring7 | 1628 | 72 | 1.0 | 2.44 | 36 | 2937.5 | 813 | 0.87 |
| squaring8 | 1101 | 72 | 1.0 | 1.77 | 36 | 5212.19 | 550 | 1.0 |
| Squaring10 | 1099 | 72 | 1.0 | 1.83 | 36 | 4521.11 | 550 | 0.5 |
| s1196a_7_4 | 708 | 32 | 1.0 | 6.9 | 16 | 833.1 | 353 | 0.37 |
| s1238a_7_4 | 704 | 32 | 1.0 | 7.26 | 16 | 1570.27 | 352 | 0.35 |
| s953a_3_2 | 515 | 45 | 0.99 | 12.48 | 23 | 22414.86 | 257 | * |
| EnqueueSeqSK | 16466 | 42 | 1.0 | 32.39 | 21 | – | – | – |
| LoginService2 | 11511 | 36 | 0.98 | 6.14 | 18 | – | – | – |
| LLReverse | 63797 | 25 | 1.0 | 33.92 | 13 | 3460.58 | 31888 | 0.63 |
| Sort | 12125 | 52 | 0.99 | 79.44 | 26 | – | – | – |
| Karatsuba | 19594 | 41 | 1.0 | 85.64 | 21 | – | – | – |
| tutorial3 | 486193 | 31 | 0.98 | 782.85 | 16 | – | – | – |

A "*" entry indicates insufficient data for estimating success probability benchmark "s953a_3_2". Unlike in UniGen, there is no way to amortize this large time over multiple runs in UniWit, while preserving the guarantee of near-uniformity.

Table 8.1 also shows that the observed success probability of UniGen is almost always 1, much higher than what Theorem 30 guarantees and better than those from UniWit. It is clear from our experiments that UniGen can scale to problems involving almost 500K variables, while preserving guarantees of almost uniformity. This goes much beyond the reach of any other random-witness generator that gives strong guarantees on the distribution of witnesses.

Theorem 30 guarantees that the probability of generation of every witness lies within a specified tolerance of the uniform probability. In practice, however, the distribution of witnesses generated by UniGen is much more closer to a uniform distribution. To illustrate this, we implemented a *uniform sampler*, henceforth called US, and compared the distributions of witnesses generated by UniGen and by US for some representative benchmarks. Given a CNF formula $F$, US first determines $|R_F|$ using an exact model counter (such as sharpSAT). To mimic generating a random witness, US simply generates a random number $i$ in $\{1 \ldots |R_F|\}$. To ensure fair comparison, we used the same source of randomness in both UniGen and US. For every problem instance on which the comparison was done, we generated a large number $N$ $(= 4 \times 10^6)$ of sample witnesses using each of US and UniGen. In each case, the number of times various witnesses were generated was recorded, yielding a distribution of the counts. Figure 8.1 shows the distributions of counts generated by UniGen and by US for one of our benchmarks (case110) with $16,384$ witnesses. The horizontal axis represents counts and the vertical axis represents the number of witnesses appearing a specified number of times. Thus, the point $(242, 450)$ represents the fact that each of 450 distinct witnesses were generated 242 times in $4 \times 10^6$ runs. Observe that the distributions resulting from UniGen and US can hardly be distinguished in practice. This holds not only for this benchmark, but for all other benchmarks we experimented with. Overall, our experiments confirm that UniGen is two to three orders of magnitude more efficient than state-of-the-art random witness generators, has probability of success almost 1, and preserves strong guarantees about the uniformity of generated witnesses. Furthermore, the distribution of generated witnesses can hardly be distinguished from that of a uniform sampler in practice.

## 8.6 Chapter Summary

Marrying scalability with strong guarantees of uniformity has been the holy grail of sampling algorithms. Despite long history of theoretical as well as practical interest, prior work in this area either provided strong theoretical guarantees at the cost of scalability, or remedy the scalabilty problem at the cost of guarantees of uniformity.

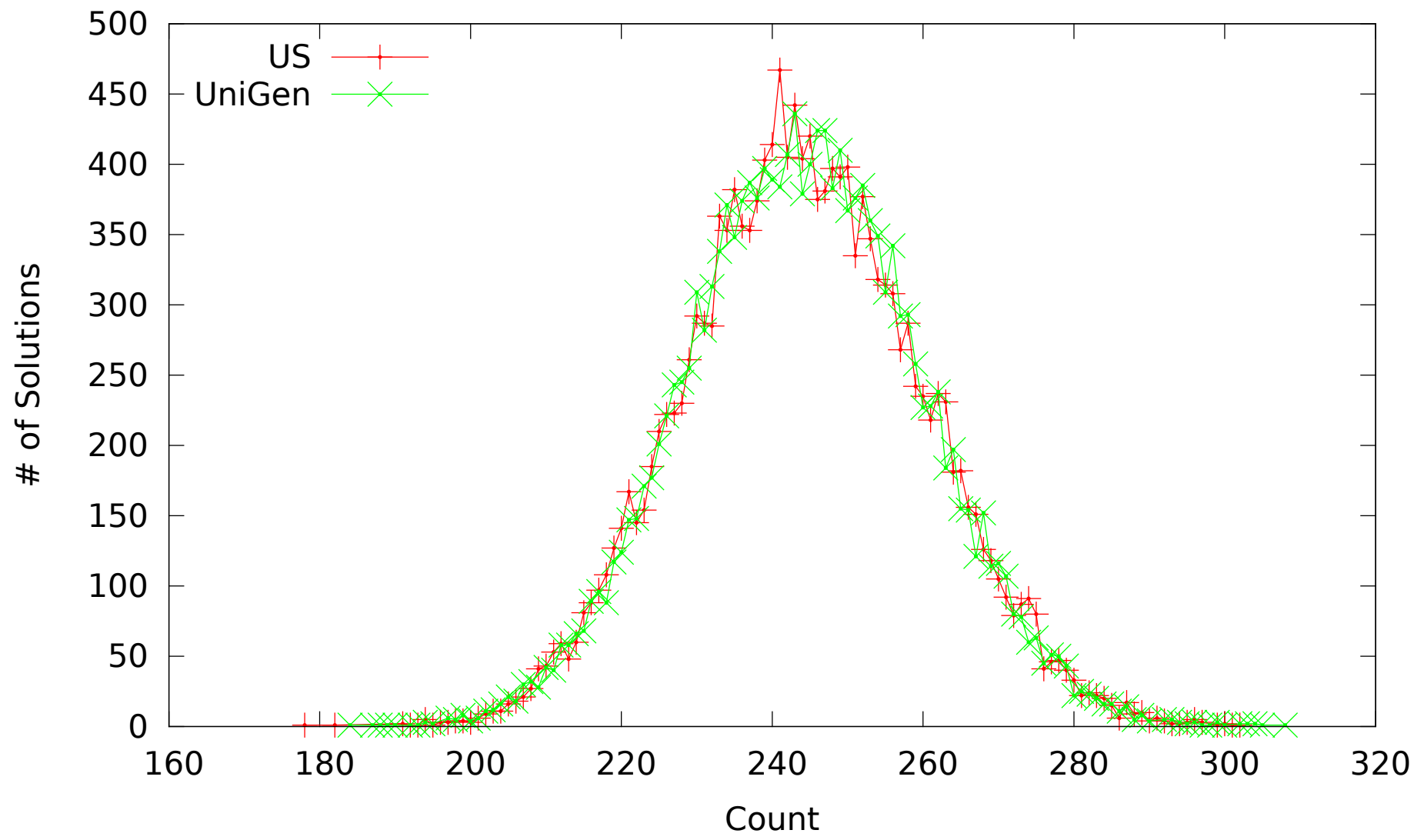

Figure 8.1 : Uniformity comparison between Uniform Sampler (US) and UniGen on benchmark 'case110'

In this chapter, we took a step towards design of scalable algorithms with strong theoretical guarantees. Building on the ideas introduced in the previous part, we designed a new hashing-based algorithm called UniGen, which is the first algorithm to provide guarantees of almost-uniformity, while scaling to the problems involving hundreds of thousands of variables. As a mark of departure from previous hashing-based approach, UniGen first invokes an approximate model counting routine to get an estimate of the number of cells that it should divide the space of solutions into. Then, UniGen employs SAT solver to enumerate all the solutions for a randomly chosen cell that passes the check for "smallness". In order to design efficient SAT queries, we introduced the notion of sampling set of the variables which allows construction of sparser hash functions. Consequently, UniGen is able to scale to problems involving hundreds of thousands of variables where the sampling set is small.

While UniGen significantly outperforms prior state of the art, it should be viewed as the first step towards design of scalable sampling techniques with rigorous formal guarantees. The next two chapters will discuss several ways to push the scalability barrier further. In particular, we employ parallelism and sacrifice of independence to obtain performance gain in the following Chapter.

# Chapter 9

# Parallelization

Since the end of Dennard scaling, there has been a strong revival of interest in parallelizing a wide variety of algorithms to achieve improved performance [75]. While simulation-based verification typically involves running in parallel many simulations with different input stimuli, the generation of these stimuli is often done sequentially. This is because existing approaches to stimulus generation are not efficiently parallelizable without degrading guarantees of uniformity. One of the main goals in parallel-algorithm design is to achieve a speedup nearly linear in the number of processors, which requires the avoidance of dependencies among different parts of the algorithm [65]. Most of the sampling algorithms used for uniform witness generation fail to meet this criterion, and are hence not easily parallelizable. For example, approaches based on random seeding of a SAT solver maintain information about which regions of the solution space have already been explored, since the random seed often is not enough to steer the solver towards new regions of the solution space [109]. Different threads generating solutions must therefore communicate with each other, impeding efficient parallelization. In MCMC-based approaches, to generate independent samples in parallel each thread has to take a walk until a stationary distribution is reached. This often takes exponential time in the case of hard combinatorial spaces with complex internal structure [69]. Heuristics to speed up MCMC-based techniques destroy guarantees of uniformity even in the sequential case [109]. Methods based on random walks on WBDDs are amenable to parallelization, but they are known not to scale beyond a few hundred variables. The lack of techniques for sampling solutions of constraints in parallel while preserving guarantees of effectiveness in finding bugs is therefore a major impediment to high-performance CRV.

The algorithm UniGen2 presented in this Chapter takes a step forward in addressing the above problem. It has an initial preprocessing step that is sequential but low-overhead, followed by inherently parallelizable sampling steps. It generates samples (stimuli) that are provably almost as effective as those generated by a uniform sampler for purposes of detecting a bug. Furthermore, our experiments demonstrate that a parallel implementation of UniGen2 achieves a near-linear speedup in the number of processor cores. Given that current practitioners are forced to trade guarantees of effectiveness in bug hunting for scalability, the above properties of UniGen2 are significant. Specifically, they enable a new paradigm of CRV wherein parallel stimulus generation and simulation can provide the required runtime performance while also providing theoretical guarantees.

## 9.1 Algorithm

Our algorithm, named UniGen2, bears some structural similarities with the UniGen algorithm proposed earlier in [39]. Nevertheless, there are key differences that allow UniGen2 to outperform UniGen significantly. Like UniGen, UniGen2 takes a CNF formula $F$, a sampling set $S$ and a tolerance $\varepsilon$ (that is chosen to be at least 6.84 for technical reasons). Note that the formula $F$ and set $S$ uniquely define the solution set $R_{F\downarrow S}$.

Similarly to UniGen, UniGen2 works by partitioning $R_{F\downarrow S}$ into "cells" using random hash functions, then randomly selecting a cell by adding appropriate constraints to $F$. If the chosen cell has the right size (where the acceptable size range depends on the desired tolerance $\varepsilon$), we can enumerate all the solutions in it and return a uniform random sample from among them. Unlike UniGen, however, UniGen2 samples multiple times from the same cell. This decreases the generation time per sample by a large factor (about $10\times$ in our experiments), while preserving strong guarantees of effectiveness of the samples in finding bugs.

**Algorithm 11** EstimateParameters($F, S, \varepsilon$)

/* Returns (hashBits, loThresh, thresh) as required by GenerateSamples */

1: Find $\kappa \in (0,1)$ such that $\varepsilon = (1+\kappa)(7.44 + \frac{0.392}{(1-\kappa)^2}) - 1$
2: pivot $\leftarrow \left\lceil 4.03\left(1+\frac{1}{\kappa}\right)^2 \right\rceil$
3: thresh $\leftarrow \lceil 1 + \sqrt{2}(1+\kappa)\text{pivot} \rceil$; loThresh $\leftarrow \left\lfloor \frac{1}{\sqrt{2}(1+\kappa)}\text{pivot} \right\rfloor$
4: $i \leftarrow 0$
5: **while** $i < n$ **do**
6: $\quad i \leftarrow i+1$
7: $\quad$ Choose $h$ at random from $H_{xor}(|S|, i)$
8: $\quad$ Choose $\alpha$ at random from $\{0,1\}^i$
9: $\quad Y \leftarrow$ BoundedSAT$(F \wedge (h(S) = \alpha), 61, S)$
10: $\quad$ **if** $1 \leq |Y| \leq 60$ **then**
11: $\quad\quad$ **return** (round $(\log|Y| + i + \log 1.8 - \log \text{pivot})$ , loThresh, thresh)
12: **return** $\perp$

UniGen2 is an algorithmic framework that operates in two stages: the first stage, EstimateParameters (Algorithm 11), performs low-overhead one-time preprocessing for a given $F$, $S$, and $\varepsilon$ to compute numerical parameters 'hashBits', 'loThresh', and 'thresh'. The quantity hashBits controls how many cells $R_{F\downarrow S}$ will be partitioned

**Algorithm 12** GenerateSamples($F, S, \text{hashBits}, \text{loThresh}, \text{thresh}$)

1: Pick an order $V$ of the values $\{\text{hashBits} - 2, \text{hashBits} - 1, \text{hashBits}\}$
2: Choose $h$ at random from $H_{xor}(|S|, \text{hashBits})$
3: Choose $\alpha$ at random from $\{0, 1\}^{\text{hashBits}}$
4: **for** $i \in V$ **do**
5: $\quad Y \leftarrow \text{BoundedSAT}(F \wedge (h_i(S) = \alpha_i), \text{thresh}, S)$
6: $\quad$ **if** $(\text{loThresh} \leq |Y| < \text{thresh})$ **then**
7: $\quad\quad$ **return** loThresh distinct random elements of $Y$
8: **return** $\bot$

into, while loThresh and thresh delineate the range of acceptable sizes for a cell. In the second stage, GenerateSamples (Algorithm 12) uses these parameters to generate loThresh samples. If more samples are required, GenerateSamples is simply called again with the same parameters. Theorem 33 below shows that invoking GenerateSamples multiple times does not cause the loss of any theoretical guarantees. We now explain the operation of the two subroutines in detail.

Lines 1–3 of EstimateParameters compute numerical parameters based on the tolerance $\varepsilon$ which are used by GenerateSamples. The variable 'pivot' can be thought of as the ideal cell size we are aiming for, while as mentioned above 'loThresh' and 'thresh' define the allowed size range around this ideal. For simplicity of exposition, we assume that $|R_{F\downarrow S}| > \max(60, \text{thresh})$. If not, there are very few solutions and we can do uniform sampling by enumerating all of them as in UniGen [39].

Lines 4–11 of EstimateParameters compute 'hashBits', an estimate of the number of hash functions required so that the corresponding partition of $R_{F\downarrow S}$ (into $2^{\text{hashBits}}$ cells) has cells of the desired size. This is done along the same lines as in UniGen, which used an approximate model counter such as ApproxMC [38]. The procedure invokes a SAT solver through the function BoundedSAT$(\phi, m, S)$. This returns a set, consisting of models of the formula $\phi$ which all differ on the set of variables $S$, that has size $m$. If there is no such set of size $m$, the function returns a maximal set. If the estimation procedure fails, EstimateParameters returns $\perp$ on line 12. In practice, it would be called repeatedly until it succeeds. Theorem 41 below shows that on average few repetitions are needed for EstimateParameters to succeed, and this is borne out in practice.

The second stage of UniGen2, GenerateSamples, begins on lines 1–4 by picking a hash count $i$ close to hashBits, then selecting a random hash function from the family $H_{xor}(|S|, i)$ on line 2. On line 3 we pick a random output value $\alpha$, so that the constraint $h(S) = \alpha$ picks out a random cell. Then, on line 5 we invoke BoundedSAT on $F$ with this additional constraint, obtaining at most hiThresh elements $Y$ of the cell. If $|Y| <$ thresh then we have enumerated every element of $R_{F\downarrow S}$ in the cell, and if $|Y| \geq$ loThresh the cell is large enough for us to get a good sample. So if loThresh $\leq |Y| <$ thresh, we randomly select loThresh elements of $Y$ and return them on line 7.

If the number of elements of $R_{F\downarrow S}$ in the chosen cell is too large or too small, we choose a new hash count on line 4. Note that line 1 can pick an arbitrary order for the three hash counts to be tried, since our analysis of UniGen2 does not depend on the order. This allows us to use an optimization where if we run GenerateSamples multiple times, we we choose an order which starts with the value of $i$ that was successful in the previous invocation of GenerateSamples. Since hashBits is only an estimate of the correct value for $i$, in many benchmarks on which we experimented, UniGen2 initially failed to generate a cell of the right size with $i = \text{hashBits} - 2$, but then succeeded with $i = \text{hashBits} - 1$. In such scenarios, beginning with $i = \text{hashBits} - 1$ in subsequent iterations saves considerable time. This heuristic is similar in spirit to "leapfrogging" in ApproxMC [38] and UniWit [37], but does not compromise the theoretical guarantees of UniGen2 in any way.

If all three hash values tried on line 4 fail to generate a correctly-sized cell, GenerateSamples fails and returns $\perp$ on line 8. We show below that this happens with probability at most 0.38. Otherwise, UniGen2 completes by returning loThresh samples.

## 9.2 Parallelization

As described above, UniGen2 operates in two stages: EstimateParameters is initially called to do one-time preprocessing, and then GenerateSamples is called to do the actual sampling. To generate $N$ samples, we can invoke EstimateParameters once, and then GenerateSamples $N/\mathsf{loThresh}$ times, since each of the latter calls generates loThresh samples (unless it fails). Furthermore, each invocation of GenerateSamples is completely independent of the others. Thus if we have $k$ processor cores, we can just perform $N/(k \cdot \mathsf{loThresh})$ invocations of GenerateSamples on each. There is no need for any inter-thread communication: the "leapfrogging" heuristic for choosing the order on line 1 can simply be done on a per-thread basis. This gives us a linear speedup in the number of cores $k$, since the per-thread work (excluding the initial preprocessing) is proportional to $1/k$. Furthermore, Theorem 33 below shows that assuming each thread has its own source of randomness, performing multiple invocations of GenerateSamples in parallel does not alter its guarantees of uniformity. This means that UniGen2 can scale to an arbitrary number of processor cores as more samples are desired, while not sacrificing any theoretical guarantees.

## 9.3 Analysis of UniGen2

**Theorem 32.** EstimateParameters *and* GenerateSamples *return* $\bot$ *with probabilities at most* 0.009 *and* 0.38 *respectively.*

*Proof.* By Lemmas 44 and 42 below respectively. □

**Theorem 33.** *For given* $F$*,* $S$*, and* $\varepsilon$*, let* $L$ *be the set of samples generated using* UniGen2 *with a single call to* GenerateSamples*. Then for each* $y \in R_{F\downarrow S}$*, we have*

$$\frac{\text{loThresh}}{(1+\varepsilon)|R_{F\downarrow S}|} \leq \mathsf{Pr}[y \in L] \leq 1.02 \cdot (1+\varepsilon)\frac{\text{loThresh}}{|R_{F\downarrow S}|}.$$

*Proof.* By Lemma 41 below. □

**Theorem 34.** *For given* $F$*,* $S$*, and* $\varepsilon$*, and for* hashBits*,* loThresh*, and* thresh *as estimated by* EstimateParameters*, let* GenerateSamples *be called* $N$ *times with these parameters in an arbitrary parallel or sequential interleaving. Let* $E_{y,i}$ *denote the event that* $y \in R_{F\downarrow S}$ *is generated in the* $i^{th}$ *call to* GenerateSamples*. Then the events* $E_{y,i}$ *are* $(l,u)$-a.a.d. *with* $l = \frac{\text{loThresh}}{(1+\varepsilon)|R_{F\downarrow S}|}$ *and* $u = \frac{1.02\cdot(1+\varepsilon)\text{loThresh}}{|R_{F\downarrow S}|}$.

*Proof.* Different invocations of GenerateSamples use independent randomness for the choices on lines 2, 3, and 7. Therefore the only part of GenerateSamples which can be affected by earlier invocations is the ordering heuristic used on line 1. But Lemma 41 shows that the probability that GenerateSamples returns a particular witness is between $l$ and $u$ regardless of the order used. Therefore $l \leq \Pr[E_{y,i}] \leq u$ even if conditioned on the results of previous invocations, and so the events $E_{y,i}$ are $(l,u)$-*a.a.d.*. □

**Theorem 35.** *There exists a fixed constant $\lambda = 40$ such that for every $F$, $S$, and $\varepsilon$, the expected number of* SAT *queries made by* UniGen2 *per generated sample is at most $\lambda$.*

*Proof.* A successful invocation of GenerateSamples produces loThresh samples and makes at most 3·hiThresh SAT queries (at most hiThresh for each call to BoundedSAT). Since by Theorem 32 GenerateSamples succeeds with probability at least 0.62, the expected number of SAT queries per generated sample is at most $(3 \cdot \text{hiThresh})/(0.62 \cdot \text{loThresh})$. Optimization shows that $\text{hiThresh}/\text{loThresh} < 8.2$, so the expected number of queries per sample is less than 40. □

Finally, we bound the probability of generating a given witness with multiple calls to GenerateSamples.

**Theorem 36.** *Given $F$, $S$, and $\varepsilon$ as above, let* UniGen2 *generate $N$ samples in a list $L$ (by running* GenerateSamples *$N/\text{loThresh}$ times). Then for each $y \in R_{F\downarrow S}$,*

$$\frac{0.93 \cdot N}{(1+\varepsilon)|R_{F\downarrow S}|} \leq \Pr[y \in L] \leq 1.02(1+\varepsilon)\frac{N}{|R_{F\downarrow S}|}.$$

*Proof.* By Theorem 33, if $R$ is the set returned by a single invocation of GenerateSamples we have

$$\frac{\text{loThresh}}{(1+\varepsilon)|R_{F\downarrow S}|} \leq \Pr[y \in R] \leq \frac{1.02 \cdot \text{loThresh}(1+\varepsilon)}{|R_{F\downarrow S}|}$$

regardless of the results of any prior invocations. Therefore

$$\Pr[y \in L] = 1 - \Pr[y \notin L] \geq 1 - \left(1 - \frac{\text{loThresh}}{(1+\varepsilon)|R_{F\downarrow S}|}\right)^{N/\text{loThresh}}.$$

Now noting that

$$\frac{\text{loThresh}}{(1+\varepsilon)|R_{F\downarrow S}|} \cdot \frac{N}{\text{loThresh}} = \frac{N}{(1+\varepsilon)|R_{F\downarrow S}|} \leq \frac{1}{7.84},$$

applying the binomial theorem and observing that the sum of the cubic and higher order terms is positive, we have

$$\Pr[y \in L] \geq \frac{N}{(1+\varepsilon)|R_{F\downarrow S}|}\left(1 - \frac{1}{2! \cdot 7.84}\right) = \frac{0.93 \cdot N}{(1+\varepsilon)|R_{F\downarrow S}|}.$$

For the upper bound, a similar argument shows that

$$\Pr[y \in L] \leq 1 - \left(1 - \frac{1.02(1+\varepsilon)\text{loThresh}}{|R_{F\downarrow S}|}\right)^{N/\text{loThresh}} \leq \frac{1.02(1+\varepsilon)N}{|R_{F\downarrow S}|}.$$

□

### 9.3.1 Analysis of GenerateSamples

Throughout this section, we use the notations $R_{F|S}$ and $R_{F|S,h,\alpha}$ introduced in Chapter 2. We denote by $U_y$ the event that witness $y \in R_{F\downarrow S}$ is output by GenerateSamples when called with the parameters calculated by EstimateParameters on inputs $F$, $S$, and $\varepsilon$. We are interested in providing lower and upper bounds for $\Pr[U_y]$. The proofs presented here follow the structure of the proofs in [39].

Let us denote $\text{round}(\log(|R_{F\downarrow S}| - 1) - \log \text{pivot})$ by $m$, where 'pivot' is the quantity computed on line 2 of EstimateParameters. The expression used for computing pivot ensures that pivot $\geq 17$. Also, as mentioned in Section 9.1, for simplicity we assume that $|R_{F\downarrow S}| > \max(60, \text{hiThresh})$ (in practice this can be checked by simply enumerating up to $\max(60, \text{hiThresh})$ witnesses). Finally, note that the expression for computing $\kappa$ on line 1 of EstimateParameters requires $\varepsilon \geq 6.84$ in order to ensure that $\kappa \in [0, 1)$ can always be found.

The next lemma provides a lower bound on the probability of generation of a witness. Let $w_{i,y,\alpha}$ denote the probability $\Pr\left[\frac{\text{pivot}}{\sqrt{2}(1+\kappa)} \leq |R_{F|S,h,\alpha}| \leq 1 + \sqrt{2}(1+\kappa)\text{pivot} \text{ and } h(y) = \alpha :\right.$ The proof of the lemma also provides a lower bound on $w_{m,y,\alpha}$. Let $p_{i,y}$ denote the probability that GenerateSamples returns on line 7 with a particular value of $i$ and with $y$ in $R_{F|S,h,\alpha}$, where $\alpha \in \{0,1\}^i$ is the value chosen on line 3. Also let $f_m = \mathsf{Pr}[q-2 \leq m \leq q]$, where $q$ is shorthand for the quantity hashBits computed by EstimateParameters.

**Lemma 37.** *Regardless of the order chosen on line 1 of* GenerateSamples*, we have* $\frac{1}{4} \cdot \frac{\text{loThresh}}{\text{hiThresh}} \cdot f_m \cdot p_{m,y} \leq \mathsf{Pr}[U_y] \leq \frac{\text{loThresh}}{|Y|} \sum_{i=q-2}^{q} p_{i,y}$ *for each* $y \in R_{F\downarrow S}$.

*Proof.* $\mathsf{Pr}[U_y] \geq \frac{\text{loThresh}}{\text{hiThresh}} \cdot \frac{1}{2^{q-m}} \cdot p_{m,y} \cdot f_m \geq \frac{\text{loThresh}}{\text{hiThresh}} \cdot \frac{1}{4} \cdot p_{m,y} \cdot f_m$. The upper bound follows from the definition of $U_y$. □

All subsequent results in this section will bound $\mathsf{Pr}[U_y]$ using Lemma 37, so they also hold regardless of the order of hash counts. For notational simplicity we do not always mention this fact in the lemma statements.

**Lemma 38.** *For every* $y \in R_{F\downarrow S}$, $\mathsf{Pr}[U_y] \geq \frac{0.7(1-e^{-3/2})}{4(1.05+\kappa)(|R_{F\downarrow S}|-1)}$

*Proof.* From Lemma 37, we have $\mathsf{Pr}[U_y] \geq \frac{\text{loThresh}}{\text{hiThresh}} \cdot f_m \cdot \frac{1}{4} \cdot p_{m,y}$. Therefore, $\mathsf{Pr}[U_y] \geq \frac{\text{loThresh}}{1+\sqrt{2}(1+\kappa)\text{pivot}} \cdot p_{m,y} \cdot f_m$. By Lemma 47, $f_m > 0.7$. The proof is now completed by showing $p_{m,y} \geq 1//2^m(1 - \frac{1}{(\kappa/(1+\kappa))^2(|R_F|-1)})$. This gives $\mathsf{Pr}[U_y] \geq \frac{0.7(1-e^{-3/2})\text{loThresh}}{4(1+\sqrt{2}(1+\kappa)\text{pivot})2^m} \geq \frac{0.7(1-e^{-3/2})\text{loThresh}}{4(1.05+\kappa)(|R_{F\downarrow S}|-1)}$. The last inequality uses the observation that $1/(\sqrt{2} \cdot \text{pivot}) \leq 0.05$.

To calculate $p_{m,y}$, we first note that since $y \in R_{F\downarrow S}$, the requirement "$y \in R_{F|S,h,\alpha}$" reduces to "$y \in h^{-1}(\alpha)$". For $\alpha \in \{0,1\}^n$, we define $w_{m,y,\alpha}$ as $\mathsf{Pr}\left[\frac{\text{pivot}}{\sqrt{2}(1+\kappa)} \leq |R_{F|S,h,\alpha}| \leq 1+ \sqrt{2}(1+\kappa) \text{ pivot and } h(y) = \alpha : h \xleftarrow{R} H_{xor}(n,m)\right]$. Therefore, $p_{m,y} = \Sigma_{\alpha \in \{0,1\}^m} (w_{m,y,\alpha} \cdot 2^{-m})$. The proof is now completed by showing that $w_{m,y,\alpha} \geq (1-e^{-3/2})/2^m$ for every $\alpha \in \{0,1\}^m$ and $y \in \{0,1\}^n$.

Towards this end, let us first fix a random $y$. Now we define an indicator variable $\gamma_{z,\alpha}$ for every $z \in R_F \setminus \{y\}$ such that $\gamma_{z,\alpha} = 1$ if $h(z) = \alpha$, and $\gamma_{z,\alpha} = 0$ otherwise. Let us fix $\alpha$ and choose $h$ uniformly at random from $H_{xor}(n,m)$. The random choice of h induces a probability distribution on $\gamma_{z,\alpha}$ such that $E[\gamma_{z,\alpha}] = \mathsf{Pr}[\gamma_{z,\alpha} = 1] = 2^{-m}$. Since we have fixed $y$, and since hash functions chosen from $H_{xor}(n, m, 3)$ are 3-wise independent, it follows that for every distinct $z_a, z_b \in R_F \setminus \{y\}$, the random variables $\gamma_{z_a,\alpha}, \gamma_{z_b,\alpha}$ are 2-wise independent. Let $\Gamma_\alpha = \sum_{z \in R_F \setminus \{y\}} \gamma_{z,\alpha}$ and $\mu_\alpha = E[\Gamma_\alpha]$. Clearly, $\Gamma_\alpha = |R_{F|S,h,\alpha}| - 1$ and $\mu_\alpha = \sum_{z \in R_F \setminus \{y\}} E[\gamma_{z,\alpha}] = \frac{|R_F|-1}{2^m}$. Also, $\mathsf{Pr}[\frac{\text{pivot}}{\sqrt{2}(1+\kappa)} \leq |R_{F|S,h,\alpha}| \leq 1 + \sqrt{2}(1+\kappa)\text{pivot}] = \mathsf{Pr}[\frac{\text{pivot}}{\sqrt{2}(1+\kappa)} - 1 \leq |R_{F|S,h,\alpha}| - 1 \leq \sqrt{2}(1+\kappa)\text{pivot}] \geq \mathsf{Pr}[\frac{\text{pivot}}{\sqrt{2}(1+\kappa)} \leq |R_{F|S,h,\alpha}| - 1 \leq \sqrt{2}(1+\kappa)\text{pivot}]$. Using the expression for pivot, we get $2 \leq \lfloor e^{-1/2}(1+1/\epsilon)^2 \cdot \frac{|R_F|-1}{2^m} \rfloor$. Therefore using Chebyshev Inequality and substituting pivot $= (|R_F| - 1)/2^m$, we get $\mathsf{Pr}[\frac{\text{pivot}}{\sqrt{2}(1+\kappa)} \leq |R_{F|S,h,\alpha}| - 1 \leq \sqrt{2}(1+\kappa)\text{pivot}] \geq 1 - \frac{1}{(\kappa/(1+\kappa))^2(|R_F|-1)/2^m}$. Therefore, $\mathsf{Pr}[\frac{\text{pivot}}{\sqrt{2}(1+\kappa)} \leq |R_{F|S,h,\alpha}| \leq 1 + \sqrt{2}(1+\kappa)\text{pivot}] \geq 1 - \frac{1}{(\kappa/(1+\kappa))^2(|R_F|-1)/2^m}$ Since $h$ is chosen at random from $H_{xor}(n, m)$, we also have $\mathsf{Pr}[h(y) = \alpha] = 1/2^m$. It follows that $w_{m,y,\alpha} \geq (1 - \frac{1}{(\kappa/(1+\kappa))^2(|R_F|-1)})/2^m$ □

The next lemma provides an upper bound on $w_{i,y,\alpha}$ and $p_{i,y}$.

**Lemma 39.** *For $i < m-1$, both $w_{i,y,\alpha}$ and $p_{i,y}$ are bounded above by* $\frac{1}{|R_{F\downarrow S}|-1} \frac{1}{\left(1 - \frac{2(1+\kappa)}{2^{m-i}}\right)^2}$.

*Proof.* We will use the terminology introduced in the proof of Lemma 38. Clearly, $\mu_\alpha = \frac{|R_{F\downarrow S}|-1}{2^i}$. Since each $\gamma_{z,\alpha}$ is a 0-1 variable, $\mathsf{V}[\gamma_{z,\alpha}] \leq \mathsf{E}[\gamma_{z,\alpha}]$. Therefore, $\sigma^2_{z,\alpha} \leq \sum_{z \neq y, z \in R_{F\downarrow S}} \mathsf{E}[\gamma_{z,\alpha}] \leq \sum_{z \in R_{F\downarrow S}} \mathsf{E}[\gamma_{z,\alpha}] = \mathsf{E}[\Gamma_\alpha] = 2^{-i}(|R_{F\downarrow S}| - 1)$. So $\mathsf{Pr}[\frac{pivot}{\sqrt{2}(1+\kappa)} \leq |R_{F|S,h,\alpha}| \leq 1 + (1+\kappa)\sqrt{2}\text{pivot}] \leq \mathsf{Pr}[|R_{F|S,h,\alpha}| - 1 \leq (1+\kappa)\sqrt{2}\text{pivot}] \leq \mathsf{Pr}[|R_{F|S,h,\alpha}| - 1 \leq 2(1+\kappa)\frac{|R_{F\downarrow S}|-1}{2^m}]$. From Chebyshev's inequality, we know that $\mathsf{Pr}[|\Gamma_\alpha - \mu_{z,\alpha}| \geq \lambda\sigma_{z,\alpha}] \leq 1/\lambda^2$ for every $\kappa > 0$. By choosing $\lambda = (1 - \frac{2(1+\kappa)}{2^{m-i}})\frac{\mu_{z,\alpha}}{\sigma_{z,\alpha}}$ (Note that $\lambda > 0$ for $i < m-1$), we have $\mathsf{Pr}[|R_{F,h,\alpha}| - 1 \leq (1+\kappa)2\frac{|R_{F\downarrow S}|-1}{2^m}] \leq \mathsf{Pr}\left[|(|R_{F,h,\alpha}| - 1) - \frac{|R_{F\downarrow S}|-1}{2^i}| \geq (1 - \frac{2(1+\kappa)}{2^{m-i}})\frac{|R_{F\downarrow S}|-1}{2^i}\right] \leq \frac{1}{\left(1 - \frac{2(1+\kappa)}{2^{m-i}}\right)^2} \cdot \frac{2^i}{|R_{F\downarrow S}|-1}$. Since $h$ is chosen at random from $H_{xor}(n, m)$, we also have $\mathsf{Pr}[h(y) = \alpha] = 1/2^i$. It follows that $w_{i,y,\alpha} \leq \frac{1}{|R_{F\downarrow S}|-1} \frac{1}{\left(1 - \frac{2(1+\kappa)}{2^{m-i}}\right)^2}$. The bound for $p_{i,y}$ is easily obtained by noting that $p_{i,y} = \Sigma_{\alpha \in \{0,1\}^i} (w_{i,y,\alpha} \cdot 2^{-i})$. □

This allows us to give an upper bound for $\mathsf{Pr}[U_y]$.

**Lemma 40.** *For every* $y \in R_{F\downarrow S}$, $\mathsf{Pr}[U_y] \leq \frac{1+\kappa}{|R_{F\downarrow S}|-1}(7.55 + \frac{0.29}{(1-\kappa)^2})$.

*Proof.* We will use the terminology introduced in the proof of Lemma 38. The proof below uses the inequality $2^m \times \text{pivot} \leq \frac{|R_{F\downarrow S}-|}{\sqrt{2}}$ at several points. Also, From Lemma 37, we have $\mathsf{Pr}[U_y] \leq \sum_{i=q-2}^{q} \frac{\text{loThresh}}{|Y|} p_{i,y} \leq \frac{\sqrt{2}(1+\kappa)\text{loThresh}}{\text{pivot}} \sum_{i=q-2}^{q} p_{i,y}$. We can sub-divide the calculation of $\mathsf{Pr}[U_y]$ into three cases based on the range of the values m can take.

**Case 1 :** $q-2 \leq m \leq q$.

Now there are three values that $m$ can take.

1. $m = q-2$. We know that $p_{i,y} \leq \mathsf{Pr}[h(y) = \alpha] = \frac{1}{2^i}$. Therefore, $\mathsf{Pr}[U_y|m = q-2] \leq \frac{\sqrt{2}(1+\kappa)\text{loThresh}}{\text{pivot}} \cdot \frac{1}{2^{q-2}} \frac{7}{4}$. Substituting the value of pivot and $m$, we get $\mathsf{Pr}[U_y|m = q-2] \leq \frac{7(1+\kappa)\text{loThresh}}{2(|R_{F\downarrow S}|-1)}$.

2. $m = q-1$. For $i \in [q-2, q]$ $p_{i,y} \leq \mathsf{Pr}[h(y) = \alpha] = \frac{1}{2^i}$. $\mathsf{Pr}[U_y|m = q-1] \leq \frac{\sqrt{2}(1+\kappa)\text{loThresh}}{\text{pivot}} \cdot \frac{1}{2^{q-2}} \frac{7}{2}$. Substituting the value of pivot and $m$, we get $\mathsf{Pr}[U_y|m = q-2] \leq \frac{7(1+\kappa)\text{loThresh}}{|R_{F\downarrow S}|-1}$.

3. $m = q$. For $i \in [q-1, q]$ $p_{i,y} \leq \mathsf{Pr}[h(y) = \alpha] = \frac{1}{2^i}$.. Using Lemma 39, we get $p_{q-2,y} \leq \frac{1}{|R_{F\downarrow S}|-1}\left(\frac{1}{(1-\frac{1+\kappa}{2})^2}\right)$. Therefore, $\mathsf{Pr}[U_y|m = q] \leq \frac{\sqrt{2}(1+\kappa)\text{loThresh}}{\text{pivot}} \left(\frac{1}{|R_{F\downarrow S}|-1}\left(\frac{1}{(1-\frac{1+\kappa}{2})^2} \quad + \frac{3}{2^q}\right)\right)$. Noting that pivot $\geq 17$ and $\kappa \leq 1$, $\mathsf{Pr}[U_y|m = q] \leq \frac{(1+\kappa)\text{loThresh}}{|R_{F\downarrow S}|-1}(6 + \frac{0.333}{(1-\kappa)^2})$.

$\mathsf{Pr}[U_y|q-2 \leq m \leq q] \leq \max_i(\mathsf{Pr}[U_y|m = i])$. Therefore, $\mathsf{Pr}[U_y|q-2 \leq m \leq q] \leq \mathsf{Pr}[U_y|m = q] \leq \frac{(1+\kappa)\text{loThresh}}{|R_{F\downarrow S}|-1}(6.667 + \frac{0.333}{(1-\kappa)^2})$.

**Case 2 :** $m < q-2$. $\mathsf{Pr}[U_y|m < q-3] \leq \frac{\sqrt{2}(1+\kappa)}{\text{pivot}} \cdot \frac{1}{2^{q-3}} \frac{7}{4}$. Substituting the value of

pivot and maximizing $m = q+3$, we get $\mathsf{Pr}[U_y | m < q-2] \leq \frac{7(1+\kappa)\text{loThresh}}{4(|R_{F\downarrow S}|-1)}$.

**Case 3 :** $m > q.\mathsf{Pr}[U_y | m > q] \leq \mathsf{Pr}[U_y | m = q+1] = \frac{\sqrt{2}(1+\kappa)\text{loThresh}}{pivot}\left(\frac{2}{2^m} + \frac{1}{|R_{F\downarrow S}|-1}\left(\sum_{i=q-2}^{q-1} \frac{1}{1-\frac{2(1+\kappa)}{2^{m-i}}}\right)\right)$. Noting that pivot $\geq 17$ and expanding the summation, $\mathsf{Pr}[U_y | m > q] \leq \frac{(1+\kappa)\text{loThresh}}{|R_{F\downarrow S}|-1}\left(4 + \frac{\sqrt{2}}{17}\left(\frac{1}{(1-\frac{2(1+\kappa)}{2^3})^2} + \frac{1}{(1-\frac{2(1+\kappa)}{2^2})^2}\right)\right)$. Using $\kappa < 1$ for the first term, $\mathsf{Pr}[U_y | m > q] \leq \frac{(1+\kappa)\text{loThresh}}{|R_{F\downarrow S}|-1}(4.333 + \frac{0.333}{(1-\kappa)^2})$

Summing up all the above cases, $\mathsf{Pr}[U_y] = \mathsf{Pr}[U_y | m < q-2] \times \mathsf{Pr}[m < q-2] + \mathsf{Pr}[U_y | q-2 \leq m \leq q] \times \mathsf{Pr}[q-2 \leq m \leq q] + \mathsf{Pr}[U_y | m > q] \times \mathsf{Pr}[m > q]$. From Lemma 46, we have $\mathsf{Pr}[m < q-1] + \mathsf{Pr}[m > q] \leq 0.177$ and $\mathsf{Pr}[q-3 \leq m \leq q] \leq 1$. Also, $\mathsf{Pr}[U_y | m < q-2] \leq \mathsf{Pr}[U_y | m > q]$. Therefore, $\mathsf{Pr}[U_y | m < q-2] \times \mathsf{Pr}[m < q-2] + \mathsf{Pr}[U_y | m > q] \times \mathsf{Pr}[m > q] \leq 0.177 \times \mathsf{Pr}[U_y | m > q]$ Therefore, $\mathsf{Pr}[U_y] \leq \frac{(1+\kappa)\text{loThresh}}{|R_{F\downarrow S}|-1}(7.44 + \frac{0.392}{(1-\kappa)^2})$. □

Combining Lemmas 38 and 40, the following lemma is obtained.

**Lemma 41.** *Regardless of the order chosen on line 1 of* GenerateSamples, *for every* $y \in R_{F\downarrow S}$ *and* $\varepsilon > 6.84$ *we have*

$$\frac{\text{loThresh}}{(1+\varepsilon)|R_{F\downarrow S}|} \leq \mathsf{Pr}[U_y] \leq 1.02(1+\varepsilon)\frac{\text{loThresh}}{|R_{F\downarrow S}|}.$$

*Proof.* The proof is completed by using Lemmas 38 and 40 and substituting $(1+\varepsilon) = (1+\kappa)(7.44 + \frac{0.392}{(1-\kappa)^2})$. To arrive at the results, we use the inequality $\frac{4(1.05+\kappa)}{0.7(1-e^{-3/2})} \leq (1+\kappa)(7.44 + \frac{0.392}{(1-\kappa)^2})$. Furthermore, we use $\frac{\text{loThresh}}{(1+\varepsilon)|R_{F\downarrow S}|} < \frac{\text{loThresh}}{(1+\varepsilon)(|R_{F\downarrow S}|-1)}$. Also, since we assume $|R_{F\downarrow S}| - 1 \geq 60$, we have $\frac{(1+\varepsilon)\text{loThresh}}{|R_{F\downarrow S}|-1} < \frac{1.02(1+\varepsilon)\text{loThresh}}{|R_{F\downarrow S}|}$. □

**Lemma 42.** GenerateSamples *succeeds (i.e. does not return* $\perp$*) with probability at least* 0.62.

*Proof.* As mentioned above, we are assuming $|R_{F\downarrow S}| > 1 + \sqrt{2}(1+\kappa)\text{pivot}$. Let $P_{\text{succ}}$ denote the probability that $\mathsf{GenerateSamples}$ succeeds. Let $p_i$ with $q-2 \leq i \leq q$ denote the conditional probability that the condition on line 6 of $\mathsf{GenerateSamples}$ evaluates to true with $\frac{\text{pivot}}{\sqrt{2}(1+\kappa)} \leq |R_{F|S,h,\alpha}| \leq 1 + \sqrt{2}(1+\kappa)\text{pivot}$, given that $|R_{F\downarrow S}| > 1 + \sqrt{2}(1+\kappa)\text{pivot}$. Let $f_m = \mathsf{Pr}[q-2 \leq m \leq q]$. Therefore as shown in Lemma 37, $P_{\text{succ}} \geq p_m f_m \geq 0.7 p_m$. The theorem is now proved by using Chebyshev's Inequality to show that $p_m \geq 1 - e^{-3/2} \geq 0.77$.

For every $y \in \{0,1\}^n$ and for every $\alpha \in \{0,1\}^m$, define an indicator variable $\nu_{y,\alpha}$ as follows: $\nu_{y,\alpha} = 1$ if $h(y) = \alpha$, and $\nu_{y,\alpha} = 0$ otherwise. Let us fix $\alpha$ and $y$ and choose $h$ uniformly at random from $H_{xor}(n,m)$. The random choice of $h$ induces a probability distribution on $\nu_{y,\alpha}$, such that $\mathsf{Pr}[\nu_{y,\alpha} = 1] = \mathsf{Pr}[h(y) = \alpha] = 2^{-m}$ and $\mathsf{E}[\nu_{y,\alpha}] = \mathsf{Pr}[\nu_{y,\alpha} = 1] = 2^{-m}$. In addition 3-wise independence of hash functions chosen from $H_{xor}(n,m)$ implies that for every distinct $y_a, y_b, y_c \in R_{F\downarrow S}$, the random variables $\nu_{y_a,\alpha}, \nu_{y_b,\alpha}$ and $\nu_{y_c,\alpha}$ are 3-wise independent.

Let $\Gamma_\alpha = \sum_{y \in R_{F\downarrow S}} \nu_{y,\alpha}$ and $\mu_\alpha = \mathsf{E}[\Gamma_\alpha]$. Clearly, $\Gamma_\alpha = |R_{F,h,\alpha}|$ and $\mu_\alpha = \sum_{y \in R_{F\downarrow S}} \mathsf{E}[\nu_{y,\alpha}] = 2^{-m}|R_{F\downarrow S}|$. Since $|R_{F\downarrow S}| > \text{pivot}$ and $i - l > 0$, using the expression for pivot we get $3 \leq \left\lfloor e^{-1/2}(1+\frac{1}{\kappa})^{-2} \cdot \frac{|R_{F\downarrow S}|}{2^m} \right\rfloor$. Therefore, by Chebyshev's Inequality, $\mathsf{Pr}\left[\frac{|R_{F\downarrow S}|}{2^m}.\left(1 - \frac{\kappa}{1+\kappa}\right) \leq |R_{F|S,h,\alpha}| \leq (1+\kappa)\frac{|R_{F\downarrow S}|}{2^m}\right] > 1 - e^{-3/2}$. Simplifying and noting that $\frac{\kappa}{1+\kappa} < \kappa$ for all $\kappa > 0$, we obtain $\mathsf{Pr}\left[(1+\kappa)^{-1} \cdot \frac{|R_{F\downarrow S}|}{2^m} \leq |R_{F|S,h,\alpha}| \leq (1+\kappa) \cdot \frac{|R_{F\downarrow S}|}{2^m}\right] > 1 - e^{-3/2}$. Also, $\frac{\text{pivot}}{\sqrt{2}(1+\kappa)} \leq \frac{1}{1+\kappa}\frac{|R_{F\downarrow S}|-1}{2^m} \leq \frac{|R_{F\downarrow S}|}{(1+\kappa)2^m}$ and $1 + \sqrt{2}(1+\kappa)\text{pivot} \geq 1 + \frac{(1+\kappa)(|R_{F\downarrow S}|-1)}{2^m} \geq \frac{(1+\kappa)|R_{F\downarrow S}|}{2^m}$. Therefore, $p_m = \mathsf{Pr}[\frac{\text{pivot}}{\sqrt{2}(1+\kappa)} \leq |R_{F|S,h,\alpha}| \leq 1 + \sqrt{2}(1+\kappa)\text{pivot}] \geq \mathsf{Pr}\left[(1+\kappa)^{-1} \cdot \frac{|R_{F\downarrow S}|}{2^m} \leq |R_{F|S,h,\alpha}| \leq (1+\kappa) \cdot \frac{|R_{F\downarrow S}|}{2^m}\right] \geq 1 - e^{-3/2}$. □

### 9.3.2 Analysis of EstimateParameters

In this section we define $\ell = \log(60) - 1$ and $\mu = \mathsf{E}\left[[]\,|R_{F|S,h,\alpha}|\right] = 2^{-i}|R_{F\downarrow S}|$. Putting $HC(x) = \mathsf{round}(\log x + \log 1.8 - \log \text{pivot})$, we show that the value hashBits computed by EstimateParameters is a good estimate of $HC(|R_{F\downarrow S}|)$ with high probability.

The following property of pairwise independent hash functions is the main tool in our analysis.

**Lemma 43.** *With $h$ and $\alpha$ chosen as in* EstimateParameters*, for each $\gamma > 0$ we have*

$$\Pr[(1-\gamma)\mu \leq |R_{F|S,h,\alpha}| \leq (1+\gamma)\mu] \geq 1 - \frac{1}{\gamma^2 \mu}.$$

*Proof.* By pairwise independence, the variance of $|R_{F|S,h,\alpha}|$ is at most $\mu$. The result then follows from Chebyshev's inequality. □

**Lemma 44.** *Given $|R_{F\downarrow S}| > 60$, the probability that* EstimateParameters *returns non-$\perp$ with $i + \ell \leq \log_2 |R_{F\downarrow S}|$, is at least* $0.991$.

*Proof.* Let us denote $\log_2 |R_{F\downarrow S}| - \ell = \log_2 |R_{F\downarrow S}| - (\lfloor \log_2(60) \rfloor - 1)$ by $m$. Since $|R_{F\downarrow S}| > 60$ as noted above and $|R_{F\downarrow S}| \leq 2^n$, we have $\ell < m + \ell \leq n$. Let $p_i$ $(\ell \leq i \leq n)$ denote the conditional probability that EstimateParameters terminates in iteration $i$ of its loop with $1 \leq |R_{F|S,h,\alpha}| \leq 60$, given $|R_{F\downarrow S}| > 60$. Since the choice of $h$ and $\alpha$ in each iteration of the loop are independent of those in previous iterations, the conditional probability that EstimateParameters returns non-$\perp$ with $i \leq \log_2 |R_{F\downarrow S}| = m + l$, given $|R_{F\downarrow S}| > 60$, is $p_\ell + (1-p_\ell)p_{\ell+1} + \cdots + (1-p_\ell)(1-p_{\ell+1})\cdots(1-p_{m+\ell-1})p_{m+\ell}$. Let us denote this sum by $P$. Thus, $P = p_\ell + \sum_{i=\ell+1}^{m+\ell} \prod_{k=\ell}^{i-1}(1-p_k)p_i \geq \left(p_\ell + \sum_{i=\ell+1}^{m+\ell-1} \prod_{k=\ell}^{i-1}(1-p_k)p_i\right) p_{m+\ell} + \prod_{s=\ell}^{m+\ell-1}(1-p_s)p_{m+\ell} = p_{m+\ell}$. The lemma is now proved by showing that $p_{m+\ell} \geq 0.991$. Applying Lemma 43 with $\gamma = 1 - 1/30$ and $i = m = \log_2 |R_{F\downarrow S}| - \ell$, and noting that $\mu = 2^{-i}|R_{F\downarrow S}| = 2^\ell = 30$, we have $\Pr[1 \leq |R_{F|S,h,\alpha}| \leq 59] \geq 0.991$. □

**Lemma 45.** *Let* EstimateParameters *return a* hashBits *value of $c$, with $i$ being the final value of its loop counter. Let $p_i$ be a shorthand for*
$1 - \mathsf{Pr}\left[HC((1.8)^{-1} \cdot |R_{F\downarrow S}|) \leq c \leq HC(1.8 \cdot |R_{F\downarrow S}|)\right]$. *Then* $p_i \leq \frac{0.169}{2^{\log_2 |R_{F\downarrow S}|-l-i}}$.

*Proof.* Since $c \neq \bot$, by line 11 of the pseudocode we have $c = HC(2^i \cdot |R_{F|S,h,\alpha}|)$, where $\alpha, i$ and $h$ denote (with abuse of notation) the values of the corresponding variables in the final iteration of the loop. As mentioned above, we are assuming that $|R_{F\downarrow S}| > 60$. We have $\mu = \frac{30}{2^{\log_2 |R_{F\downarrow S}|-l-i}}$. Applying Lemma 43 with $\gamma = 0.8/(1 + 0.8) < 0.8$, we obtain $\Pr[(1.8)^{-1} \cdot 2^{-i}|R_{F\downarrow S}| \leq |R_{F|S,h,\alpha}| \leq (1.8) \cdot 2^{-i}|R_{F\downarrow S}|] \geq 1 - \frac{5.0625}{\mu} \geq 1 - \frac{0.169}{2^{\log_2 |R_{F\downarrow S}|-l-i}}$. □

Now we can establish that EstimateParameters provides a good estimate of $HC(|R_{F\downarrow S}|)$.

**Lemma 46.** *With* hashBits *computed by* EstimateParameters, *we have*

$$\mathsf{Pr}\left[HC((1.8)^{-1} \cdot |R_{F\downarrow S}|) \leq \text{hashBits} \leq HC((1.8) \cdot |R_{F\downarrow S}|)\right] > 0.823.$$

*Proof.* Let $k^* = \log_2 |R_{F\downarrow S}| - l$ It follows that $\mathsf{Pr}\left[HC((1.8)^{-1} \cdot |R_{F\downarrow S}|) \leq \text{hashBits} \leq HC((1.8) \cdot |R_{F\downarrow S}|)\right] \geq 1 - p_{k^*} - p_{k^*-1} - p_{k^*-2} \geq 1 - 0.169 - \frac{0.169}{2} - \frac{0.169}{4} \geq 0.7$ □

This in turn means that hashBits is a good estimate of the quantity $m$ used in the analysis of GenerateSamples.

**Lemma 47.** *Let* $m = \mathsf{round}(\log(|R_{F\downarrow S}| - 1) - \log \text{pivot})$ *be defined as in Section 9.3.1. For the value* hashBits *computed by* EstimateParameters, *we have*

$$\mathsf{Pr}[\text{hashBits} - 2 \leq m \leq \text{hashBits}] > 0.7.$$

*Proof.* Straightforward computation from Lemma 46, noting that $|R_{F\downarrow S}| > 60$. □

## 9.4 Evaluation

To evaluate the performance of UniGen2, we built a prototype implementation in C++ that employs the solver CryptoMiniSAT [1] to handle CNF-SAT augmented with XORs efficiently *. We conducted an extensive set of experiments on diverse public domain benchmarks, seeking to answer the following questions:

1. How does UniGen2's runtime performance compare to that of UniGen, a state-of-the-art almost-uniform SAT sampler?

2. How does the performance of parallel UniGen2 scale with the # of cores?

3. How does the distribution of samples generated by UniGen2 compare with the ideal distribution?

4. Does parallelization affect the uniformity of the distribution of the samples?

Our experiments showed that UniGen2 outperforms UniGen by a factor of about 20× in terms of runtime. The distribution generated by UniGen2 is statistically indistinguishable from that generated by an ideal uniform sampler. Finally, the runtime performance of parallel UniGen2 scales linearly with the number of cores, while its output distribution continues to remain uniform.

### 9.4.1 Experimental Setup

We conducted experiments on a heterogeneous set of benchmarks used in earlier related work [39]. The benchmarks consisted of ISCAS89 circuits augmented with parity conditions on randomly chosen subsets of outputs and next-state variables, constraints arising in bounded model checking, bit-blasted versions of SMTLib benchmarks, and problems arising from automated program synthesis. For each benchmark, the sampling set $S$ was either taken to be the independent support of the formula or was provided by the corresponding source. Experiments were conducted on a total of 200+ benchmarks. We present results for only a subset of representative benchmarks here. A detailed list of all the benchmarks is available in the Appendix.

*The tool (with source code) is available at https://bitbucket.org/kuldeepmeel/unigen

For purposes of comparison, we also ran experiments with UniGen [39], a state-of-the-art almost-uniform SAT witness generator. We employed the Mersenne Twister to generate pseudo-random numbers, and each thread was seeded independently using the C++ class `random_device`. Both tools used an overall timeout of 20 hours, and a BoundedSAT timeout of 2500 seconds. All experiments used $\varepsilon = 16$, corresponding to loThresh $= 11$ and hiThresh $= 64$. The experiments were conducted on a high-performance computer cluster, where each node had a 12-core, 2.83 GHz Intel Xeon processor, with 4GB of main memory per core.

### 9.4.2 Runtime performance

We compared the runtime performance of UniGen2 with that of UniGen for all our benchmarks. For each benchmark, we generated between 1000 and 10000 samples (depending on the size of the benchmark) and computed the average time taken to generate a sample on a single core. The results of these experiments for a representative subset of benchmarks are shown in Table 9.1. The columns in this table give the benchmark name, the number of variables and clauses, the size of the sampling set, the success probability of UniGen2, and finally the average runtime per sample for both UniGen2 and UniGen in seconds. The success probability of UniGen2 was computed as the fraction of calls to GenerateSamples that successfully generated samples.

Table 9.1 clearly shows that UniGen2 significantly outperforms UniGen on all types of benchmarks, even when run on a single core[†]. Over the entire set of 200+ benchmarks, UniGen2's runtime performance was about 20× better than that of UniGen on average (using the geometric mean). The observed performance gain can be attributed to two factors. First, UniGen2 generates loThresh (11 in our experiments) samples from every cell instead of just 1 in the case of UniGen. This provides a speedup of about 10×. Second, as explained in Section 9.1, UniGen2 uses "leapfrogging" to optimize the order in which the values of $i$ in line 4 of Algorithm 12 are chosen. In contrast, UniGen uses a fixed order. This provides an additional average speedup of 2× in our experiments. Note also that the success probability of UniGen2 is consistently very close to 1 across the entire set of benchmarks.

| | | | | UniGen2 | | UniGen |
|---|---|---|---|---|---|---|
| **Benchmark** | **#vars** | **#clas** | **\|S\|** | **Succ. Prob** | **Runtime(s)** | **Runtime(s)** |
| s1238a_3_2 | 686 | 1850 | 32 | 1.0 | 0.3 | 7.17 |
| s1196a_3_2 | 690 | 1805 | 32 | 1.0 | 0.23 | 4.54 |
| s832a_15_7 | 693 | 2017 | 23 | 1.0 | 0.04 | 0.51 |
| case_1_b12_2 | 827 | 2725 | 45 | 1.0 | 0.24 | 6.77 |
| squaring16 | 1627 | 5835 | 72 | 1.0 | 4.16 | 79.12 |
| squaring7 | 1628 | 5837 | 72 | 1.0 | 0.79 | 21.98 |
| doublyLinkedList | 6890 | 26918 | 37 | 1.0 | 0.04 | 1.23 |
| LoginService2 | 11511 | 41411 | 36 | 1.0 | 0.05 | 0.55 |
| Sort | 12125 | 49611 | 52 | 1.0 | 4.15 | 82.8 |
| 20 | 15475 | 60994 | 51 | 1.0 | 19.08 | 270.78 |
| enqueue | 16466 | 58515 | 42 | 1.0 | 0.87 | 14.67 |
| Karatsuba | 19594 | 82417 | 41 | 1.0 | 5.86 | 80.29 |
| lltraversal | 39912 | 167842 | 23 | 1.0 | 0.18 | 4.86 |
| llreverse | 63797 | 257657 | 25 | 1.0 | 0.73 | 7.59 |
| diagStencil_new | 94607 | 2838579 | 78 | 1.0 | 3.53 | 60.18 |
| tutorial3 | 486193 | 2598178 | 31 | 1.0 | 58.41 | 805.33 |
| demo2_new | 777009 | 3649893 | 45 | 1.0 | 3.47 | 40.33 |

Table 9.1 : Runtime performance comparison of UniGen2 and UniGen (on a single core).

[†]The full version of Table 9.1 is available in Appendix as Table A5.

### 9.4.3 Parallel speedup

To measure the effect of parallelization on runtime performance, we ran the parallel version of UniGen2 with 1 to 12 processor cores on our benchmarks. In each experiment with $C$ cores, we generated 2500 samples per core, and computed the $C$-core resource usage as the ratio of the average individual core runtime to the total number of samples (i.e. $C \times 2500$). We averaged our computations over 7 identical runs. The speedup for $C$ cores was then computed as the ratio of 1-core resource usage to $C$-core resource usage. Figure 9.1 shows how the speedup varies with the number of cores for a subset of our benchmarks. The figure illustrates that parallel UniGen2 generally scales almost linearly with the number of processor cores.

To obtain an estimate of how close UniGen2's performance is to real-world requirements (roughly 10× slowdown compared to a simple SAT call), we measured the slowdown of UniGen2 (and UniGen) running on a single core relative to a simple SAT call on the input formula. The (geometric) mean slowdown for UniGen2 turned out to be 21 compared to 470 for UniGen. This shows that UniGen2 running in parallel on 2–4 cores comes close to matching the requirements of CRV in industrial practice.

### 9.4.4 Uniformity comparison

To measure the quality of the distribution generated by UniGen2 and parallel UniGen2 in practice, we implemented an *ideal sampler*, henceforth denoted as IS. Given a formula $F$, the sampler IS first enumerates all witnesses in $R_{F\downarrow S}$, and then picks an element of $R_{F\downarrow S}$ uniformly at random. We compared the distribution generated by IS with that generated by UniGen2 run sequentially, and with that generated by UniGen2 run in parallel on 12 cores. In the last case, the samples generated by all the cores were aggregated before comparing the distributions. We had to restrict the experiments for comparing distributions to a small subset of our benchmarks, specifically those which had less than $100,000$ solutions. We generated a large number $N$ $(\geq 4 \times 10^6)$ of samples for each benchmark using each of IS, sequential UniGen2, and parallel UniGen2. Since we chose $N$ much larger than $|R_{F\downarrow S}|$, all witnesses occurred multiple times in the list of samples. We then computed the frequency of generation of individual witnesses, and grouped witnesses appearing the same number of times together. Plotting the distribution of frequencies — that is, plotting points $(x, y)$ to indicate that each of $x$ distinct witnesses were generated $y$ times — gives a convenient way to visualize the distribution of the samples. Figure 9.2 depicts this for one representative benchmark (case110, with 16,384 solutions).

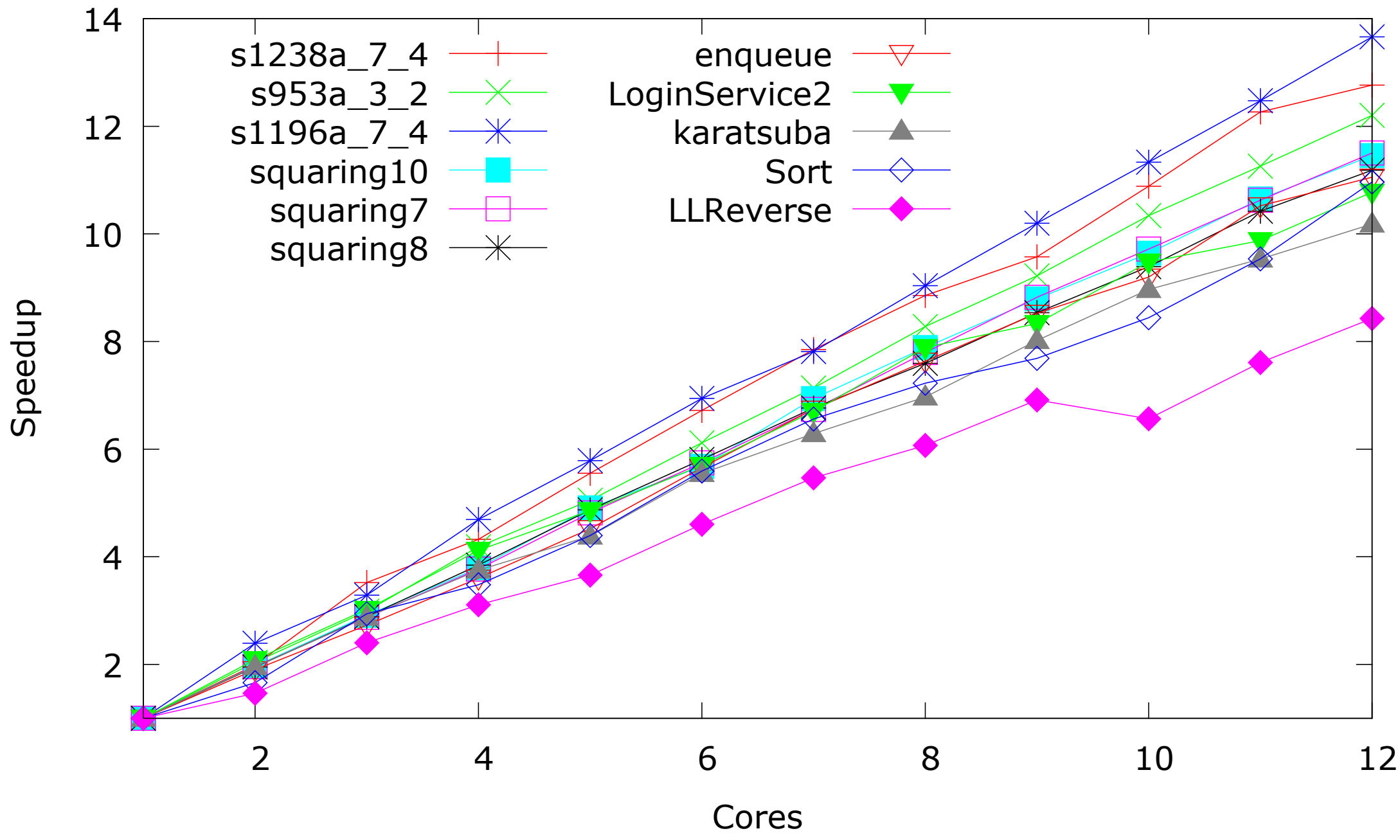

Figure 9.1 : Effect of parallelization on the runtime performance of UniGen2.

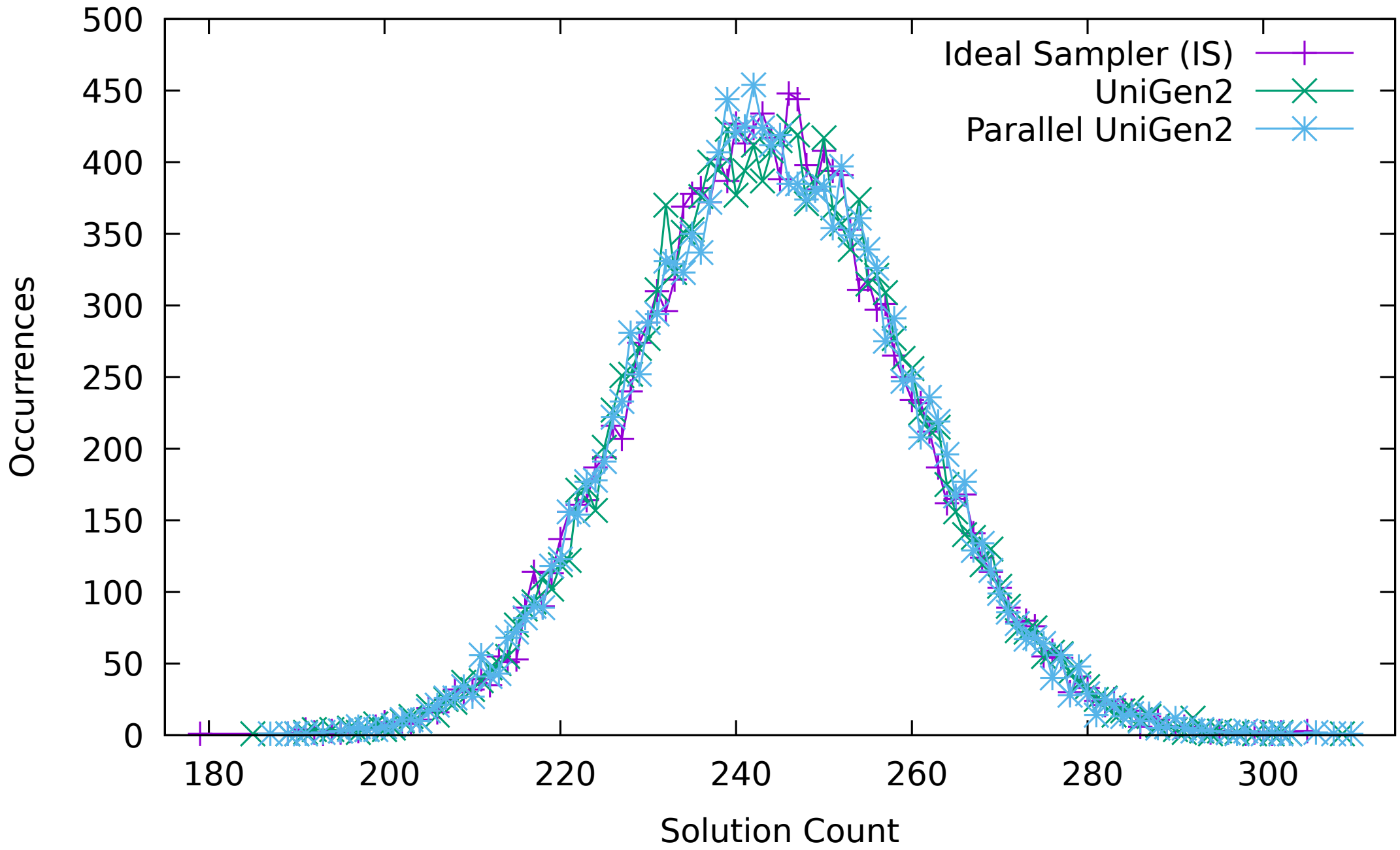

Figure 9.2 : Uniformity comparison between an ideal sampler (IS), UniGen2, and parallel UniGen2. Results from benchmark 'case110' with $N = 4 \cdot 10^6$.

It is clear from Figure 9.2 that the distribution generated by UniGen2 is practically indistinguishable from that of IS. Furthermore, the quality of the distribution is not affected by parallelization. Similar observations also hold for the other benchmarks for which we were able to enumerate all solutions. For the example shown in Fig. 9.2, the Jensen-Shannon distance between the distributions from sequential UniGen2 and IS is 0.049, while the corresponding figure for parallel UniGen2 and IS is 0.052. These small Jensen-Shannon distances make the distribution of UniGen2 (whether sequential or parallel) indistinguishable from that of IS.

## 9.5 Chapter Summary

In this Chapter, we introduced an adaptation of UniGen, UniGen2, that addresses key performance deficiencies of UniGen. Significantly, we showed that UniGen2 achieves a near-linear speedup with the number of cores, without any degradation of uniformity either in theory or in practice. This suggests a new high-performance paradigm for generating (near-)uniformly distributed solutions of a system of constraints. Specifically, it is no longer necessary to gain performance by sacrificing uniformity in a sequential sampler.

In this part, we have introduced a hashing-based paradigm that provides rigorous guarantees of almost-uniformity while scaling to large formulas. Furthermore, the approach is highly parallelizing. Now, let us not lose sight of the forest for the trees. Remember, the problem of sampling as defined in Chapter 2 included a weight function and that weight function was not necessarily uniform!. Can the sampling framework introduced in this part be generalized to handle general weight? We will find out in the next Chapter.

# Chapter 10

# Handling Weighted Distributions for Sampling

In this chapter, we seek to answer whether the hashing-based framework introduced in the previous two chapters can be extended to handle general sampling problems where the weight distribution of interest is not necessarily uniform. To this end, we introduce a novel parameter, *tilt*, which is the ratio of the maximum weight of a satisfying assignment to the minimum weight of a satisfying assignment, to categorize hardness of sampling problem and to provide an affirmative answer to the above question when *tilt* is small. Specifically, we show that UniGen can be adapted to work in the setting of weighted assignments, using only a SAT solver (NP-oracle) and a black-box weight function $w(\cdot)$ when *tilt* is small.

Our assumption about tilt being bounded by a small number is reasonable in several practical situations. For example, when solving probabilistic inference with evidence by reduction to weighted model counting [42], every satisfying assignment of the CNF formula corresponds to an assignment of values to variables in the underlying probabilistic graphical model that is consistent with the evidence. Furthermore, the weight of a satisfying assignment is the joint probability of the corresponding assignment of variables in the probabilistic graphical model. A large tilt would therefore mean existence of two assignments that are consistent with the evidence, but one of which is overwhelmingly more likely than the other. In several real-world problems (see, e.g. Sec 8.3 of [59]), this is considered unlikely given that numerical conditional probability values are often obtained from human experts providing qualitative and rough quantitative data. The algorithms presented in this section require an upper bound for $\rho$ as the input. It is worth noting that although better estimation of upper bounds improve the performance, the algorithms are sound with respect to any upper bound estimate. While an algorithm solution to estimation of upper bound for $\rho$ is beyond the scope of this work, such an estimate can be easily obtained from the designers of probabilistic models. It is easy for designers to estimate upper bound for $\rho$ than accurate estimation of $\mathrm{w}_{\mathrm{max}}$ as the former does not require precise knowledge of probabilities of all the models.

The remainder of the Chapter is organized as follows. Section 10.1 presents an adaptation of UniGen, called WeightGen, that works with small-tilt weight functions. Results of experimenting with a suite of large benchmarks are presented in Section 10.3. Finally, we conclude in Section 10.4.

## 10.1 Algorithm

We assume access to a subroutine called BoundedWeightSAT that takes a CNF formula $F$, a "pivot", an upper bound $r$ of the tilt and an upper bound $w_{max}$ of the maximum weight of a satisfying assignment in the sampling set set $S$. It returns a set of satisfying assignments of $F$ such that the total weight of the returned assignments scaled by $1/w_{max}$ exceeds pivot. It also updates the minimum weight of a satisfying assignment seen so far and returns the same. We discussed BoundedWeightSAT in Chapter 5 but for ease of readability, we repeat the discussion here. BoundedWeightSAT accesses a subroutine AddBlockClause that takes as inputs a formula $F$ and a projected assignment $\sigma|_S$, computes a blocking clause for $\sigma|_S$, and returns the formula $F'$ obtained by conjoining $F$ with the blocking clause thus obtained. Finally, the algorithms assume access to an NP-oracle, which in particular can decide SAT. Both algorithms also accept as input a positive real-valued parameter $r$ which is an upper bound on $\rho$.

The pseudo-code for WeightGen is presented in Algorithm 14. WeightGen takes in a CNF formula $F$, tolerance $\varepsilon > 6.84$, tilt upper bound $r$, and sampling set $S$ and returns a random (approximately weighted-uniform) satisfying assignment. WeightGen can be viewed as adaptation of UniGen to weighted domain.

**Algorithm 13** BoundedWeightSAT$(F, \text{pivot}, r, \mathrm{w}_{\max}, S)$

$\mathrm{w}_{\min} \leftarrow \mathrm{w}_{\max}/r; \mathrm{w}_{\text{total}} \leftarrow 0; Y = \{\};$
**repeat**
    $y \leftarrow$ SolveSAT$(F)$;
    **if** y == UNSAT **then**
        break;
    $Y = Y \cup y$;
    $F =$ AddBlockClause$(F, y|_S)$;
    $\mathrm{w}_{\text{total}} \leftarrow \mathrm{w}_{\text{total}} + W(y)$;
    $\mathrm{w}_{\min} \leftarrow min(\mathrm{w}_{\min}, W(y))$;
**until** $\mathrm{w}_{\text{total}}/(\mathrm{w}_{\min} \cdot r) > pivot$;
**return** $(Y, \mathrm{w}_{\min} \cdot r)$;

**Algorithm 14** WeightGen$(F, \varepsilon, r, S)$

/*Assume $\varepsilon > 6.84$ */

$\mathrm{w_{max}} \leftarrow 1$; Samples $= \{\}$;

$(\kappa, \mathrm{pivot}) \leftarrow$ ComputeKappaPivot$(\varepsilon)$;

$\mathrm{hiThresh} \leftarrow 1 + \sqrt{2}(1+\kappa)\mathrm{pivot}$;

$\mathrm{loThresh} \leftarrow \frac{1}{\sqrt{2}(1+\kappa)}\mathrm{pivot}$;

$(Y, \mathrm{w_{max}}) \leftarrow$ BoundedWeightSAT$(F, \mathrm{hiThresh}, r, \mathrm{w_{max}}, S)$;

**if** $(W(Y)/\mathrm{w_{max}} \leq \mathrm{hiThresh})$ **then**

  Choose $y$ weighted-uniformly at random from $Y$;

  **return** $y$;

**else**

  $(C, \mathrm{w_{max}}) \leftarrow$ WeightMC$(F, 0.8, 0.2)$;

  $q \leftarrow \lceil \log C - \log \mathrm{w_{max}} + \log 1.8 - \log \mathrm{pivot} \rceil$;

  $i \leftarrow q - 4$;

  Choose $h$ at random from $H_{xor}(|S|, q)$

  Choose $\alpha$ at random from $\{0,1\}^q$

  **repeat**

    $i \leftarrow i + 1$;

    $(Y, \mathrm{w_{max}}) \leftarrow$ BoundedWeightSAT$(F \wedge (h_i(S) = \alpha_i), \mathrm{hiThresh}, r, \mathrm{w_{max}}, S)$;

    $W \leftarrow W(Y)/\mathrm{w_{max}}$

  **until** $(\mathrm{loThresh} \leq W \leq \mathrm{hiThresh})$ or $(i = q)$

  **if** $(W > \mathrm{hiThresh})$ or $(W < \mathrm{loThresh})$ **then**

    **return** $\perp$

  **else** Choose $y$ weighted-uniformly at random from $Y$;

      **return** $y$;

**Algorithm 15** ComputeKappaPivot($\varepsilon$)

1: Find $\kappa \in [0, 1)$ such that $\varepsilon = (1 + \kappa)(7.55 + \frac{0.29}{(1-\kappa)^2}) - 1$

2: pivot $\leftarrow \lceil 4.03 \left(1 + \frac{1}{\kappa}\right)^2 \rceil$; **return** ($\kappa$, pivot)

WeightGen first computes $\kappa$ and pivot and uses them to compute hiThresh and loThresh, which quantify the size of a "small" cell. The easy case of the weighted count being less than hiThresh is handled in lines 6–9. Otherwise, WeightMC is called to estimate the weighted model count, which is used to estimate the range of candidate values for $m$. The choice of parameters for WeightMC is motivated by technical reasons. The loop in 15– 19 terminates when a small cell is found and a sample is picked weighted-uniformly at random. Otherwise, the algorithm reports a failure.

### Implementation Details

Similar to WeightMC, our implementation of WeightGen, BoundedWeightSAT is implemented using CryptoMiniSAT [1], a SAT solver that handles xor clauses efficiently. CryptoMiniSAT uses *blocking clauses* to prevent already generated witnesses from being generated again. Since we are interested in only the assignments to sampling set $S$, blocking clauses can be restricted to only variables in the set $S$. We used "random_device" implemented in C++11 as source of pseudo-random numbers to make random choices in WeightGen.

## 10.2 Analysis of WeightGen

For convenience of analysis, we assume that $\log(W(F \downarrow S) - 1) - \log pivot$ is an integer, where *pivot* is the quantity computed by algorithm ComputeKappaPivot. A more careful analysis removes this assumption by scaling the probabilities by constant factors. Let us denote $\log(W(F \downarrow S) - 1) - \log pivot$ by $m$. The expression used for computing $pivot$ in algorithm ComputeKappaPivot ensures that pivot $\geq 17$. Therefore, if an invocation of WeightGen does not return from line 8 of the pseudocode, then $W(F \downarrow S) \geq 18$. Note also that the expression for computing $\kappa$ in algorithm ComputeKappaPivot requires $\varepsilon \geq 1.71$ in order to ensure that $\kappa \in [0, 1)$ can always be found.

In the case where $W(F \downarrow S) \leq 1 + (1 + \kappa)pivot$, BoundedWeightSAT returns all witnesses of $F$ and WeightGen returns a perfect weighted-uniform sample on line 8. So we restrict our attention in the lemmas below to the other case, where as noted above we have $W(F \downarrow S) \geq 18$. The following lemma shows that $q$, computed in line 11 of the pseudocode, is a good estimator of $m$.

**Lemma 48.** $\mathsf{Pr}[q - 3 \leq m \leq q] \geq 0.8$

*Proof.* Recall that in line 10 of the pseudocode, an approximate weighted model counter is invoked to obtain an estimate, $C$, of $W(R_F)$ with tolerance 0.8 and confidence 0.8. By the definition of approximate weighted model counting, we have $\mathsf{Pr}[\frac{C}{1.8} \leq W(R_F) \leq (1.8)C] \geq 0.8$. Defining $c = C/w_{max}$, we have $\mathsf{Pr}[\log c - \log(1.8) \leq \log W(F \downarrow S) \leq \log c + \log(1.8)] \geq 0.8$. It follows that $\mathsf{Pr}[\log c - \log(1.8) - \log pivot - \log(\frac{1}{1-1/W(F\downarrow S)}) \leq \log(W(F \downarrow S) - 1) - \log pivot \leq \log c - \log pivot + \log(1.8) - \log(\frac{1}{1-1/W(F\downarrow S)})] \geq 0.8$. Substituting $q = \lceil \log C - \log \mathrm{w_{max}} + \log 1.8 - \log pivot \rceil = \lceil \log c + \log 1.8 - \log pivot \rceil$, and using the bounds $\mathrm{w_{max}} \leq 1$, $\log 1.8 \leq 0.85$, and $\log(\frac{1}{1-1/W(F\downarrow S)}) \leq 0.12$ (since $W(F \downarrow S) \geq 18$ at line 10 of the pseudocode, as noted above), we have $\mathsf{Pr}[q - 3 \leq m \leq q] \geq 0.8$. □

The next lemma provides a lower bound on the probability of generation of a wit-ness. Let $w_{i,y,\alpha}$ denote the probability $\mathsf{Pr}\left[\frac{pivot}{1+\kappa} \leq \mathcal{W}\left(R_{F,h,\alpha}\right) \leq 1 + (1+\kappa)pivot \wedge h(y) = \alpha\right]$, with $h \xleftarrow{R} H_{xor}(n, i, 3)$. The proof of the lemma also provides a lower bound on $w_{m,y,\alpha}$.

**Lemma 49.** *For every witness $y \in R_F$, $\mathsf{Pr}[y \text{ is output}] \geq \frac{0.8(1-e^{-3/2})\mathcal{W}(y)}{(1.06+\kappa)(W(F{\downarrow}S)-1)}$*

*Proof.* Let $U$ denote the event that witness $y \in R_F$ is output by WeightGen on inputs $F$, $\varepsilon$, $r$, and $X$. Let $p_{i,y}$ denote the probability that we exit the loop at line 19 with a particular value of $i$ and $y \in R_{F,h,\alpha}$, where $\alpha \in \{0,1\}^i$ is the value chosen on line 14. Then, $\mathsf{Pr}[U] = \cup_{i=q-3}^{q} \frac{\mathcal{W}(y)}{\mathcal{W}(Y)} p_{i,y}$, where $Y$ is the set returned by BoundedWeightSAT on line 17. Let $f_m = \mathsf{Pr}[q{-}3 \leq m \leq q]$. From Lemma 48, we know that $f_m \geq 0.8$. From line 20, we also know that $\frac{1}{1+\kappa}pivot \leq \mathcal{W}(Y) \leq 1+(1+\kappa)pivot$. Therefore, $\mathsf{Pr}[U] \geq \frac{\mathcal{W}(y)}{1+(1+\kappa)pivot} \cdot p_{m,y} \cdot f_m$. The proof is now completed by showing $p_{m,y} \geq \frac{1}{2^m}(1 - e^{-3/2})$, as then we have $\mathsf{Pr}[U] \geq \frac{0.8(1-e^{-3/2})}{(1+(1+\kappa)pivot)2^m} \geq \frac{0.8(1-e^{-3/2})}{(1.06+\kappa)(W(F{\downarrow}S)|-1)}$. The last inequality uses the observation that $1/pivot \leq 0.06$.

To calculate $p_{m,y}$, we first note that since $y \in R_F$, the requirement "$y \in R_{F,h,\alpha}$" reduces to "$y \in h^{-1}(\alpha)$". For $\alpha \in \{0,1\}^n$, we define $w_{m,y,\alpha} = \mathsf{Pr}\left[\frac{pivot}{1+\kappa} \leq \mathcal{W}\left(R_{F,h,\alpha}\right) \leq 1 + (1+\kappa)\right.$ $\left. pivot \wedge h(y) = \alpha : h \xleftarrow{R} H_{xor}(n, m, 3)\right]$. Then we have $p_{m,y} = \Sigma_{\alpha \in \{0,1\}^m}\left(w_{m,y,\alpha} \cdot 2^{-m}\right)$. So to prove the desired bound on $p_{m,y}$ it suffices to show that $w_{m,y,\alpha} \geq (1-e^{-3/2})/2^m$ for every $\alpha \in \{0,1\}^m$ and $y \in \{0,1\}^n$.

Towards this end, let us first fix a random $y$. Now we define an indicator variable $\gamma_{z,\alpha}$ for every $z \in R_F \setminus \{y\}$ such that $\gamma_{z,\alpha} = \mathcal{W}(z)$ if $h(z) = \alpha$, and $\gamma_{z,\alpha} = 0$ otherwise. Let us fix $\alpha$ and choose $h$ uniformly at random from $H_{xor}(n, m, 3)$. The random choice of $h$ induces a probability distribution on $\gamma_{z,\alpha}$ such that $\mathsf{E}[\gamma_{z,\alpha}] = \mathcal{W}(z)\,\mathsf{Pr}[\gamma_{z,\alpha} = \mathcal{W}(z)] = \mathcal{W}(z)\,\mathsf{Pr}[h(z) = \alpha] = \mathcal{W}(z)/2^m$. Since we have fixed $y$, and since hash functions chosen from $H_{xor}(n, m, 3)$ are 3-wise independent, it follows that for every distinct $z_a, z_b \in R_F \setminus \{y\}$, the random variables $\gamma_{z_a,\alpha}, \gamma_{z_b,\alpha}$ are 2-wise independent. Let $\Gamma_\alpha = \sum_{z \in R_F \setminus \{y\}} \gamma_{z,\alpha}$ and $\mu_\alpha = \mathsf{E}[\Gamma_\alpha]$. Clearly, $\Gamma_\alpha = \mathcal{W}(R_{F,h,\alpha}) - \mathcal{W}(y)$ and $\mu_\alpha = \sum_{z \in W(F\downarrow S)\setminus\{y\}} \mathsf{E}[\gamma_{z,\alpha}] = (W(F \downarrow S) - \mathcal{W}(y))/2^m$. Since *pivot* $= (W(F \downarrow S) - 1)/2^m \leq (W(F \downarrow S) - \mathcal{W}(y))/2^m$, we have $\mathsf{Pr}[\frac{pivot}{1+\kappa} \leq \mathcal{W}(R_{F,h,\alpha}) \leq 1 + (1+\kappa)pivot] \geq \mathsf{Pr}[\frac{W(F\downarrow S)-\mathcal{W}(y)}{(1+\kappa)2^m} \leq \mathcal{W}(R_{F,h,\alpha}) \leq 1 + (1+\kappa)\frac{W(F\downarrow S)-1}{2^m}] \geq \mathsf{Pr}[\frac{W(F\downarrow S)-\mathcal{W}(y)}{2^m(1+\kappa)} \leq \mathcal{W}(R_{F,h,\alpha}) - \mathcal{W}(y) \leq (1+\kappa)\frac{(W(F\downarrow S)-\mathcal{W}(y))}{2^m}]$. Since *pivot* $= \lceil e^{3/2}(1+1/\kappa)^2 \rceil$ and the variables $\gamma_{z,\alpha}$ are 2-wise independent and in the range $[0,1]$, we apply Chebyshev's Inequality to obtain $\mathsf{Pr}[\frac{\text{pivot}}{1+\kappa} \leq \mathcal{W}(R_{F,h,\alpha}) \leq 1 + (1+\kappa)\text{pivot}] \geq 1 - e^{-3/2}$. Since $h$ is chosen at random from $H_{xor}(n,m,3)$, we also have $\mathsf{Pr}[h(y) = \alpha] = 1/2^m$. It follows that $w_{m,y,\alpha} \geq (1 - e^{-3/2})/2^m$. □

The next lemma provides an upper bound of $w_{i,y,\alpha}$ and $p_{i,y}$.

**Lemma 50.** *For $i < m$, both $w_{i,y,\alpha}$ and $p_{i,y}$ are bounded above by* $\frac{1}{W(F\downarrow S)-1}\frac{1}{\left(1-\frac{1+\kappa}{2^{m-i}}\right)^2}$.

*Proof.* We will use the terminology introduced in the proof of Lemma 49. Clearly, $\mu_\alpha = \frac{W(F\downarrow S)-\mathcal{W}(y)}{2^i}$. Since each $\gamma_{z,\alpha}$ takes values in $[0,1]$, $\mathsf{V}[\gamma_{z,\alpha}] \leq \mathsf{E}[\gamma_{z,\alpha}]$. Therefore, $\sigma^2_{z,\alpha} \leq \sum_{z\neq y, z\in R_F} \mathsf{E}[\gamma_{z,\alpha}] \leq \sum_{z\in R_F} \mathsf{E}[\gamma_{z,\alpha}] = \mathsf{E}[\Gamma_\alpha] \leq 2^{-m}(W(F \downarrow S) - \mathcal{W}(y))$. So $\mathsf{Pr}[\frac{pivot}{1+\kappa} \leq \mathcal{W}(R_{F,h,\alpha}) \leq 1 + (1+\kappa)\text{pivot}] \leq \mathsf{Pr}[\mathcal{W}(R_{F,h,\alpha}) - \mathcal{W}(y) \leq (1+\kappa)\text{pivot}]$. From Chebyshev's inequality, we know that $\mathsf{Pr}[|\Gamma_\alpha - \mu_{z,\alpha}| \geq \lambda\sigma_{z,\alpha}] \leq 1/\lambda^2$ for every $\lambda > 0$. $\mathsf{Pr}[\mathcal{W}(R_{F,h,\alpha}) - \mathcal{W}(y) \leq (1+\kappa)\frac{(W(F\downarrow S)-\mathcal{W}(y))}{2^i}] \leq \mathsf{Pr}\left[|(\mathcal{W}(R_{F,h,\alpha}) - \mathcal{W}(y)) - \frac{W(F\downarrow S)-1}{2^i}| \geq (1 - \frac{1+\kappa}{2^{m-i}})\frac{W(F\downarrow S)-\mathcal{W}(y)}{2^i}\right] \leq \frac{1}{\left(1-\frac{(1+\kappa)}{2^{m-i}}\right)^2} \cdot \frac{2^i}{W(F\downarrow S)-1}$. Since $h$ is chosen at random from $H_{xor}(n,m,3)$, we also have $\mathsf{Pr}[h(y) = \alpha] = 1/2^i$. It follows that $w_{i,y,\alpha} \leq \frac{1}{W(F\downarrow S)-1}\frac{1}{\left(1-\frac{1+\kappa}{2^{m-i}}\right)^2}$. The bound for $p_{i,y}$ is easily obtained by noting that $p_{i,y} = \Sigma_{\alpha\in\{0,1\}^i}(w_{i,y,\alpha} \cdot 2^{-i})$. □

**Lemma 51.** *For every witness $y \in R_F$,* $\mathsf{Pr}[y \textit{ is output}] \leq \frac{(1+\kappa)\mathcal{W}(y)}{W(F\downarrow S)-1}(2.23 + \frac{0.48}{(1-\kappa)^2})$

*Proof.* We will use the terminology introduced in the proof of Lemma 49. Using $\frac{pivot}{1+\kappa} \leq \mathcal{W}(Y)$, we have $\mathsf{Pr}[U] = \cup_{i=q-3}^{q}\frac{\mathcal{W}(y)}{\mathcal{W}(Y)}p_{i,y} \leq \frac{1+\kappa}{pivot}\mathcal{W}(y)\sum_{i=q-3}^{q} p_{i,y}$. Now we subdivide the calculation of $\mathsf{Pr}[U]$ into three cases depending on the value of $m$.

**Case 1 :** $q - 3 \leq m \leq q$.

Now there are four values that $m$ can take.

1. $m = q - 3$. We know that $p_{i,y} \leq \mathsf{Pr}[h(y) = \alpha] = \frac{1}{2^i}$, so $\mathsf{Pr}[U|m = q-3] \leq \frac{1+\kappa}{pivot} \cdot \frac{\mathcal{W}(y)}{2^{q-3}} \frac{15}{8}$. Substituting the values of *pivot* and $m$ gives $\mathsf{Pr}[U|m = q-3] \leq \frac{15(1+\kappa)\mathcal{W}(y)}{8(W(F{\downarrow}S)-1)}$.

2. $m = q - 2$. For $i \in [q-2, q]$ $p_{i,y} \leq \mathsf{Pr}[h(y) = \alpha] = \frac{1}{2^i}$ Using Lemma 50, we get $p_{q-3,y} \leq \frac{1}{W(F{\downarrow}S)-1} \frac{1}{\left(1-\frac{1+\kappa}{2}\right)^2}$. Therefore, $\mathsf{Pr}[U\,|m = q-2] \leq \frac{1+\kappa}{pivot}\mathcal{W}(y) \frac{1}{W(F{\downarrow}S)-1} \frac{4}{(1-\kappa)^2} + \frac{1+\kappa}{pivot}\mathcal{W}(y) \frac{1}{2^{q-2}} \frac{7}{4}$. Noting that $pivot = \frac{W(F{\downarrow}S)-1}{2^m} > 10$, we obtain $\mathsf{Pr}[U|m = q-2] \leq \frac{(1+\kappa)\mathcal{W}(y)}{W(F{\downarrow}S)-1}(\frac{7}{4} + \frac{0.4}{(1-\kappa)^2})$

3. $m = q - 1$. For $i \in [q-1, q]$, $p_{i,y} \leq \mathsf{Pr}[h(y) = \alpha] = \frac{1}{2^i}$. Using Lemma 50, we get $p_{q-3,y} + p_{q-2,y} \leq \frac{1}{W(F{\downarrow}S)-1}\left(\frac{1}{\left(1-\frac{1+\kappa}{2^2}\right)^2} + \frac{1}{\left(1-\frac{1+\kappa}{2}\right)^2}\right) = \frac{1}{W(F{\downarrow}S)-1}\left(\frac{16}{(3-\kappa)^2} + \frac{4}{(1-\kappa)^2}\right)$. Therefore, $\mathsf{Pr}[U|m = q-1] \leq \frac{1+\kappa}{pivot}\mathcal{W}(y)\left(\frac{1}{W(F{\downarrow}S)-1}\left(\frac{16}{(3-\kappa)^2} + \frac{4}{(1-\kappa)^2}\right) + \frac{1}{2^{q-1}}\frac{3}{2}\right)$. Since $pivot = \frac{W(F{\downarrow}S)-1}{2^m} > 10$ and $\kappa \leq 1$, $\mathsf{Pr}[U|m = q-1] \leq \frac{(1+\kappa)\mathcal{W}(y)}{W(F{\downarrow}S)-1}(1.9 + \frac{0.4}{(1-\kappa)^2})$.

4. $m = q$. We have $p_{q,y} \leq \mathsf{Pr}[h(y) = \alpha] = \frac{1}{2^q}$, and using Lemma 50 we get $p_{q-3,y} + p_{q-2,y} + p_{q-1,y} \leq \frac{1}{W(F{\downarrow}S)-1}\left(\frac{1}{\left(1-\frac{1+\kappa}{2^3}\right)^2} + \frac{1}{\left(1-\frac{1+\kappa}{2^2}\right)^2} + \frac{1}{\left(1-\frac{1+\kappa}{2}\right)^2}\right) = \frac{1}{W(F{\downarrow}S)-1}\left(\frac{64}{(7-\kappa)^2} + \frac{16}{(3-\kappa)^2} + \frac{4}{(1-\kappa)^2}\right)$. So $\mathsf{Pr}[U|m = q] \leq \frac{1+\kappa}{pivot}\mathcal{W}(y)\left(\frac{1}{W(F{\downarrow}S)-1}\left(\frac{64}{(7-\kappa)^2} + \frac{16}{(3-\kappa)^2} +\right.\right.$ Using $pivot = \frac{W(F{\downarrow}S)-1}{2^m} > 10$ and $\kappa \leq 1$, we obtain $\mathsf{Pr}[U|m = q] \leq \frac{(1+\kappa)\mathcal{W}(y)}{W(F{\downarrow}S)-1}(1.58 + \frac{0.4}{(1-\kappa)^2})$.

Since $\mathsf{Pr}[U|q-3 \leq m \leq q] \leq \max_{q-3 \leq i \leq q}(\mathsf{Pr}[U|m = i])$, we have $\mathsf{Pr}[U|q-3 \leq m \leq q] \leq \frac{1+\kappa}{W(F{\downarrow}S)-1}(1.9 + \frac{0.4}{(1-\kappa)^2})$ from the $m = q-1$ case above.

**Case 2 :** $m < q - 3$. Since $p_{i,y} \leq \mathsf{Pr}[h(y) = \alpha] = \frac{1}{2^i}$, we have $\mathsf{Pr}[U|m < q-3] \leq \frac{1+\kappa}{pivot}\mathcal{W}(y) \cdot \frac{1}{2^{q-3}}\frac{15}{8}$. Substituting the value of *pivot* and maximizing $m - q + 3$, we get

$\mathsf{Pr}[U|m < q-3] \leq \frac{15(1+\kappa)\mathcal{W}(y)}{16(W(F{\downarrow}S)-1)}$.

**Case 3 :** $m > q$. Using Lemma 50, we know that $\mathsf{Pr}[U|m > q] \leq \frac{1+\kappa}{pivot}\frac{\mathcal{W}(y)}{W(F{\downarrow}S)-1}$ $\sum_{i=q-3}^{q}\frac{1}{\left(1-\frac{1+\kappa}{2^{m-i}}\right)^2}$. The R.H.S. is maximized when $m = q+1$. Hence $\mathsf{Pr}[U|m > q] \leq \frac{1+\kappa}{pivot}\ \frac{\mathcal{W}(y)}{W(F{\downarrow}S)-1}\times\sum_{i=q-3}^{q}\frac{1}{\left(1-\frac{1+\kappa}{2^{q+1-i}}\right)^2}$. Noting that $pivot = \frac{W(F{\downarrow}S)-1}{2^m} > 10$ and expanding the above summation we have $\mathsf{Pr}[U|m > q] \leq \frac{(1+\kappa)\mathcal{W}(y)}{W(F{\downarrow}S)-1}\frac{1}{10}\left(\frac{256}{(15-\kappa)^2}+\frac{64}{(7-\kappa)^2}+\frac{16}{(3-\kappa)^2}+\frac{2}{(1-\kappa)^2}\right)$. Using $\kappa \leq 1$ for the first three summation terms, we obtain $\mathsf{Pr}[U|m > q] \leq \frac{(1+\kappa)\mathcal{W}(y)}{W(F{\downarrow}S)-1}(0.71+\frac{0.4}{(1-\kappa)^2})$

Summing up all the above cases, $\mathsf{Pr}[U] = \mathsf{Pr}[U|m < q-3]\times\mathsf{Pr}[m < q-3] + \mathsf{Pr}[U|q-3 \leq m \leq q]\times\mathsf{Pr}[q-3 \leq m \leq q] + \mathsf{Pr}[U|m > q]\times\mathsf{Pr}[m > q]$. From Lemma 48 we have $\mathsf{Pr}[m < q-1] \leq 0.2$ and $\mathsf{Pr}[m > q] \leq 0.2$, so $\mathsf{Pr}[U] \leq \frac{(1+\kappa)\mathcal{W}(y)}{W(F{\downarrow}S)-1}(2.23+\frac{0.48}{(1-\kappa)^2})$

□

Combining Lemmas 49 and 51, the following lemma is obtained.

**Lemma 52.** *For every witness $y \in R_F$, if $\varepsilon > 1.71$, then*

$\frac{W(y)}{(1+\varepsilon)W(R_F)} \leq \mathsf{Pr}\left[\mathsf{WeightGen}(F,\varepsilon,r,X) = y\right] \leq (1+\varepsilon)\frac{W(y)}{W(R_F)}$.

*Proof.* In the case where $W(F\downarrow S) \leq 1+(1+\kappa)pivot$, the result holds because WeightGen returns a perfect weighted-uniform sample. Otherwise, using Lemmas 49 and 51 and substituting $(1+\varepsilon) = (1+\kappa)(2.36+\frac{0.51}{(1-\kappa)^2}) = \frac{18}{17}(1+\kappa)(2.23+\frac{0.48}{(1-\kappa)^2})$, via the inequality $\frac{1.06+\kappa}{0.8(1-e^{-3/2})} \leq \frac{18}{17}(1+\kappa)(2.23+\frac{0.48}{(1-\kappa)^2})$ we have the bounds $\frac{\mathcal{W}(y)}{(1+\varepsilon)(W(F{\downarrow}S)-1)} \leq \mathsf{Pr}\left[\mathsf{WeightGen}(F,\varepsilon,r,X) = y\right] \leq \frac{18}{17}(1+\varepsilon)\frac{\mathcal{W}(y)}{W(F{\downarrow}S)-1}$. Using $W(F\downarrow S) \geq 18$, we obtain the desired result. □

**Lemma 53.** *Algorithm* WeightGen *succeeds (i.e. does not return $\bot$) with probability at least* 0.62.

*Proof.* If $W(F\downarrow S) \leq 1+(1+\kappa)$pivot, the theorem holds trivially. Suppose $W(F\downarrow S) > 1+(1+\kappa)$pivot and let $P_{\text{succ}}$ denote the probability that a run of the algorithm suc-ceeds. Let $p_i$ with $q-3 \leq i \leq q$ denote the conditional probability that $\mathsf{WeightGen}$ $(F,\ \varepsilon,\ r,\ X)$ terminates in iteration $i$ of the repeat-until loop (lines 15–19) with $\frac{pivot}{1+\kappa} \leq \mathcal{W}\left(R_{F,h,\alpha}\right) \leq 1+(1+\kappa)pivot$, given that $W(F \downarrow S) > 1+(1+\kappa)pivot$. Then $P_{\text{succ}} = \sum_{i=q-3}^{q} p_i \prod_{j=q-3}^{i}(1-p_j)$. Letting $f_m = \mathsf{Pr}[q-3 \leq m \leq q]$, by Lemma 48 we have $P_{\text{succ}} \geq p_m f_m \geq 0.8 p_m$. The theorem is now proved by using Chebyshev's Inequality to show that $p_m \geq 1-e^{-3/2} \geq 0.776$.

For every $y \in \{0,1\}^n$ and $\alpha \in \{0,1\}^m$, define an indicator variable $\nu_{y,\alpha}$ as follows: $\nu_{y,\alpha} = \mathcal{W}\left(y\right)$ if $h(y) = \alpha$, and $\nu_{y,\alpha} = 0$ otherwise. Let us fix $\alpha$ and $y$ and choose $h$ uniformly at random from $H_{xor}(n,m,3)$. The random choice of $h$ induces a probability distribution on $\nu_{y,\alpha}$, such that $\mathsf{Pr}[\nu_{y,\alpha} = \mathcal{W}\left(y\right)] = \mathsf{Pr}[h(y) = \alpha] = 2^{-m}$ and $\mathsf{E}[\nu_{y,\alpha}] = \mathcal{W}\left(y\right)\mathsf{Pr}[\nu_{y,\alpha} = 1] = 2^{-m}\mathcal{W}\left(y\right)$. In addition 3-wise independence of hash functions chosen from $H_{xor}(n,m,3)$ implies that for every distinct $y_a, y_b, y_c \in R_F$, the random variables $\nu_{y_a,\alpha}, \nu_{y_b,\alpha}$ and $\nu_{y_c,\alpha}$ are 3-wise independent.

Let $\Gamma_\alpha = \sum_{y \in R_F} \nu_{y,\alpha}$ and $\mu_\alpha = \mathsf{E}\left[\Gamma_\alpha\right]$. Clearly, $\Gamma_\alpha = \mathcal{W}\left(R_{F,h,\alpha}\right)$ and $\mu_\alpha = \sum_{y \in R_F} \mathsf{E}\left[\nu_{y,\alpha}\right] = 2^{-m} W(F \downarrow S)$. Since $pivot = \lceil e^{3/2}(1+1/\epsilon)^2 \rceil$, we have $2^{-m}W(F \downarrow S) \geq e^{3/2}(1+1/\varepsilon)^2$, and so using Chebyshev's Inequality with $\beta = \kappa/(1+\kappa)$ we obtain $\mathsf{Pr}\left[\frac{W(F\downarrow S)}{2^m}.\left(1-\frac{\kappa}{1+\kappa}\right) \leq \mathcal{W}\left(R_{F,h,\alpha}\right) \leq (1+\frac{\kappa}{1+\kappa})\frac{W(F\downarrow S)}{2^m}\right] > 1-e^{-3/2}$. Simplifying and noting that $\frac{\kappa}{1+\kappa} < \kappa$ for all $\kappa > 0$, we have $\mathsf{Pr}\left[(1+\kappa)^{-1} \cdot \frac{W(F\downarrow S)}{2^m} \leq \mathcal{W}\left(R_{F,h,\alpha}\right) \leq (1+\kappa) \cdot \frac{W(F\downarrow S)}{2^m}\right] > 1-e^{-3/2}$. Also, $\frac{pivot}{1+\kappa} = \frac{1}{1+\kappa}\frac{W(F\downarrow S)-1}{2^m} \leq \frac{W(F\downarrow S)}{(1+\kappa)2^m}$ and $1+(1+\kappa)pivot = 1+\frac{(1+\kappa)(W(F\downarrow S)-1)}{2^m} \geq \frac{(1+\kappa)W(F\downarrow S)}{2^m}$. Therefore, $p_m = \mathsf{Pr}[\frac{pivot}{1+\kappa} \leq \mathcal{W}\left(R_{F,h,\alpha}\right) \leq 1+(1+\kappa)pivot] \geq \mathsf{Pr}\left[(1+\kappa)^{-1} \cdot \frac{W(F\downarrow S)}{2^m} \leq \mathcal{W}\left(R_{F,h,\alpha}\right) \leq (1+\kappa) \cdot \frac{W(F\downarrow S)}{2^m}\right] \geq 1 - e^{-3/2}$. □

By combining Lemmas 52 and 53, we get the following:

**Theorem 54.** . *Given a CNF formula $F$, tolerance $\varepsilon > 1.71$, tilt bound $r$, and sam-*

*pling set $S$, for every $y \in R_F$ we have* $\frac{W(y)}{(1+\varepsilon)W(R_F)} \leq \mathsf{Pr}\left[\mathsf{WeightGen}(F, \varepsilon, r, X) = y\right] \leq (1+\varepsilon)\frac{W(y)}{W(R_F)}$. *Also,* WeightGen *succeeds (i.e. does not return* $\perp$*) with probability at least* 0.62.

**Theorem 55.** *Given an oracle for* SAT*,* WeightGen$(F, \varepsilon, r, S)$ *runs in time polynomial in* $r, |F|$ *and* $1/\varepsilon$ *relative to the oracle.*

*Proof.* Referring to the pseudocode for WeightGen, the runtime of the algorithm is bounded by the runtime of the constant number (at most 5) of calls to BoundedWeightSAT and one call to WeightMC (with parameters $\delta = 0.2, \varepsilon = 0.8$). As shown in Theorem 9, the call to WeightMC can be done in time polynomial in $|F|$ and $r$ relative to the oracle. Every invocation of BoundedWeightSAT can be implemented by at most $(r \cdot pivot) + 1$ calls to a SAT oracle (as in the proof of Theorem 10), and the total time taken by all calls to BoundedWeightSAT is polynomial in $|F|, r$ and *pivot* relative to the oracle. Since $pivot = \mathcal{O}(1/\varepsilon^2)$, the runtime of WeightGen is polynomial in $r$, $|F|$ and $1/\varepsilon$ relative to the oracle. □

## 10.3 Experimental Results

To evaluate the performance of WeightGen, we built prototype implementation and conducted an extensive set of experiments [*]. The suite of benchmarks was made up of problems arising from various practical domains as well as problems of theoretical interest. Specifically, we used bit-level unweighted versions of constraints arising from grid networks, plan recognition, DQMR networks, bounded model checking of circuits, bit-blasted versions of SMT-LIB [2] benchmarks, and ISCAS89 [26] circuits with parity conditions on randomly chosen subsets of outputs and next-state variables [139,

[*]The tool (with source code) is available at https://bitbucket.org/kuldeepmeel/weightgen

102]. While WeightGen is agnostic to the weight function, other tools that we used for comparison require the weight of an assignment to be the product of the weights of its literals. Consequently, to create weighted problems with tilt at most some bound $r$, we randomly selected $m = \max(15, n/100)$ of the variables and assigned them the weight $w$ such that $(w/(1-w))^m = r$, their negations the weight $1-w$, and all other literals the weight 1. To illustrate agnostic nature of our algorithms w.r.t. to weight oracle, we also evaluated WeightGen with non-factored representation of the weights. In our implementation of weight oracle without factored representation, we first randomly chose a range of minimum ($w_{min}$) and maximum ($w_{max}$) possible weights and then randomly selected 20 variables of the input formula. We now compute weight of an assignment as $w_{min}+(w_{max} - w_{min} * \frac{x}{2^20})$, where $x$ is the integer value of binary representation of assignment to our randomly selected 20 variables. Unless mentioned otherwise, our experiments used $r = 5$ and $\epsilon = 16$.

To facilitate performing multiple experiments in parallel, we used a high performance cluster, each experiment running on its own core. Each node of the cluster had two quad-core Intel Xeon processors with 4GB of main memory. We used 2500 seconds as the timeout of each invocation of BoundedWeightSAT and 20 hours as the overall timeout for WeightGen. If an invocation of BoundedWeightSAT timed out in line 17 (WeightGen), we repeated the execution of the corresponding loops without incrementing the variable $i$ (in both algorithms). With this setup, WeightGen was able to successfully generate weighted random samples for formulas with close to 64,000 variables.

Since a probabilistic generator is likely to be invoked many times with the same formula and weights, it is useful to perform the counting on line 10 of WeightGen only once, and reuse the result for every sample. Reflecting this, column 6 in Table 5.1 lists the time, averaged over a large number of runs, taken by WeightGen to generate one sample given that the weighted model count on line 10 has already been found. It is clear from Table 5.1 that WeightGen scales to formulas with thousands of variables.

Table 10.1 : SDD and WeightGen runtimes in seconds.

| **Benchmark** | **vars** | **#clas** | **SDD** | **Weight-Gen** |
|---|---|---|---|---|
| or-50 | 100 | 266 | 0.38 | 0.14 |
| or-70 | 140 | 374 | 0.83 | 13.37 |
| s526_3_2 | 365 | 943 | 29.54 | 0.85 |
| s526a_3_2 | 366 | 944 | 12.16 | 1.1 |
| s953a_3_2 | 515 | 1297 | 355.7 | 21.14 |
| s1238a_7_4 | 704 | 1926 | mem | 19.52 |
| s1196a_15_7 | 777 | 2165 | 2275 | 19.59 |
| Squaring9 | 1434 | 5028 | mem | 110.37 |
| Squaring7 | 1628 | 5837 | mem | 113.12 |
| ProcessBean | 4768 | 14458 | mem | 418.29 |
| LoginService2 | 11511 | 41411 | mem | 3.45 |
| Sort | 12125 | 49611 | T | 140.19 |
| EnqueueSeq | 16466 | 58515 | mem | 165.64 |
| Karatsuba | 19594 | 82417 | mem | 193.11 |
| TreeMax | 24859 | 103762 | T | 2.0 |
| LLReverse | 63797 | 257657 | mem | 88.0 |

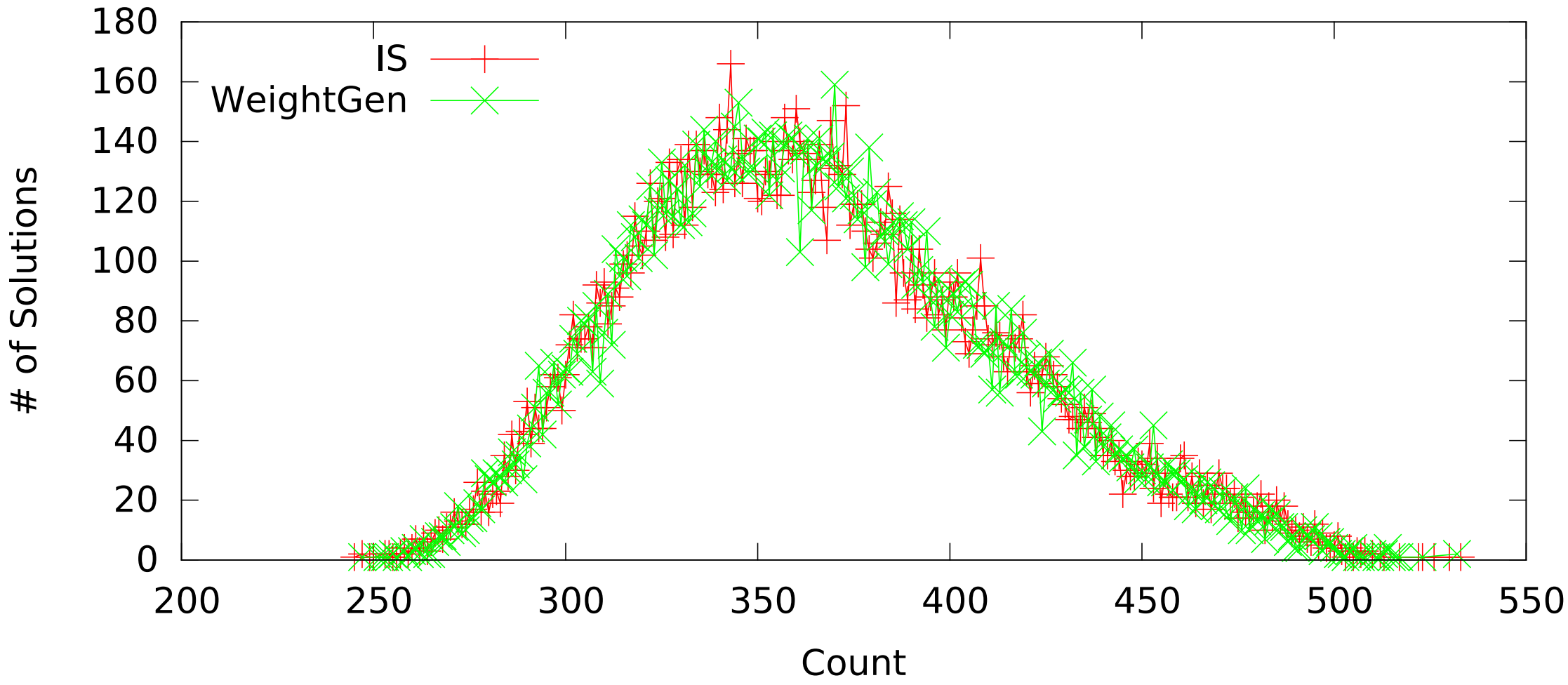

Figure 10.1 : Uniformity comparison for case110

To measure the accuracy of WeightGen, we implemented an *Ideal Sampler*, henceforth called IS, and compared the distributions generated by WeightGen and IS for a representative benchmark. Given a CNF formula $F$, IS first generates all the satisfying assignments, then computes their weights and uses these to sample the ideal distribution. We then generated a large number $N$ $(= 6 \times 10^5)$ of sample witnesses using both IS and WeightGen. In each case, the number of times various witnesses were generated was recorded, yielding a distribution of the counts. Figure 10.1 shows the distributions generated by WeightGen and IS for one of our benchmarks (case110) with 16,384 solutions. The almost perfect match between the distribution generated by IS and WeightGen held also for other benchmarks. Thus, as was the case for WeightMC, the accuracy of WeightGen is better in practice than that established by

Theorem 54.

## 10.4 Chapter Summary

In this Chapter, we considered adaptation of UniGen to handle the problem of distribution aware sampling. For approximation techniques that provide strong theoretical two-way bounds, a major limitation is the reliance on potentially-expensive most probable explanation (MPE) queries. We identify a novel parameter, *tilt*, to categorize weighted counting and sampling problems for SAT. We showed how to remove this reliance on MPE queries, while retaining strong theoretical guarantees. Experimental results demonstrate the effectiveness of this approach in practice when the tilt is small.

The experimental results presented in this Chapter are promising but indicate that this work is still a first step in the design of scalable sampling algorithms that can handle arbitrary distribution while providing rigorous formal guarantees. Chakraborty et al [34] present an extension of WeightGen that can handle large tilt if the weight function is white box. The resulting algorithm, however, employs Pseudo Boolean solvers, which are less scalable as compared to SAT solvers. As a result, the proposed algorithm faces significant scalability hurdles. An interesting direction of future work would be to propose extension of WeightGen that requires SAT solvers instead of Pseudo-Boolean solvers.

This is the final chapter of Part III. Let us take a moment to summarize what we accomplished in this Part. Given the practical as well as theoretical significance of the problem of sampling, the design of scalable sampling algorithms with rigorous formal guarantees have been a central problem for past three decades. The prior work, however, allows the end user to choose only one between theoretical guarantees and scalability. In this Part, we introduced a hashing-based paradigm that provides strong theoretical guarantees and promises scalability. Furthermore, the approach is highly parallelizable and achieves near linear speedup in practice. UniGen framework introduced in this part opens up several interesting directions of future research, which are discussed in detail in Chapter 12.

# Part IV

# Epilogue

# Chapter 11

# On Computing Minimal Independent Support

The hashing-based techniques for sampling and counting presented in this thesis crucially relies on choose a hash function randomly from $H_{xor}(n, m)$, which consists of conjunction of parity constraints, i.e. XOR, each of average density of 1/2. Consequently, a cell is represented as a conjunction of input constraints and XOR constraints. Since combinatorial reasoning tools are invoked to search for solutions of the conjunction of input constraints and XOR-based universal hash functions, one might wonder if there is any relationship between runtime performance of combinatorial search techniques and features of XOR-constraints? It has been observed that lower density XORs are easy to reason in practice and runtime performance of solvers greatly enhances with the decrease in the density of XOR-constraints [85]. This has led to recent work focused on designing hash functions with low density XOR-constraints [72] but such hash functions provide very weak guarantees of universality that did not translate to scalable algorithms for counting and sampling.

In Chapter 2, we introduced the notion of an *independent support* of a Boolean formula [39]: a subset of variables whose values uniquely determine the values of the remaining variables in any satisfying assignment to the formula. Formally, let $\mathcal{I} \subseteq X$ be a subset of the support such that if two satisfying assignments $\sigma_1$ and $\sigma_2$ agree on $\mathcal{I}$, then $\sigma_1 = \sigma_2$. In other words, in every satisfying assignment, the truth values of variables in $\mathcal{I}$ uniquely determine the truth value of every variable in $X \setminus \mathcal{I}$. The set $\mathcal{I}$ is called an *independent support* of $F$, and $\mathcal{D} = X \setminus \mathcal{I}$ is referred to as *dependent* *support*. Note that there is a one-to-one correspondence between $R_F$ and $R_{F\downarrow I}$. There may be more than one independent support: $(a \vee \neg b) \wedge (\neg a \vee b)$ has three, namely $\{a\}$, $\{b\}$ and $\{a, b\}$. Clearly, if $\mathcal{I}$ is an independent support of $F$, so is every superset of $\mathcal{I}$.

Next, note that $H_{xor}(X, \cdot)$ can be constructed by picking variables from the $\mathcal{I}$ alone. In particular, The hashing-based algorithms for sampling and counting presented in this paper take in sampling set $S$ as parameter. Therefore, when $S = X$ is supplied, we substitute $S$ with $\mathcal{I}$ if $\mathcal{I}$ is unknown. The importance of this observation comes from the fact that for many important classes of problems the size of an independent support is typically one to two orders of magnitude smaller than the number of all variables, which in turn leads to XOR constraints of typical density of 1/200 to 1/20, i.e. one to two orders of magnitude smaller than that of the traditional hash functions. We emphasize that unlike recent work of Ermon et al. [72], these hash functions still preserve the strong guarantees of universality and therefore can be used as replacement for traditional hash functions in recent hashing-based techniques for sampling and counting.

The notion of independent support is closely related to the concept of functional dependency in the context of relational databases; it is essentially equivalent to the concept of a *key* in a relation [120]. The difference is that in the context of relational databases, relations are represented *explicitly*, while here the relation $R_F$ is represented *implicitly* by means of the formula $F$. Thus, algorithmic techniques from relational-database theory do not scale to the setting considered here. In the context of combinational logic circuits, there has been some work that constructs a logic circuit whose Tseitin-encoding corresponds to the given Boolean formula [78]. The primary inputs of this constructed circuit form the independent support of the given

Boolean formula. This construction is based on pattern matching the formula to find sub-formulas corresponding to commonly used gates. This technique is not guaranteed to be complete and unlikely to succeed if the formulas did not originate from combinational circuits.

The variables that are not part of an independent support can be considered as *redundant*, and there is extensive research related to redundancy in propositional logic in general [114, 18]. In our context, a particular problem of importance is that of computing a concise reason of inconsistency of an over-constrained Boolean formula. Significant recent research focuses on efficiently computing a *minimal unsatisfiable subformula* (MUS) of a Boolean formula [122] and its extension on computing a *group-oriented* (also called *high-level*) minimal unsatisfiable subformula (GMUS) of an explicitly partitioned Boolean formula [116, 127]. In addition, there are highly optimized algorithms and off-the-shelf implementations for computing MUSes and GMUSes, such as MUSer2 [19]. Even more recent research focuses on computing a *smallest* (i.e., minimum-sized) MUS of a Boolean formula (SMUS) [115, 96], which in general is a significantly more computationally-intensive task. Similarly, one can consider a smallest GMUS of an explicitly partitioned Boolean formula (SGMUS). The tool Forqes described in [96] can compute SMUSes and SGMUSes.

In this chapter, we present an algorithmic procedure to determine minimal and minimum independent supports. The key idea of this algorithmic procedure is the reduction of the problem of minimizing an independent support of a Boolean formula to (S)GMUS. In this reduction, each independent subset of variables naturally corresponds to an unsatisfiable subformula of the total formula, and in particular the problems of finding a minimal independent support, or a minimum-sized independent support, or all minimal or minimum-sized independent supports, can be naturally translated to the corresponding problems in the MUS framework. For future reference, we denote by MIS the tool that computes a minimal independent support of a Boolean formula by employing the above translation and MUSer2, and we denote by SMIS the tool that computes a minimum independent support that uses Forqes instead.

An experimental comparison of MIS and SMIS highlights an important tradeoff between the performance and the sizes of computed independent supports. In particular, while MIS scales to larger formulas, SMIS computes even smaller independent supports for a subset of benchmarks that are within its reach. We illustrate the practical gains of MIS by augmenting the state of the art sampler and counter with the new hashing scheme that uses the computed minimal independent supports.

## 11.1 Computing Minimal/Minimum Independent Supports

In this section, we first discuss how computation of minimal/minimum independent supports can be reduced to computation of minimal/minimum unsatisfiable subsets. Building on our reduction, we propose the first algorithmic procedure, MIS, to compute a minimal independent support for a given formula. We then discuss how MIS can make efficient usage of information from users. We also discuss a variant SMIS that computes a minimum independent support. Finally, we discuss how minimal and minimum independent supports computed by MIS and SMIS can be applied to hashing-based approximate techniques for counting and sampling.

### 11.1.1 Reduction to Group-oriented Minimal Unsatisfiable Subsets

For a given Boolean formula $F$ and $S \subseteq X$, we know that $S$ is an independent support of $F$ whenever every two satisfying assignments $\sigma_1, \sigma_2$ to $F$ that agree on $S$, must be identical. We formalize this as follows. We introduce additional variables $Y = \{y_1, \ldots, y_n\}$, and let $F(y_1, \ldots, y_n)$ be obtained from $F(x_1, \ldots, x_n)$ by replacing every occurrence of a variable in $X$ by the corresponding variable in $Y$. The definition of independence is captured by the following formula:

$$F(x_1, \ldots, x_n) \wedge F(y_1, \ldots, y_n) \wedge \bigwedge_{i \in Ind(S)} (x_i = y_i) \implies \bigwedge_{j \in Ind(X \setminus S)} (x_j = y_j),$$

where $Ind(S)$ and $Ind(X \setminus S)$, respectively, denote the index sets of $S$ and $X \setminus S$. Since it obviously holds that $\bigwedge_{i \in Ind(S)}(x_i = y_i) \Rightarrow \bigwedge_{i \in Ind(S)}(x_i = y_i)$, we can replace the right-hand side of the above formula by $\bigwedge_{j \in Ind(X)}(x_j = y_j)$. Finally, define the Boolean function $Q_{F,S}(x_1, \ldots, x_n, y_1, \ldots, y_n)$ by

$$Q_{F,S} = F(x_1, \ldots, x_n) \wedge F(y_1, \ldots, y_n) \wedge \bigwedge_{i \in Ind(S)} (x_i = y_i) \wedge \neg \left( \bigwedge_{j \in Ind(X)} (x_j = y_j) \right).$$

**Proposition 1.** *$S$ in an independent support for $F$ if and only if $Q_{F,S}$ is unsatisfiable.*

From Proposition 1 we obtain the following upper bound.

**Theorem 56.** *The problem of deciding whether $S$ is a minimal independent support of $F$ is in $\mathsf{D^P}$, where $\mathsf{D^P} = \{A - B | A, B \in \mathsf{NP}\}$.*

*Proof.* Checking that $S$ is independent support of $F$ is reducible to unsatisfiability of $Q_{F,S}$, which is in co-NP. To check minimality, we can select each variable $x \in S$ and check that $Q_{F,S-\{x\}}$ is satisfiable. □

We offer the following lower bound conjecture:

**Conjecture 1.** *The problem of deciding whether $S$ is a minimal independent support of $F$ is $\mathsf{D^P}$-complete.*

Proposition 1 leads to algorithms for computing a minimal independent support of $F$. One possible approach is to start with $S = X$ and the obviously unsatisfiable formula $Q_{F,X}$, and then remove variables $x_i$ from $S$ (corresponding to conjuncts $x_i = y_i$ in $Q_{F,S}$) as long as $Q_{F,S}$ remains unsatisfiable. Instead, we observe that the problem of minimizing independent support can be restated as the problem of minimizing unsatisfiable subsets, and hence we can benefit from the full variety of different algorithms and various important optimizations developed in the latter context. We now pursue this direction.

Using notation from Section 2.4, define $H_1, \ldots, H_n$ and $\Omega$ as follows:

$$H_1 = \{x_1 = y_1\},\ \ldots,\ H_n = \{x_n = y_n\},$$

$$\Omega = F(x_1, \ldots, x_n) \wedge F(y_1, \ldots, y_n) \wedge \bigvee_{i \in Ind(X)} (x_i \neq y_i).$$

To obtain a CNF representation, suppose that the original formula $F$ is given in CNF. Then we let $H_i = \{(\neg x_i \vee y_i) \wedge (x_i \vee \neg y_i)\}$ for $i = 1, \ldots, n$. For $\Omega$, the terms $F(x_1, \ldots, x_n)$ and $F(y_1, \ldots, y_n)$ are already in CNF. To encode $\bigvee_{i=1}^{n}(x_i \neq y_i)$, we introduce additional variables $b_1, \ldots, b_n$, add clauses $(\neg x_i \vee \neg y_i \vee b_i)$, $(x_i \vee y_i \vee b_i)$ for $i = 1, \ldots, n$, and add the clause $(\neg b_1 \vee \cdots \vee \neg b_n)$.

The following proposition follows immediately from the construction and Proposition 1:

**Proposition 2.** *The formula $H_1 \wedge \cdots \wedge H_n \wedge \Omega$ is unsatisfiable. Moreover, for a subset $S \subseteq X$: $S$ is an independent support of $F$ if and only if $\{H_i | i \in Ind(S)\}$ is a group-oriented unsatisfiable subset of $\{H_1, \ldots, H_n\}$.*

It immediately follows that problems of computing independent support can be reduced to analogous problems of finding group oriented unsatisfiable subsets. Specifi-cally, computing a minimal independent support can be reduced to computing a minimal unsatisfiable subset; computing a minimum independent support can be reduced to computing a minimum unsatisfiable subset; computing all minimal independent supports can be reduced to computing all minimal unsatisfiable subsets; and so on.

### 11.1.2 Handling Under- and Over- Approximations

In Section 11.1.3 we describe a light-weight technique for detecting a set of variables that is dependent on the remaining variables in the formula, thus allowing us to restrict the search for a minimal independent support by excluding the dependent variables. Furthermore, in some of our applications (see Section 11.1.6), the user has the additional freedom to specify which variables should or should not be in the independent support. In both cases, we can think of the set of variables that should to be included as specifying an *under-approximation* of the independent support, and we can think of complement of the set of variables that should be excluded as specifying an *over-approximation* of the independent support.

Due to these considerations, we introduce the following extension of the independent-support problem. Let $U \subseteq V \subseteq X$ and suppose that $V$ is an independent support of $F$. Let us say that an independent support of $F$ relative to an *under-approximation* $U$ and an *over-approximation* $V$ is a set $S$ such that $U \subseteq S \subseteq V$ and $S$ is an independent support of $F$. Further, let us say that a *minimal independent support of $F$ relative to $U$ and $V$* is a minimal $S$ with these properties. Note that $S$ does not need to be a minimal independent support of $F$ (as $U$ itself might have dependent variables). Also note the explicit requirement that $V$ is an independent support (if $V$ is not an independent support, then no subset of $V$ is an independent support).

The reduction to group-oriented unsatisfiable subset described in Section 11.1.1

can be easily extended to handle this more general problem. Given $F$, $U$ and $V$ as above, let $H_i = \{x_i = y_i\}$ for $i \in Ind(V \setminus U)$, and let $\Omega = F(x_1, \ldots, x_n) \wedge F(y_1, \ldots, y_n) \wedge \bigwedge_{i \in Ind(U)} (x_i = y_i) \wedge \bigvee_{i \in Ind(X)} (x_i \neq y_i)$.

**Proposition 3.** *The following statements are true:*

1. *The formula* $\Omega \wedge \bigwedge_{i \in Ind(V \setminus U)} H_i$ *is unsatisfiable.*

2. *For a subset* $W \subseteq V \setminus U$: $\{H_i \mid i \in Ind(W)\}$ *is a group-oriented unsatisfiable subset of* $\{H_i \mid i \in Ind(V \setminus U)\}$ *if and only if* $U \cup W$ *is an independent support of* $F$ *relative to* $U$ *and* $V$.

3. $\{H_i \mid i \in Ind(W)\}$ *is a minimal group-oriented unsatisfiable subset of* $\{H_i \mid i \in Ind(V \setminus U)\}$ *if and only if* $U \cup W$ *is a minimal independent support of* $F$ *relative to* $U$ *and* $V$.

We, henceforth, denote this reduction as $\mathsf{TranslateToGMUS}(F, U, V)$. Note that when $U = \emptyset$ and $V = X$ the definition of an independent support relative to $U$ and $V$ corresponds to the standard definition of independent support, and $\mathsf{TranslateToGMUS}(F, U, V)$ coincides with the reduction given in Section 11.1.1. In what follows, we omit "relative to an under-approximation $U$" when $U = \emptyset$, and we omit "relative to an over-approximation $V$" when $V = X$.

### 11.1.3 Exploiting Local Dependencies

In various important contexts, a variable $x \in X$ can be shown to be dependent on other variables, either purely syntactically or by analyzing only a small subset of all clauses. An especially important case is when the formula $F$ encodes a circuit, in which case many variables can be detected to be dependent simply from their defining clauses.

**Example 1** Suppose that $F$ contains the following clauses (among others): $(\neg x \vee y \vee b), (x \vee \neg y), (x \vee \neg b)$. It can be readily seen that in every satisfying assignment to $F$ we have that $x = y \vee b$, and so $x$ is dependent on $\{y, b\}$.

Intuitively, the variables that are locally dependent on other variables do not need be considered for independent support. We need, however, to avoid *cyclic reasoning*, such as when $F := (\neg x \vee y) \wedge (x \vee \neg y)$, $x$ depends on $y$ and that $y$ also depends on $x$.

**Algorithm 16** FindLocalDependencies(F, V)

**Input:** CNF formula $F$; set $V \subseteq X$

**Output:** A subset $Z \subseteq V$ of dependent variables.

1: $Z = \emptyset$
2: **for** $x \in V$ **do**
3: $\quad G = \mathsf{SelectLocalClauses}(F, x)$
4: $\quad W = \mathsf{Vars}(G)$ /*$\mathsf{Vars}(G)$ denotes the support of $G$ */
5: $\quad$ **if** $Q_{G,W \setminus \{x\}}$ is UNSAT **then**
6: $\quad\quad Z = Z \cup \{x\}$
7: $\quad\quad F = F \setminus G$
8: **return** $Z$

We propose Algorithm 16 to detect a set of non-cyclic locally dependent variables. The algorithm accepts a formula $F$ in CNF and a set $V$ of candidate variables to consider, and returns a set $Z \subseteq V$ of variables that are (non-cyclically) dependent on the remaining variables. Initially, $Z$ is empty. In the algorithm we iteratively select a variable $x \in V$ and call SelectLocalClauses to select a set of clauses of $F$

"around" $x$. These should include at least all the clauses of $F$ involving $x$, but more generally can correspond to a larger neighborhood of $x$ in the *primal graph* (the graph with vertexes $\mathsf{Vars}(F)$, and an edge between $x_1$ and $x_2$ whenever $F$ contains a clause involving both $x_1$ and $x_2$). Next we check whether $x$ can be shown to be dependent on the remaining variables in $G$: this could be either a purely syntactic check or involve a SAT invokation. When $x$ is indeed dependent, then $x$ is added to $Z$, and moreover all the clauses involved into showing this dependency are removed from $F$ (for simplicity in the algorithm we remove all clauses of $G$, but a more refined analysis is also possible). This step is important to avoid cyclic dependencies.

**Proposition 4.** *Let $Z$ be an outcome of Algorithm 16. Then $X \setminus Z$ is an independent support for $F$. Moreover, let $S$ be a minimal independent support of $F$ relative to the over-approximation $X \setminus Z$. Then $S$ is also a minimal independent support of $F$.*

The first part of Proposition 4 summarizes the correctness of Algorithm 16. The second part shows that the output of the algorithm can be used to obtain an over-approximation of a minimal independent support – and thus it can be viewed as a preprocessing step for computing minimal independent support.

### 11.1.4 Combined Algorithm

Algorithm MIS (Algorithm 17) presents our combined approach to compute a minimal independent support. The algorithm accepts a formula $F$ in CNF, and both an under-approximation $U$ and an over-approximation $V$. We require that $U \subseteq V$ and that $V$ is an independent support for $F$. As the first step, we call FindLocalDependencies described in Section 11.1.3 to compute a set of (locally) dependent variables, which is essentially used to further refine the over-approximation $V$. Next, following the description in Section 11.1.2, we translate the problem into a GMUS computation.

The call to ComputeGMUS refers to a state-of-the-art algorithm to compute GMUSes (in our experiments, we use MUSer2). The independent support returned by the algorithm consists of the variables in the under-approximation $U$ and the variables that correspond to the groups in the minimal group-unsatisfiable subset. The correctness of this algorithm follows from Proposition 3.

**Algorithm 17** MIS(F, U, V)

**Input:** CNF formula $F$; sets $U$,$V$ such that $U \subseteq V \subseteq \mathsf{Vars}(F)$ and $V$ is independent support for $F$

**Output:** Minimal $S$ with the property that $U \subseteq S \subseteq V$ and $S$ is an independent support for $F$

1: Z = FindLocalDependencies$(F, V)$

2: $\{\Omega, H_1, \ldots, H_n\}$ = TranslateToGMUS$(F, U, V \setminus Z)$

3: $\{H_{i_1}, \ldots, H_{i_n}\}$ = ComputeGMUS$(\{\Omega, H_1, \ldots, H_n\})$

4: $S = U \cup \{x_{i_1}, \ldots, x_{i_n}\}$

5: **return** $S$

Given the computationally expensive nature of GMUS computation, it may happen that ComputeGMUS exceeds a specified time-limit. However, it is important to note that MUSer2 still returns a sound over-approximation of a minimal group-unsatisfiable subset in case of a time-out (as it employs a variant of the *deletion-based* approach described in [122]). In this case the support consisting of the variables in $U$ and the variables in the computed over-approximation returned by ComputeGMUS is still an independent support. Therefore, MIS behaves as an *anytime* algorithm; that is, it always returns a sound independent support for a given time budget. Our experiments indicate that this anytime behavior is useful in computing independent

supports – even if these are not minimal, they are significantly smaller than the support of $F$ and improve performance of sampling and counting tools by 2-3 orders of magnitude.

### 11.1.5 Computation of Minimum Independent Support

Since the problem of computing a minimum-sized independent support can be reduced to that of computing minimum-sized group-unsatisfiable subset, we can extend MIS to compute a minimum-sized independent support, by following the two modifications below. First, we remove the call to FindLocalDependencies – as this is a greedy heuristic that provide guarantees of minimality but not of mimum size. Second, we replace the call to compute minimal group-unsatisfiable subset with the call to compute minimum group-unsatisfiable subset. We use SMIS to denote the resulting algorithm. Our experimental comparison of MIS and SMIS, discussed in Section 11.2, shows that MIS scales to larger formulas, while SMIS computes even smaller sized independent supports for a subset of benchmarks that are within its reach.

### 11.1.6 Handling User Input

In some of our applications the user is allowed to additionally provide a set of variables $W$ that is believed to form an independent support of $F$, and the task is to minimize this set. There are two interesting scenarios associated with this. If $W$ is indeed an independent support of $F$, as can be checked by checking satisfiability of $Q_{F,W}$, then $W$ can be used an as over-approximation of an independent support, that is, one can look for a minimal independent support relative to the over-approximation prescribed by $W$. It is possible, however, that $W$ is not really an over-approximation. In our experience, in these cases the user input is still "close" to being correct, and so we suggest the following two-step approach. First, we treat $W$ as an under-approximation and find a minimal set $U$ such that $W \cup U$ forms an independent support. Second, we treat $W \cup U$ as an over-approximation and find a minimal subset of $W \cup U$. In our experience, not only does this scheme results in a minimal independent support that is close to the user input, but is also significantly faster than computing a minimal independent support from scratch

## 11.2 Evaluation

To evaluate the performance and impact of MIS, we built a prototype implementation* in C++ and conducted an extensive set of experiments on diverse set of public-domain problem instances. In these experiments, a typical instance is a formula $F$, with set of support $X$, and independent support $\mathcal{I}$ computed by MIS. The main objectives of our experimental set up was to seek answers for the following questions:

1. How do MIS and SMIS scale to large formulas and how do sizes of $\mathcal{I}$ computed by MIS and SMIS compare to $X$?

2. How does the performance and size of computed $\mathcal{I}$ vary with the user input?

3. How does employing $H_{xor}$ on $\mathcal{I}$ instead of $X$ affect the performance of ApproxMC, the state-of-the-art counting tool?

4. How do new provable bounds on the size of XORs required for approximate model counting techniques compare with previously known bounds?

In summary, we observe that MIS scales to large formulas with tens of thousands of variables, and the minimal independent support computed by MIS are typically of

*The tool along with source code is available at http://bitbucket.org/kuldeepmeel/mis

1/10 to 1/100 the size of support of the formulas. Furthermore, utilizing user input even when the initial user input is only an under-approximation, MIS can compute minimal independent supports significantly faster than without user input. Finally, by utilizing $\mathcal{I}$ computed by MIS and SMIS, we provide the first theoretically proven bounds on size of XOR constraints that are close to empirically observed bounds.

### 11.2.1 Experimental Setup

We conducted experiments on a heterogeneous suite of benchmarks used in earlier works on sampling and counting [39]. The benchmark suite employed in the experiments consisted of problems arising from probabilistic inference in grid networks, synthetic grid-structured random interaction Ising models, plan recognition, DQMR networks, bit-blasted versions of SMTLIB benchmarks, ISCAS89 combinational circuits with weighted inputs, and program synthesis examples. We employed MUSer2 [19] for group minimal group-unsatisfiable subset computation and forqes [96] for group minimum-unsatisfiable subset computation. We used a high-performance cluster to conduct multiple experiments in parallel. Each node of the cluster had a 12-core 2.83 GHz Intel Xeon processor, with 4GB of main memory, and each of our experiments was run on a single core. We employed the Mersenne Twister to generate pseudo-random numbers, and each thread was seeded independently using the C++ class `random_device`. Since different runs of MIS compute different minimal independent supports depending on the input from pseudo-random generator, we compute up to five independent supports for each benchmark and report the median of corresponding statistics.

### 11.2.2 Runtime Performance of MIS and SMIS

Table 11.1 presents the runtime of MIS and SMIS for our benchmark suite. The names of the benchmarks are specified in column 1, while columns 2 and 3 list the number of variables and clauses for each benchmark. Column 4 and 6 list the median runtime and median size of minimal independent supports ($\mathcal{I}$) computed by MIS. Column 6 lists the ratio of the number of variables to $|\mathcal{I}|$. Column 7 and 8 list the runtime and size of a minimum-sized independent support ($\mathsf{I}_m$). The ratio of $|\mathsf{I}_m|$ to $|\mathcal{I}|$ is presented in column 9. The results demonstrate that MIS scales to fairly large formulas, and the minimal independent supports computed by MIS are one to two orders of magnitude compared to the overall support. The comparison of MIS vis-a-vis SMIS highlights a tradeoff in performance. In particular, while MIS scales to larger formulas, SMIS computes even smaller independent supports for a subset of benchmarks that are within its reach (and in some cases removes up to 40% additional variables).

### 11.2.3 Impact of User Input on MIS

To study the impact of user input on MIS, we experimented with the suite of benchmarks for which independent support was provided by the sources. Table 11.2 presents the result of our experiments. Column 1 lists the benchmark while columns 2 and 3 list the number of variables and clauses for each benchmark. Columns 4 and 5 list the runtime and the median size of computed $\mathcal{I}$ by MIS without user input. Columns 6–9 report statistics when the user input is provided to MIS. Column 6 lists the size of $\mathcal{I}$ provided by the user while column 7 and 8 present the runtime and the size of computed $\mathcal{I}$ by MIS. Column 9 lists the fraction of ratio of intersection of computed $\mathcal{I}$ and user-provided $\mathcal{I}$ to the computed $\mathcal{I}$. We use “U” and ”O” to denote that the input provided by user was an under-approximation and over-approximation of an independent support respectively.

| **Benchmark** | **#vars** | **#clas** | MIS **time(s)** | $\|\mathcal{I}\|$ | $\frac{\#vars}{\|\mathcal{I}\|}$ | SMIS **time(s)** | $\|\mathsf{I}_m\|$ | $\frac{\|\mathsf{I}_m\|}{\|\mathcal{I}\|}$ |
|---|---|---|---|---|---|---|---|---|
| squaring4 | 891 | 2839 | 868.71 | 55 | 16.05 | 1174.46 | 36 | 0.65 |
| s953a_15_7 | 602 | 1657 | 7.48 | 48 | 12.41 | 11.03 | 45 | 0.93 |
| squaring30 | 1031 | 3693 | 192.14 | 30 | 34.37 | 144.82 | 29 | 0.97 |
| case_2_b12_1 | 427 | 1385 | 1.42 | 34 | 12.56 | 16.52 | 30 | 0.88 |
| scenarios_llreverse | 1096 | 4217 | 59.8 | 81 | 13.45 | 205.0 | 46 | 0.56 |
| squaring10 | 1099 | 3632 | 3321.29 | 56 | 19.45 | 1609.63 | 40 | 0.71 |
| TR_ptb_1_linear | 1969 | 6288 | 1297.77 | 122 | 16.07 | 768.37 | 106 | 0.87 |
| s1488_7_4 | 872 | 2499 | 11.38 | 24 | 36.33 | – | – | – |
| s5378a_15_7 | 3766 | 8732 | 1990.1 | 227 | 16.59 | – | – | – |
| lssBig | 12438 | 149909 | 536.88 | 46 | 270.39 | – | – | – |
| blockmap_10_02.net | 12562 | 26022 | 2637.74 | 78 | 161.05 | – | – | – |
| lss | 13373 | 156208 | 971.24 | 45 | 297.18 | – | – | – |
| blockmap_10_03.net | 13786 | 28826 | 13442.28 | 125 | 110.29 | – | – | – |
| 20 | 13887 | 60046 | 40.29 | 51 | 272.29 | 14.6 | 50 | 0.98 |
| scenarios_tree_insert_search | 16573 | 61922 | 18000 | 943 | 17.57 | – | – | – |
| blockmap_15_01.net | 33035 | 67424 | 781.94 | 49 | 674.18 | – | – | – |
| blockmap_20_01.net | 78650 | 160055 | 2513.32 | 67 | 1173.88 | – | – | – |

Table 11.1 : Runtime performance of MIS and SMIS

Table 11.2 shows that user-provided input are not necessarily minimal and are sometimes under approximation of an independent support. Since several minimal independent supports exist, it does not necessarily imply that size of an under-approximation would be smaller than every minimal independent support; e.g., for benchmark "Pollard", while oen of the independent supports is of size 48, the inpt with size 50 is not an independent support and is, therefore, an under-approximation of some other independent support. Table 11.2 clearly demonstrates that MIS is able to take advantage of user input, even when the initial user input is only an under approximation, and can compute $\mathcal{I}$ significantly faster than without user input. Since initial user input is only an under approximation in several cases and therefore, algorithmic techniques such as MIS are required to compute a sound independent support.

### 11.2.4 Impact on Performance of Sampling and Counting Techniques

Since $H_{xor}$ constructed over an independent support $\mathcal{I}$, denoted $H_{xor}^{\mathcal{I}}$, is 3-universal. Therefore, hashing-based counting techniques can be augmented with $H_{xor}^{\mathcal{I}}$. We compared the performance of ApproxMC with IApproxMC, where IApproxMC is ApproxMC augmented with $H_{xor}^{\mathcal{I}}$. We used an overall timeout of 5 hours, and the tolerance ($\varepsilon$) and confidence ($1 - \delta$) were set to 0.8 and 0.8, respectively, for ApproxMC and IApproxMC. The parameter values were chosen to match the corresponding values in previously published works on ApproxMC [38]. In summary, ApproxMC timed out on 36 out of 112 benchmarks, IApproxMC were able to count respectively on all the benchmarks. Since we computed up to five independent supports for each benchmark, we also computed range of runtime for IApproxMC.

| | | | Without User Input | | With User Input | | | |
|---|---|---|---|---|---|---|---|---|
| **Benchmark** | **#vars** | **#clas** | MIS **time(s)** | $\lvert\mathcal{I}\rvert$ | **User** $\lvert\mathcal{I}\rvert$ | MIS **time(s)** | **Computed** $\lvert\mathcal{I}\rvert$ | **Type** |
| TR_b14_2_linear | 1570 | 4963 | 243.65 | 136 | 204 | 234.0 | 103 | O |
| squaring7 | 1628 | 5837 | 12329.2 | 58 | 72 | 4404.22 | 40 | O |
| 55 | 1874 | 8384 | 0.1 | 38 | 46 | 0.24 | 38 | U |
| TR_b12_1_linear | 1914 | 6619 | 5963.92 | 73 | 99 | 1559.43 | 60 | U |
| TR_b12_2_linear | 2426 | 8373 | 15505.02 | 79 | 107 | 1779.25 | 64 | O |
| TR_device_1_even_linear | 2447 | 7612 | 612.19 | 176 | 281 | 338.06 | 158 | O |
| case_1_b12_even1 | 2681 | 8492 | 4507.71 | 155 | 150 | 1534.94 | 147 | O |
| case_2_b12_even1 | 2681 | 8492 | 4249.56 | 149 | 150 | 2008.88 | 147 | O |
| scenarios_tree_insert_insert | 2797 | 10427 | 837.14 | 101 | 84 | 725.08 | 85 | U |
| Pollard | 2800 | 49543 | 1211.4 | 179 | 50 | 543.94 | 48 | U |
| 56 | 2801 | 9965 | 2.23 | 37 | 38 | 1.84 | 37 | U |
| ProcessBean | 3130 | 11689 | 172.64 | 305 | 166 | 92.44 | 156 | U |
| scenarios_tree_delete2 | 3411 | 12783 | 444.61 | 179 | 138 | 389.79 | 137 | U |
| lss_harder | 3465 | 62713 | 1727.77 | 116 | 21 | 1690.61 | 22 | U |
| s5378a_15_7 | 3766 | 8732 | 1990.1 | 227 | 214 | 559.06 | 214 | O |
| listReverseEasy | 4092 | 15867 | 16715.34 | 144 | 121 | 1959.57 | 99 | U |
| reverse | 9485 | 535676 | 25.03 | 201 | 262 | 24.2 | 195 | U |
| lss | 13373 | 156208 | 971.24 | 45 | 20 | 665.22 | 20 | U |
| 110 | 15316 | 60974 | 9.2 | 80 | 88 | 9.08 | 80 | U |

Table 11.2 : Impact of User Input on MIS. "U" and "O" denote that the input provided by user was an under-approximation and over-approximation of an independent support respectively.

Table 11.3 presents the comparison of runtimes of ApproxMC and IApproxMC for a subset of the benchmarksColumn 1 lists the benchmarks, while column 2 report the number of variables for each benchmark. Column 3 lists the runtime of MIS to compute $\mathcal{I}$. Column 4 lists the runtime of ApproxMC, while the median runtime and range of runtimes for IApproxMC are listed in columns 5 and 6. (We generated 100 samples for each benchmark, and sampling time is amortized per sample.) We use '–' to denote the timeout (5 hours).

Table 11.3 clearly demonstrates that employing 3-universal hash functions $H_{xor}$ over $\mathcal{I}$ resulted in 2-3 orders of magnitude performance improvement for both counting and sampling. It is worth noting that for the case of "squaring14", MIS times out, but the over-approximation returned by MIS still allows IApproxMC to sample and count, while UniGen2 and ApproxMC timed out. Furthermore, the considerably smaller range of runtimes for most of the benchmarks illustrate the dominating effect of minimal independent supports on the runtime performance. This observation is, however, not always true and we observe that there are cases where the range is considerably large – a detailed analysis is beyond the scope of this work and is left for future work.

### 11.2.5 Impact on XOR Size Bounds for Model Counting Techniques

Since approximation techniques for model counting only requires weaker guarantees of universality [38], several techniques have been proposed on employing shorter XORs for model counting [85, 72]. The investigations into shorter XORs [85, 72] empirically demonstrated that short XORs, surprisingly, perform quite well for wide variety of benchmarks, even without a theoretical guarantee, but have failed to obtain provable bounds on adequate size of XOR constraints that are close to empirical observations. By computing the size of XOR constraints based on the size of minimal indepen-dent support and then applying Theorem 3 of [72], we provide the first theoretically proven bounds on adequate size of XOR constraints that are very close to empirically observed bounds.

| | | MIS | ApproxMC | IApproxMC | |
|---|---|---|---|---|---|
| **Benchmark** | **#vars** | **time(s)** | **time (s)** | **Median time (s)** | **Range time(s)** |
| squaring4 | 891 | 868.71 | – | 1550.04 | 986.47 |
| s953a_15_7 | 602 | 7.48 | – | 1221.22 | 250.72 |
| squaring30 | 1031 | 192.14 | 29974.19 | 89.82 | 42.23 |
| case_2_b12_1 | 427 | 1.42 | 1449.15 | 212.82 | 82.42 |
| squaring10 | 1099 | 3321.29 | – | 3135.01 | 3800.54 |
| s1196a_7_4 | 708 | 35.44 | – | 314.21 | 167.29 |
| s1238a_7_4 | 704 | 47.59 | – | 404.54 | 93.32 |
| case_0_b12_2 | 827 | 34.87 | – | 1528.17 | 4418.18 |
| case_1_b12_2 | 827 | 23.87 | – | 1541.06 | 399.75 |
| scenarios_llreverse | 1096 | 59.8 | – | 17109.1 | 10040.38 |
| case_2_b12_2 | 827 | 25.32 | – | 1228.28 | 872.1 |
| lss_harder | 3465 | 1727.77 | 13116.78 | 120.46 | 301.58 |
| BN_57 | 1154 | 103.22 | – | 517.27 | 1118.89 |
| BN_59 | 1112 | 104.85 | – | 484.79 | 236.62 |
| BN_65 | 925 | 29.64 | – | 1322.17 | 261.33 |
| squaring1 | 891 | 718.78 | – | 1480.99 | 296.37 |
| squaring8 | 1101 | 3453.48 | – | 2061.31 | 4970.9 |

Table 11.3 : Runtime comparison of ApproxMC vis-a-vis IApproxMC

Table 11.4 presents the comparison of new theoretical bounds with previously known best theoretical and empirical bounds for benchmarks reported in previous works [85, 72]. Column 1 lists the benchmarks, while column 2 and 3 report the number of variables and clauses for each benchmark. Column 4 and 5 present previously known theoretical and empirical bounds on size of XORs [72]. Finally, the new theoretical bounds based on computation of independent supports is presented in column 6. Table 11.4 clearly shows that new bounds obtained based on minimal independent supports computed by MIS greatly improve on the previously reported theoretical bounds. Furthermore, the bounds are very close to empirically observed bounds. In fact, in one case we obtain theoretical bound that is better than the best known empirical bounds. (It is worth noting that previous results [85, 72] on shorter XORs do not extend to sampling techniques as sampling requires stronger guarantees of universality.)

| | | | **Previous Bounds** | | **New Bounds** |
|---|---|---|---|---|---|
| **Benchmark** | **#vars** | **#clas** | **Theoretical** | **Empirical** | **Theoretical** |
| ls7R34med | 119 | 622 | 46 | 3 | 12 |
| ls7R35med | 136 | 745 | 53 | 3 | 16 |
| ls7R36med | 149 | 870 | 56 | 3 | 18 |
| log.c.red | 352 | 1933 | 112 | 28 | 9 |
| 2bitmax_6 | 252 | 766 | 26 | 8 | 21 |
| blk-50-3-10-20 | 50 | 30 | 10 | 5 | 5 |
| blk-50-10-3-20 | 50 | 30 | 8 | 3 | 5 |

Table 11.4 : Comparison of bounds on shorter XORs for model counting

## 11.3 Chapter Summary

The performance of hashing-based techniques presented in this thesis is primarily affected by the runtime of combinatorial solvers for the queries that are typically expressed as conjunction of CNF and constraints from hash functions. Furthermore, it has been observed that lower density XORs are easy to reason in practice and runtime performance of solvers greatly enhances with the decrease in the density of XOR-constraints [85]. In this context, it is important to note that hash functions constructed over Independent Support still retains the same theoretical guarantees with respect to universality. The importance of this observation comes from the fact that for many important classes of problems the size of an independent support is typically one to two orders of magnitude smaller than the number of all variables, which in turn leads to XOR constraints of typical density of 1/200 to 1/20, i.e. one to two orders of magnitude smaller than that of the traditional hash functions.

In this Chapter, we presented the first algorithmic procedure and corresponding tool, MIS, to determine minimal independent support via reduction to Group MUS. The experimental evaluation over an extensive suite of benchmarks demonstrate that MIS scales to large formulas. Furthermore, the minimal independent supports computed by MIS lead to 2-3 orders of magnitude improvement in the performance of UniGen2 and ApproxMC. Finally, construction of XORs over independent support allows us to obtain tight theoretical bounds on the size of XOR constraints for approximate model counting – in some cases, even better than previously observed empirical bounds.

# Chapter 12

# Conclusion and Future Work

Constrained sampling and counting are two fundamental problems in artificial intelligence. In constrained sampling, the task is to sample randomly from the set of solutions of input constraints while the problem of constrained counting is to count the number of solutions. Both problems have numerous applications, including in probabilistic reasoning, machine learning, planning, statistical physics, inexact computing, and constrained-random verification [10, 99, 129, 135]. For example, probabilistic inference over graphical models can be reduced to constrained counting for propositional formulas [48, 135]. In addition, approximate probabilistic reasoning relies heavily on sampling from high-dimensional probabilistic spaces encoded as sets of constraints [69, 99]. Both constrained sampling and counting can be viewed as aspects of one of the most fundamental problems in artificial intelligence: exploring the structure of the solution space of a set of constraints [137].

Constrained sampling and counting are known to be computationally hard [150, 100, 148]. To bypass these hardness results, approximate versions of the problems have been investigated. Despite strong theoretical and practical interest in approximation techniques over the years, there is still an immense gap between theory and practice in this area. Theoretical algorithms offer guarantees on the quality of approximation, but do not scale in practice, whereas practical tools achieve scalability at the cost of offering weaker or no guarantees.

The hashing-based approach introduced in this thesis has yielded significant progress in this area. By combining the ideas of using constraint solvers as an oracle and the reduction of the solution space via universal hashing, we have developed *highly scalable* algorithms that offer *rigorous* approximation guarantees.

In the context of constrained counting, we presented a new approach to hashing-based unweighted counting, which allows out-of-order-search with dependent hash functions, dramatically reducing the number of SAT solver calls from linear to logarithmic in the size of the support of interest. This is achieved while retaining strong theoretical guarantees and without increasing the complexity of each SAT solver call. We then discussed how our hashing-based techniques can be lifted to handle weighted counting. Prior hashing-based approaches to WMC employed computationally expensive MPE oracle. In contrast, we only employ NP oracle. We introduced a novel parameter, *t*ilt, to capture the hardness of benchmarks with respect to hashing-based approach. We then presented a complementary approach wherein we propose an efficient reduction of weighted to unweighted counting if the weight function is expressed using *literal-weighted* representation. To handle word-level constraints, we presented, $\mathcal{H}_{SMT}$, a word-level hash function and employed it to build SMTApproxMC, an approximate word-level model counter.

We employed hashing-based counting framework to estimate reliability of power-grid networks, which is crucial for decision making to ensure availability and resilience of critical facilities. Our counting-based reliability estimation framework, RelNet, unlike the current state of the art techniques, can scale to real world networks arising from cities across U.S., especially when exact reliability computations are not affordable.

In the context of constrained sampling, we designed a new hashing-based algorithm called UniGen, which is the first algorithm to provide guarantees of almost-uniformity, while scaling to the problems involving hundreds of thousands of variables. As a mark of departure from previous hashing-based approach, UniGen first invokes an approximate model counting routine to get an estimate of the number of cells that it should divide the space of solutions into. Then, UniGen employs NP oracle to enumerate all the solutions for a randomly chosen cell that passes the check for "smallness". In order to design efficient NP queries, we introduced the notion of sampling set of the variables which allows construction of sparser hash functions. Consequently, UniGen is able to scale to problems involving hundreds of thousands of variables where the sampling set is small. We then introduced an adaptation of UniGen, UniGen2, that addresses key performance deficiencies of UniGen. Significantly, we showed that UniGen2 achieves a near-linear speedup with the number of cores, without any degradation of uniformity either in theory or in practice. This suggests a new high-performance paradigm for generating (near-)uniformly distributed solutions of a system of constraints. Specifically, it is no longer necessary to gain performance by sacrificing uniformity in a sequential sampler.

Finally, we considered adaptation of UniGen to handle the problem of distribution aware sampling. For approximation techniques that provide strong theoretical two-way bounds, a major limitation is the reliance on potentially-expensive most probable explanation (MPE) queries. We identify a novel parameter, *tilt*, to categorize weighted counting and sampling problems for SAT. We showed how to remove this reliance on MPE queries, while retaining strong theoretical guarantees. Experimental results demonstrate the effectiveness of this approach in practice when the tilt is small.

The performance of hashing-based techniques presented in this thesis is primarily affected by the runtime of combinatorial solvers for the queries that are typically expressed as conjunction of CNF and constraints from hash functions. Furthermore, it has been observed that lower density XORs are easy to reason in practice and runtime performance of solvers greatly enhances with the decrease in the density of XOR-constraints [85]. In this context, it is important to note that hash functions constructed over Independent Support still retains the same theoretical guarantees with respect to universality. The importance of this observation comes from the fact that for many important classes of problems the size of an independent support is typically one to two orders of magnitude smaller than the number of all variables, which in turn leads to XOR constraints of typical density of 1/200 to 1/20, i.e. one to two orders of magnitude smaller than that of the traditional hash functions.

In Chapter 11, we presented the first algorithmic procedure and corresponding tool, MIS, to determine minimal independent support via reduction to Group MUS. The experimental evaluation over an extensive suite of benchmarks demonstrate that MIS scales to large formulas. Furthermore, the minimal independent supports computed by MIS lead to 2-3 orders of magnitude improvement in the performance of UniGen2 and ApproxMC. Finally, construction of XORs over independent support allows us to obtain tight theoretical bounds on the size of XOR constraints for approximate model counting – in some cases, even better than previously observed empirical bounds.

Overall, we were able to take the first step in bridging the gap between theory and practice in constrained sampling and counting.

## 12.1 Future Work

The hashing-based framework introduced in this thesis is just the first step in bridging the gap between theory and practice in constrained counting and sampling. Inspired by the success of SAT solving and in the hope of creating a similar counting and sampling revolution, we end with a list of several open directions that, we believe, would be crucial to achieve the "promised" revolution:

### 12.1.1 Dependence on $\varepsilon$

The hashing-based counting algorithms proposed in this thesis provide $(\varepsilon, \delta)$ guarantees and make $\mathcal{O}(\log n \log(\frac{1}{\delta})(\frac{1}{\varepsilon^2}))$ calls to SAT oracle. While in practice the quality of approximations are significantly better than the theoretical guarantees, we are still gazing at a wide gap between theory and practice (c.f. discussion in Section 4.2.2). In particular, invoking the hashing-based algorithms with very small $\varepsilon$ would imply impractical running times. ApproxMC2 requires $\mathcal{O}(\frac{1}{\varepsilon^2})$ invocation of the underlying NP oracle. This presents significant challenges when $\varepsilon$ is very small. Approaches based on Stockmeyer's hashing-based approach [144] lead to $\mathcal{O}(\frac{1}{\varepsilon})$ dependence but fail to scale to large instances. Therefore, a promising direction of future research would be to design techniques that would provide $\mathcal{O}(\frac{1}{\varepsilon})$ dependence on $\varepsilon$ and scale to large instances.

### 12.1.2 Weighted to Unweighted Reductions

The proposed reductions in Chapter 5 open up new research directions. While we focused on exact WMC in Chapter 5, the computational difficulty of exact inferencing in complex graphical models has led to significant recent interest in approximate WMC. In this context, it is worth noting that the reduction proposed in Theorem 13a allows us to lift approximation guarantees from the unweighted to the weighted setting for CNF formulas. Unfortunately, this is not the case for the reduction proposed in Theorem 13b, which is required for DNF formulas. The question of whether there exists an approximation-preserving reduction from WMC to UMC that also preserves DNF is open. The practical feasibility of solving approximate WMC problems by reducing them to their unweighted counterpart, even in the case of CNF formulas, requires further detailed investigation. This is particularly challenging since the current reductions introduce extra variables, which is known to adversely affect XOR-hashing-based state-of-the-art approximation techniques [39, 72].

Another interesting direction of research is CNF/DNF-preserving reductions from ConstraintWMC to UMC. Investigations in this direction can lead to improvements in both modeling and inferencing techniques in probabilistic programming frameworks. The design of practically efficient unweighted DNF model counters is also a fruitful line of research, since our reduction allows us to transform any such tool to a weighted DNF model counter.

### 12.1.3 Eager and Lazy SMT

The fundamental idea underlying our approach is the use of 3-universal hash functions (see ealier discussion) to partition the space of solutions into small cells. We express hashing constraints by means of random XOR constraints over a subset of sampling variables. The resulting formula is, consequently, the input CNF formula augmented with random XOR constraints. While XOR constraints by themselves are solvable efficiently using Gaussian elimination, CNF formulas augmented with random XOR clauses are very hard to solve by traditional SAT solvers. In our work, we use CryptoMiniSAT, a specialized solver for CNF formulas augmented with XOR constraints. CryptoMiniSAT can be viewed as a *lazy* SMT solver. This means that solving CNF constraints and XOR constraints is separated. Generally, lazy SMT solvers interleave iterations focusing on CNF constraints, using standard SAT techniques, with iterations focusing on theory constraints, using theory-specific techniques. Unlike lazy solvers, *eager* SMT solvers do not separate propositional and theory constraints. For example, eager SMT(BV) (bit-vector constraints) solvers rely on efficient CNF encoding of the input problem, leveraging solely the power of SAT solvers, e.g. [28]. It is not, however, clear that the lazy approach is necessarily superior to eager approach for solving CNF formulas augmented with *random* XOR constraints. While random XOR constraints on their own have been thoroughly studied, cf. [6], the combination of CNF constraints with random XOR constraints is yet to be studied. Therefore, an interesting direction of future research would be to study the tradeoff between eager and lazy SMT solving in this context; particularly, in our study of sampling and counting for SMT(BV), building on recent comparisons of the eager and lazy approaches in SMT solving [92].

### 12.1.4 Sampling for SMT

The problems of sampling of propositional formulas generalize naturally to the corresponding problems for formulas in richer first-order theories. Of particular interest are SMT (Satisfiability Modulo Theories) formulas that arise in program verification and testing, probabilistic-program analysis, quantitative-information flow, word-level hardware verification, constrained-random verification, probabilistic databases, and the like [50, 56, 90, 128, 132, 158].

The hash function, $H_{SMT}$ only provides guarantees of 2-universality while the proposed hashing-based approach, UniGen2, requires 3-universal hash fucntions. Therefore, an extension of hashing-based technique would requires us to design 3-universal hash functions. While XOR constraints are excellent choices for 3-universal hash functions when reasoning about propositional constraints, the existence of richer operators and domains in first-order theories provides an opportunity for using alternative hash functions, which lend themselves to increased computational efficiency in practice, while still providing guarantees of 3-universality.

### 12.1.5 Application to PAC Learning

We expect our efficient constrained sampling and counting algorithms to have broad applications. For example, our techniques may have an interesting application in PAC (probably approximately correct) learning [105, 151]. Bshouty et al. [32] showed that any class of Boolean functions that can be learned from membership queries with unlimited computational power can also be learned in probabilistic polynomial time with an NP oracle ($\mathsf{BPP}^{\mathsf{NP}}$) using membership queries. The technique relies on the probabilistic estimation of threshold functions, which employs almost-uniform sampling. So far, this widely cited result has been considered of pure theoretical interest due to the perceived infeasibility of almost-uniform sampling, but it may now become practical. More generally, sampling is a very fundamental basic operation in many computational-learning algorithms, cf. [152]. Therefore, an interesting direction of future research would be to study how hashing-based framework can enable reduction to practice of PAC-learning algorithms that have been studied only theoretically until now.

# APPENDIX

Table A1 : Extended Table of Performance Comparison of ApproxMC vis-a-vis ApproxMC2

| Benchmark | Vars | Clauses | ApproxMC2 Time | ApproxMC Time | ApproxMC2 SATCalls | ApproxMC SATCalls |
|---|---|---|---|---|---|---|
| case106 | 204 | 509 | 133.92 | – | 2377 | – |
| case35 | 400 | 1414 | 215.35 | – | 1809 | – |
| case146 | 219 | 558 | 4586.26 | – | 1986 | – |
| tutorial3 | 486193 | 2598178 | 12373.99 | – | 1744 | – |
| case202 | 200 | 544 | 149.56 | – | 1839 | – |
| case203 | 214 | 580 | 165.17 | – | 1800 | – |
| case205 | 214 | 580 | 300.11 | – | 1793 | – |
| s953a_15_7 | 602 | 1657 | 161.41 | – | 1648 | – |
| s953a_7_4 | 533 | 1373 | 16218.67 | – | 1832 | – |
| case_1_b14_1 | 238 | 681 | 132.47 | – | 1814 | – |
| case_2_b14_1 | 238 | 681 | 129.95 | – | 1805 | – |
| case119 | 267 | 787 | 906.88 | – | 2044 | – |
| case133 | 211 | 615 | 18502.44 | – | 2043 | – |
| case_3_b14_1 | 238 | 681 | 125.69 | – | 1831 | – |
| case204 | 214 | 580 | 166.2 | – | 1808 | – |
| case136 | 211 | 615 | 9754.08 | – | 2026 | – |
| llreverse | 63797 | 257657 | 1938.1 | 4482.94 | 1219 | 2801 |
| lltraversal | 39912 | 167842 | 151.33 | 450.57 | 1516 | 4258 |
| karatsuba | 19594 | 82417 | 23553.73 | 28817.79 | 1378 | 13360 |
| enqueueSeqSK | 16466 | 58515 | 192.96 | 2036.09 | 2207 | 23321 |
| 20 | 15475 | 60994 | 1778.45 | 20557.24 | 2308 | 34815 |
| 77 | 14535 | 27573 | 88.36 | 1529.34 | 2054 | 24764 |
| sort | 12125 | 49611 | 209.0 | 3610.4 | 1605 | 27731 |
| LoginService2 | 11511 | 41411 | 26.04 | 110.77 | 1533 | 10653 |
| 81 | 10775 | 38006 | 158.93 | 10555.13 | 2220 | 33954 |
| 17 | 10090 | 27056 | 100.76 | 4874.39 | 1810 | 28407 |
| 29 | 8866 | 31557 | 87.78 | 3569.25 | 1712 | 28630 |
| LoginService | 8200 | 26689 | 21.77 | 101.15 | 1498 | 12520 |
| 19 | 6993 | 23867 | 126.23 | 11051.95 | 1827 | 31352 |
| Pollard | 7815 | 41258 | 12.8 | 16.55 | 1023 | 695 |
| 7 | 6683 | 24816 | 84.1 | 5332.76 | 2062 | 31195 |
| doublyLinkedList | 6890 | 26918 | 17.05 | 75.45 | 1615 | 10647 |
| tree_delete | 5758 | 22105 | 8.87 | 33.84 | 1455 | 7647 |
| 35 | 4915 | 10547 | 77.53 | 6074.75 | 2028 | 32096 |
| 80 | 4969 | 17060 | 76.88 | 5039.37 | 2389 | 30294 |
| ProcessBean | 4768 | 14458 | 213.78 | 15558.75 | 2296 | 33493 |
| 56 | 4842 | 17828 | 126.96 | 1024.36 | 2218 | 22988 |

Continued on next page

| Benchmark | Vars | Clauses | ApproxMC2 Time | ApproxMC Time | ApproxMC2 SATCalls | ApproxMC SATCalls |
|---|---|---|---|---|---|---|
| 70 | 4670 | 15864 | 68.18 | 1026.99 | 2307 | 23902 |
| ProjectService3 | 3175 | 11019 | 190.98 | 19626.24 | 1715 | 36762 |
| 32 | 3834 | 13594 | 49.86 | 1102.68 | 1882 | 21835 |
| 55 | 3128 | 12145 | 90.33 | 7623.13 | 1810 | 28322 |
| 51 | 3708 | 14594 | 86.9 | 1538.87 | 2091 | 22115 |
| 109 | 3565 | 14012 | 77.69 | 917.19 | 1752 | 21104 |
| NotificationServiceImpl2 | 3540 | 13425 | 22.2 | 74.76 | 2265 | 15186 |
| aig_insertion2 | 2592 | 10156 | 13.18 | 120.56 | 2412 | 16729 |
| 53 | 2586 | 10747 | 32.29 | 248.26 | 1885 | 17680 |
| ConcreteActivityService | 2481 | 9011 | 6.01 | 33.56 | 1619 | 13072 |
| 111 | 2348 | 5479 | 42.49 | 567.25 | 1884 | 20383 |
| aig_insertion1 | 2296 | 9326 | 24.91 | 127.94 | 2416 | 16779 |
| case_3_b14_2 | 270 | 805 | 90.88 | 18114.84 | 2028 | 31194 |
| ActivityService2 | 1952 | 6867 | 2.74 | 13.09 | 1542 | 9700 |
| IterationService | 1896 | 6732 | 3.39 | 16.74 | 1572 | 10570 |
| squaring7 | 1628 | 5837 | 323.58 | 8774.17 | 1791 | 29298 |
| ActivityService | 1837 | 5968 | 2.39 | 11.62 | 1633 | 9606 |
| 10 | 1494 | 2215 | 135.04 | 4759.18 | 2020 | 30270 |
| case_2_b14_2 | 270 | 805 | 90.17 | 13479.3 | 2002 | 31179 |
| PhaseService | 1686 | 5655 | 2.45 | 12.03 | 1617 | 9649 |
| squaring9 | 1434 | 5028 | 308.34 | 6131.25 | 1718 | 29324 |
| case_1_b12_2 | 827 | 2725 | 129.03 | 9964.91 | 1808 | 29328 |
| UserServiceImpl | 1509 | 5009 | 1.49 | 7.1 | 1480 | 7707 |
| 27 | 1509 | 2707 | 34.96 | 130.23 | 1885 | 17489 |
| squaring8 | 1101 | 3642 | 250.2 | 9963.56 | 1784 | 29386 |
| case_2_b12_2 | 827 | 2725 | 122.64 | 7967.12 | 1803 | 29342 |
| case_1_b14_2 | 270 | 805 | 89.69 | 10777.71 | 2038 | 31187 |
| case_0_b12_2 | 827 | 2725 | 134.65 | 8362.19 | 1808 | 29340 |
| IssueServiceImpl | 1393 | 4319 | 2.48 | 13.37 | 1589 | 10469 |
| squaring10 | 1099 | 3632 | 290.64 | 6208.98 | 1773 | 29391 |
| squaring11 | 966 | 3213 | 324.63 | 11111.49 | 1795 | 29280 |
| s953a_3_2 | 515 | 1297 | 165.81 | 11968.07 | 1826 | 33920 |
| squaring29 | 1141 | 4248 | 135.4 | 1290.88 | 2002 | 18662 |
| squaring3 | 885 | 2809 | 281.29 | 8836.68 | 1802 | 27618 |
| squaring28 | 1060 | 3839 | 129.46 | 1164.31 | 2091 | 18685 |
| squaring6 | 885 | 2809 | 233.72 | 5799.3 | 1753 | 27580 |
| s1196a_15_7 | 777 | 2165 | 73.26 | 2577.71 | 1938 | 23097 |
| squaring30 | 1031 | 3693 | 117.53 | 1134.18 | 2006 | 18668 |
| squaring1 | 891 | 2839 | 227.03 | 5145.1 | 1787 | 27557 |
| squaring4 | 891 | 2839 | 274.71 | 6094.24 | 1774 | 27646 |
| squaring2 | 885 | 2809 | 240.35 | 5112.72 | 1805 | 27577 |
| squaring5 | 885 | 2809 | 352.17 | 6477.17 | 1819 | 27559 |
| GuidanceService | 988 | 3088 | 3.59 | 17.08 | 1632 | 13115 |
| case_1_b14_3 | 304 | 941 | 109.46 | 7432.67 | 1829 | 28444 |
| s1488_15_7 | 941 | 2783 | 1.57 | 5.02 | 1553 | 5867 |
| squaring26 | 894 | 3187 | 102.08 | 787.16 | 1997 | 17569 |
| case_3_b14_3 | 304 | 941 | 104.65 | 6821.33 | 1815 | 28424 |

Continued on next page

| Benchmark | Vars | Clauses | ApproxMC2 Time | ApproxMC Time | ApproxMC2 SATCalls | ApproxMC SATCalls |
|---|---|---|---|---|---|---|
| case201 | 200 | 544 | 221.78 | 16171.04 | 1814 | 32970 |
| squaring25 | 846 | 2947 | 110.25 | 791.63 | 2074 | 17437 |
| tree_delete3 | 795 | 2734 | 46.39 | 562.39 | 1595 | 20763 |
| s1488_7_4 | 872 | 2499 | 1.46 | 5.43 | 1523 | 6891 |
| squaring27 | 837 | 2901 | 110.1 | 714.37 | 2028 | 17337 |
| s1488_3_2 | 854 | 2423 | 1.8 | 6.51 | 1501 | 5527 |
| case_2_b14_3 | 304 | 941 | 114.36 | 6643.4 | 1815 | 28443 |
| s1238a_15_7 | 773 | 2210 | 66.87 | 713.17 | 1841 | 22792 |
| case_0_b11_1 | 340 | 1026 | 123.65 | 6398.95 | 1777 | 29323 |
| s1196a_7_4 | 708 | 1881 | 76.44 | 917.27 | 1800 | 22442 |
| s1196a_3_2 | 690 | 1805 | 62.64 | 827.91 | 1711 | 22177 |
| s1238a_7_4 | 704 | 1926 | 66.48 | 716.53 | 1813 | 22545 |
| case_1_b11_1 | 340 | 1026 | 124.08 | 5754.05 | 1810 | 29352 |
| s1238a_3_2 | 686 | 1850 | 77.88 | 895.66 | 1848 | 23171 |
| GuidanceService2 | 715 | 2181 | 2.37 | 15.56 | 1605 | 13252 |
| squaring23 | 710 | 2268 | 74.37 | 429.83 | 2358 | 15911 |
| squaring22 | 695 | 2193 | 71.75 | 466.91 | 2357 | 15891 |
| squaring20 | 696 | 2198 | 78.24 | 466.67 | 2357 | 15813 |
| squaring21 | 697 | 2203 | 81.89 | 460.94 | 2451 | 15877 |
| squaring24 | 695 | 2193 | 80.76 | 462.12 | 2363 | 15849 |
| s832a_15_7 | 693 | 2017 | 6.01 | 29.68 | 1608 | 14808 |
| s820a_15_7 | 685 | 1987 | 2.52 | 12.0 | 1483 | 12488 |
| s832a_7_4 | 624 | 1733 | 2.47 | 11.66 | 1543 | 12713 |
| s832a_3_2 | 606 | 1657 | 1.26 | 6.71 | 1717 | 11449 |
| s820a_7_4 | 616 | 1703 | 2.41 | 9.83 | 1435 | 12328 |
| s820a_3_2 | 598 | 1627 | 1.19 | 5.75 | 1646 | 10746 |
| case34 | 409 | 1597 | 124.7 | 2665.47 | 1818 | 27561 |
| s420_15_7 | 366 | 994 | 81.34 | 2011.14 | 2060 | 24871 |
| case6 | 329 | 996 | 113.94 | 3233.94 | 2043 | 25750 |
| s420_new_15_7 | 351 | 934 | 73.18 | 1897.5 | 2054 | 24885 |
| case131 | 432 | 1830 | 76.96 | 1293.21 | 1852 | 24230 |
| s420_7_4 | 312 | 770 | 82.7 | 2373.55 | 2049 | 24887 |
| s420_new1_15_7 | 366 | 994 | 79.42 | 1732.28 | 2053 | 24868 |
| case121 | 291 | 975 | 112.0 | 3046.07 | 1809 | 29418 |
| case_0_b12_1 | 427 | 1385 | 67.81 | 914.84 | 1880 | 22212 |
| squaring50 | 500 | 1965 | 31.92 | 190.39 | 2388 | 16703 |
| squaring51 | 496 | 1947 | 37.45 | 230.85 | 2094 | 16804 |
| case_1_b12_1 | 427 | 1385 | 66.94 | 866.66 | 1894 | 22152 |
| case_2_b12_1 | 427 | 1385 | 63.55 | 797.71 | 1882 | 22206 |
| s420_new1_7_4 | 312 | 770 | 85.19 | 2045.89 | 2061 | 24869 |
| case125 | 393 | 1555 | 86.17 | 1324.85 | 2306 | 23975 |
| case123 | 267 | 980 | 58.88 | 1625.83 | 2250 | 23066 |
| case143 | 427 | 1592 | 71.83 | 696.46 | 2139 | 19449 |
| s420_new_7_4 | 312 | 770 | 74.5 | 1485.23 | 2054 | 24887 |
| case105 | 170 | 407 | 227.36 | 7361.33 | 2330 | 32045 |
| case114 | 428 | 1851 | 24.83 | 151.71 | 1854 | 17679 |
| case115 | 428 | 1851 | 29.09 | 173.42 | 1888 | 17659 |

Continued on next page

| Benchmark | Vars | Clauses | ApproxMC2 Time | ApproxMC Time | ApproxMC2 SATCalls | ApproxMC SATCalls |
|---|---|---|---|---|---|---|
| case116 | 438 | 1881 | 31.59 | 156.56 | 1897 | 17636 |
| s526a_15_7 | 453 | 1304 | 20.35 | 67.56 | 1887 | 15811 |
| s526_15_7 | 452 | 1303 | 17.69 | 58.43 | 1898 | 15861 |
| case126 | 302 | 1129 | 74.05 | 1312.09 | 2316 | 23068 |
| s420_new_3_2 | 294 | 694 | 88.48 | 1577.85 | 2052 | 24925 |
| s420_new1_3_2 | 294 | 694 | 93.87 | 1590.05 | 2053 | 24485 |
| s420_3_2 | 294 | 694 | 97.18 | 1399.45 | 2052 | 24933 |
| s526a_7_4 | 384 | 1020 | 13.39 | 46.53 | 1805 | 15711 |
| case57 | 288 | 1158 | 57.97 | 703.78 | 1647 | 21193 |
| s444_15_7 | 377 | 1072 | 8.43 | 26.74 | 1634 | 14897 |
| case62 | 291 | 1165 | 71.35 | 833.88 | 1973 | 22174 |
| s526_7_4 | 383 | 1019 | 20.55 | 44.41 | 1820 | 15200 |
| s526_3_2 | 365 | 943 | 7.66 | 24.51 | 1964 | 14977 |
| s526a_3_2 | 366 | 944 | 12.45 | 26.09 | 1772 | 15219 |
| s382_15_7 | 350 | 995 | 22.29 | 67.46 | 1763 | 16207 |
| registerlesSwap | 372 | 1493 | 0.42 | 0.33 | 1018 | 685 |
| s510_15_7 | 340 | 948 | 20.59 | 56.42 | 1840 | 16558 |
| s510_7_4 | 316 | 844 | 18.06 | 73.38 | 1842 | 16622 |
| case117 | 309 | 1367 | 0.75 | 3.44 | 1712 | 8665 |
| case122 | 314 | 1258 | 17.59 | 67.53 | 1963 | 16806 |
| case111 | 306 | 1358 | 0.62 | 2.86 | 1519 | 7686 |
| case118 | 309 | 1367 | 0.84 | 3.44 | 1933 | 8650 |
| case113 | 309 | 1367 | 0.93 | 3.77 | 1972 | 8624 |
| s510_3_2 | 298 | 768 | 15.14 | 74.08 | 1871 | 16667 |
| s349_15_7 | 285 | 829 | 13.18 | 76.65 | 1906 | 15850 |
| s444_7_4 | 308 | 788 | 18.25 | 62.97 | 1766 | 16260 |
| s298_15_7 | 292 | 870 | 0.86 | 4.08 | 1756 | 9569 |
| case2 | 296 | 1116 | 10.08 | 39.7 | 1662 | 14956 |
| s344_15_7 | 284 | 824 | 12.76 | 60.94 | 1887 | 15837 |
| case3 | 294 | 1110 | 11.52 | 40.84 | 1648 | 14935 |
| case110 | 287 | 1263 | 0.69 | 2.74 | 1776 | 7771 |
| s444_3_2 | 290 | 712 | 6.48 | 18.76 | 1601 | 14909 |
| s382_7_4 | 281 | 711 | 7.83 | 28.86 | 1538 | 14832 |
| s382_3_2 | 263 | 635 | 5.33 | 21.47 | 1624 | 14915 |
| case109 | 241 | 915 | 5.53 | 24.43 | 1711 | 13172 |
| case132 | 236 | 708 | 22.7 | 94.67 | 1683 | 14076 |
| s298_7_4 | 223 | 586 | 0.67 | 3.42 | 1690 | 9492 |
| case135 | 236 | 708 | 19.74 | 68.81 | 1659 | 13858 |
| case56 | 202 | 722 | 1.84 | 10.02 | 1676 | 13176 |
| s298_3_2 | 205 | 510 | 0.59 | 2.92 | 1747 | 8670 |
| case108 | 205 | 800 | 0.87 | 4.15 | 1731 | 9554 |
| s344_7_4 | 215 | 540 | 14.53 | 47.24 | 1875 | 15887 |
| case54 | 203 | 725 | 2.49 | 10.56 | 1679 | 13197 |
| case5 | 176 | 518 | 72.42 | 474.93 | 2103 | 18572 |
| case1 | 187 | 681 | 0.73 | 3.8 | 1726 | 10331 |
| case46 | 176 | 660 | 0.64 | 3.53 | 1726 | 9572 |
| case44 | 173 | 651 | 0.61 | 3.52 | 1754 | 9548 |

Continued on next page

| Benchmark | Vars | Clauses | ApproxMC2 Time | ApproxMC Time | ApproxMC2 SATCalls | ApproxMC SATCalls |
|---|---|---|---|---|---|---|
| case124 | 133 | 386 | 66.62 | 653.36 | 1730 | 20333 |
| s344_3_2 | 197 | 464 | 12.37 | 38.68 | 1896 | 15915 |
| s349_7_4 | 216 | 545 | 41.0 | 39.31 | 1893 | 15854 |
| case68 | 178 | 553 | 1.12 | 5.25 | 1744 | 10430 |
| s349_3_2 | 198 | 469 | 18.26 | 40.07 | 1862 | 15841 |
| case8 | 160 | 525 | 9.68 | 37.17 | 1883 | 15874 |
| case53 | 132 | 395 | 0.67 | 4.18 | 1741 | 11410 |
| case55 | 149 | 442 | 2.18 | 8.88 | 1667 | 13128 |
| case51 | 132 | 395 | 0.66 | 3.76 | 1740 | 11220 |
| case38 | 143 | 568 | 0.34 | 1.31 | 1641 | 5956 |
| case112 | 137 | 520 | 0.5 | 2.17 | 1975 | 8668 |
| case52 | 132 | 395 | 0.85 | 3.83 | 1743 | 11357 |
| case22 | 126 | 411 | 0.27 | 1.22 | 1516 | 6856 |
| case21 | 126 | 411 | 0.28 | 1.2 | 1526 | 6808 |
| case47 | 118 | 328 | 1.11 | 5.47 | 1756 | 11378 |
| case45 | 116 | 421 | 0.29 | 1.49 | 1496 | 7662 |
| case7 | 116 | 365 | 0.57 | 2.83 | 1739 | 10475 |
| case43 | 116 | 421 | 0.31 | 1.54 | 1517 | 7726 |
| case11 | 105 | 371 | 0.28 | 1.48 | 1458 | 7719 |
| case4 | 103 | 316 | 0.37 | 1.71 | 1900 | 8515 |
| case63 | 96 | 299 | 0.36 | 1.75 | 1630 | 8621 |
| case64 | 93 | 285 | 0.4 | 1.85 | 1927 | 8748 |
| case58 | 96 | 299 | 0.42 | 1.79 | 1884 | 8704 |
| case59 | 93 | 285 | 0.39 | 1.75 | 1927 | 8723 |
| case59_1 | 93 | 285 | 0.39 | 1.69 | 1972 | 8642 |
| case134 | 60 | 146 | 0.37 | 2.34 | 1710 | 11336 |
| case101 | 72 | 178 | 2.12 | 10.02 | 1666 | 14100 |
| case100 | 72 | 178 | 2.0 | 8.73 | 1675 | 14072 |
| case23 | 77 | 235 | 0.22 | 0.7 | 1604 | 5034 |
| case17 | 77 | 235 | 0.22 | 0.69 | 1608 | 5069 |
| case137 | 60 | 146 | 0.52 | 2.43 | 1779 | 11219 |
| case32 | 52 | 146 | 0.15 | 0.76 | 1372 | 4106 |
| case25 | 68 | 195 | 0.18 | 0.44 | 1323 | 3266 |
| case30 | 68 | 195 | 0.18 | 0.43 | 1341 | 3259 |
| case26 | 53 | 148 | 0.16 | 0.55 | 1352 | 4120 |
| case36 | 64 | 208 | 0.15 | 0.34 | 1338 | 2426 |
| case27 | 52 | 146 | 0.15 | 0.51 | 1369 | 4156 |
| case31 | 53 | 148 | 0.16 | 0.52 | 1374 | 4125 |
| case29 | 65 | 190 | 0.15 | 0.28 | 1181 | 2360 |
| case24 | 65 | 190 | 0.17 | 0.28 | 1227 | 2267 |
| case33 | 51 | 143 | 0.18 | 0.52 | 1369 | 4199 |
| case28 | 51 | 143 | 0.18 | 0.48 | 1316 | 4153 |
| case103 | 32 | 86 | 0.12 | 0.24 | 1233 | 2349 |
| case102 | 34 | 92 | 0.15 | 0.25 | 1215 | 2357 |
| squaring12 | 1507 | 5210 | – | 8419.06 | 423 | 31880 |
| squaring16 | 1627 | 5835 | – | 9926.56 | 423 | 31778 |
| squaring14 | 1458 | 5009 | – | 13892.48 | 423 | 31842 |

Table A2 : Extended Table of Runtime Performance comparison of sharpWeightSAT vis-a-vis SDD

| | | | sharpSAT + Reduction = sharpWeightSAT | | | | DSharp | SDD |
|---|---|---|---|---|---|---|---|---|
| **Benchmark** | **Orig #vars** | **Orig #clas** | **Final #vars** | **Final #claus** | **Transform time(s)** | **Counting time(s)** | **Counting time(s)** | **time(s)** |
| fs-01.net | 32 | 38 | 242 | 278 | 0.38 | 0.04 | 0.01 | 0.01 |
| or-50-10-10-UC-40 | 100 | 272 | 310 | 512 | 0.12 | 0.06 | 0.05 | 0.2 |
| or-50-10-10-UC-30 | 100 | 264 | 190 | 384 | 0.02 | 0.06 | 0.05 | 0.49 |
| or-50-10-1-UC-40 | 100 | 273 | 310 | 513 | 0.02 | 0.06 | 0.05 | 0.32 |
| or-50-20-10-UC-30 | 100 | 267 | 310 | 507 | 0.03 | 0.15 | 0.45 | 3.92 |
| or-50-20-10-UC-40 | 100 | 274 | 190 | 394 | 0.02 | 0.02 | 0.14 | 1.05 |
| or-50-10-1-UC-30 | 100 | 266 | 190 | 386 | 0.02 | 0.07 | 0.06 | 0.39 |
| or-50-20-1-UC-40 | 100 | 272 | 190 | 392 | 0.02 | 0.08 | 0.07 | 0.98 |
| or-50-10-9-UC-40 | 100 | 264 | 190 | 384 | 0.04 | 0.15 | 0.08 | 0.67 |
| or-50-10-1-UC-20 | 100 | 262 | 190 | 382 | 0.03 | 0.08 | 0.09 | 0.98 |
| or-50-10-10-UC-20 | 100 | 261 | 310 | 501 | 0.16 | 0.09 | 0.04 | 0.78 |
| cliquen10 | 110 | 200 | 276 | 411 | 0.06 | 0.34 | – | 11.26 |
| or-60-5-2-UC-40 | 120 | 323 | 330 | 563 | 0.03 | 0.01 | 0.05 | 0.53 |
| or-70-10-3-UC-40 | 140 | 383 | 230 | 503 | 0.03 | 0.01 | 0.06 | – |
| or-70-10-3-UC-30 | 140 | 374 | 350 | 614 | 0.05 | 0.09 | 0.07 | 0.85 |
| or-70-5-7-UC-40 | 140 | 381 | 350 | 621 | 0.06 | 0.09 | 0.05 | 0.57 |
| or-70-20-9-UC-30 | 140 | 374 | 350 | 614 | 0.03 | 0.02 | 0.05 | 1.16 |
| or-70-5-2-UC-30 | 140 | 371 | 350 | 611 | 0.11 | 0.06 | 0.05 | 1.43 |
| or-70-5-7-UC-30 | 140 | 370 | 350 | 610 | 0.03 | 0.02 | 0.06 | 1.05 |
| or-70-20-9-UC-40 | 140 | 383 | 230 | 503 | 0.02 | 0.07 | 0.07 | 0.73 |
| or-70-5-2-UC-40 | 140 | 378 | 350 | 618 | 0.02 | 0.01 | 0.05 | 0.59 |
| or-70-10-6-UC-40 | 140 | 391 | 230 | 511 | 0.02 | 0.01 | 0.05 | 0.37 |
| or-70-20-6-UC-40 | 140 | 375 | 350 | 615 | 0.05 | 0.02 | 0.15 | 1.28 |
| or-70-10-6-UC-30 | 140 | 379 | 230 | 499 | 0.02 | 0.06 | 0.08 | 1.21 |
| 5step | 177 | 475 | 267 | 595 | 0.02 | 0.09 | 0.02 | 4.49 |
| or-100-20-9-UC-50 | 200 | 557 | 290 | 677 | 0.04 | 0.03 | 0.07 | 1.17 |
| or-100-20-9-UC-60 | 200 | 561 | 410 | 801 | 0.05 | 0.11 | 0.05 | 1.34 |
| or-100-20-6-UC-60 | 200 | 564 | 290 | 684 | 0.04 | 0.06 | 0.05 | 0.95 |
| cliquen15 | 240 | 450 | 617 | 932 | 0.02 | 11.29 | – | 536.85 |
| case121 | 291 | 975 | 381 | 1095 | 0.04 | 0.12 | 12.46 | 6.6 |
| BN_104 | 294 | 537 | 914 | 1307 | 0.03 | 0.07 | 0.31 | 0.73 |
| case_1_b11_1 | 340 | 1026 | 550 | 1266 | 0.03 | 92.16 | 1059.82 | 64.3 |

Continued on next page

| | | | sharpSAT + Reduction = sharpWeightSAT | | | | DSharp | SDD |
|---|---|---|---|---|---|---|---|---|
| **Benchmark** | **Orig #vars** | **Orig #clas** | **Final #vars** | **Final #claus** | **Transform time(s)** | **Counting time(s)** | **Counting time(s)** | **time(s)** |
| s526_3_2 | 365 | 943 | 455 | 1063 | 0.02 | 0.22 | 0.62 | – |
| s526a_3_2 | 366 | 944 | 456 | 1064 | 0.04 | 0.17 | 1.79 | – |
| case35 | 400 | 1414 | 490 | 1534 | 0.03 | 0.54 | 18.66 | 76.35 |
| cliquen20 | 420 | 800 | 1750 | 2320 | 0.62 | 21.49 | – | – |
| s526_15_7 | 452 | 1303 | 542 | 1423 | 0.03 | 0.94 | 5.75 | – |
| s953a_3_2 | 515 | 1297 | 605 | 1417 | 0.03 | 0.1 | 1.08 | – |
| BN_112 | 541 | 1443 | 2187 | 3489 | 0.05 | 0.08 | 0.6 | 3.46 |
| lang12 | 576 | 13584 | 786 | 13824 | 0.06 | 334.5 | 1276.3 | – |
| BN_110 | 620 | 1568 | 3966 | 5392 | 0.46 | 0.08 | 0.48 | 9.74 |
| BN_106 | 630 | 1692 | 2607 | 4155 | 0.07 | 0.14 | 0.25 | 3.6 |
| cliquen25 | 650 | 1250 | 1742 | 2642 | 0.05 | 260.06 | – | – |
| s1238a_3_2 | 686 | 1850 | 896 | 2090 | 0.04 | 0.45 | 6.94 | – |
| s1196a_3_2 | 690 | 1805 | 780 | 1925 | 0.04 | 0.34 | 10.53 | – |
| s1238a_7_4 | 704 | 1926 | 914 | 2166 | 0.11 | 0.63 | 6.54 | – |
| s1196a_7_4 | 708 | 1881 | 798 | 2001 | 0.07 | 0.61 | 7.28 | – |
| s1238a_15_7 | 773 | 2210 | 863 | 2330 | 0.05 | 0.56 | 10.2 | – |
| s1196a_15_7 | 777 | 2165 | 867 | 2285 | 0.06 | 0.54 | 8.88 | – |
| case_2_b12_2 | 827 | 2725 | 917 | 2845 | 0.06 | 34.11 | 714.37 | 735.68 |
| squaring1 | 891 | 2839 | 981 | 2959 | 0.04 | 10.02 | 97.86 | – |
| BN_67 | 925 | 2063 | 1240 | 2423 | 0.07 | 0.38 | 2.11 | 3239.31 |
| BN_65 | 925 | 2063 | 1150 | 2333 | 0.05 | 0.11 | 1.52 | – |
| cliquen30 | 930 | 1800 | 2517 | 3821 | 0.11 | 300.86 | – | – |
| rbm_20 | 960 | 1760 | 4546 | 6226 | 0.1 | 1231.3 | – | – |
| squaring10 | 1099 | 3632 | 1189 | 3752 | 0.1 | 43.08 | 998.32 | – |
| squaring8 | 1101 | 3642 | 1191 | 3762 | 0.06 | 21.41 | 392.11 | – |
| BN_59 | 1112 | 2661 | 1272 | 2853 | 0.21 | 0.68 | 12.07 | 820.8 |
| BN_63 | 1112 | 2661 | 1272 | 2853 | 0.04 | 0.68 | 8.68 | – |
| BN_53 | 1154 | 2692 | 1314 | 2884 | 0.07 | 1.23 | 8.5 | 1425.19 |
| BN_57 | 1154 | 2692 | 1378 | 2948 | 0.05 | 0.64 | 10.89 | 523.0 |
| BN_55 | 1154 | 2692 | 1314 | 2884 | 0.1 | 1.11 | – | – |
| BN_47 | 1336 | 3376 | 1406 | 3460 | 0.11 | 0.11 | 1.49 | 170.92 |
| BN_51 | 1336 | 3376 | 1434 | 3488 | 0.09 | 0.09 | 1.08 | 185.71 |
| BN_49 | 1336 | 3376 | 1434 | 3488 | 0.09 | 0.24 | – | 1296.78 |
| BN_61 | 1348 | 3388 | 1418 | 3472 | 0.05 | 0.2 | 1.77 | 157.88 |

Continued on next page

| | | | sharpSAT + Reduction = sharpWeightSAT | | | | DSharp | SDD |
|---|---|---|---|---|---|---|---|---|
| **Benchmark** | **Orig #vars** | **Orig #clas** | **Final #vars** | **Final #claus** | **Transform time(s)** | **Counting time(s)** | **Counting time(s)** | **time(s)** |
| blockmap_05_01.net | 1411 | 2737 | 1501 | 2857 | 0.04 | 0.13 | 1.05 | 11.75 |
| squaring9 | 1434 | 5028 | 1524 | 5148 | 0.07 | 32.68 | 721.14 | – |
| squaring14 | 1458 | 5009 | 1548 | 5129 | 0.07 | 68.75 | 2152.41 | – |
| squaring12 | 1507 | 5210 | 1597 | 5330 | 0.06 | 79.02 | 1305.58 | – |
| squaring16 | 1627 | 5835 | 1723 | 5963 | 0.07 | – | 2623.12 | – |
| squaring7 | 1628 | 5837 | 1724 | 5965 | 0.13 | 40.31 | 1383.83 | – |
| blockmap_05_02.net | 1738 | 3452 | 1976 | 3724 | 0.09 | 0.08 | 2.35 | 15.45 |
| BN_43 | 1820 | 3806 | 2240 | 4286 | 0.34 | 8393.12 | – | – |
| BN_108 | 2289 | 8218 | 11028 | 19105 | 0.27 | 2.14 | 8.66 | 270.31 |
| smokers_20 | 2580 | 3740 | 6840 | 8860 | 0.33 | 224.25 | – | – |
| BN_38 | 3938 | 7661 | 8027 | 12800 | – | – | – | 5760.54 |
| treemax | 24859 | 103762 | 26353 | 105754 | 1.5 | 3.93 | 338.16 | – |
| BN_26 | 50470 | 93870 | 276675 | 352390 | 244.29 | 68.99 | 259.42 | 693.09 |

Table A3 : Extended Runtime performance of SMTApproxMC vis-a-vis CDM

| Benchmark | Total Bits | Variable Types | # of Operations | SMTApproxMC time(s) | CDM time(s) |
|---|---|---|---|---|---|
| squaring27 | 59 | {1: 11, 16: 3} | 10 | – | 2998.97 |
| 1159708 | 64 | {8: 4, 32: 1} | 12 | 14793.93 | – |
| 1159472 | 64 | {8: 4, 32: 1} | 8 | 16308.82 | – |
| 1159115 | 64 | {8: 4, 32: 1} | 12 | 23984.55 | – |
| 1159520 | 64 | {8: 4, 32: 1} | 1388 | 114.53 | 155.09 |
| 1160300 | 64 | {8: 4, 32: 1} | 1183 | 44.02 | 71.16 |
| 1159005 | 64 | {8: 4, 32: 1} | 213 | 28.88 | 105.6 |
| 1159751 | 64 | {8: 4, 32: 1} | 681 | 143.32 | 193.84 |
| 1159391 | 64 | {8: 4, 32: 1} | 681 | 57.03 | 91.62 |
| case1 | 17 | {1: 13, 4: 1} | 13 | 17.89 | 65.12 |
| 1159870 | 64 | {8: 4, 32: 1} | 164 | 17834.09 | 9152.65 |
| 1160321 | 64 | {8: 4, 32: 1} | 10 | 117.99 | 265.67 |
| 1159914 | 64 | {8: 4, 32: 1} | 8 | 230.06 | 276.74 |
| 1159064 | 64 | {8: 4, 32: 1} | 10 | 69.58 | 192.36 |
| 1160493 | 64 | {8: 4, 32: 1} | 8 | 317.31 | 330.47 |
| 1159197 | 64 | {8: 4, 32: 1} | 8 | 83.22 | 176.23 |
| 1160487 | 64 | {8: 4, 32: 1} | 10 | 74.92 | 149.44 |
| 1159606 | 64 | {8: 4, 32: 1} | 686 | 431.23 | 287.85 |
| case100 | 22 | {1: 6, 16: 1} | 8 | 32.62 | 89.69 |
| 1160397 | 64 | {8: 4, 32: 1} | 70 | 126.08 | 172.24 |

Continued on next page

| Benchmark | Total Bits | Variable Types | # of Operations | SMTApproxMC time(s) | CDM time(s) |
|---|---|---|---|---|---|
| 1160475 | 64 | {8: 4, 32: 1} | 67 | 265.58 | 211.16 |
| case108 | 24 | {1: 20, 4: 1} | 7 | 37.33 | 100.2 |
| case101 | 22 | {1: 6, 16: 1} | 12 | 44.74 | 90 |
| 1159244 | 64 | {8: 4, 32: 1} | 1474 | 408.63 | 273.57 |
| case46 | 20 | {1: 8, 4: 3} | 12 | 16.95 | 76.4 |
| case44 | 20 | {1: 8, 4: 3} | 8 | 13.69 | 72.05 |
| case134 | 19 | {1: 3, 16: 1} | 8 | 5.36 | 54.22 |
| case137 | 19 | {1: 3, 16: 1} | 9 | 10.98 | 56.12 |
| case68 | 26 | {8: 3, 1: 2} | 7 | 34.9 | 67.48 |
| case54 | 20 | {1: 16, 4: 1} | 8 | 50.73 | 103.91 |
| 1160365 | 64 | {8: 4, 32: 1} | 286 | 98.38 | 99.74 |
| 1159418 | 32 | {8: 2, 16: 1} | 7 | 3.73 | 43.68 |
| 1160877 | 32 | {8: 2, 16: 1} | 8 | 2.57 | 44.01 |
| 1160988 | 32 | {8: 2, 16: 1} | 8 | 4.4 | 44.64 |
| 1160521 | 32 | {8: 2, 16: 1} | 7 | 4.96 | 44.52 |
| 1159789 | 32 | {8: 2, 16: 1} | 13 | 6.35 | 43.09 |
| 1159117 | 32 | {8: 2, 16: 1} | 13 | 5.55 | 43.18 |
| 1159915 | 32 | {8: 2, 16: 1} | 11 | 7.02 | 45.62 |
| 1160332 | 32 | {8: 2, 16: 1} | 12 | 3.94 | 44.35 |
| 1159582 | 32 | {8: 2, 16: 1} | 8 | 5.37 | 43.98 |
| 1160530 | 32 | {8: 2, 16: 1} | 12 | 2.01 | 43.28 |
| 1160482 | 64 | {8: 4, 32: 1} | 36 | 153.99 | 120.55 |

Continued on next page

| Benchmark | Total Bits | Variable Types | # of Operations | SMTApproxMC time(s) | CDM time(s) |
|---|---|---|---|---|---|
| 1159564 | 32 | {8: 2, 16: 1} | 12 | 7.36 | 41.77 |
| 1159990 | 64 | {8: 4, 32: 1} | 34 | 71.17 | 97.25 |
| case7 | 18 | {1: 10, 8: 1} | 12 | 17.93 | 51.96 |
| case56 | 20 | {1: 16, 4: 1} | 12 | 41.54 | 109.3 |
| case43 | 15 | {1: 11, 4: 1} | 12 | 8.6 | 37.63 |
| case45 | 15 | {1: 11, 4: 1} | 12 | 9.3 | 35.77 |
| case53 | 19 | {1: 7, 8: 1, 4: 1} | 9 | 53.66 | 69.96 |
| case4 | 16 | {1: 12, 4: 1} | 12 | 8.42 | 35.49 |
| 1160438 | 64 | {8: 4, 32: 1} | 2366 | 199.08 | 141.84 |
| case109 | 29 | {1: 21, 4: 2} | 12 | 171.51 | 179.98 |
| case38 | 13 | {1: 9, 4: 1} | 7 | 6.21 | 30.27 |
| case11 | 15 | {1: 11, 4: 1} | 8 | 7.26 | 33.75 |
| 1158973 | 64 | {8: 4, 32: 1} | 94 | 366.6 | 270.17 |
| case22 | 14 | {1: 10, 4: 1} | 12 | 5.46 | 26.03 |
| case21 | 14 | {1: 10, 4: 1} | 12 | 5.57 | 24.59 |
| case52 | 19 | {1: 7, 8: 1, 4: 1} | 9 | 45.1 | 70.72 |
| case23 | 12 | {1: 8, 4: 1} | 11 | 2.29 | 12.84 |
| case51 | 19 | {1: 7, 8: 1, 4: 1} | 12 | 40 | 67.22 |
| case17 | 12 | {1: 8, 4: 1} | 12 | 2.75 | 11.09 |
| case33 | 11 | {1: 7, 4: 1} | 12 | 1.7 | 9.66 |
| case30 | 13 | {1: 5, 4: 2} | 13 | 1.41 | 8.69 |
| case28 | 11 | {1: 7, 4: 1} | 12 | 1.66 | 8.73 |

Continued on next page

| Benchmark | Total Bits | Variable Types | # of Operations | SMTApproxMC time(s) | CDM time(s) |
|---|---|---|---|---|---|
| case25 | 13 | {1: 5, 4: 2} | 12 | 1.39 | 8.27 |
| case27 | 11 | {1: 7, 4: 1} | 12 | 1.69 | 8.57 |
| case26 | 11 | {1: 7, 4: 1} | 12 | 1.68 | 8.35 |
| case32 | 11 | {1: 7, 4: 1} | 12 | 1.46 | 8.16 |
| case31 | 11 | {1: 7, 4: 1} | 12 | 1.64 | 7.64 |
| case29 | 12 | {1: 4, 4: 2} | 8 | 0.67 | 5.16 |
| case24 | 12 | {1: 4, 4: 2} | 12 | 0.77 | 4.94 |
| 1160335 | 64 | {8: 4, 32: 1} | 216 | 0.31 | 0.54 |
| 1159940 | 64 | {8: 4, 32: 1} | 94 | 0.17 | 0.04 |
| 1159690 | 32 | {8: 2, 16: 1} | 8 | 0.12 | 0.04 |
| 1160481 | 32 | {8: 2, 16: 1} | 12 | 0.13 | 0.03 |
| 1159611 | 64 | {8: 4, 32: 1} | 73 | 0.2 | 0.09 |
| 1161180 | 32 | {8: 2, 16: 1} | 12 | 0.11 | 0.04 |
| 1160849 | 32 | {8: 2, 16: 1} | 7 | 0.1 | 0.03 |
| 1159790 | 64 | {8: 4, 32: 1} | 113 | 0.15 | 0.04 |
| 1160315 | 64 | {8: 4, 32: 1} | 102 | 0.17 | 0.04 |
| 1159720 | 64 | {8: 4, 32: 1} | 102 | 0.17 | 0.05 |
| 1159881 | 64 | {8: 4, 32: 1} | 102 | 0.16 | 0.04 |
| 1159766 | 64 | {8: 4, 32: 1} | 73 | 0.15 | 0.03 |
| 1160220 | 64 | {8: 4, 32: 1} | 681 | 0.17 | 0.03 |
| 1159353 | 64 | {8: 4, 32: 1} | 113 | 0.16 | 0.04 |
| 1160223 | 64 | {8: 4, 32: 1} | 102 | 0.17 | 0.04 |

Continued on next page

| Benchmark | Total Bits | Variable Types | # of Operations | SMTApproxMC time(s) | CDM time(s) |
|---|---|---|---|---|---|
| 1159683 | 64 | {8: 4, 32: 1} | 102 | 0.17 | 0.03 |
| 1159702 | 64 | {8: 4, 32: 1} | 102 | 0.19 | 0.04 |
| 1160378 | 64 | {8: 4, 32: 1} | 476 | 0.17 | 0.04 |
| 1159183 | 64 | {8: 4, 32: 1} | 172 | 0.17 | 0.03 |
| 1159747 | 64 | {8: 4, 32: 1} | 322 | 0.18 | 0.03 |
| 1159808 | 64 | {8: 4, 32: 1} | 539 | 0.17 | 0.03 |
| 1159849 | 64 | {8: 4, 32: 1} | 322 | 0.18 | 0.03 |
| 1159449 | 64 | {8: 4, 32: 1} | 540 | 0.3 | 0.05 |
| case47 | 26 | {1: 6, 8: 2, 4: 1} | 11 | 81.5 | 80.25 |
| case2 | 24 | {1: 20, 4: 1} | 10 | 273.91 | 194.33 |
| 1159239 | 64 | {8: 4, 32: 1} | 238 | 1159.32 | 449.21 |
| case8 | 24 | {1: 12, 8: 1, 4: 1} | 8 | 433.2 | 147.35 |
| 1159936 | 64 | {8: 4, 32: 1} | 238 | 5835.35 | 1359.9 |
| squaring51 | 40 | {1: 32, 4: 2} | 7 | 3285.52 | 607.22 |
| 1159431 | 64 | {8: 4, 32: 1} | 12 | 36406.4 | – |
| 1160191 | 64 | {8: 4, 32: 1} | 12 | 40166.1 | – |

Table A4 : Extended Table of Runtime performance comparison of UniGen and UniWit

| | | | UniGen | | | UniWit | |
|---|---|---|---|---|---|---|---|
| Benchmark | #Variables | $\|S\|$ | Succ Prob | Avg Run Time (s) | Avg XOR len | Avg Run Time (s) | Avg XOR len |
| Case121 | 291 | 48 | 1.0 | 0.19 | 24 | 56.09 | 145 |
| Case1_b11_1 | 340 | 48 | 1.0 | 0.2 | 24 | 755.97 | 170 |
| Case2_b12_2 | 827 | 45 | 1.0 | 0.33 | 22 | – | – |
| Case35 | 400 | 46 | 0.99 | 11.23 | 23 | 666.14 | 199 |
| Squaring1 | 891 | 72 | 1.0 | 0.38 | 36 | – | – |
| Squaring8 | 1101 | 72 | 1.0 | 1.77 | 36 | 5212.19 | 550 |
| Squaring10 | 1099 | 72 | 1.0 | 1.83 | 36 | 4521.11 | 550 |
| Squaring7 | 1628 | 72 | 1.0 | 2.44 | 36 | 2937.5 | 813 |
| Squaring9 | 1434 | 72 | 1.0 | 4.43 | 36 | 4054.42 | 718 |
| Squaring14 | 1458 | 72 | 1.0 | 24.34 | 36 | 2697.42 | 728 |
| Squaring12 | 1507 | 72 | 1.0 | 31.88 | 36 | 3421.83 | 752 |
| Squaring16 | 1627 | 72 | 1.0 | 41.08 | 36 | 2852.17 | 812 |
| s526_3_2 | 365 | 24 | 0.98 | 0.68 | 12 | 51.77 | 181 |
| s526a_3_2 | 366 | 24 | 1.0 | 0.97 | 12 | 84.04 | 182 |
| s526_15_7 | 452 | 24 | 0.99 | 1.68 | 12 | 23.04 | 225 |
| s1196a_7_4 | 708 | 32 | 1.0 | 6.9 | 16 | 833.1 | 353 |
| s1196a_3_2 | 690 | 32 | 1.0 | 7.12 | 16 | 451.03 | 345 |
| s1238a_7_4 | 704 | 32 | 1.0 | 7.26 | 16 | 1570.27 | 352 |
| s1238a_15_7 | 773 | 32 | 1.0 | 7.94 | 16 | 136.7 | 385 |
| s1196a_15_7 | 777 | 32 | 0.97 | 8.98 | 16 | 133.45 | 388 |
| s1238a_3_2 | 686 | 32 | 0.99 | 10.85 | 16 | 1416.28 | 342 |
| s953a_3_2 | 515 | 45 | 0.99 | 12.48 | 23 | 22414.86 | 257 |
| TreeMax | 24859 | 19 | 1.0 | 0.52 | 10 | 49.78 | 12423 |
| LLReverse | 63797 | 25 | 1.0 | 33.92 | 13 | 3460.58 | 31888 |
| LoginService2 | 11511 | 36 | 0.98 | 6.14 | 18 | – | – |
| EnqueueSeqSK | 16466 | 42 | 1.0 | 32.39 | 21 | – | – |
| ProjectService3 | 3175 | 55 | 1.0 | 71.74 | 28 | – | – |
| Sort | 12125 | 52 | 0.99 | 79.44 | 26 | – | – |
| Karatsuba | 19594 | 41 | 1.0 | 85.64 | 21 | – | – |
| ProcessBean | 4768 | 64 | 0.98 | 123.52 | 32 | – | – |
| tutorial3_4_31 | 486193 | 31 | 0.98 | 782.85 | 16 | – | – |

Table A5 : Extended Runtime performance comparison of UniGen2 and UniGen (on a single core)

| | | | | UniGen2 | | UniGen |
|---|---|---|---|---|---|---|
| **Benchmark** | **#vars** | **#clas** | **\|S\|** | **Succ. Prob** | **Runtime(s)** | **Runtime(s)** |
| 109_new | 60 | 55 | 36 | 1.0 | 0.14 | 19.35 |
| 32_new | 60 | 49 | 38 | 1.0 | 0.12 | 22.66 |
| 70_new | 62 | 49 | 40 | 1.0 | 0.13 | 16.34 |
| 29_new | 69 | 55 | 45 | 1.0 | 0.12 | 6.05 |
| case100 | 72 | 178 | 24 | 1.0 | 0.01 | 0.2 |
| case101 | 72 | 178 | 24 | 1.0 | 0.01 | 0.2 |
| 10_new | 103 | 135 | 46 | 1.0 | 0.14 | 2.18 |
| case47 | 118 | 328 | 28 | 1.0 | 0.01 | 0.08 |
| case124 | 133 | 386 | 31 | 1.0 | 0.12 | 3.43 |
| case55 | 149 | 442 | 26 | 1.0 | 0.01 | 0.15 |
| case8 | 160 | 525 | 26 | 1.0 | 0.04 | 0.96 |
| lltraversal_new | 163 | 359 | 41 | 1.0 | 0.19 | 9.73 |
| case105 | 170 | 407 | 59 | 1.0 | 0.3 | 7.07 |
| case5 | 176 | 518 | 36 | 1.0 | 0.65 | 5.09 |
| treemin_new | 177 | 451 | 29 | 1.0 | 0.12 | 2.55 |
| s344_3_2 | 197 | 464 | 24 | 1.0 | 0.12 | 1.38 |
| s349_3_2 | 198 | 469 | 24 | 1.0 | 0.12 | 1.46 |
| case201 | 200 | 544 | 45 | 1.0 | 0.17 | 5.46 |
| case202 | 200 | 544 | 45 | 1.0 | 0.18 | 5.44 |
| case56 | 202 | 722 | 23 | 1.0 | 0.01 | 0.17 |
| case54 | 203 | 725 | 23 | 1.0 | 0.01 | 0.17 |
| case106 | 204 | 509 | 60 | 1.0 | 0.35 | 8.61 |
| 19_new | 211 | 594 | 48 | 1.0 | 0.11 | 6.76 |
| case133 | 211 | 615 | 42 | 0.98 | 136.04 | 1330.62 |
| case136 | 211 | 615 | 42 | 0.98 | 128.91 | 1710.95 |
| case203 | 214 | 580 | 49 | 1.0 | 0.13 | 5.22 |
| case205 | 214 | 580 | 49 | 1.0 | 0.13 | 5.24 |
| case204 | 214 | 580 | 49 | 1.0 | 0.13 | 5.23 |
| tree_delete3_new | 215 | 521 | 44 | 0.99 | 0.27 | 2.81 |
| s344_7_4 | 215 | 540 | 24 | 1.0 | 0.2 | 1.66 |
| s349_7_4 | 216 | 545 | 24 | 1.0 | 0.2 | 1.58 |
| case146 | 219 | 558 | 64 | 1.0 | 24.22 | 386.47 |

Continued on next page

| | | | | UniGen2 | | UniGen |
|---|---|---|---|---|---|---|
| **Benchmark** | **#vars** | **#clas** | **\|S\|** | **Succ. Prob** | **Runtime(s)** | **Runtime(s)** |
| case145 | 219 | 558 | 64 | 1.0 | 17.49 | 478.73 |
| case132 | 236 | 708 | 41 | 1.0 | 0.14 | 2.04 |
| case135 | 236 | 708 | 41 | 1.0 | 0.15 | 2.02 |
| case_1_b14_1 | 238 | 681 | 45 | 1.0 | 0.16 | 5.47 |
| case_2_b14_1 | 238 | 681 | 45 | 1.0 | 0.15 | 5.28 |
| case_3_b14_1 | 238 | 681 | 45 | 1.0 | 0.18 | 5.26 |
| case109 | 241 | 915 | 31 | 1.0 | 0.03 | 0.39 |
| case14 | 247 | 649 | 67 | 1.0 | 108.11 | 2675.16 |
| s382_3_2 | 263 | 635 | 24 | 1.0 | 0.04 | 0.41 |
| case123 | 267 | 980 | 34 | 1.0 | 0.4 | 5.22 |
| case119 | 267 | 787 | 59 | 1.0 | 1.21 | 39.86 |
| case_1_b14_2 | 270 | 805 | 43 | 1.0 | 0.17 | 5.38 |
| case_2_b14_2 | 270 | 805 | 43 | 1.0 | 0.16 | 5.34 |
| case_3_b14_2 | 270 | 805 | 43 | 1.0 | 0.18 | 5.59 |
| case9 | 279 | 753 | 67 | 1.0 | 91.03 | 4632.83 |
| s382_7_4 | 281 | 711 | 24 | 1.0 | 0.04 | 0.47 |
| case61 | 282 | 753 | 66 | 1.0 | 103.77 | 1964.81 |
| s344_15_7 | 284 | 824 | 24 | 1.0 | 0.08 | 1.27 |
| case120 | 284 | 851 | 61 | 1.0 | 4.18 | 198.08 |
| s349_15_7 | 285 | 829 | 24 | 1.0 | 0.09 | 1.23 |
| case57 | 288 | 1158 | 32 | 1.0 | 0.85 | 4.88 |
| s444_3_2 | 290 | 712 | 24 | 1.0 | 0.04 | 0.4 |
| case121 | 291 | 975 | 48 | 1.0 | 0.17 | 5.5 |
| case62 | 291 | 1165 | 33 | 1.0 | 0.21 | 5.59 |
| s420_3_2 | 294 | 694 | 34 | 1.0 | 0.36 | 7.38 |
| s420_new1_3_2 | 294 | 694 | 34 | 1.0 | 0.35 | 7.41 |
| s420_new_3_2 | 294 | 694 | 34 | 1.0 | 0.24 | 6.2 |
| case3 | 294 | 1110 | 26 | 1.0 | 0.04 | 0.84 |
| case2 | 296 | 1116 | 26 | 1.0 | 0.05 | 0.83 |
| s510_3_2 | 298 | 768 | 25 | 1.0 | 0.1 | 1.27 |
| case126 | 302 | 1129 | 34 | 1.0 | 0.24 | 4.88 |
| case_1_b14_3 | 304 | 941 | 40 | 1.0 | 0.18 | 5.61 |
| case_2_b14_3 | 304 | 941 | 40 | 1.0 | 0.18 | 5.62 |
| case_3_b14_3 | 304 | 941 | 40 | 1.0 | 0.18 | 5.76 |

Continued on next page

| | | | | UniGen2 | | UniGen |
|---|---|---|---|---|---|---|
| **Benchmark** | **#vars** | **#clas** | **\|S\|** | **Succ. Prob** | **Runtime(s)** | **Runtime(s)** |
| s444_7_4 | 308 | 788 | 24 | 0.99 | 0.13 | 1.28 |
| s420_new1_7_4 | 312 | 770 | 34 | 1.0 | 0.22 | 5.79 |
| s420_new_7_4 | 312 | 770 | 34 | 1.0 | 0.17 | 5.81 |
| s420_7_4 | 312 | 770 | 34 | 1.0 | 0.24 | 5.79 |
| case122 | 314 | 1258 | 27 | 1.0 | 0.07 | 1.46 |
| s510_7_4 | 316 | 844 | 25 | 1.0 | 0.1 | 1.21 |
| case6 | 329 | 996 | 52 | 1.0 | 0.2 | 6.96 |
| case_0_b11_1 | 340 | 1026 | 48 | 1.0 | 0.21 | 6.06 |
| case_1_b11_1 | 340 | 1026 | 48 | 1.0 | 0.23 | 6.16 |
| s510_15_7 | 340 | 948 | 25 | 1.0 | 0.09 | 1.23 |
| s382_15_7 | 350 | 995 | 24 | 1.0 | 0.14 | 1.41 |
| s420_new_15_7 | 351 | 934 | 34 | 1.0 | 0.16 | 5.31 |
| s526_3_2 | 365 | 943 | 24 | 1.0 | 0.04 | 0.87 |
| s420_15_7 | 366 | 994 | 34 | 1.0 | 0.18 | 5.3 |
| s420_new1_15_7 | 366 | 994 | 34 | 1.0 | 0.19 | 5.37 |
| s526a_3_2 | 366 | 944 | 24 | 1.0 | 0.06 | 0.6 |
| s444_15_7 | 377 | 1072 | 24 | 1.0 | 0.06 | 0.84 |
| s526_7_4 | 383 | 1019 | 24 | 1.0 | 0.09 | 0.86 |
| 77_new | 384 | 2171 | 44 | 1.0 | 0.12 | 26.92 |
| s526a_7_4 | 384 | 1020 | 24 | 1.0 | 0.08 | 0.98 |
| case125 | 393 | 1555 | 35 | 1.0 | 0.44 | 6.48 |
| case35 | 400 | 1414 | 46 | 1.0 | 0.27 | 8.88 |
| case34 | 409 | 1597 | 39 | 1.0 | 0.2 | 5.78 |
| case143 | 427 | 1592 | 48 | 1.0 | 0.2 | 5.24 |
| case_0_b12_1 | 427 | 1385 | 37 | 1.0 | 0.17 | 4.41 |
| case_2_b12_1 | 427 | 1385 | 37 | 1.0 | 0.18 | 4.39 |
| case_1_b12_1 | 427 | 1385 | 37 | 1.0 | 0.19 | 4.57 |
| case115 | 428 | 1851 | 28 | 1.0 | 0.11 | 2.44 |
| case114 | 428 | 1851 | 28 | 1.0 | 0.1 | 2.43 |
| case131 | 432 | 1830 | 36 | 1.0 | 0.26 | 2.97 |
| case116 | 438 | 1881 | 28 | 1.0 | 0.09 | 2.4 |
| s526_15_7 | 452 | 1303 | 24 | 1.0 | 0.07 | 1.4 |
| s526a_15_7 | 453 | 1304 | 24 | 1.0 | 0.07 | 1.37 |
| isolateRightmost_new | 483 | 1498 | 64 | 1.0 | 0.28 | 18.56 |

Continued on next page

| | | | | UniGen2 | | UniGen |
|---|---|---|---|---|---|---|
| **Benchmark** | **#vars** | **#clas** | **\|S\|** | **Succ. Prob** | **Runtime(s)** | **Runtime(s)** |
| squaring51 | 496 | 1947 | 42 | 1.0 | 0.13 | 2.86 |
| squaring50 | 500 | 1965 | 42 | 1.0 | 0.14 | 2.86 |
| s953a_3_2 | 515 | 1297 | 45 | 1.0 | 0.67 | 11.6 |
| s953a_7_4 | 533 | 1373 | 45 | 1.0 | 22.42 | 1303.7 |
| s953a_15_7 | 602 | 1657 | 45 | 1.0 | 0.49 | 10.64 |
| s820a_7_4 | 616 | 1703 | 23 | 1.0 | 0.01 | 0.15 |
| s832a_7_4 | 624 | 1733 | 23 | 1.0 | 0.01 | 0.12 |
| s820a_15_7 | 685 | 1987 | 23 | 1.0 | 0.01 | 0.11 |
| s1238a_3_2 | 686 | 1850 | 32 | 1.0 | 0.3 | 7.17 |
| s1196a_3_2 | 690 | 1805 | 32 | 1.0 | 0.23 | 4.54 |
| s832a_15_7 | 693 | 2017 | 23 | 1.0 | 0.04 | 0.51 |
| squaring24 | 695 | 2193 | 61 | 1.0 | 0.29 | 6.98 |
| squaring22 | 695 | 2193 | 61 | 1.0 | 0.28 | 6.89 |
| squaring20 | 696 | 2198 | 61 | 1.0 | 0.29 | 6.96 |
| squaring21 | 697 | 2203 | 61 | 1.0 | 0.27 | 6.78 |
| s1238a_7_4 | 704 | 1926 | 32 | 1.0 | 0.23 | 3.11 |
| s1196a_7_4 | 708 | 1881 | 32 | 1.0 | 0.29 | 3.3 |
| squaring23 | 710 | 2268 | 61 | 1.0 | 0.28 | 7.05 |
| GuidanceService2 | 715 | 2181 | 27 | 1.0 | 0.02 | 0.34 |
| s1238a_15_7 | 773 | 2210 | 32 | 1.0 | 0.21 | 3.55 |
| s1196a_15_7 | 777 | 2165 | 32 | 1.0 | 0.18 | 4.87 |
| tree_delete3 | 795 | 2734 | 32 | 1.0 | 0.2 | 3.64 |
| case_0_b12_2 | 827 | 2725 | 45 | 1.0 | 0.25 | 6.74 |
| case_2_b12_2 | 827 | 2725 | 45 | 1.0 | 0.24 | 6.72 |
| case_1_b12_2 | 827 | 2725 | 45 | 1.0 | 0.24 | 6.77 |
| squaring27 | 837 | 2901 | 61 | 1.0 | 0.36 | 6.39 |
| squaring25 | 846 | 2947 | 61 | 1.0 | 0.35 | 6.66 |
| squaring3 | 885 | 2809 | 72 | 1.0 | 0.58 | 15.94 |
| squaring2 | 885 | 2809 | 72 | 1.0 | 0.6 | 17.15 |
| squaring6 | 885 | 2809 | 72 | 1.0 | 0.76 | 15.81 |
| squaring5 | 885 | 2809 | 72 | 1.0 | 0.58 | 15.49 |
| squaring1 | 891 | 2839 | 72 | 1.0 | 0.69 | 16.0 |
| squaring4 | 891 | 2839 | 72 | 1.0 | 0.66 | 15.49 |
| squaring26 | 894 | 3187 | 61 | 1.0 | 0.4 | 6.92 |

Continued on next page

| | | | | UniGen2 | | UniGen |
|---|---|---|---|---|---|---|
| **Benchmark** | **#vars** | **#clas** | **\|S\|** | **Succ. Prob** | **Runtime(s)** | **Runtime(s)** |
| squaring11 | 966 | 3213 | 72 | 1.0 | 0.85 | 18.53 |
| GuidanceService | 988 | 3088 | 27 | 1.0 | 0.02 | 0.23 |
| squaring30 | 1031 | 3693 | 61 | 1.0 | 0.55 | 15.3 |
| squaring28 | 1060 | 3839 | 61 | 1.0 | 0.46 | 15.67 |
| llreverse_new | 1096 | 4217 | 47 | 1.0 | 0.18 | 10.1 |
| squaring10 | 1099 | 3632 | 72 | 1.0 | 0.73 | 20.65 |
| squaring8 | 1101 | 3642 | 72 | 1.0 | 0.76 | 19.64 |
| squaring29 | 1141 | 4248 | 61 | 1.0 | 0.65 | 19.42 |
| 79_new | 1217 | 4034 | 40 | 1.0 | 2.93 | 21.24 |
| IssueServiceImpl | 1393 | 4319 | 30 | 1.0 | 0.01 | 0.1 |
| squaring9 | 1434 | 5028 | 72 | 1.0 | 1.03 | 20.35 |
| squaring14 | 1458 | 5009 | 72 | 1.0 | 2.62 | 48.73 |
| 10 | 1494 | 2215 | 46 | 1.0 | 0.33 | 85.45 |
| squaring12 | 1507 | 5210 | 72 | 1.0 | 3.25 | 62.44 |
| 27 | 1509 | 2707 | 32 | 1.0 | 0.22 | 6.37 |
| squaring16 | 1627 | 5835 | 72 | 1.0 | 4.16 | 79.12 |
| squaring7 | 1628 | 5837 | 72 | 1.0 | 0.79 | 21.98 |
| PhaseService | 1686 | 5655 | 27 | 1.0 | 0.01 | 0.18 |
| 27_new | 1792 | 6717 | 32 | 1.0 | 0.38 | 13.45 |
| ActivityService | 1837 | 5968 | 27 | 1.0 | 0.01 | 0.17 |
| 55_new | 1874 | 8384 | 46 | 1.0 | 3.05 | 146.83 |
| IterationService | 1896 | 6732 | 27 | 1.0 | 0.01 | 0.23 |
| ActivityService2 | 1952 | 6867 | 27 | 1.0 | 0.01 | 0.19 |
| aig_insertion1 | 2296 | 9326 | 60 | 1.0 | 0.18 | 3.57 |
| 111 | 2348 | 5479 | 36 | 1.0 | 0.48 | 15.79 |
| ConcreteActivityService | 2481 | 9011 | 28 | 1.0 | 0.02 | 0.36 |
| 53 | 2586 | 10747 | 32 | 1.0 | 0.26 | 6.96 |
| aig_insertion2 | 2592 | 10156 | 60 | 1.0 | 0.18 | 3.54 |
| 55 | 3128 | 12145 | 46 | 1.0 | 31.11 | 178.17 |
| ProjectService3 | 3175 | 11019 | 55 | 1.0 | 0.68 | 17.32 |
| NotificationServiceImpl2 | 3540 | 13425 | 36 | 1.0 | 0.12 | 1.34 |
| 109 | 3565 | 14012 | 36 | 1.0 | 0.88 | 12.99 |
| 51 | 3708 | 14594 | 38 | 1.0 | 0.52 | 18.77 |
| 32 | 3834 | 13594 | 38 | 1.0 | 0.47 | 19.39 |

Continued on next page

| | | | | UniGen2 | | UniGen |
|---|---|---|---|---|---|---|
| **Benchmark** | **#vars** | **#clas** | **\|S\|** | **Succ. Prob** | **Runtime(s)** | **Runtime(s)** |
| 70 | 4670 | 15864 | 40 | 1.0 | 0.78 | 24.58 |
| ProcessBean | 4768 | 14458 | 64 | 1.0 | 0.8 | 32.2 |
| 56 | 4842 | 17828 | 38 | 1.0 | 0.61 | 15.98 |
| 35 | 4915 | 10547 | 52 | 1.0 | 1.33 | 65.12 |
| 80 | 4969 | 17060 | 48 | 1.0 | 0.98 | 181.87 |
| tree_delete | 5758 | 22105 | 30 | 1.0 | 0.02 | 0.35 |
| 7 | 6683 | 24816 | 50 | 1.0 | 1.69 | 160.65 |
| doublyLinkedList | 6890 | 26918 | 37 | 1.0 | 0.04 | 1.23 |
| 19 | 6993 | 23867 | 48 | 1.0 | 3.34 | 52.28 |
| LoginService | 8200 | 26689 | 34 | 1.0 | 0.08 | 0.9 |
| 29 | 8866 | 31557 | 45 | 1.0 | 8.19 | 100.46 |
| 17 | 10090 | 27056 | 45 | 1.0 | 35.0 | 526.58 |
| parity_new | 10137 | 44830 | 50 | 1.0 | 4.09 | 41.08 |
| 81 | 10775 | 38006 | 51 | 1.0 | 15.19 | 285.7 |
| LoginService2 | 11511 | 41411 | 36 | 1.0 | 0.05 | 0.55 |
| Sort | 12125 | 49611 | 52 | 1.0 | 4.15 | 82.8 |
| 77 | 14535 | 27573 | 44 | 1.0 | 11.33 | 38.54 |
| 20 | 15475 | 60994 | 51 | 1.0 | 19.08 | 270.78 |
| enqueue | 16466 | 58515 | 42 | 1.0 | 0.87 | 14.67 |
| Karatsuba | 19594 | 82417 | 41 | 1.0 | 5.86 | 80.29 |
| lltraversal | 39912 | 167842 | 23 | 1.0 | 0.18 | 4.86 |
| LLReverse | 63797 | 257657 | 25 | 1.0 | 0.73 | 7.59 |
| diagStencil_new | 94607 | 2838579 | 78 | 1.0 | 3.53 | 60.18 |
| demo4_new | 381129 | 1801463 | 45 | 1.0 | 4.01 | 74.68 |
| tutorial3 | 486193 | 2598178 | 31 | 1.0 | 58.41 | 805.33 |
| demo2_new | 777009 | 3649893 | 45 | 1.0 | 3.47 | 40.33 |
| demo3_new | 865935 | 3509158 | 45 | 1.0 | 6.36 | 87.12 |

Table A6 : Comparison of MIS vs SMIS. “TO” indicates timeout after 18000 seconds

| Benchmark | #vars | #clas | MIS time(s) | $\mathcal{I}$ | Min Time | MinSize | Ratio size (s) |
|---|---|---|---|---|---|---|---|
| s298_3_2 | 205 | 510 | 0.29 | 17 | 2.13 | 15 | 0.88 |
| s298_7_4 | 223 | 586 | 0.26 | 18 | 2.19 | 16 | 0.89 |
| blasted_case132 | 236 | 708 | 0.44 | 21 | 1.36 | 21 | 1.0 |
| s382_7_4 | 281 | 711 | 0.46 | 23 | 3.36 | 22 | 0.94 |
| s344_15_7 | 284 | 824 | 0.61 | 25 | 4.28 | 24 | 0.94 |
| s349_15_7 | 285 | 829 | 0.64 | 27 | 5.26 | 24 | 0.89 |
| blasted_case110 | 287 | 1263 | 0.91 | 15 | 2.01 | 14 | 0.9 |
| s444_3_2 | 290 | 712 | 0.66 | 26 | 5.95 | 22 | 0.85 |
| s298_15_7 | 292 | 870 | 0.45 | 17 | 6.63 | 16 | 0.94 |
| blasted_case3 | 294 | 1110 | 1.81 | 25 | 5.49 | 22 | 0.88 |
| blasted_case2 | 296 | 1116 | 1.7 | 25 | 7.85 | 22 | 0.88 |
| blasted_case111 | 306 | 1358 | 0.84 | 16 | 1.3 | 14 | 0.85 |
| s444_7_4 | 308 | 788 | 0.67 | 27 | 5.12 | 24 | 0.89 |
| blasted_case117 | 309 | 1367 | 0.99 | 22 | 1.56 | 15 | 0.68 |
| blasted_case118 | 309 | 1367 | 0.97 | 23 | 1.63 | 15 | 0.65 |
| blasted_case122 | 314 | 1258 | 0.89 | 26 | 1.8 | 24 | 0.92 |
| s510_7_4 | 316 | 844 | 0.52 | 28 | 2.82 | 25 | 0.89 |
| s510_15_7 | 340 | 948 | 0.57 | 29 | 4.64 | 25 | 0.85 |
| s526_3_2 | 365 | 943 | 1.61 | 26 | 17.32 | 22 | 0.85 |
| s526a_3_2 | 366 | 944 | 1.29 | 30 | 7.06 | 24 | 0.8 |
| s420_new1_15_7 | 366 | 994 | 0.86 | 33 | 7.48 | 33 | 0.99 |
| s420_15_7 | 366 | 994 | 1.33 | 34 | 7.29 | 33 | 0.97 |
| registerlesSwap | 370 | 1090 | 0.5 | 24 | 1.9 | 15 | 0.62 |
| s444_15_7 | 377 | 1072 | 1.48 | 27 | 13.46 | 23 | 0.85 |
| s526_7_4 | 383 | 1019 | 1.53 | 30 | 9.6 | 24 | 0.8 |
| s526a_7_4 | 384 | 1020 | 1.47 | 28 | 8.85 | 24 | 0.86 |
| blasted_case125 | 393 | 1555 | 1.93 | 34 | 3.81 | 32 | 0.94 |
| blasted_case34 | 409 | 1597 | 2.38 | 37 | 8.41 | 36 | 0.97 |
| blasted_case143 | 427 | 1592 | 4.71 | 38 | 12.4 | 27 | 0.71 |
| blasted_case_0_b12_1 | 427 | 1385 | 1.89 | 34 | 17.11 | 30 | 0.87 |
| blasted_case_1_b12_1 | 427 | 1385 | 2.18 | 33 | 16.97 | 30 | 0.91 |
| blasted_case_2_b12_1 | 427 | 1385 | 1.42 | 34 | 16.52 | 30 | 0.88 |
| blasted_case115 | 428 | 1851 | 3.6 | 35 | 8.53 | 25 | 0.71 |

Continued on next page

| Benchmark | #vars | #clas | MIS time(s) | $\mathcal{I}$ | Min Time | MinSize | Ratio size (s) |
|---|---|---|---|---|---|---|---|
| blasted_case131 | 432 | 1830 | 2.51 | 35 | 7.37 | 33 | 0.94 |
| blasted_case116 | 438 | 1881 | 3.47 | 33 | 10.03 | 25 | 0.76 |
| scenarios_treemax | 452 | 1637 | 0.66 | 32 | 2.52 | 26 | 0.81 |
| s526_15_7 | 452 | 1303 | 2.78 | 27 | 14.4 | 23 | 0.85 |
| s526a_15_7 | 453 | 1304 | 3.04 | 28 | 16.48 | 24 | 0.84 |
| isolateRightmost | 483 | 1498 | 1.37 | 47 | 6.79 | 45 | 0.96 |
| blasted_squaring51 | 496 | 1947 | 50.82 | 38 | 45.19 | 24 | 0.62 |
| blasted_squaring50 | 500 | 1965 | 50.04 | 39 | 47.45 | 24 | 0.61 |
| s953a_3_2 | 515 | 1297 | 5.89 | 47 | 10.78 | 45 | 0.95 |
| s953a_7_4 | 533 | 1373 | 8.64 | 49 | 10.48 | 45 | 0.92 |
| s641_15_7 | 576 | 1399 | 1.93 | 54 | 20.6 | 54 | 1.0 |
| s713_15_7 | 596 | 1477 | 2.64 | 54 | 18.58 | 54 | 1.0 |
| s820a_3_2 | 598 | 1627 | 0.46 | 20 | 0.95 | 19 | 0.95 |
| s953a_15_7 | 602 | 1657 | 7.48 | 48 | 11.03 | 45 | 0.93 |
| s832a_3_2 | 606 | 1657 | 0.64 | 20 | 1.41 | 19 | 0.93 |
| s832a_7_4 | 624 | 1733 | 0.92 | 23 | 3.82 | 20 | 0.87 |
| blasted_case130 | 644 | 2056 | 7.27 | 53 | 24.05 | 49 | 0.92 |
| s820a_15_7 | 685 | 1987 | 1.23 | 25 | 3.47 | 21 | 0.84 |
| s1238a_3_2 | 686 | 1850 | 41.93 | 48 | 65.22 | 32 | 0.67 |
| s1196a_3_2 | 690 | 1805 | 27.3 | 41 | 2522.76 | 32 | 0.78 |
| s832a_15_7 | 693 | 2017 | 3.94 | 31 | 48.62 | 23 | 0.73 |
| blasted_squaring22 | 695 | 2193 | 11.09 | 27 | 25.57 | 23 | 0.85 |
| blasted_squaring24 | 695 | 2193 | 11.08 | 25 | 27.05 | 23 | 0.9 |
| blasted_squaring20 | 696 | 2198 | 12.81 | 26 | 29.37 | 23 | 0.87 |
| blasted_squaring21 | 697 | 2203 | 11.62 | 25 | 32.23 | 23 | 0.9 |
| s1238a_7_4 | 704 | 1926 | 47.59 | 49 | 232.6 | 32 | 0.65 |
| s1196a_7_4 | 708 | 1881 | 35.44 | 45 | TO | – | – |
| blasted_squaring23 | 710 | 2268 | 12.64 | 27 | 36.03 | 23 | 0.85 |
| blasted_case12 | 737 | 2310 | 97.48 | 65 | TO | – | – |
| s1238a_15_7 | 773 | 2210 | 90.89 | 52 | 181.59 | 32 | 0.61 |
| s1196a_15_7 | 777 | 2165 | 67.1 | 47 | TO | – | – |
| blasted_case_2_b12_2 | 827 | 2725 | 25.32 | 42 | TO | – | – |
| blasted_case_0_b12_2 | 827 | 2725 | 34.87 | 43 | TO | – | – |
| blasted_case_1_b12_2 | 827 | 2725 | 23.87 | 43 | TO | – | – |
| blasted_squaring27 | 837 | 2901 | 34.83 | 27 | 82.78 | 27 | 0.98 |

Continued on next page

| Benchmark | #vars | #clas | MIS time(s) | $\mathcal{I}$ | Min Time | MinSize | Ratio size (s) |
|---|---|---|---|---|---|---|---|
| blasted_case50 | 843 | 3288 | 19.88 | 65 | 40.55 | 62 | 0.95 |
| blasted_squaring25 | 846 | 2947 | 42.86 | 28 | 71.81 | 27 | 0.95 |
| s1488_3_2 | 854 | 2423 | 5.6 | 20 | TO | – | – |
| blasted_case211 | 869 | 2929 | 25.21 | 81 | 55.19 | 80 | 0.98 |
| blasted_case210 | 872 | 2937 | 28.15 | 83 | 48.99 | 80 | 0.96 |
| s1488_7_4 | 872 | 2499 | 11.38 | 24 | TO | – | – |
| blasted_squaring5 | 885 | 2809 | 957.48 | 62 | TO | – | – |
| blasted_squaring6 | 885 | 2809 | 782.16 | 58 | TO | – | – |
| blasted_squaring2 | 885 | 2809 | 739.63 | 57 | 2916.58 | 36 | 0.63 |
| blasted_squaring3 | 885 | 2809 | 796.04 | 56 | 3590.09 | 36 | 0.64 |
| blasted_squaring1 | 891 | 2839 | 718.78 | 60 | TO | – | – |
| blasted_squaring4 | 891 | 2839 | 868.71 | 55 | 1174.46 | 36 | 0.65 |
| blasted_squaring26 | 894 | 3187 | 85.11 | 30 | 70.17 | 27 | 0.9 |
| BN_65 | 925 | 2063 | 29.64 | 48 | 80.24 | 45 | 0.94 |
| s1488_15_7 | 941 | 2783 | 8.65 | 22 | TO | – | – |
| blasted_case_2_ptb_1 | 963 | 3027 | 32.35 | 79 | 51.0 | 77 | 0.97 |
| blasted_case_1_ptb_1 | 966 | 3035 | 29.81 | 79 | 52.26 | 77 | 0.97 |
| blasted_squaring30 | 1031 | 3693 | 192.14 | 30 | 144.82 | 29 | 0.97 |
| BN_46 | 1039 | 2265 | 28.13 | 41 | TO | – | – |
| blasted_squaring28 | 1060 | 3839 | 217.91 | 29 | 133.42 | 29 | 0.98 |
| scenarios_llreverse | 1096 | 4217 | 59.8 | 81 | 205.0 | 46 | 0.56 |
| blasted_squaring10 | 1099 | 3632 | 3321.29 | 56 | 1609.63 | 40 | 0.71 |
| blasted_squaring8 | 1101 | 3642 | 3453.48 | 54 | 799.63 | 40 | 0.74 |
| BN_59 | 1112 | 2661 | 104.85 | 48 | 593.78 | 32 | 0.67 |
| BN_63 | 1112 | 2661 | 115.64 | 44 | 887.17 | 32 | 0.72 |
| BN_55 | 1154 | 2692 | 126.32 | 43 | 1643.1 | 32 | 0.74 |
| BN_57 | 1154 | 2692 | 103.22 | 43 | TO | – | – |
| BN_53 | 1154 | 2692 | 120.55 | 42 | 1966.96 | 32 | 0.75 |
| blasted_case209 | 1189 | 3477 | 57.83 | 94 | 79.62 | 88 | 0.93 |
| blasted_case212 | 1189 | 3477 | 56.15 | 97 | 77.34 | 88 | 0.91 |
| GuidanceService2 | 1192 | 4362 | 22.94 | 91 | 103.44 | 65 | 0.71 |
| scenarios_aig_insertion2 | 1194 | 4304 | 10.86 | 92 | 11.56 | 73 | 0.79 |
| scenarios_aig_insertion1 | 1195 | 4301 | 8.96 | 93 | 13.8 | 73 | 0.78 |
| 79 | 1217 | 4034 | 0.91 | 40 | 1.08 | 39 | 0.97 |
| blasted_TR_device_1_linear | 1249 | 3927 | 121.44 | 100 | TO | – | – |

Continued on next page

| **Benchmark** | **#vars** | **#clas** | **MIS time(s)** | $\mathcal{I}$ | **Min Time** | **MinSize** | **Ratio size (s)** |
|---|---|---|---|---|---|---|---|
| SetTest | 1252 | 5411 | 4.52 | 23 | 95.4 | 19 | 0.83 |
| blasted_case_2_b14_even | 1304 | 4057 | 24.66 | 118 | TO | – | – |
| blasted_case_1_b14_even | 1304 | 4057 | 20.33 | 118 | TO | – | – |
| blasted_case3_b14_even3 | 1304 | 4057 | 22.06 | 120 | TO | – | – |
| blasted_case1_b14_even3 | 1318 | 4093 | 22.26 | 123 | TO | – | – |
| BN_49 | 1336 | 3376 | 37.0 | 26 | TO | – | – |
| BN_47 | 1336 | 3376 | 30.65 | 25 | TO | – | – |
| BN_51 | 1336 | 3376 | 33.92 | 25 | TO | – | – |
| BN_61 | 1348 | 3388 | 37.18 | 27 | TO | – | – |
| blockmap_05_01.net | 1411 | 2737 | 0.39 | 14 | 432.48 | 14 | 1.0 |
| blasted_squaring9 | 1434 | 5028 | 6396.02 | 60 | 1319.6 | 40 | 0.67 |
| blasted_squaring14 | 1458 | 5009 | 18000 | 100 | TO | – | – |
| blasted_squaring12 | 1507 | 5210 | 18000 | 102 | TO | – | – |
| blasted_case_0_ptb_1 | 1507 | 4621 | 212.13 | 95 | 299.17 | 88 | 0.92 |
| blasted_case49 | 1510 | 6505 | 233.45 | 68 | 228.76 | 61 | 0.9 |
| blasted_case_3_4_b14_even | 1532 | 4761 | 32.45 | 139 | TO | – | – |
| blasted_case_1_4_b14_even | 1532 | 4761 | 36.5 | 140 | TO | – | – |
| blasted_TR_b14_2_linear | 1570 | 4963 | 243.65 | 136 | TO | – | – |
| blasted_squaring16 | 1627 | 5835 | 18000 | 142 | TO | – | – |
| blasted_squaring7 | 1628 | 5837 | 12329.2 | 58 | TO | – | – |
| blockmap_05_02.net | 1738 | 3452 | 11.68 | 39 | 2.63 | 37 | 0.95 |
| 27 | 1792 | 6717 | 0.62 | 32 | 0.15 | 32 | 1.0 |
| NotificationServiceImpl2 | 1816 | 6614 | 87.68 | 126 | 218.0 | 88 | 0.7 |
| BN_44 | 1820 | 3806 | 150.49 | 61 | TO | – | – |
| BN_43 | 1820 | 3806 | 149.31 | 62 | TO | – | – |
| BN_45 | 1820 | 3806 | 118.24 | 63 | TO | – | – |
| 55 | 1874 | 8384 | 0.1 | 38 | 0.09 | 38 | 1.0 |
| blasted_TR_b12_1_linear | 1914 | 6619 | 5963.92 | 73 | TO | – | – |
| blasted_TR_ptb_1_linear | 1969 | 6288 | 1297.77 | 122 | 768.37 | 106 | 0.87 |
| tutorial2 | 2022 | 17764 | 0.84 | 31 | 3.15 | 12 | 0.38 |
| LoginService2 | 2024 | 7382 | 111.7 | 148 | 3284.69 | 102 | 0.69 |
| doublyLinkedList | 2038 | 7962 | 455.48 | 117 | TO | – | – |
| blockmap_05_03.net | 2055 | 4143 | 16.33 | 62 | 14.35 | 60 | 0.97 |
| compress2 | 2134 | 9106 | 12.17 | 188 | 186.61 | 71 | 0.38 |
| blasted_TR_b12_2_linear | 2426 | 8373 | 15505.02 | 79 | TO | – | – |

Continued on next page

| Benchmark | #vars | #clas | MIS time(s) | $\mathcal{I}$ | Min Time | MinSize | Ratio size (s) |
|---|---|---|---|---|---|---|---|
| blasted_TR_device_1_even_linear | 2447 | 7612 | 612.19 | 176 | TO | – | – |
| blasted_case_1_b12_even2 | 2669 | 8460 | 4440.45 | 149 | TO | – | – |
| blasted_case_2_b12_even2 | 2669 | 8460 | 4459.44 | 147 | TO | – | – |
| blasted_case_0_b12_even2 | 2669 | 8460 | 4418.07 | 150 | TO | – | – |
| blasted_case_1_b12_even1 | 2681 | 8492 | 4507.71 | 155 | TO | – | – |
| blasted_case_2_b12_even1 | 2681 | 8492 | 4249.56 | 149 | TO | – | – |
| scenarios_tree_insert_insert | 2797 | 10427 | 837.14 | 101 | TO | – | – |
| Pollard | 2800 | 49543 | 1211.4 | 179 | TO | – | – |
| 56 | 2801 | 9965 | 2.23 | 37 | 2.97 | 37 | 1.0 |
| ProcessBean | 3130 | 11689 | 172.64 | 305 | TO | – | – |
| scenarios_tree_delete2 | 3411 | 12783 | 444.61 | 175 | 2352.19 | 137 | 0.78 |
| lss_harder | 3465 | 62713 | 1727.77 | 116 | 1212.93 | 22 | 0.19 |
| s5378a_3_2 | 3679 | 8372 | 945.05 | 225 | TO | – | – |
| s5378a_7_4 | 3697 | 8448 | 1599.65 | 228 | TO | – | – |
| s5378a_15_7 | 3766 | 8732 | 1990.1 | 227 | TO | – | – |
| scenarios_tree_delete | 4038 | 16142 | 56.94 | 27 | 76.69 | 21 | 0.78 |
| listReverseEasy | 4092 | 15867 | 16715.34 | 121 | 3397.65 | 99 | 0.81 |
| 71 | 5314 | 11254 | 11.48 | 66 | 35.35 | 62 | 0.94 |
| 36 | 5627 | 24717 | 0.61 | 72 | 0.37 | 72 | 1.0 |
| scenarios_tree_delete4 | 6198 | 23509 | 18000 | 574 | TO | – | – |
| 107 | 7679 | 36225 | 14.29 | 83 | 13.08 | 80 | 0.96 |
| reverse | 9485 | 535676 | 25.03 | 200 | 7.1 | 195 | 0.97 |
| 54 | 9691 | 39993 | 422.65 | 99 | 2560.42 | 93 | 0.94 |
| blockmap_10_01.net | 11328 | 23175 | 44.56 | 35 | TO | – | – |
| 30 | 12022 | 50532 | 42.85 | 76 | 69.6 | 74 | 0.97 |
| lssBig | 12438 | 149909 | 536.88 | 46 | TO | – | – |
| blockmap_10_02.net | 12562 | 26022 | 2637.74 | 78 | TO | – | – |
| lss | 13373 | 156208 | 971.24 | 45 | TO | – | – |
| blockmap_10_03.net | 13786 | 28826 | 13442.28 | 125 | TO | – | – |
| 20 | 13887 | 60046 | 40.29 | 51 | 14.6 | 50 | 0.98 |
| 110 | 15316 | 60974 | 9.2 | 80 | 8.71 | 80 | 1.0 |
| scenarios_tree_insert_search | 16573 | 61922 | 18000 | 943 | TO | – | – |
| blockmap_15_01.net | 33035 | 67424 | 781.94 | 49 | TO | – | – |
| blockmap_20_01.net | 78650 | 160055 | 2513.32 | 67 | TO | – | – |

# Bibliography

[1] CryptoMiniSAT. http://www.msoos.org/cryptominisat2/.

[2] SMTLib. http://goedel.cs.uiowa.edu/smtlib/.

[3] The SDD package. http://reasoning.cs.ucla.edu/sdd/, 2014.

[4] System Verilog. http://www.systemverilog.org, 2015.

[5] Jacob A. Abraham. An Improved Algorithm for Network Reliability. *IEEE Transactions on Reliability*, R-28(1):58–61, apr 1979.

[6] Dimitris Achlioptas and Michael Molloy. The solution space geometry of random linear equations. *Random Structures & Algorithms*, 2013.

[7] K.K. Aggarwal, Krishna .B. Misra, and J.S. Gupta. A Fast Algorithm for Reliability Evaluation. *IEEE Transactions on Reliability*, R-24(1):83–85, apr 1975.

[8] Christos Alexopoulos. State space partitioning methods for stochastic shortest path problems. *Networks*, 30(1):9–21, aug 1997.

[9] Sanjeev Arora and Boaz Barak. *Computational Complexity: A Modern Approach*. Cambridge Univ. Press, 2009.

[10] Fahiem Bacchus, Shannon Dalmao, and Toniann Pitassi. Algorithms and complexity results for #SAT and Bayesian inference. In *Proc. of FOCS*, pages

340–351, 2003.

[11] Clark W Barrett, Leonardo Mendonça de Moura, and Aaron Stump. SMT-COMP: Satisfiability Modulo Theories Competition. In *Proc. of CAV*, pages 20–23, 2005.

[12] Clark W. Barrett, M. Deters, L. de Moura, A. Oliveras, and A. Stump. 6 Years of SMT-COMP. *Journal of Automated Reasoning*, pages 1–35, 2012.

[13] Clark W. Barrett, Pascal Fontaine, and Cesare Tinelli. The SMT-LIB standard - Version 2.5. http://smtlib.cs.uiowa.edu/, 2010.

[14] Roberto J Bayardo Jr and Robert Schrag. Using CSP look-back techniques to solve real-world SAT instances. In *Proc. of AAAI*, pages 203–208, 1997.

[15] Mihir Bellare, Oded Goldreich, and Erez Petrank. Uniform generation of NP-witnesses using an NP-oracle. *Information and Computation*, 163(2):510–526, 2000.

[16] Vaishak Belle, Andrea Passerini, and Guy Van den Broeck. Probabilistic inference in hybrid domains by weighted model integration. In *Proc. of IJCAI*, pages 2770–2776, 2015.

[17] Vaishak Belle, Guy Van den Broeck, and Andrea Passerini. Hashing-based approximate probabilistic inference in hybrid domains. In *Proc. of UAI*, 2015.

[18] Anton Belov and João Marques-Silva. Generalizing redundancy in propositional logic: Foundations and hitting sets duality. *CoRR*, abs/1207.1257, 2012.

[19] Anton Belov and João Marques-Silva. Muser2: An efficient MUS extractor. *JSAT*, 8(3/4):123–128, 2012.

[20] Lionel Bening and Harry Foster. *Principles of verifiable RTL design – a functional coding style supporting verification processes.* Springer, 2001.

[21] Michelle Bensi, Armen Der Kiureghian, and Daniel Straub. Efficient Bayesian network modeling of systems. *Reliability Engineering & System Safety*, 112:200–213, apr 2013.

[22] Armin Biere, Marijn Heule, Hans van Maaren, and Toby Walsh. *Handbook of Satisfiability.* IOS Press, 2009.

[23] Roy Billinton and Wenyuan Li. *Reliability Assessment of Electric Power Systems Using Monte Carlo Methods.* 1994.

[24] Elazar Birnbaum and Eliezer L. Lozinskii. The good old Davis-Putnam procedure helps counting models. *Journal of Artificial Intelligence Research*, 10(1):457–477, June 1999.

[25] Rodrigo de Salvo Braz, Ciaran O'Reilly, Vibhav Gogate, and Rina Dechter. Probabilistic inference modulo theories. In *Workshop on Hybrid Reasoning at IJCAI*, 2015.

[26] Franc Brglez, David Bryan, and Krzysztof Kozminski. Combinational profiles of sequential benchmark circuits. In *ISCAS*, 1989.

[27] Björn Bringmann, Siegfried Nijssen, Nikolaj Tatti, Jilles Vreeken, and A Zimmerman. Mining sets of patterns. *Tutorial at ECMLPKDD*, 2010.

[28] Robert Brummayer and Armin Biere. Boolector: An efficient SMT solver for bit-vectors and arrays. In *Proc. of TACAS*, volume 5505 of *Lecture Notes in Computer Science*, pages 174–177. Springer, 2009.

[29] Michel Bruneau, Stephanie E. Chang, Ronald T. Eguchi, George C. Lee, Thomas D. O'Rourke, Andrei M. Reinhorn, Masanobu Shinozuka, Kathleen Tierney, William A. Wallace, and Detlof von Winterfeldt. A Framework to Quantitatively Assess and Enhance the Seismic Resilience of Communities. *Earthquake Spectra*, 19(4):733–752, nov 2003.

[30] Roberto Bruttomesso. *RTL Verification: From SAT to SMT(BV)*. PhD thesis, DIT, University of Trento/FBK - Fondazione Bruno Kessler, Feb 2008.

[31] Roberto Bruttomesso, Alessandro Cimatti, Anders Franzén, Alberto Griggio, Ziyad Hanna, Alexander Nadel, Amit Palti, and Roberto Sebastiani. A lazy and layered smt(bv) solver for hard industrial verification problems. In *Proc. of CAV*, pages 547–560, 2007.

[32] Nader H. Bshouty, Richard Cleve, Ricard Gavaldà, Sampath Kannan, and Christino Tamon. Oracles and queries that are sufficient for exact learning. *J. Comput. Syst. Sci.*, 52(3):421–433, 1996.

[33] Toon Calders, Christophe Rigotti, and Jean-François Boulicaut. A survey on condensed representations for frequent sets. In *Constraint-based mining and inductive databases*, pages 64–80. Springer, 2006.

[34] Supratik Chakraborty, Daniel J. Fremont, Kuldeep S. Meel, Sanjit A. Seshia, and Moshe Y. Vardi. Distribution-aware sampling and weighted model counting for SAT. In *Proc. of AAAI*, pages 1722–1730, 2014.

[35] Supratik Chakraborty, Daniel J. Fremont, Kuldeep S. Meel, Sanjit A. Seshia, and Moshe Y. Vardi. On parallel scalable uniform sat witness generation. In *Proc. of TACAS*, pages 304–319, 2015.

[36] Supratik Chakraborty, Kuldeep S. Meel, Rakesh Mistry, and Moshe Y. Vardi. Approximate probabilistic inference via word-level counting. In *Proc. of AAAI*, 2016.

[37] Supratik Chakraborty, Kuldeep S. Meel, and Moshe Y. Vardi. A Scalable and Nearly Uniform Generator of SAT Witnesses. In *Proc. of CAV*, pages 608–623, 2013.

[38] Supratik Chakraborty, Kuldeep S. Meel, and Moshe Y. Vardi. A scalable approximate model counter. In *Proc. of CP*, pages 200–216, 2013.

[39] Supratik Chakraborty, Kuldeep S. Meel, and Moshe Y. Vardi. Balancing scalability and uniformity in SAT witness generator. In *Proc. of DAC*, pages 1–6, 2014.

[40] Supratik Chakraborty, Kuldeep S. Meel, and Moshe Y. Vardi. Algorithmic improvements in approximate counting for probabilistic inference: From linear to logarithmic SAT calls. In *Proc. of IJCAI*, 2016.

[41] Yi Chang and Yasuhiro Mori. A study on the relaxed linear programming bounds method for system reliability. *Structural Safety*, 41:64–72, mar 2013.

[42] Mark Chavira and Adnan Darwiche. On probabilistic inference by weighted model counting. *Artificial Intelligence*, 172(6):772–799, 2008.

[43] Dmitry Chistikov, Rayna Dimitrova, and Rupak Majumdar. Approximate Counting for SMT and Value Estimation for Probabilistic Programs. In *Proc. of TACAS*, pages 320–334, 2015.

[44] Arthur Choi and Adnan Darwiche. Dynamic minimization of sentential decision diagrams. In *Proc. of AAAI*, pages 187–194, 2013.

[45] David Clark, Sebastian Hunt, and Pasquale Malacaria. Quantitative analysis of the leakage of confidential data. *Electr. Notes Theor. Comput. Sci.*, 59(3):238–251, 2001.

[46] Peter Clote and Anton Setzer. On PHP, st-connectivity and odd charged graphs. *Proof Complexity and Feasible Arithmetics*, 39:93–117, 1998.

[47] Stephen A. Cook. The complexity of theorem-proving procedures. In *Proceedings of the third annual ACM symposium on Theory of computing*, STOC, pages 151–158, 1971.

[48] Gregory F. Cooper. The computational complexity of probabilistic inference using bayesian belief networks. *Artificial intelligence*, 42(2):393–405, 1990.

[49] Thomas M. Cover and Joy A. Thomas. *Elements of Information Theory*. John Wiley & Sons, Inc., 1991.

[50] Nilesh Dalvi, Christopher Ré, and Dan Suciu. Probabilistic databases: diamonds in the dirt. *Commun. ACM*, 52(7):86–94, 2009.

[51] Nilesh Dalvi and Dan Suciu. Efficient query evaluation on probabilistic databases. *The VLDB Journal*, 16(4):523–544, 2007.

[52] Adnan Darwiche. New advances in compiling CNF to decomposable negation normal form. In *Proc. of ECAI*, pages 328–332. Citeseer, 2004.

[53] Adnan Darwiche. SDD: A new canonical representation of propositional knowledge bases. In *Proc. of IJCAI*, volume 22, page 819, 2011.

[54] Martin Davis, George Logemann, and Donald Loveland. A machine program for theorem-proving. *Communications of the ACM*, 5(7):394–397, 1962.

[55] Martin Davis and Hilary Putnam. A computing procedure for quantification theory. *Journal of the ACM (JACM)*, 7(3):201–215, 1960.

[56] Leonardo De Moura and Nikolaj Bjørner. Z3: An efficient SMT solver. In *Proc. of TACAS*, pages 337–340. Springer, 2008.

[57] Rina Dechter, K. Kask, E. Bin, and R. Emek. Generating random solutions for constraint satisfaction problems. In *AAAI*, pages 15–21, 2002.

[58] Shujun Deng, Zhiqiu Kong, Jinian Bian, and Yanni Zhao. Self-adjusting constrained random stimulus generation using splitting evenness evaluation and xor constraints. In *Proc. of ASP-DAC*, pages 769–774, 2009.

[59] F Javier Dıez and Marek J Druzdzel. Canonical probabilistic models for knowledge engineering. Technical report, Technical Report CISIAD-06-01, UNED, Madrid, Spain, 2006.

[60] Carmel Domshlak and Jörg Hoffmann. Probabilistic planning via heuristic forward search and weighted model counting. *Journal of Artificial Intelligence Research*, 30(1):565–620, 2007.

[61] William Dotson and J. Gobien. A new analysis technique for probabilistic graphs. *IEEE Transactions on Circuits and Systems*, 26(10):855–865, 1979.

[62] Leonardo Dueñas-Osorio, Kuldeep S. Meel, Roger Paredes, and Moshe Y. Vardi. Sat-based connectivity reliability estimation for power transmission grids. Technical report, Rice University, 2017.

[63] Leonardo Dueñas-Osorio and Javier Rojo. Reliability Assessment of Lifeline Systems with Radial Topology. *Computer-Aided Civil and Infrastructure Engineering*, 26(2):111–128, 2011.

[64] Vladimir Dzyuba, Matthijs van Leeuwen, and Luc De Raedt. Flexible constrained sampling with guarantees for pattern mining. *Data Mining and Knowledge Discovery*, pages 1–28, 2017.

[65] Derek L Eager, John Zahorjan, and Edward D Lazowska. Speedup versus efficiency in parallel systems. *IEEE Trans. on Computers*, 38(3):408–423, 1989.

[66] Niklas Eén and Niklas Sörensson. Translating pseudo-boolean constraints into sat. *JSAT*, 2(1-4):1–26, 2006.

[67] Kevin Ellis, Armando Solar-Lezama, and Josh Tenenbaum. Sampling for bayesian program learning. In *Advances in Neural Information Processing Systems*, pages 1297–1305, 2016.

[68] Paul Erdos and Joel Spencer. Probabilistic methods in combinatorics. 1974.

[69] Stefano Ermon, Carla P Gomes, Ashish Sabharwal, and Bart Selman. Embed and project: Discrete sampling with universal hashing. In *Proc. of NIPS*, pages 2085–2093, 2013.

[70] Stefano Ermon, Carla P. Gomes, Ashish Sabharwal, and Bart Selman. Optimization with parity constraints: From binary codes to discrete integration. In *Proc. of UAI*, 2013.

[71] Stefano Ermon, Carla P. Gomes, Ashish Sabharwal, and Bart Selman. Taming the curse of dimensionality: Discrete integration by hashing and optimization.

In *Proc. of ICML*, pages 334–342, 2013.

[72] Stefano Ermon, Carla P. Gomes, Ashish Sabharwal, and Bart Selman. Low-density parity constraints for hashing-based discrete integration. In *Proc. of ICML*, pages 271–279, 2014.

[73] Stefano Ermon, Carla P. Gomes, and Bart Selman. Uniform solution sampling using a constraint solver as an oracle. In *Proc. of UAI*, 2012.

[74] Stefano Ermon, Carla P. Gomes, and Bart Selman. Uniform solution sampling using a constraint solver as an oracle. In *Proc. of UAI*, pages 255–264, 2012.

[75] Hadi Esmaeilzadeh, Emily Blem, Renee St Amant, Karthikeyan Sankaralingam, and Doug Burger. Dark silicon and the end of multicore scaling. In *Proc. of ISCA*, pages 365–376. IEEE, 2011.

[76] George S. Fishman. A Monte Carlo Sampling Plan for Estimating Network Reliability. *Operations Research*, 34(4):581–594, aug 1986.

[77] Alan M Frisch, Brahim Hnich, Zeynep Kiziltan, Ian Miguel, and Toby Walsh. Propagation algorithms for lexicographic ordering constraints. *Artificial Intelligence*, 170(10):803–834, 2006.

[78] Zhaohui Fu and Sharad Malik. Extracting logic circuit structure from conjunctive normal form descriptions. In *Prof. of VLSID*, pages 37–42, 2007.

[79] Marco Gavanelli. The log-support encoding of CSP into SAT. In *Proc. of CP*, pages 815–822. Springer, 2007.

[80] Vibhav Gogate and Rina Dechter. A new algorithm for sampling CSP solutions uniformly at random. In *CP*, pages 711–715. Springer, 2006.

[81] Vibhav Gogate and Rina Dechter. Approximate Solution Sampling (and Counting) on AND/OR Spaces. In *Proc. of CP*, pages 534–538, 2008.

[82] Vibhav Gogate and Rina Dechter. SampleSearch: Importance sampling in presence of determinism. *Artificial Intelligence*, 175(2):694–729, 2011.

[83] Oded Goldreich. The Counting Class #P. Lecture notes of course on "Introduction to Complexity Theory", Weizmann Institute of Science, 1999.

[84] Carla P. Gomes, J. Hoffmann, Ashish Sabharwal, and Bart Selman. From sampling to model counting. In *Proc. of IJCAI*, pages 2293–2299, 2007.

[85] Carla P Gomes, Joerg Hoffmann, Ashish Sabharwal, and Bart Selman. From sampling to model counting. In *Proc. of IJCAI*, pages 2293–2299, 2007.

[86] Carla P. Gomes, Joerg Hoffmann, Ashish Sabharwal, and Bart Selman. Short XORs for Model Counting; From Theory to Practice. In *SAT*, pages 100–106, 2007.

[87] Carla P. Gomes, Ashish Sabharwal, and Bart Selman. Model counting: A new strategy for obtaining good bounds. In *Proc. of AAAI*, volume 21, pages 54–61, 2006.

[88] Carla P. Gomes, Ashish Sabharwal, and Bart Selman. Near-uniform sampling of combinatorial spaces using XOR constraints. In *Proc. of NIPS*, pages 670–676, 2007.

[89] Camilo Gómez, M Sánchez-Silva, and Leonardo Dueñas-Osorio. Clustering methods for risk assessment of infrastructure network systems. In *International*

*Conference on Applications of Statistics and Probability in Civil Engineering*, pages 1389–1397, 2011.

[90] Andrew D Gordon, Thomas A Henzinger, Aditya V Nori, and Sriram K Rajamani. Probabilistic Programming. In *Prof. of ICSE*, 2014.

[91] Sumit Gulwani. Automating string processing in spreadsheets using input-output examples. In *Proc. of POPL*, pages 317–330. ACM, 2011.

[92] Liana Hadarean, Kshitij Bansal, Dejan Jovanović, Clark Barrett, and Cesare Tinelli. A Tale of Two Solvers: Eager and Lazy Approaches to Bit-Vectors. In *Proc. of CAV*, volume 8559 of *Lecture Notes in Computer Science*, pages 680–695, 2014.

[93] HAZUS. Multi-hazard Loss Estimation Methodology, Earthquake Model, HAZUS-MH MR4 Technical Manual., 2003.

[94] Jinbo Huang. Universal booleanization of constraint models. In *Proc. of CP*, pages 144–158, 2008.

[95] Jorge Eduardo Hurtado. *Structural reliability: statistical learning perspectives*, volume 17. Springer Science & Business Media, 2013.

[96] Alexey Ignatiev, Alessandro Previti, Mark Liffiton, and Joao Marques-Silva. Smallest mus extraction with minimal hitting set dualization. In *Proc. of International Conference on Principles and Practice of Constraint Programming*, 2015.

[97] Alexander Ivrii, Sharad Malik, Kuldeep S. Meel, and Moshe Y. Vardi. On computing minimal independent support and its applications to sampling and

counting. *Constraints*, pages 1–18, 2015.

[98] Mahesh A. Iyer. RACE: A word-level ATPG-based constraints solver system for smart random simulation. In *ITC*, pages 299–308. Citeseer, 2003.

[99] Mark R. Jerrum and Alistair Sinclair. The Markov Chain Monte Carlo method: an approach to approximate counting and integration. *Approximation algorithms for NP-hard problems*, pages 482–520, 1996.

[100] Mark R. Jerrum, Leslie G. Valiant, and Vijay V. Vazirani. Random generation of combinatorial structures from a uniform distribution. *Theoretical Computer Science*, 43(2-3):169–188, 1986.

[101] Susmit Jha, Rhishikesh Limaye, and Sanjit Seshia. Beaver: Engineering an Efficient SMT Solver for Bit-Vector Arithmetic. In *Computer Aided Verification*, pages 668–674, 2009.

[102] Ajith K. John and Supratik Chakraborty. A quantifier elimination algorithm for linear modular equations and disequations. In *Proc. of CAV*, pages 486–503, 2011.

[103] David R Karger. A Randomized Fully Polynomial Time Approximation Scheme for the All-Terminal Network Reliability Problem. *SIAM Review*, 43(3):499–522, 2001.

[104] Richard M Karp, Michael Luby, and Neal Madras. Monte-Carlo approximation algorithms for enumeration problems. *Journal of Algorithms*, 10(3):429–448, 1989.

[105] Michael J Kearns and Umesh Virkumar Vazirani. *An introduction to computational learning theory*. MIT press, 1994.

[106] Kristian Kersting. Lifted probabilistic inference. In *Proc. of ECAI*, pages 33–38, 2012.

[107] Scott Kirkpatrick, C. Daniel Gelatt, and Mario P. Vecchi. Optimization by simulated annealing. *Science*, 220(4598):671–680, 1983.

[108] Nathan Kitchen. *Markov Chain Monte Carlo Stimulus Generation for Constrained Random Simulation*. PhD thesis, University of California, Berkeley, 2010.

[109] Nathan Kitchen and Andreas Kuehlmann. Stimulus generation for constrained random simulation. In *Proc. of ICCAD*, pages 258–265, 2007.

[110] Lukas Kroc, Ashish Sabharwal, and Bart Selman. Leveraging belief propagation, backtrack search, and statistics for model counting. In *Proc. of CPAIOR*, pages 127–141, 2008.

[111] Daniel Kroening and Ofer Strichman. *Decision Procedures: An Algorithmic Point of View*. Springer, 1 edition, 2008.

[112] James H Kukula and Thomas R Shiple. Building circuits from relations. In *CAV*, pages 113–123. Springer, 2000.

[113] Jie Li and Jun He. A recursive decomposition algorithm for network seismic reliability evaluation. *Earthquake Engineering & Structural Dynamics*, 31(8):1525–1539, aug 2002.

[114] Paolo Liberatore. Redundancy in logic I: CNF propositional formulae. *Artif. Intell.*, 163(2):203–232, 2005.

[115] Mark H. Liffiton, Maher N. Mneimneh, Inês Lynce, Zaher S. Andraus, João Marques-Silva, and Karem A. Sakallah. A branch and bound algorithm for extracting smallest minimal unsatisfiable subformulas. *Constraints*, 14(4):415–442, 2009.

[116] Mark H. Liffiton and Karem A. Sakallah. Algorithms for computing minimal unsatisfiable subsets of constraints. *J. Autom. Reasoning*, 40(1):1–33, 2008.

[117] Hyun-woo Lim and Junho Song. Efficient risk assessment of lifeline networks under spatially correlated ground motions using selective recursive decomposition algorithm. *Earthquake Engineering & Structural Dynamics*, 41(13):1861–1882, 2012.

[118] Martin Löbbing and Ingo Wegener. The number of knight's tours equals 33,439,123,484,294 – counting with binary decision diagrams. *The Electronic Journal of Combinatorics*, 3(1):R5, 1996.

[119] Neal Madras. Lectures on monte carlo methods, fields institute monographs 16. *American Mathematical Society*, 2002.

[120] David Maier. *The Theory of Relational Databases.* Computer Science Press, Rockville, Md., 1983.

[121] Sharad Malik and Lintao Zhang. Boolean satisfiability from theoretical hardness to practical success. *Commun. ACM*, 52(8):76–82, 2009.

[122] João Marques-Silva. Computing minimally unsatisfiable subformulas: State of the art and future directions. *Multiple-Valued Logic and Soft Computing*, 19(1-3):163–183, 2012.

[123] Kuldeep S. Meel. *Sampling Techniques for Boolean Satisfiability*. Rice University, 2014. M.S. Thesis.

[124] Kuldeep S. Meel, Moshe Y. Vardi, Supratik Chakraborty, Daniel J Fremont, Sanjit A Seshia, Dror Fried, Alexander Ivrii, and Sharad Malik. Constrained sampling and counting: Universal hashing meets sat solving. In *Proc. of Beyond NP Workshop*, 2016.

[125] S. Minato. Zero-Suppressed BDDs for Set Manipulation in Combinatorial Problems. In *Proc. of Design Automation Conference*, pages 272–277, 1993.

[126] Matthew W Moskewicz, Conor F Madigan, Ying Zhao, Lintao Zhang, and Sharad Malik. Chaff: Engineering an efficient sat solver. In *Proceedings of the 38th annual Design Automation Conference*, pages 530–535. ACM, 2001.

[127] Alexander Nadel. Boosting minimal unsatisfiable core extraction. In *Proc. of FMCAD*, pages 221–229, 2010.

[128] Yehuda Naveh and Roy Emek. Random Stimuli Generation for Functional Hardware Verification as a CP Application. In *Proc. of CP*, volume 3709 of *Lecture Notes in Computer Science*, pages 882–882. Springer, 2005.

[129] Yehuda Naveh, Michal Rimon, Itai Jaeger, Yoav Katz, Michael Vinov, Eitan s Marcu, and Gil Shurek. Constraint-based random stimuli generation for hardware verification. In *Proc of IAAI*, pages 1720–1727, 2006.

[130] Lavon B Page and Jo Ellen Perry. A practical implementation of the factoring theorem for network reliability. *IEEE Transactions on Reliability*, 37(3):259–267, 1988.

[131] Avi Pfeffer. Figaro: An object-oriented probabilistic programming language. *Charles River Analytics Technical Report*, page 137, 2009.

[132] Quoc-Sang Phan, Pasquale Malacaria, Oksana Tkachuk, and Corina S Păsăreanu. Symbolic quantitative information flow. *ACM SIGSOFT Software Engineering Notes*, 37(6):1–5, 2012.

[133] Stephen M Plaza, Igor L Markov, and Valeria Bertacco. Random stimulus generation using entropy and xor constraints. In *Proc. of DAC*, pages 664–669, 2008.

[134] Suresh Rai and Arun Kumar. Recursive Technique For Computing System Reliability. *IEEE Transactions on Reliability*, R-36(1):38–44, apr 1987.

[135] Dan Roth. On the hardness of approximate reasoning. *Artificial Intelligence*, 82(1):273–302, 1996.

[136] Reuven Rubinstein. Stochastic Enumeration Method for Counting NP-Hard Problems. *Methodology and Computing in Applied Probability*, pages 1–43, 2012.

[137] Stuart J. Russell and Peter Norvig. *Artificial Intelligence: A Modern Approach (3rd Ed.).* Prentice Hall, 2009.

[138] Tian Sang, Fahiem Bacchus, Paul Beame, Henry A Kautz, and Toniann Pitassi. Combining component caching and clause learning for effective model counting. In *Proc. of SAT*, 2004.

[139] Tian Sang, Paul Beame, and Henry Kautz. Performing bayesian inference by weighted model counting. In *Prof. of AAAI*, pages 475–481, 2005.

[140] Rishabh Singh, Sumit Gulwani, and Armando Solar-Lezama. Automated feedback generation for introductory programming assignments. In *Proc. of POPL*. ACM, 2013.

[141] Michael Sipser. A complexity theoretic approach to randomness. In *Proc. of STOC*, pages 330–335, 1983.

[142] Armando Solar-Lezama. The sketching approach to program synthesis. In *APLAS*, volume 5904, pages 4–13. Springer, 2009.

[143] Mate Soos, Karsten Nohl, and Claude Castelluccia. Extending SAT Solvers to Cryptographic Problems. In *Proc. of SAT*, 2009.

[144] Larry Stockmeyer. The complexity of approximate counting. In *Proc. of STOC*, pages 118–126, 1983.

[145] Stefano Teso, Roberto Sebastiani, and Andrea Passerini. Structured learning modulo theories. *Artificial Intelligence*, 244:166–187, 2017.

[146] The White House, Office of the Press Secretary,. FACT SHEET: Obama Administration Announces Public and Private Sector Efforts to Increase Community Resilience through Building Codes and Standards. 2016.

[147] Marc Thurley. SharpSAT: counting models with advanced component caching and implicit BCP. In *Proc. of SAT*, pages 424–429, 2006.

[148] Seinosuke Toda. On the computational power of PP and (+)P. In *Proc. of FOCS*, pages 514–519. IEEE, 1989.

[149] Celina G. Val, Michael A. Enescu, Sam Bayless, William Aiello, and Alan J. Hu. Precisely measuring quantitative information flow: 10K lines of code and beyond. In *Proc. of EuroS&P*, pages 31–46, 2016.

[150] Leslie G. Valiant. The complexity of enumeration and reliability problems. *SIAM Journal on Computing*, 8(3):410–421, 1979.

[151] Leslie G. Valiant. *Probably Approximately Correct: Nature's Algorithms for Learning and Prospering in a Complex World.* Basic Books, 2013.

[152] Santosh S. Vempala. A random sampling based algorithm for learning the intersection of half-spaces. In *Proc. of FOCS*, pages 508–513. IEEE, 1997.

[153] Ronald E Walpole, Raymond H Myers, Sharon L Myers, and Keying Ye. *Probability and statistics for engineers and scientists*, volume 8. Prentice Hall Upper Saddle Riverˆ eNJ NJ, 1993.

[154] Toby Walsh. SAT v CSP. In *Proc. of CP*, pages 441–456. Springer, 2000.

[155] Wei Wei, Jordan Erenrich, and Bart Selman. Towards Efficient Sampling: Exploiting Random Walk Strategies. In *Proc. of AAAI*, pages 670–676, 2004.

[156] Wei Wei and Bart Selman. A new approach to model counting. In *Proc. of SAT*, pages 2293–2299. Springer, 2005.

[157] Daniel Wilhelm and Jehoshua Bruck. Stochastic switching circuit synthesis. In *Proc. of ISIT*, pages 1388–1392, 2008.

[158] Robert Wille, Daniel Große, Finn Haedicke, and Rolf Drechsler. SMT-based stimuli generation in the SystemC verification library. In *Advances in Design*

*Methods from Modeling Languages for Embedded Systems and SoCs*, pages 227–244. Springer, 2010.

[159] Yexiang Xue, Arthur Choi, and Adnan Darwiche. Basing decisions on sentences in decision diagrams. In *Proc. of AAAI*, 2012.

[160] Chao Yin and Ahsan Kareem. Computation of failure probability via hierarchical clustering. *Structural Safety*, 61:67–77, jul 2016.

[161] Jun Yuan, Adnan Aziz, Carl Pixley, and Ken Albin. Simplifying boolean constraint solving for random simulation-vector generation. *IEEE Trans. on CAD of Integrated Circuits and Systems*, 23(3):412–420, 2004.

[162] Shengjia Zhao, Sorathan Chaturapruek, Ashish Sabharwal, and Stefano Ermon. Closing the gap between short and long xors for model counting. In *Proc. of AAAI*, 2016.

[163] Michael Zhu and Stefano Ermon. A hybrid approach for probabilistic inference using random projections. In *Proc. of ICML*, pages 2039–2047, 2015.

[164] Albrecht Zimmermann and Siegfried Nijssen. Supervised pattern mining and applications to classification. In *Frequent pattern mining*, pages 425–442. Springer, 2014.

[165] Enrico Zio. Reliability engineering: Old problems and new challenges. *Reliability Engineering & System Safety*, 94(2):125–141, 2009.

[166] Konstantin M. Zuev, Stephen Wu, and James L. Beck. General network reliability problem and its efficient solution by subset simulation. *Probabilistic Engineering Mechanics*, 40:25–35, 2015.